%% file: main.tex
 \documentclass[3p,11pt]{elsarticle}

\input{style.tex}

\begin{document}

\title{Beyond-mean-field approaches for nuclear neutrinoless double beta decay in the standard mechanism}

\author[1,2]{J. M. Yao \footnote{yaojm8@mail.sysu.edu.cn}}
\address[1]{School of Physics and Astronomy, Sun Yat-sen University,  Zhuhai, 519082 China}
\address[2]{Facility for Rare Isotope Beams, Michigan State University, East Lansing, Michigan 48824-1321, USA}

\author[3,4]{J. Meng 
\footnote{mengj@pku.edu.cn}}
\address[3]{State Key Laboratory of Nuclear Physics and Technology, School of Physics,  Peking University, Beijing 100871, China}
\address[4]{Yukawa Institute for Theoretical Physics, Kyoto University, Kyoto 606-8502, Japan}

\author[5]{Y. F. Niu 
\footnote{niuyf@lzu.edu.cn}}
\address[5]{School of Nuclear Science and Technology, Lanzhou University, Lanzhou, 730000 P. R. China }

\author[6]{P. Ring 
\footnote{ring@ph.tum.de}}
\address[6]{Physik-Department der Technischen Universität München, D-85748 Garching, Germany}

\begin{abstract} 
 Nuclear weak decays provide important probes  to  fundamental symmetries in nature. A precise description of these  processes in atomic nuclei requires comprehensive knowledge on both the strong and weak interactions in the nuclear medium and on the dynamics of quantum many-body systems. In particular, an observation of the hypothetical double beta decay without emission of neutrinos ($\znubb$) would unambiguously demonstrate the Majorana nature of neutrinos and the existence of the lepton-number-violation process. It would also provide unique information on the ordering and absolute scale of neutrino masses. The next-generation tonne-scale experiments with sensitivity up to $10^{28}$ years after a few years of running will probably provide a definite answer to these fundamental questions based on our current knowledge on the nuclear matrix element (NME), the precise determination of which is a challenge to nuclear theory. Beyond-mean-field  approaches have been frequently adapted for the study of nuclear structure and decay throughout the nuclear chart for several decades. In this review, we summarize  the status of  beyond-mean-field calculations of the NMEs of $\znubb$ decay assuming the standard mechanism of an exchange of light Majorana neutrinos. The challenges and prospects in the extension and application of beyond-mean-field approaches for $\znubb$ decay are discussed.
\end{abstract}

 \maketitle

\setcounter{tocdepth}{4}
\setcounter{secnumdepth}{4}
\tableofcontents


 \section{Introduction}

\input{0introduction}

  \section{Nuclear beyond-mean-field  approaches}
  \label{sec:BMF}

\input{1BMF-mod.tex}

   \section{The theory of $\znubb$ decay in atomic nuclei}
    \label{sec:theory4dbd}

\input{2theory4dbd.tex}

   \section{The NME of $\znubb$ decay}
    \label{sec:NME}

\input{3nuclear_structure_calculations.tex}

  \section{Summary and outlook} 
   \label{sec:summary}
   \input{4summary.tex}

\section*{Acknowledgements}

We thank D. L. Fang,  G. Li,  Y. Liao, W. L. Lv, L. J. Wang,  Y. K. Wang, S. Zhou for their careful readings of the manuscript and fruitful discussions.   
J.M.Y is grateful to B. Bally, R. A. M. Basili, A. Belley,  J. Engel, Z. C. Gao, K. Hagino, H. Hergert,  J. D. Holt,  C. F. Jiao,   S. Novario, T. R. Rodriguez, A. Marquez Romero,  S. R. Stroberg,  L. S. Song,   and R. Wirth for collaboration and discussions on this topic and all participants for the discussions on the `` 2021 workshop of neutrinoless double beta decay" held in Zhuhai, China.  J.M.Y. was partially supported by the Fundamental Research Funds for the Central Universities, Sun Yat-sen University, \jmyr{the National Natural Science Foundation of China (Grant No. 12141501)} and  U.S.  Department of Energy, Office of Science, Office of Nuclear Physics under Awards No. DE-SC0017887 and No. DE-SC0015376 (the DBD  Topical Theory Collaboration). J.M. was partly supported by the National Key R\&D Program of China (Contracts No. 2018YFA0404400 and 2017YFE0116700) and the National Natural Science Foundation of China (Grants No. 12141501, 12070131001, 11935003). Y.F.N. was supported by the National Natural Science Foundation of China (Grant No.  12075104).  P.R. acknowledges partial support from the Deutsche Forschungsgemeinschaft (DFG, German Research Foundation) under Germany’s Excellence Strategy-EXC-2094-390783311.






%

\end{document}

%% file: style.tex
\usepackage{amsfonts,amsmath,amssymb} 

\usepackage{graphicx}
\usepackage{xcolor}
\usepackage{color}
\usepackage{xspace}
\usepackage{isotope}
\usepackage{float}
\usepackage{placeins} 

\usepackage{verbatim}{}{}
\usepackage{comment} 
\usepackage{soul} 
 
\usepackage{enumitem}
\usepackage{wrapfig}

\usepackage{epsfig}
\usepackage{cancel}
  
\usepackage[colorlinks,linkcolor=blue,anchorcolor=blue,citecolor=blue]{hyperref} 
\hypersetup{
   colorlinks=true, 
   linkcolor=blue,  
   citecolor=blue,  
   filecolor=blue,  
   urlcolor=blue    
}

\usepackage{enumitem}

\usepackage[T1]{fontenc}
\usepackage[utf8]{inputenc}
\usepackage{graphicx,color,rotating,pifont}
\usepackage{amsmath,amssymb}  
\usepackage{bbold,textcomp}

\usepackage{mathtools}
\usepackage{siunitx}
\usepackage{bm}
\usepackage{ae}
\usepackage{dcolumn}
\usepackage{txfonts}
\usepackage{tensor}
\usepackage{braket}
\usepackage{booktabs}
\usepackage{xspace}
\usepackage{soul}
\setstcolor{red}
\setul{}{.2ex} 

\newcommand{\znubb}{\ensuremath{0\nu\beta\beta}}

 \newcommand{\unity}{1\!\!1}

\DeclareMathSymbol{\NS}{\mathord}{AMSb}{"4E}

\DeclareSIUnit{\fm}{\femto\meter}
\DeclareSIUnit{\MeVc}{\MeV\per\text{\ensuremath{c}}}

\newcommand{\dd}{\ensuremath{\mathrm{d}}}

\renewcommand{\vec}[1]{\ensuremath{\bm{#1}}}

\newcommand{\tj}[6]{ \begin{pmatrix}
   #1 & #2 & #3 \\
   #4 & #5 & #6 
  \end{pmatrix}}
  
\newcommand{\Gj}[6]{ \begin{Bmatrix}
   #1 & #2 & #3 \\
   #4 & #5 & #6 
  \end{Bmatrix}}

\newcommand{\nj}[9]{ \begin{Bmatrix}
   #1 & #2 & #3 \\
   #4 & #5 & #6 \\
   #7 & #8 & #9 
  \end{Bmatrix}}

\newcommand{\beq}{\begin{equation}}
\newcommand{\eeq}{\end{equation}}
\newcommand{\beqn}{\begin{eqnarray}}
\newcommand{\eeqn}{\end{eqnarray}}
\newcommand{\bsub}{\begin{subequations}}
\newcommand{\esub}{\end{subequations}}
\newcommand{\bpm}{\begin{pmatrix}}
\newcommand{\epm}{\end{pmatrix}}

\newcommand{\te}{\mathrm{e}}
\newcommand{\ti}{\mathrm{i}}

\newcommand{\jmy}[1]{\textcolor{black}{#1}}
\newcommand{\jmyr}[1]{\textcolor{black}{#1}}

\newcolumntype{C}{>{$\displaystyle} c <{$}}

%% file: 0introduction.tex
Since the beginning of nuclear physics, atomic nuclei have served as a natural laboratory to explore fundamental interactions and symmetries in nature. In particular, nuclear single-$\beta$ decay governed by the weak interaction has been one of the major subjects of nuclear and particle physics as it turns out that the weak interaction is the only fundamental interaction that breaks parity symmetry \cite{Lee:1956PR,Wu:1957PR}. Historically, it was the study of nuclear single-$\beta$ decay that provided the first physical evidence of neutrinos -- neutral elementary particles that are the key to helping us better understand the nature of our universe. In 1914, Chadwick established the continuous feature of the kinetic energy spectrum of the $\beta$-decay electron \cite{Chadwick:1914}. To explain this observation without the violation of energy and momentum conservation,  in 1930, Pauli conjectured tentatively the existence of a spin-1/2, weakly interacting neutral particle with a small mass emitted together with the electron in the single-$\beta$ decay. This particle (called neutrino afterward by Fermi) was first detected by Reines and Cowan with inverse $\beta$-decay reactions \cite{Reines:1953,Cowan:1956Science}. Thereafter, nuclear single-$\beta$ decay has been studied extensively to search for new physics beyond the standard model of particle physics~\cite{Herczeg:2001PPNP,Severijns:2006,Otten:2008RPP,Falkowski:2021JHEP}.

\begin{figure}[tb]
\centering    
\includegraphics[width=10cm]{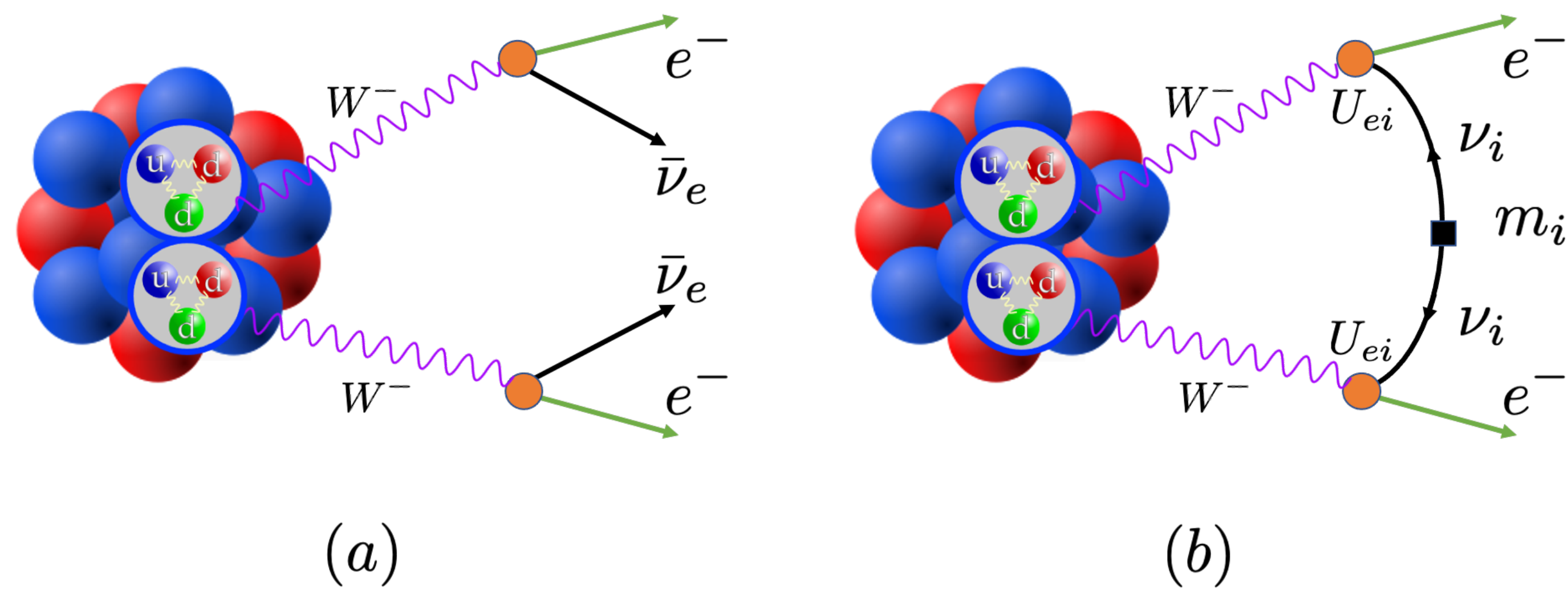}  
\caption{\label{fig:Introd_cartoon_dbd} Cartoon pictures for (a) $2\nu\beta\beta$ decay and (b) $0\nu\beta\beta$ decay from an atomic nucleus $(A, Z)$ to the neighboring one $(A, Z+2)$ via the emission of two electrons with and without the emission of two electron antineutrinos, respectively.  In (b), only the standard mechanism of an exchange of a light Majorana neutrino is plotted for demonstration, even though there are many other possible non-standard mechanisms responsible for $\znubb$ decay, such as those with the emission of Majorons~\cite{Chikashige:1981PLB,Aulakh:1982PLB,Berezhiani:1992,Hirsch:1996PLB,Faessler:1998JPG,Cepedello:2018PRL}.}
\end{figure}

  Nuclear double-$\beta$ decay is a second-order weak process in which two neutrons (or protons) in a parent nucleus ($A, Z$) are simultaneously transforming into two protons (or neutrons) in a daughter nucleus ($A, Z\pm2$).  There are typically four  types of double-$\beta$ decay processes, including double-electron emission ($2\beta^-$), double-positron emission ($2\beta^+$), single-positron emission with single-electron capture ($\epsilon\beta^+$), and double-electron capture ($2\epsilon$), and two kinds of decay modes ($0\nu, 2\nu$) depending on if there is accompanying two-(anti)neutrino emission. In the present Review, we concentrate mainly  on the following decay processes, namely, the $2\nu\beta^-\beta^-$ decay process
 \begin{align}
 \label{eq:Introd_2nudbd}
    (A, Z) \to  (A, Z+2)  + 2e^-  + 2\bar\nu_e,
 \end{align} 
 and the $0\nu\beta^-\beta^-$ decay process
 \begin{align}
 \label{eq:Introd_0nudbd}
    (A, Z) &\to  (A, Z+2)  + 2e^-,
 \end{align} 
 which are depicted schematically in Fig.~\ref{fig:Introd_cartoon_dbd}(a) and (b), respectively. Hereafter, we simply call these two processes $2\nu\beta\beta$ decay and $\znubb$ decay, respectively, for convenience.

\begin{figure}[t]
\centering  
\includegraphics[width=14cm]{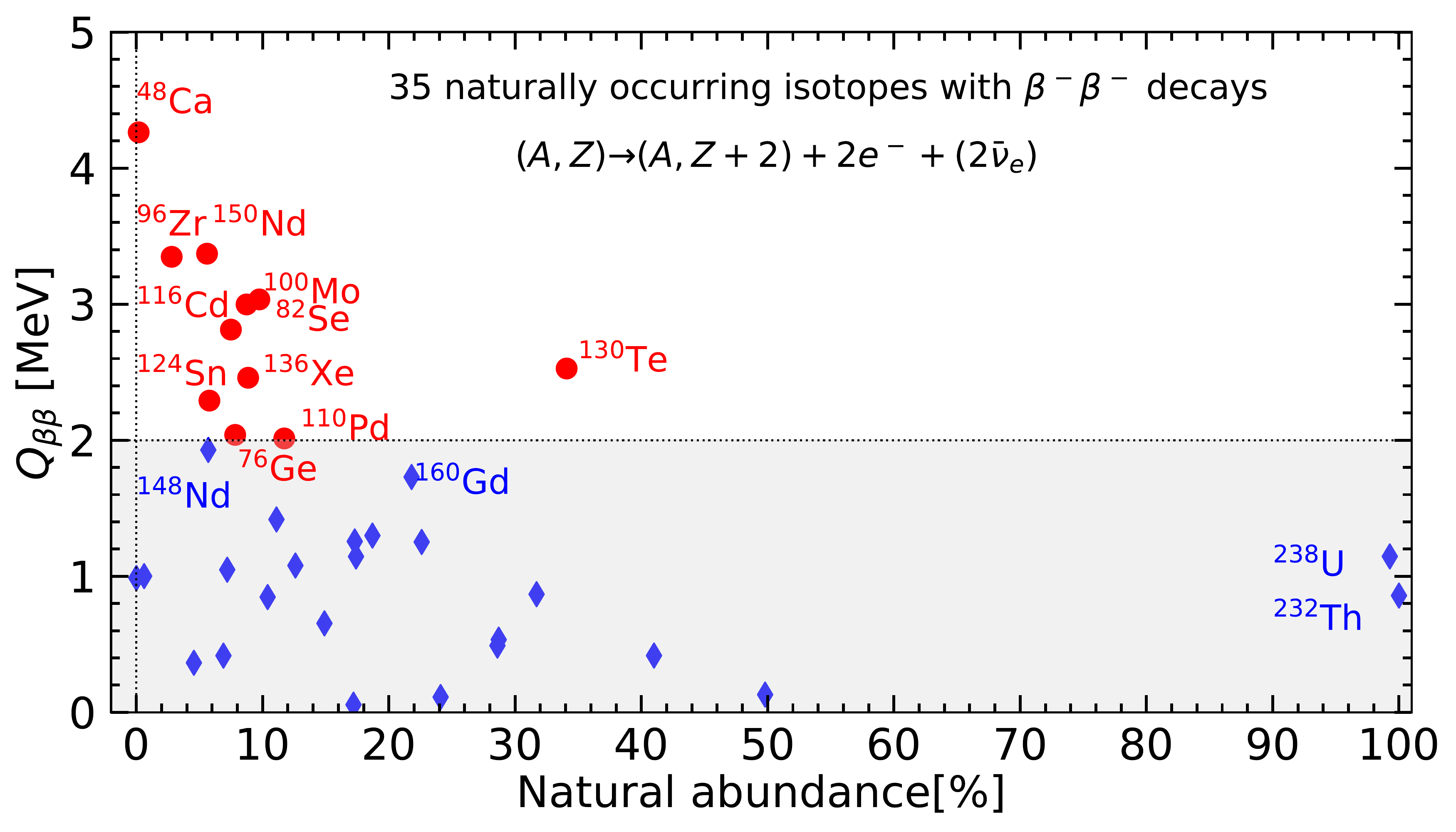}  
\caption{\label{fig:Introd_Qbb_mass_35} The $Q_{\beta\beta}$ value of 35 natural isotopes fulfilling the energy condition for undergoing $\beta^-\beta^-$ decay. The isotopes with $Q_{\beta\beta}>2.0$ MeV are candidates of $0\nu\beta\beta$ decay with experimental interest.  }
\end{figure}

  The $2\nu\beta\beta$ decay is allowed in the standard model and  was first considered by Goeppert-Mayer in 1935~\cite{Goeppert-Mayer:1935} based on the effective theory of single-$\beta$ decay proposed by Fermi in 1934 \cite{Fermi:1934ZP}. Note that double-$\beta$ decay is  different from two sequential single-$\beta$ decays and \jmyr{it can only be detected in the nuclei} where the  single-$\beta$ decay is energetically forbidden or at least strongly suppressed by spin change.  In nature, only 35 isotopes fulfill the energy condition $B(A,Z+2)+2\Delta_{npe} > B(A, Z) > B(A, Z+1)+\Delta_{npe}$ allowing for $2\nu\beta^-\beta^-$ decay~\cite{Tomoda:1991}, \footnote{Some isotopes can undergo the $2\nu\beta^+\beta^+$ decay, including $^{78}$Kr(0.355\%), $^{96}$Ru(5.54\%), $^{106}$Cd (1.25\%), $^{124}$Xe (0.095\%), $^{130}$Ba (0.106\%), and $^{136}$Ce (0.185 \%) \cite{Tretyak:2002} with the numbers in the parenthesis indicating their natural abundances. Note that the abundances of these isotopes are generally very low, except for $^{96}$Ru and $^{106}$Cd whose abundances are comparable to some of $0\nu\beta\beta$-decay candidates but with smaller $Q_{\beta\beta}$ values.} where $B(A,Z)$ is the binding energy of the isotope $(A, Z)$~\footnote{It is noted that the impacts of electron binding energies, small corrections due to atomic physics, nuclear recoil, and a possible nonzero neutrino mass are usually neglected.}   and $\Delta_{npe}=(m_n-m_p-m_e)c^2=0.782$~MeV.  Fig.~\ref{fig:Introd_Qbb_mass_35} shows the $Q_{\beta\beta}=B(A,Z+2)-B(A,Z)+2\Delta_{npe}$ value of these isotopes.  Since the first experiments on $2\nu\beta\beta$ decay \cite{Fireman:1949PR,Inghram:1950PR}, a dozen of isotopes have already been measured with half-lives $T^{2\nu}_{1/2}$ ranging from $10^{18}$ to $10^{21}$ years~\cite{Tretyak:2002,Saakyan:2013,Pritychenko:2014NDS,Barabash:2020nck}.

  The $0\nu\beta\beta$ decay mode was first considered by Furry~\cite{Furry:1939} who was motivated by the seminal papers of Majorana \cite{Majorana:1937} and Racah \cite{Racah:1937} which suggested that the neutrino may coincide with its own antiparticle.  Even though $0\nu\beta\beta$ decay is forbidden in the standard model because it violates lepton number conservation, it is preferred by many extensions of the standard model \cite{Georgi:1974PRL,Fritzsch:1974AP,Pati:1974PRD,Mohapatra:1975PRD}.
  According to the black-box theorem proposed by Schechter and Valle \cite{Schechter:1982PRD}, the observation of $0\nu\beta\beta$ decay  would confirm the existence of a Majorana mass term for neutrinos, even though the black-box induced Majorana mass could be only an extremely tiny correction  $\delta m_\nu={\cal O}(10^{-28})$ eV \cite{Duerr:2011JHEP,Liu:2016PLB}. It implies that other lepton-number-violating operators (such as seesaw mechanisms) must provide a leading contribution to the Majorana neutrino mass term. In any case, the search of $0\nu\beta\beta$ decay provides a direct way to disclose the physics of neutrinos and the existence of a lepton-number-violation process which has an important implication for the matter-antimatter asymmetry \cite{Fukugita:1986PLB} in the Universe as indicated from the ratio of baryon-to-photon number density $n_b/n_\gamma \sim 10^{-10}$ \cite{Nakamura:2017IJMPE,PDG:2018}.

 The research on $0\nu\beta\beta$ decay has gone through several stages since the 1930s, keeping being alive with both depression and encouragements coming from discoveries, which continue updating our understanding of neutrinos and the weak interaction. See the comprehensive review paper on the history of this research by Barabash \cite{Barabash:2011}
or the review paper by Vergados et al.~\cite{Vergados:2012RPP}.  In recent decades, the discoveries of neutrino oscillations \cite{Super-Kamiokande:1998,SNO:2001,KamLAND:2003,Dayabay:2012}   and their implications that neutrinos are  massive particles provide compelling arguments for the $0\nu\beta\beta$ decay~\cite{PDG:2018}  and revive the experimental search with increased sensitivity~\cite{US2015,EU2019,China2020}. The neutrino oscillation experiments have already provided a precise measurement on many neutrino parameters \cite{Xing:2020PR}, but they cannot provide the absolute mass $m_i$ for each neutrino as the signals are only sensitive to the neutrino mixing angles $\theta_{ij}$ and to the difference of squared neutrino masses $\Delta m^2_{ij}\equiv m^2_i-m^2_j$. Complementary to neutrino oscillation experiments, the single-$\beta$ decay of tritium  being measured in the Karlsruhe Tritium Neutrino (KATRIN) experiment~\cite{KATRIN:2019PRL}, the Planck  measurements of the cosmic microwave background anisotropies \cite{Planck:2018} and worldwide $0\nu\beta\beta$-decay experiments \cite{Dolinski:2019} are running to provide direct constraints on different combinations of the neutrino masses $m_j$.

 If the $0\nu\beta\beta$ decay is driven by the standard mechanism of exchanging light Majorana neutrinos, the effective electron neutrino mass $ \langle m_{\beta\beta}\rangle$ as a linear combination of all the three neutrino masses
 \begin{equation}
 \label{eq:effective-mass}
     \langle m_{\beta\beta}\rangle\equiv  \sum^3_{j=1} U^2_{ej} m_j
 \end{equation} 
 can be determined from its half-life $T^{0\nu}_{1/2}$ as follows
\begin{equation}
\label{half-life}
 |\langle m_{\beta\beta}\rangle|
=\left[\dfrac{m^2_e}{g^4_A(0) G_{0\nu}T^{0\nu}_{1/2}
 \left\lvert M^{0\nu}\right\rvert^2}\right]^{1/2},
\end{equation}
where $m_e\simeq0.511$ MeV is electron mass.  The axial-vector coupling strength $g_A(0)$ is factorized out from the nuclear matrix element (NME) $M^{0\nu}$. Its free-space value is around 1.27. This value is expected to be renormalized in nuclear medium and will be discussed later. The phase-space factor $G_{0\nu}\approx 10^{-14}$ yr$^{-1}$ can be evaluated rather precisely. The elements $U_{ej}$ in the quantity $\langle m_{\beta\beta}\rangle$ form the first row of the Pontecorvo-Maki-Nakagawa-Sakata (PMNS) matrix \cite{Pontecorvo:1957,MNS:1962} that is connecting the three light flavor neutrinos $\ket{\nu_{\alpha}}$ ($\alpha=e, \mu, \tau$) to the three neutrino mass eigenstates $\ket{\nu_{j}}$ ($j=1, 2, 3$)~\cite{Bilenky:1987},
        \beq
        \label{eq:neutrino-mixing}
         \ket{\nu_\alpha} = \sum^3_{j=1} U^\ast_{\alpha j} \ket{\nu_j}.
        \eeq
    The PMNS matrix $U$ is usually parametrized as \cite{PDG:2018}
    \beqn
    \label{eq:Introd_PMNS}
     U  
     &=&\begin{pmatrix}
      1 & 0 & 0\\
      0 & c_{23} & s_{23} \\
     0 & -s_{23} & c_{23}
    \end{pmatrix}
    \begin{pmatrix}
     c_{13} & 0 & s_{13}e^{-i\delta}\\
     0 & 1 & 0 \\
     -s_{13}e^{i\delta} & 0 & c_{13}
    \end{pmatrix}
    \begin{pmatrix}
      c_{12} & s_{12} & 0\\
       -s_{12} & c_{12} &0 \\
      0 & 0 & 1
    \end{pmatrix}
     \cdot {\cal P},
    \eeqn
where $c_{ij}\equiv \cos\theta_{ij}$, and $s_{ij}\equiv \sin\theta_{ij}$ with three mixing angles $\theta_{12}, \theta_{23}, \theta_{13}$. The Dirac phase $\delta$ is nonzero if there is a yet-unobserved charge-parity (CP) symmetry violation in neutrino oscillations. The values of the diagonal matrix ${\cal P}$ depend on the nature of neutrinos:
\beqn
{\cal P}=\left\{
\begin{array}{ll}
{\rm diag}(1, 1, 1), & \quad {\rm Dirac},\\
{\rm diag}(1, e^{i\alpha_{21}/2}, e^{i\alpha_{31}/2}), & \quad {\rm Majorana}\\
\end{array}\right.
\eeqn 
 where $\alpha_{21}, \alpha_{31}$ are the CP-violating Majorana phases which do not affect neutrino oscillations but influence the rate of $0\nu\beta\beta$ decay if it exists.

  The isotopes in Fig.~\ref{fig:Introd_Qbb_mass_35} with $Q_{\beta\beta}>2.0$ MeV are selected as candidates for detecting $0\nu\beta\beta$ decay as a larger $Q_{\beta\beta}$-value usually corresponds to a larger phase-space factor and lower background in the energy region of interest.    Even though the $0\nu\beta\beta$ decay is not observed yet, the null $0\nu\beta\beta$ decay signal from current experiments provides a constraint on the upper limits of the effective neutrino mass $\langle m_{\beta\beta}\rangle$. So far, the best half-life lower limit is achieved in the experiments with $^{76}$Ge ($T^{0\nu}_{1/2}>1.8\times 10^{26}$ yr~\cite{Agostini:2020}),  $^{130}$Te ($T^{0\nu}_{1/2}>3.2\times 10^{25}$ yr~\cite{CUORE2020}), and $^{136}$Xe ( $T^{0\nu}_{1/2}>1.07\times 10^{26}$ yr~\cite{Gando:2016PRL,EXO-200:2019}),  which transform into constraints on the effective neutrino mass as follows,
  \begin{itemize}
     \item $^{76}$Ge: $|\langle m_{\beta\beta}\rangle| < 0.079-0.180$ eV (90\% C.L.) \cite{Agostini:2020}.
     \item $^{130}$Te: $|\langle m_{\beta\beta}\rangle| < 0.075-0.350$ eV (90\% C.L.) \cite{CUORE2020}.
     \item $^{136}$Xe : $|\langle m_{\beta\beta}\rangle| < 0.061 - 0.165$ eV (90\% C.L.). \cite{Gando:2016PRL} 
 \end{itemize}
 The above constraints on neutrino masses are complementary to other measurements~\cite{Formaggio:2021PR}, including
  \begin{itemize}
     \item Cosmological observations from the Planck experiments: $\sum^3_{i=1} m_i <0.12-0.66$ eV (95\% C.L.) \cite{Vagnozzi:2017,Planck:2018}. 
     \item The KATRIN experiment: $m_\beta=\sqrt{\sum^3_{i=1} \vert U_{ei}\vert^2m^2_i}\leq 0.8$ eV (90\% C.L.) \cite{KATRIN:2021}. \footnote{In contrast to the other two measurements, the experiment on the tritium $\beta$ decay $^3$H $\to ^3$He $+ e^- + \bar\nu_e$ provides a model-independent direct measurement of neutrino masses via a precision measurement of the electron spectrum in the end-point region. Note that the decay rate is contributed incoherently from the sum of each mass eigenstate~\cite{Bilenky:1987,Formaggio:2021PR}, i.e.,
     $d\Gamma/d\epsilon_e \propto \sum^3_{i=1}\left|U_{ei}\right|^{2}\sqrt{\left(\epsilon_{0}-\epsilon_e\right)^{2}-m_{i}^{2}}\Theta(\epsilon_0-\epsilon_e-m_i)$, where $\epsilon_e$ denotes the electron’s kinetic energy and $\epsilon_0$ corresponds to the maximum value of $\epsilon_e$ in the absence of neutrino mass. $\Theta(\epsilon_0-\epsilon_e-m_i)$ is the step function that ensures energy conservation.}
     \item  \jmyr{The $^{163}$Ho electron-capture (ECHo) experiment: $m_\beta\leq 225$ eV (95\% C.L.) \cite{Springer:1987}.}
 \end{itemize}

The next-generation tonne-scale experimental searches of $0\nu\beta\beta$ decay are expected to reach a discovery potential of half-life about $10^{28}$ years after a few years of running \cite{Dolinski:2019}, which may provide a definite answer on the mass hierarchy of neutrinos based on the light-Majorana neutrino exchange mechanism and  current knowledge on the NME $M^{0\nu}$ for $0\nu\beta\beta$ decay. These experiments will also test other possible mechanisms, including a large uncharted region of the allowed parameter space assuming that neutrino masses follow the normal-ordering mass spectra~\cite{Agostini:2021}. 

For a given $g_A(0)$ value, the uncertainty in the upper limit of effective neutrino mass $|\langle m_{\beta\beta}\rangle|$ is mainly originated from the uncertainty in the NME, 
\begin{equation}
    M^{0\nu}=\langle \Psi_{\rm F}\vert O^{0\nu}\vert \Psi_{\rm I}\rangle,
\end{equation} 
which can in principle be computed straightforwardly if the transition operator  $O^{0\nu}$ for the $0\nu\beta\beta$ decay  and the wave functions $\vert\Psi_{I/F}\rangle$ for  the initial (I) and final (F) nuclear states are known. Since the NME  cannot be measured directly \footnote{We note that there are ongoing indirect experimental measurements on the NME, including heavy-ion double charge-exchange reactions \cite{Cappuzzello:2016EPJA} and $(p, t)$ reactions  \cite{Rebeiro:2020PLB}.},  a precise calculation of the NME is of particular importance. Achieving it is a challenge for nuclear theory as it requires a consistent treatment of both the strong and weak interactions and a systematically improvable solving of nuclear many-body systems with controllable uncertainties.

Atomic nuclei are quantum many-body systems that have been modeled in terms of different degrees of freedom confronted with different levels of difficulty.  The dynamics of nucleons inside the nucleus are governed by the strong interaction described fundamentally by quantum chromodynamics (QCD) in terms of  quarks and gluons. Due to the non-perturbative nature of the strong interaction at the nuclear energy scale, the application of QCD directly to the atomic nucleus is computationally challenging. With high-performance computing facilities, one can exploit lattice QCD to determine the nucleon-nucleon interaction \cite{Inoue:2011PRL,Lyu:2021PRL}, or to compute the nuclear systems directly~\cite{Beane:2013PRD}. However, the precision is currently far from satisfactory and it is formidable to apply the lattice QCD to atomic nuclei with the mass number $A>4$ because of the fermion sign problem and the exponential growth in noise with increasing nucleon number, and the factorial growth in the number of quark-level Wick contractions~\cite{Drischler:2021PPNP}. As discussed later, lattice QCD may provide a unique tool to determine the low-energy constants (LECs) of $0\nu\beta\beta$-decay operators due to the absence of corresponding data. See the recent review papers~\cite{Cirigliano:2020PPNP,Davoudi:2021PR}. 

Alternatively, by freezing quarks and gluons inside \jmy{colorless hadrons} (nucleons and pions), which are the relevant degrees of freedom in atomic nuclei, Weinberg proposed a chiral effective field theory (EFT) for nuclear force \cite{Weinberg:1991}. In this theory, one constructs effective Lagrangians that consist of interactions consistent with the symmetries of QCD and are organized by an expansion in the power of $(Q,m_\pi)/\Lambda_\chi$, where $Q$ ($\sim$0.14 GeV) is a typical momentum of the interacting nuclear system and $\Lambda_\chi$ (ranging from 700 MeV to 1 GeV) is the breakdown scale of the chiral EFT, which is associated with physics that is not explicitly resolved. The power counting scheme in chiral EFT provides a convenient scheme to quantify the uncertainty from nuclear forces~\cite{Machleidt:2011PR,Epelbaum:2009RMP,Hebeler:2021PR} and currents~\cite{Park:1993,Park:2003,Krebs:2020EPJA,Baroni:2021}. With the nuclear forces derived from chiral EFT up to next-to-next-to-next-to-leading order (N$^3$LO)~\cite{Entem:2003}, one can predict nuclear structure properties \cite{Morris:2018} and single-$\beta$ decay rates \cite{Gysbers:2019df} of atomic nuclei up to mass number $A=100$ or even beyond \cite{Arthuis:2020,Miyagi:2021} using many-body approaches employing some systematically improvable truncation schemes. \jmy{In recent years, motivated by the successes of relativistic theories in studies of nuclear systems and the need for a relativistic nuclear force in nuclear structure studies, a new relativistic scheme to construct the nuclear force in the framework of covariant chiral EFT was proposed~\cite{Ren:2016CPC}. Within this framework, a high-precision nucleon-nucleon interaction up to the N$^2$LO order was obtained very recently~\cite{Lu:2021}, providing an essential input for relativistic {\em ab initio} studies of nuclear structure and reactions in the near future.}

Despite the above achievements,  the calculation of the NMEs of candidates for the $0\nu\beta\beta$ decay from first-principles is still challenging due to the difficulties discussed below:  
\begin{itemize}
    \item The candidates for $0\nu\beta\beta$ decay involve medium-mass deformed nuclei with strong collective correlations for which the usual truncation schemes of nuclear many-body methods are ill-suited \cite{Hergert:2020}. An efficient way to capture nuclear collective correlations needs to be implemented into the {\em ab intio} methods applicable for heavy nuclei. This is one of the key points to be discussed in this Review.    
    
    \item The $0\nu\beta\beta$-decay operators consistent with the chiral expansion order of the employed nuclear forces are not available yet. Inspired by the first derivation of $0\nu\beta\beta$-decay operators mediated by heavy particles within the EFT~\cite{Prezeau:2003PRD}, Cirigliano et al. exploited the EFT to derive the transition operators up to N$^2$LO in the standard mechanism~\cite{Cirigliano:2018JHEP,Cirigliano2018PRL,Cirigliano:2019PRC}. It was shown that the use of their derived full transition operators up to N$^2$LO requires the knowledge of excitation energies for intermediate nuclear states, which are challenges to  many-body approaches. Besides, it was realized there based on the renormalizability argument that  a contact transition operator needs to be promoted from the N$^2$LO to the leading order. Due to the absence of  data to determine the corresponding LEC, a more fundamental theoretical method was proposed to estimate the LEC  \cite{Cirigliano:2021qko,Cirigliano:2021PRL} with which one is able to estimate its impact on the NMEs of $0\nu\beta\beta$ decay with {\em ab initio} methods~\cite{Pastore:2018,Wirth:2021,Jokiniemi:2021} and this result will hopefully be confirmed in future lattice QCD calculation~\cite{Davoudi:2020gxs}. 
    
    \item The contribution of higher-body weak currents to the NME  might become significant in the calculations of nuclear models with low-momentum nuclear forces. In the chiral EFT,  two-body weak currents contribute to the  $0\nu\beta\beta$ decay at N$^3$LO.   In most of the nuclear {\em ab initio} studies, however, the nuclear force from chiral EFT is usually preprocessed with the renormalization technique of  $V_{\rm low-k}$~\cite{Bogner:PR2003} or  similarity renormalization group (SRG)~\cite{Bogner:2010} that reduces the coupling between low and high momenta to speed up the convergence rate of nuclear many-body calculations. Using the nuclear wave functions obtained in this framework for the NME of $\znubb$ decay, the currents have to be transformed accordingly to the same energy scale. This transformation might induce sizable higher-body currents, adding coherently to the original currents. The calculation of these currents in candidate nuclei is challenging. 
  
\end{itemize}

Compared to {\it ab initio} methods, conventional nuclear models based on  either in-medium nucleon-nucleon effective interactions or energy density functionals (EDFs) offer more economic tools to modeling atomic nuclei and they have been extensively applied to compute different nuclear properties with good precision in the past decades. One of the most important conventional nuclear models is the valence-space interacting shell model (ISM) which represents the nuclear wave function as a superposition of all possible configurations described by different Slater determinants and it thus contains all types of correlations allowed within the model space.  The shell model has been frequently chosen to understand the structure of atomic nuclei with great success~\cite{Caurier:2005RMP,Otsuka:2020RMP}. However, the application of shell models is strongly limited to nuclei in particular mass regions because of the exponential growth of the Hilbert space with 
nuclear size. Therefore, in practical calculations, the shell models are restricted to a valence space  composed of a few orbits. Nucleons in this small model space interact  with each other via a phenomenological residual interaction tailored to that model space. Complementary to the shell model, self-consistent mean-field (SCMF) or EDF approaches exploiting the mean-field approximation and the mechanism of symmetry-breaking to take into account the predominant parts of many-body correlations (such as pairing correlations and deformations) have achieved great success in the description of nuclear bulk properties for nuclei over the entire nuclear chart with sets of universal parameters~\cite{Bender:2003RMP,Vretenar:2005PR,Meng:2005PPNP}. In this framework, it is well known that the mean-field approximation suffers from the drawback that the resulting wave function does not have good quantum numbers and, therefore, many results cannot be compared directly to the experimental data. To solve this problem, beyond mean-field (BMF) techniques are usually implemented to take into account the dynamic correlations associated with symmetry restoration and quantum fluctuations. This procedure is usually carried out within the symmetry-projected generator coordinate method (PGCM), in which the nuclear wave function is constructed as a linear combination of nonorthogonal mean-field wave functions projected onto the correct quantum numbers. The framework of implementing the BMF techniques of PGCM into EDF calculations is also known as multi-reference(MR)-EDF, which has become the state-of-the-art tool of choice for nuclear spectroscopic studies. The most advanced MR-EDF incorporates nuclear triaxial deformation~\cite{Bender:2008,Yao:2010,Rodriguez:2010,Yao:2014PRC} and octupole deformation~\cite{Yao:2015Octupole,Bernard:2016PRC,Marevic:2020PRL}, where the major quantum numbers (particle numbers, parity, and angular momentum) are restored in the wave function of nuclear state. See for instance the review papers~\cite{Bender:2003RMP,Niksic:2011PPNP,Egido:2016PS,Robledo:2018JPG,Sheikh2021_JPG48-123001}. In particular, the BMF techniques provide powerful tools to improve the approximations employed in various many-body approaches to the NMEs of $0\nu\beta\beta$ decay.

With the same transition operator derived from the standard mechanism, previous studies have shown that the NMEs of $0\nu\beta\beta$ decay calculated by either the  shell models \cite{Menendez:2009,Caurier:2008PRL,Horoi:2016} or by MR-EDF approaches \cite{Rodriguez:2010,Song:2014,Yao:2015,Song:2017} are not much sensitive to the choice of effective interactions or EDFs as long as the same model space and the same type of many-body approximations are employed. However, the NMEs derived from these two types of conventional nuclear models differ from each other by  a factor of about three, causing an uncertainty of an order of magnitude in the half-life for a given value of the effective neutrino mass $\langle m_{\beta\beta}\rangle$. In other words,  the uncertainty in the effective Majorana neutrino mass is  a factor of about three, which is {\em huge} compared to the uncertainties (around the percentage level \cite{PDG:2018}) in other neutrino parameters.  Resolving the discrepancy between different model predictions has been one of the major tasks in the nuclear theory community \cite{Engel:2017}.  A great deal of effort~\cite{Horoi:2013PRL,Menendez:2014,Brown:2015compare,Menendez:2016PRC,Iwata:2016,Yao:2016PRC,Wang:2021} has been devoted to understanding how the choices of EDFs or effective interactions and truncation schemes in the model space affect the predicted NMEs. Unfortunately, it is difficult to reduce the discrepancy between these models because each model has its own phenomenology and uncontrolled approximations. Besides, there are many studies based on nuclear models other than the above two, including the proton-neutron quasiparticle random-phase approximation~\cite{Simkovic:2009PRC,Mustonen:2013,Hyvarinen:2015PRC,Fang:2018,Simkovic:2018,Jokiniemi:2021,Terasaki:2021}, the interacting boson model (IBM) \cite{Barea:2009PRC,Barea:2012PRL,Barea:2013,Barea:2015,Deppisch:2020ztt}, or the projected Hartree-Fock-Bogoliubov method using a pairing-plus-quadrupole Hamiltonian~\cite{Rath:2010PRC,Rath:2013,Rath:2019}. See the recent discussions on the comparison of the quasiparticle random-phase approximation with IBM~\cite{Kotila:2021} and with the shell model~\cite{Brown:2015compare,Terasaki:2021}.

The modeling of the NMEs of $0\nu\beta\beta$ decay is now standing at the crossroads. One direction is to understand the discrepancy existing in the NMEs by the various conventional nuclear models and to improve the corresponding descriptions. The other direction is to develop an advanced nuclear {\em ab initio} method that is capable to describe the $0\nu\beta\beta$ decay of medium-mass and heavy nuclei. Thanks to the application of high-performance computing and new implementations of many-body techniques into nuclear physics in the past decades, there are lots of new exciting developments in  conventional nuclear models and {\em ab initio} methods \cite{Hergert:2020}, both of which have made significant achievements. In particular, it has been shown recently that marrying the techniques of conventional nuclear models with {\em ab initio} methods can provide a promising way to compute the  NMEs of $0\nu\beta\beta$ decay starting from realistic nuclear forces~\cite{Coraggio:2020,Yao:2020PRL,Belley2021PRL}. Nevertheless, there is still plenty of work to be done towards a precise determination of the NMEs~\cite{Yao:2020Science}. This Review aims to shed some light on the way to this end.

There are many excellent review papers on the topics related to $0\nu\beta\beta$ decay, covering the status and plans of experimental searches \cite{Dolinski:2019,Ejiri:2019PR} and the theory of $0\nu\beta\beta$ decay~\cite{Vergados:2012RPP,Bilenky:2015}, its implication on neutrino physics \cite{DellOro:2016} and extensions of the standard model \cite{Faessler:1998JPG,Deppisch:2012JPG}, and also on the status of modeling nuclear wave functions ~\cite{Engel:2017} and determination of effective transition operators with chiral EFT and lattice QCD~\cite{Cirigliano:2020PPNP,Davoudi:2021PR} for the NMEs. Besides, there are also some short review papers on the applications of a specific many-body approach to the NMEs, including the projected Hartree-Fock-Bogoliubov theory~\cite{Rath:2019} and the shell model~\cite{Coraggio:2020}. To the best of our knowledge, a comprehensive review on the status of nuclear BMF approaches for $0\nu\beta\beta$ decay is still missing. In this Review, we  highlight the recent developments along this direction and the remaining issues to be addressed.

The structure of this Review is arranged as follows. In Section \ref{sec:BMF}, \jmy{we present the general ideas and recent developments of techniques and algorithms for the BMF studies of  nuclear structure properties and the NMEs of $0\nu\beta\beta$ decay. In particular, we highlight the recent efforts towards {\em ab initio} studies of heavy nuclei exploiting the BMF techniques.}  In Section \ref{sec:theory4dbd}, we present a comprehensive derivation of the half-life of $0\nu\beta\beta$ decay based on the standard mechanism. Recent progress in the understanding of transition operators is also discussed. In Section \ref{sec:NME}, we  overview the status of BMF calculations of the NMEs of $0\nu\beta\beta$ decay starting from either a phenomenological Hamiltonian, EDF, or a realistic nuclear force. In Section \ref{sec:summary} we summarize the main content of this Review, and present an outlook on the future developments towards a precise calculation of the NMEs.

%% file: 1BMF-mod.tex
The concept of BMF is very broad and therefore it is unrealistic to cover all the works  within this framework in this Review. Given this situation, we will  overview the main ideas of some typical BMF approaches that have been actively employed in nuclear physics.  By definition, an approach that goes BMF approximation can be classified as a BMF approach, in which, the wave function of a fermionic quantum many-body system is represented as a superposition of more than one Slater determinant.   The dynamic correlation included in the correlated wave function  usually increases with the number of Slater determinants, but at the cost of increasing the complexity of the calculations. 

The starting point of BMF approaches is usually a reference state obtained with the mean-field approach, like Hartree-Fock or Hartree-Fock-Bogoliubov. The wave function of thus obtained reference state is a single Slater determinant of single-particle or quasiparticle basis functions and it is in many cases a good approximation to the exact wave function. In the mean-field approximation, the many-body problem is reduced to an effective one-body problem that can be easily solved. The interactions among all the nucleons are approximated as effective one-body potentials composed of  a long-range particle-hole field $\Gamma$ and a short-range particle-particle pairing field $\Delta$, both of which are treated in a self-consistent manner \cite{Ring:1980}. The Hartree-Fock or Hartree-Fock-Bogoliubov approaches with  effective potentials derived from  phenomenologically parametrized nuclear effective interactions or EDFs turn out to be a great success in the description of global nuclear properties \cite{Bender:2003RMP,Meng:2016Book}.

 The mean-field approaches mentioned above are also  referred to as the single-reference framework relying on the variational principle which usually, because of its non-linear character, leads to solutions of the many-body system violating the symmetries of the Hamiltonian, such as translational symmetry, rotational SO(3) symmetry, and gauge U(1) symmetries \cite{Sheikh2021_JPG48-123001}. It results in a wave function without definite momentum, angular momentum, and particle numbers, respectively. In finite nuclei, this spontaneous symmetry breaking is an artificial effect and can cause sharp phase transitions, for instance, the sudden change from spherical to deformed shapes or from normal to superfluidity state. The deficiencies of the single-reference framework can be remedied
after being extended to a  multi-reference framework with the help of the  symmetry-projected generator coordinate method \cite{Ring:1980}, a commonly used BMF approach in nuclear physics. Symmetry restoration with projection operators can be carried out either before or after variation depending on if the single-particle wave functions are determined from the projected or unprojected energy \cite{Sheikh2021_JPG48-123001}.

In this section, we will introduce the framework of BMF approaches, including the quasiparticle random-phase approximation and beyond, symmetry restoration, and the generator coordinate method, focusing on recent progress in development. In particular, in the recent decade, significant progress has been made in the {\em ab initio} studies of atomic nuclei starting from realistic nuclear interactions. How to implement these interactions into BMF approaches is an interesting topic. We will come back to this topic at the end of this section.

\subsection{The Hartree-Fock-Bogoliubov (HFB) theory}
\label{sec:HFB}

Let us start from a general Hamiltonian composed of one-, two-, and three-body terms,
\beq
\label{eq:Hamiltoian_2N_3N}
H=\sum_{pq} t^p_{q}c^\dagger_p c_{q} + \dfrac{1}{4}\sum_{pqrs}  v^{pr}_{qs}c^\dagger_p c^\dagger_{r}c_{s}c_q
+\dfrac{1}{36}\sum_{pqrstu} w^{pqr}_{stu} c^\dagger_p c^\dagger_q c^\dagger_r c_uc_tc_s,
\eeq
where $t^p_q$ is the matrix element of the kinetic energy\footnote{This applies usually for nuclear interactions or EDFs  defined in a full single-particle space. For effective Hamiltonians in restricted spaces it is just the single-particle part of this operator.}, $v^{pr}_{qs}$ and $w^{pqr}_{stu}$ stand for the antisymmetrized matrix elements of two- and three-body interactions in a selected orthonormal single-particle basis, such as the eigenstates of a harmonic oscillator (HO) potential, with the particle creation and annihilation  operators $c^\dagger_p$ and $c_p$ ($p=1, 2, \ldots, M$). The index $p$ stands for a set of quantum numbers $(n_p, l_p, j_p, m_p)$.   In the HFB theory, the nuclear wave function is approximated as a quasiparticle vacuum 
\beq
\label{eq:HFB_wf}
\ket{\Phi(\bm{q})}
=\prod^M_{k=1} \beta_k(\bm{q}) \ket{0},\quad \beta_k(\bm{q})\ket{\Phi(\bm{q})}=0,
\eeq
where $\ket{0}$ denotes the particle vacuum. The quasiparticle operators $\beta_k(\bm{q}), \beta^\dagger_k(\bm{q})$ are linear combinations of the particle operators $c_p, c^\dagger_p$,
 \beq
 \label{eq:Bogoliubov_Ttransformation}
 \begin{pmatrix}
    \beta(\bm{q})\\
    \beta^{\dagger}(\bm{q})
    \end{pmatrix}
       =
{\cal W}(\bm{q})
\begin{pmatrix}
    c\\
    c^\dagger
 \end{pmatrix}
    =
 \begin{pmatrix}
    U^\dagger(\bm{q}) & V^\dagger(\bm{q}) \\
    V^T(\bm{q})       & U^T(\bm{q})
 \end{pmatrix}
\begin{pmatrix}
    c\\
    c^\dagger
 \end{pmatrix},
 \eeq
where $\bm{q}$ labels different HFB states.
The transformation (\ref{eq:Bogoliubov_Ttransformation}) is known as Bogoliubov transformation with ${\cal W}(\bm{q})$ being a $2M\times2M$ matrix. Its unitarity requires that the  matrices $(U, V)$ satisfy the orthogonalization relations
$$
\begin{aligned}
U^{\dagger} U+V^{\dagger} V &=1, \quad U U^{\dagger}+V^{*} V^{\mathrm{T}}=1, \\
U^{\mathrm{T}} V+V^{\mathrm{T}} U &=0, \quad    U V^{\dagger}+V^{*} U^{\mathrm{T}}=0.
\end{aligned}
$$
The matrices $(U, V)$ are determined by the variational principle which leads to the HFB equation,
 \beq
 \label{eq:HFB_equation}
 \begin{pmatrix}
   h-\lambda & \Delta\\
   -\Delta^\ast & -h^\ast-\lambda
    \end{pmatrix}
    \begin{pmatrix}
    U_k\\
    V_k
 \end{pmatrix}
    =E_k
  \begin{pmatrix}
    U_k\\
    V_k
 \end{pmatrix}.
 \eeq
 In the above equation, $E_k$ is the so-called quasiparticle energy and the columns $U_k(\bm{q}), V_k(\bm{q})$ correspond to the expansion coefficients of the $k$-th quasiparticle wave function in the single-particle $(c_p, c^\dagger_p)$ basis.  The particle-hole field $h$ and particle-particle field $\Delta$ are formulated in the single-particle basis by the matrix elements given by
 \begin{align}
 \label{eq:HFB_mean_fields_ph}
 h^p_{q}&=t^p_{q}+\sum_{rs} v^{pr}_{qs}\rho^r_s+\Gamma^p_{q}(3N), \\
 \label{eq:HFB_mean_fields_pp}
 \Delta^{pq} &= \dfrac{1}{2}\sum_{rs} v^{pq}_{rs}\kappa_{rs}+\Delta^{pq}(3N),
 \end{align}
where the contribution from the three-body interaction term to \jmy{the particle-hole field} $h^p_q$ is
\beqn
\Gamma^p_q(3N)
&=&\dfrac{1}{2}\sum_{rstu} w^{prs}_{qtu}\rho^r_t\rho^s_u
+\dfrac{1}{4}\sum_{rstu} w^{prs}_{qtu}\kappa^{rs}\kappa_{tu},
\eeqn
and its contribution to \jmy{the pairing field} $\Delta^{pq}$ is
\beqn
\Delta^{pq}(3N)
&=& \dfrac{1}{2}\sum_{rstu}w^{pqr}_{stu}\kappa_{st}\rho^r_u.
\eeqn
In the above equations, we have introduced the one-body {\em normal} and {\em abnormal}  density matrices  $\rho$ and $\kappa$ (also known as pairing tensor) of the HFB state $\ket{\Phi(\bm{q})}$
\bsub\beqn
\rho^p_q(\bm{q};\bm{q}) &=& \langle\Phi(\bm{q}) \vert c^\dagger_p c_q\vert\Phi(\bm{q})\rangle,\\
\kappa^{pq}(\bm{q};\bm{q}) &=& \langle\Phi(\bm{q}) \vert c^\dagger_p c^\dagger_q\vert\Phi(\bm{q})\rangle,\\
\kappa_{ts}(\bm{q};\bm{q}) &=&  \langle \Phi(\bm{q})\vert c_sc_t\vert\Phi(\bm{q})\rangle.
\eeqn
\esub

According to the Bloch-Messiah theorem \cite{Block:1962NP}, the Bogoliubov transformation matrix ${\cal W}(\bm{q})$ in (\ref{eq:Bogoliubov_Ttransformation})  can always be decomposed into the product of three special matrices~\cite{Ring:1980}
\beqn
\label{eq:Bloch-Messiah}
{\cal W}
=\left(\begin{array}{cc}
D & 0 \\
0 & D^{*}
\end{array}\right)\left(\begin{array}{cc}
\bar{U} & \bar{V} \\
\bar{V} & \bar{U}
\end{array}\right)\left(\begin{array}{cc}
C & 0 \\
0 & C^{*}
\end{array}\right).
\eeqn
In comparison with Eq.(\ref{eq:Bogoliubov_Ttransformation}), one finds the relations $U=D\bar U C$ and $V=D\bar V C^*$, where $D$ and $C$ are $M\times M$ unitary matrices. The matrix $D_{pl}$ transforms from the original basis $p$ to the so-called canonical basis consisting of pairs ($l,\bar{l}$) of canonically conjugate states\footnote{In case of time-reversal symmetry of the Hamiltonian (\ref{eq:Hamiltoian_2N_3N}), the canonical conjugate states are connected by the time-reversal operator.}. The second transformation given by the real matrices $\bar U$ and $\bar V$
     \beqn
     \label{eq:uv-coefficients}
      \bar U_{ll'}=\left(\begin{array}{cc}
      u_l & 0 \\
      0 & u_l  \\
     \end{array}\right)\delta_{ll'},\quad
      \bar V_{ll'}=\left(\begin{array}{cc}
      0 & v_l \\
      -v_l & 0  \\
     \end{array}\right)\delta_{ll'},
     \eeqn
 is the Bogoliubov-Valatin transformation defining quasiparticle operators ($\alpha^\dagger_l, \alpha^\dagger_{\bar l}$) with $l>0$,
 \bsub\begin{align}
 \alpha^\dag_l&= u_l c^\dag_l - v_l c^\dag_{\bar{l}}, \\
 \alpha^\dag_{\bar{l}}&=u_l c^\dag_{\bar{l}} + v_l c^\dag_l,
 \label{eq:Bogoliubov-Valatin}
 \end{align}
 \esub
 where the index $l=1, 2, \ldots, M/2$ and $u_{l}^{2}+v_{l}^{2}=1$. The $v^2_l \in[0,1]$ is interpreted as the occupation probability of the $l$-th single-particle state in canonical basis, in which the one-body density $\rho$ is diagonal. The third matrix $C$ in Eq. (\ref{eq:Bloch-Messiah}) describes a transformation in quasiparticle space from the operators $\alpha$ to the operators $\beta$ in Eq. (\ref{eq:Bogoliubov_Ttransformation}).

 In the nucleus $(A, Z)$ with large shell gaps around the Fermi surfaces, the nuclear interactions might be not strong enough to scatter a pair of two nucleons across the shell gaps. In this case, the pairing field $\Delta$ in (\ref{eq:HFB_mean_fields_pp}) tends to vanish in the procedure of iterations. The HFB equation (\ref{eq:HFB_equation}) collapses to an HF equation, and the HFB state (\ref{eq:HFB_wf}) becomes an HF state,
 \beq
\label{eq:HF_wf}
\ket{\Phi(\bm{q})}
=  \prod^N_{n=1} d^\dagger_n(\bm{q}) \ket{0}\otimes\prod^Z_{p=1} d^\dagger_p(\bm{q}) \ket{0}
\eeq
where $Z$ is the number of protons and $N$ the number of neutrons in the nucleus. $d^\dagger_n$ and $d^\dagger_p$ are single-particle creation operators for neutrons and protons, respectively, and can be written as a linear combination of $c^\dagger$ operators with the weight determined by the $D$ matrix in (\ref{eq:Bloch-Messiah}).

 Note that in most applications, the Hamiltonian (\ref{eq:Hamiltoian_2N_3N}) is truncated up to the two-body term and all the terms related to the three-body term vanish. The effect of three-body interaction is mimicked by introducing a density-dependent interaction.  It can be viewed as the use of the normal-ordering technique with respect to a reference state
 \beqn
 \label{eq:NO2B}
 \dfrac{1}{36}\sum_{pqrstu} w^{pqr}_{stu} c^\dagger_p c^\dagger_q c^\dagger_r c_uc_tc_s 
 &\to&
 \dfrac{1}{6} \sum_{pqrstu}w^{pqr}_{stu}\rho^p_s\rho^q_t\rho^r_u +
 \dfrac{1}{2}\sum_{ps}  \left(\sum_{qrut}w^{pqr}_{stu}\rho^q_t\rho^r_u\right) \{c^\dagger_pc_s\}
 +\dfrac{1}{4} \sum_{pqst}  \left(\sum_{ru}w^{pqr}_{stu}\rho^r_u\right) \{c^\dagger_p c^\dagger_q  c_tc_s\}\nonumber\\
 &&+ \dfrac{1}{36}\sum_{pqrstu} w^{pqr}_{stu} \{c^\dagger_p c^\dagger_q c^\dagger_r c_uc_tc_s\},
 \eeqn
 where $\{\cdots\}$ denotes normal ordering of operators.  Recent {\em ab initio} studies employing the normal-ordering technique have demonstrated that with a properly selected reference state  most of the three-body effect can be captured with the normal-ordering two-body approximation~\cite{Roth:2012,Hergert:2016jk,Miyagi:2021,Hebeler:2021PR}, where the last term in (\ref{eq:NO2B}) is neglected.   In EDF-based approaches, the Hamiltonian (\ref{eq:Hamiltoian_2N_3N}) is replaced with an energy functional $E[\rho, \nabla^2\rho, \ldots]$ of various nuclear intrinsic densities and their derivatives. In most of 
the shell models, i.e. configuration-interaction (CI) calculations~\cite{Caurier:2005RMP},
an effective one-plus-two-body effective Hamiltonian defined in a valence space is employed with the interaction matrix elements as free parameters optimized to nuclear low-lying states. In these cases, the three-body effect should have been taken into account partially through the adjustment of the free parameters. See a comprehensive discussion on the determination of three-body nuclear force~\cite{Hebeler:2021PR} in chiral EFT and its role in shell-model calculation~\cite{Stroberg:2019}. Hereafter, the Hamiltonian (\ref{eq:Hamiltoian_2N_3N}) will be written explicitly up to the two-body term.

 \subsection{The quasiparticle random-phase approximation (QRPA)}
 \label{subsubsec:QRPA}

 The proton-neutron quasiparticle random-phase approximation (pn-QRPA)  is one of the most reliable nuclear structure models used for describing the structure of the intermediate nuclear states virtually excited in $\beta\beta$ decay.  Important BMF correlations in the nuclear ground state  are naturally accounted for within the QRPA approach through the introduction of a backward amplitude~\footnote{In some applications QRPA is not considered as BMF, because it can be derived from a time-dependent mean-field approach. In this work, we consider QRPA as beyond mean-field because of the correlations in the ground state through the backward going diagrams.}. The formalism of QRPA model has been introduced in detail in textbooks~\cite{Ring:1980,Suhonen:1988} and in a recent review paper~\cite{Schuck:2021PR}. Here we present a sketch of this approach for nuclear charge-exchange excitations.

 In the QRPA approach, nuclear excited states $|\nu \rangle$ are generated by an operator $Q^\dagger_{\nu}$ on top of the QRPA vacuum state $\ket{{\rm QRPA}}$, where the $Q^\dagger_{\nu}$ operator is defined as
 \beq
  Q^\dagger_{\nu} = \sum_{pn}  X^{\nu}_{pn} \beta^\dagger_p \beta^\dagger_n - Y^{\nu}_{pn} \beta_n \beta_p.
\eeq
The $\beta^\dagger_p$ and $\beta_n$ are the quasiparticle creation and annihilation operators for protons and neutrons defined in (\ref{eq:Bogoliubov_Ttransformation}). We first consider the pn-QRPA for the charge-exchange channel. The QRPA vacuum state $\ket{{\rm QRPA}}$ is defined as follows
\beq
\label{eq:qrpa-gs}
 Q_{\nu} | {\rm QRPA} \rangle = 0.
\eeq
If the backward amplitude $Y^{\nu}_{pn}$ in $Q^\dagger_{\nu} $ were set to be zero, the QRPA vacuum state would be a pure HFB state, and the QRPA would reduce to the quasiparticle Tamm-Dancoff approach in which the wave function of a nuclear excited state is expanded in terms of two-quasiparticle excitations on top of the ground state approximated with a HFB state. Therefore, the magnitude of the amplitude $Y^{\nu}_{pn}$ reflects the amount of BMF correlations included in the QRPA for the ground state.

For convenience, the nuclear Hamiltonian is usually written in terms of the quasiparticle operators~\cite{Ring:1980},
\beqn
 H &=& H^0 + \sum_{k_1k_2}H^{11}_{k_1k_2} \beta^\dagger_{k_1}\beta_{k_2} + \frac{1}{2} \sum_{k_1k_2} ( H^{20}_{k_1k_2} \beta^\dagger_{k_1} \beta^\dagger_{k_2} + h.c. )
  + \sum_{k_1k_2k_3k_4 } (H^{40}_{k_1k_2k_3k_4} \beta^\dagger_{k_1}\beta^\dagger_{k_2} \beta^\dagger_{k_3} \beta^\dagger_{k_4}  + h.c. ) \nonumber\\
 && +   \sum_{k_1k_2k_3k_4 } (H^{31}_{k_1k_2k_3k_4} \beta^\dagger_{k_1}\beta^\dagger_{k_2} \beta^\dagger_{k_3} \beta_{k_4}  + h.c. )
  +  \frac{1}{4} \sum_{k_1k_2k_3k_4 } H^{22}_{k_1k_2k_3k_4} \beta^\dagger_{k_1}\beta^\dagger_{k_2} \beta_{k_3} \beta_{k_4},
\eeqn
where $H^0=\bra{\Phi(\bm{q})}H\ket{\Phi(\bm{q})}$ corresponds to the energy of the HFB state $\ket{\Phi(\bm{q})}$. In the canonical basis, the matrix $H^{11}$ has the form,
 \beq
 H^{11}_{kk'} = (u_k u_{k'} - v_k v_{k'} ) h_{kk'} - (u_k v_{k'} + v_k u_{k'} ) \Delta_{kk'},
 \eeq
where the particle-hole $h$ and pairing field $\Delta$ have been defined  in Eqs. \eqref{eq:HFB_mean_fields_ph} and  \eqref{eq:HFB_mean_fields_pp}.

The unknown amplitudes $X^{\nu}_{pn}$ and $Y^{\nu}_{pn}$ are determined by the QRPA equation which can be derived with equations-of-motion method~\cite{Rowe:1968RMP},
\beqn
	\langle {\rm QRPA} | \Big[ \beta_{n}\beta_{p}, \big[ H, Q_\nu^\dagger\big] \Big] | {\rm QRPA}\rangle &=& \hbar \Omega^\nu \langle  {\rm QRPA}| \Big[ \beta_{n} \beta_{p}, Q_\nu^\dagger \Big] | {\rm QRPA}\rangle  ,\\
\langle  {\rm QRPA}| \Big[ \beta_{p}^\dagger \beta_{n}^\dagger , \big[ H, Q_\nu^\dagger \big] \Big] | {\rm QRPA}\rangle &=& \hbar \Omega^\nu \langle  {\rm QRPA}| \Big[ \beta_{p}^\dagger \beta_{n}^\dagger, Q_\nu^\dagger \Big]  | {\rm QRPA}\rangle.
 \eeqn
The above QRPA equation can be rewritten into a compact form,
 \beq
 \label{eq:QRPAeq}
 \left( \begin{array}{cc} A & B \\ -B^* & -A^* \end{array} \right) \left( \begin{array}{c} X^\nu \\ Y^\nu \end{array} \right) = \hbar \Omega^\nu  \left( \begin{array}{c} X^\nu \\ Y^\nu \end{array} \right) ,
 \eeq
 where the matrix elements of $A$ and $B$ are defined as
 \begin{align}
 \label{eq:QRPAmatrixelementA}
 A_{p_1n_1,p_2n_2} &= \langle \Phi(q)| \Big[ \beta_{n_1}\beta_{p_1}, \big[ H,   \beta^\dagger_{p_2} \beta^\dagger_{n_2}\big] \Big] | \Phi(q)\rangle
 = H^{11}_{p_1p_2} \delta_{n_1 n_2} + H^{11}_{n_1n_2} \delta_{p_1 p_2}  + H^{22}_{p_1n_1p_2n_2}, \\
  \label{eq:QRPAmatrixelementB}
 B_{p_1n_1,p_2n_2} &= -  \langle \Phi(q) | \Big[ \beta_{n_1}\beta_{p_1}, \big[ H,   \beta_{n_2} \beta_{p_2}\big] \Big] | \Phi(q)\rangle  = 4!  H^{40} _{p_1n_1p_2n_2}.
 \end{align}
 In the above derivation, use is made of the so-called quasi-boson approximation (QBA) where expectation values in the QRPA ground state are approximated by their values in the uncorrelated HFB state.  Nevertheless, the BMF correlation in the ground state has already been taken into account in terms of the $Y^{\nu}_{pn}$.
 
 
 The QRPA equation (\ref{eq:QRPAeq}) is usually expressed in the canonical basis in which the $A$ and $B$ matrix elements are given by~\footnote{We note that these expressions look different from those of Ref.~\cite{Suhonen:1988} because we are presenting the full QRPA equations  based on the HFB theory,  instead of the HF+BCS theory. Actually it can be proved that these two expressions are identical  for monopole pairing forces, where there is no difference between HFB and HF+BCS, and where $H^{11}$ is diagonal. }
 \beqn
 A_{p_1 n_1, p_2 n_2} &=& H^{11}_{p_1p_2} \delta_{n_1 n_2} + H^{11}_{n_1n_2} \delta_{p_1 p_2} + V^{pp}_{p_1n_1p_2n_2}(u_{p_1} u_{n_1} u_{p_2} u_{n_2} + v_{p_1} v_{n_1} v_{p_2} v_{n_2}) \nonumber\\
 && + V^{ph}_{p_1n_2n_1p_2}(u_{p_1} v_{n_1} u_{p_2} v_{n_2} + v_{p_1} u_{n_1} v_{p_2} u_{n_2}), \\
 B_{p_1 n_1, p_2 n_2} &=& - V^{pp}_{p_1n_1p_2n_2}(u_{p_1} u_{n_1} v_{p_2} v_{n_2} + v_{p_1} v_{n_1} u_{p_2} u_{n_2}) + V^{ph}_{p_1n_2n_1p_2}(u_{p_1} v_{n_1} v_{p_2} u_{n_2} + v_{p_1} u_{n_1} u_{p_2} v_{n_2}),
 \eeqn
 where  $V^{pp}$ and $V^{ph}$ are the particle-particle and particle-hole two-body residual interactions, respectively. \begin{itemize}
     
\item For the particle-hole interaction, either a phenomenological schematic separable force~\cite{Sarriguren:2003,Simkovic:2004NPA}, an effective interaction derived from the CD-Bonn potential~\cite{Yousef:2009PRC,Fang:2010,Fang:2011}, or an effective interaction derived from EDFs  self-consistently~\cite{Terasaki:2019,Mustonen:2013,Niu:2017PRC} has been employed. For the latter, it is determined in such a way, 
 \beq
 V^{ph}_{p_1n_2n_1p_2} =  \frac{\partial^2 E }{\partial \rho_{n_1p_1}\partial \rho_{p_2n_2}},
 \eeq
 that it is fully consistent with  that used in the HFB calculation.
 
 \item For the particle-particle interaction, it can in principle be derived from the particle-hole interaction via Pandya transformation~\cite{Pandya:PR1956}. In the QRPA based on either a relativistic EDF or Skyrme EDF, however, the  particle-particle residual interaction takes a different form from that of the particle-hole part, but it has to be consistent with that used in the underlying HFB calculation. For charge-exchange excitations, the particle-particle interaction can be separated into  isoscalar ($T=0$) channel and isovector ($T=1$) channel.\footnote{If a contact interaction is employed for the particle-particle channel, the isoscalar part acts only for non-natural parity excitations, such as the Gamow-Teller(GT)-type excitation, while the isovector part acts only for natural-parity excitations, such as the Fermi-type excitation.} For the isovector channel, one can use the same form as that used in the HFB calculation, and its strength can be determined by fitting to ground-state properties.  However, except for $N=Z$ nuclei, the isoscalar particle-particle interaction cannot be well determined by the ground-state properties. Therefore, the isoscalar part is usually chosen to be the same form as that of the isovector part, but with an additional renormalized strength parameter $g_{\mathrm{pp}}^{T=0}$ adjusted to experimental data of GT excitations~\cite{Bai:2013PLB,Bai:2014PRC} or the half-lives of single-$\beta$ decay~\cite{Suhonen:1994PRC,Niu:2013PLB, Niu:2013PRC}. \jmyr{In the application to $0\nu\beta\beta$ decay, the half-lives of $2\nu\beta\beta$ decay were adopted to determine the renormalized strength parameters $g_{\mathrm{pp}}^{T=0}$ and $g_{\mathrm{pp}}^{T=1}$ separately in more sophisticated ways~\cite{Rodin:2003,Simkovic:2004NPA,Faessler:2008JPG_gA,Lisi:2015PRD,Suhonen:2017PRC,Terasaki:2018PRC,Simkovic:2018}. For example, the isovector strength parameter $g_{\mathrm{pp}}^{T=1}$ can be adjusted such that the Fermi NME of $2 \nu \beta \beta$  decay vanishes~\cite{Rodin:2011,Simkovic:2013} as required by the isospin-symmetry conservation. Keeping the obtained value of $g_{\mathrm{pp}}^{T=1}$ fixed, the $g_{\mathrm{pp}}^{T=0}$ is further determined by requiring either the half-lives of $2\nu \beta \beta$ decay are reproduced~\cite{Suhonen:2017PRC,Fang:2018} or the NMEs of $2\nu \beta \beta$ decay under closure approximation vanish~\cite{Simkovic:2018}. 
 }

  \end{itemize}

 For spherical nuclei, it is convenient to rewrite the QRPA equation into the angular momentum ($J$)-coupled form.  In the Condon-Shortly phase convention~\cite{Edmonds:1955}, the $J$-coupled QRPA matrix elements are given by \footnote{We note that this definition is consistent with that in Ref.~\cite{Suhonen:1988} where the Baranger's notation~\cite{Baranger:1960} was employed,
 \bsub\begin{align}
 G\left(p_1 n_1 p_2 n_2 J\right)
 &=-\frac{1}{2}\left\langle p_1 n_1 (J)|V^{pp}| p_2n_2 (J)\right\rangle,\\
 F\left(p_1 n_1 p_2n_2 J\right)
 &=-\frac{1}{2}\left\langle p_1 n^{-1}_1 (J)|V^{pp}| p_2 n^{-1}_2 (J)\right\rangle.
 \end{align}
 \esub}
 \beqn
 A^J_{p_1 n_1, p_2 n_2} &=& H^{11}_{p_1p_2} \delta_{n_1 n_2} + H^{11}_{n_1n_2} \delta_{p_1 p_2} + V^{ppJ}_{p_1n_1p_2n_2}(u_{p_1} u_{n_1} u_{p_2} u_{n_2} + v_{p_1} v_{n_1} v_{p_2} v_{n_2}) \nonumber\\
 && + V^{phJ}_{p_1n_2n_1p_2}(u_{p_1} v_{n_1} u_{p_2} v_{n_2} + v_{p_1} u_{n_1} v_{p_2} u_{n_2}), \\
 B^J_{p_1 n_1, p_2 n_2} &=& - V^{ppJ}_{p_1n_1p_2n_2}(u_{p_1} u_{n_1} v_{p_2} v_{n_2} + v_{p_1} v_{n_1} u_{p_2} u_{n_2}) + V^{phJ}_{p_1n_2n_1p_2}(u_{p_1} v_{n_1} v_{p_2} u_{n_2} + v_{p_1} u_{n_1} u_{p_2} v_{n_2}),
 \eeqn
 where the $J$-coupled particle-particle two-body interaction matrix element is 
 \beqn
 V^{pp J}_{p_1n_1p_2n_2} 
 &\equiv & \langle p_1 n_1  (J)| V^{pp} | p_2n_2(J)\rangle \nonumber\\
 &=&\sum_{m_{p_1}m_{n_1}m_{p_2}m_{n_2}}  \langle j_{p_1} m_{p_1} j_{n_1} m_{n_1} | JM \rangle  \langle j_{p_2} m_{p_2} j_{n_2} m_{n_2} | JM \rangle  \langle j_{p_1} m_{p_1} j_{n_1} m_{n_1} | V^{pp} | j_{p_2} m_{p_2} j_{n_2} m_{n_2} \rangle,
 \eeqn
 and the particle-hole two-body interaction matrix element 
  \beqn
 V^{ph J}_{p_1n_2n_1p_2} 
 &\equiv & \langle p_1 n_1^{-1} (J) | V^{ph} | p_2 n_2^{-1} (J)\rangle \nonumber\\
 &= & -\sum_{J'} (2J'+1) 
 \left\{\begin{array}{ccc}
     j_{p_1} & j_{n_1} & J  \\
     j_{p_2} & j_{n_2} & J' 
 \end{array}
 \right\}
 \langle {p_1} {n_2} (J') | V^{ph} | {p_2} {n_1} (J')\rangle \nonumber\\
 &=&\sum_{m_{p_1}m_{n_1}m_{p_2}m_{n_2}} (-1)^{j_{n_2}-m_{n_2} +j_{n_1} -m_{n_1} }  \langle j_{p_1} m_{p_1}  j_{n_1} -m_{n_1} | JM \rangle  
 \langle j_{p_2} m_{p_2} j_{n_2}  -m_{n_2} | JM \rangle  \nonumber\\
 &&\times\langle  j_{p_1} m_{p_1} j_{n_2} m_{n_2} | V^{ph} | j_{n_1} m_{n_1} j_{p_2} m_{p_2} \rangle. 
 \eeqn
 
 In this case, the  strength of the transition from the QRPA ground state (\ref{eq:qrpa-gs}) to the excited state $\ket{JM}$ connected by a tensor transition operator $O^{JM}$  is given by
\beq
B = \sum_M   \Bigg| \langle JM | O^{JM} | {\rm QRPA} \rangle  \Bigg|^2.
\eeq 
 For the isospin-raising $T^+$ channel,
 \beqn
 \langle JM | \hat O^{JM} | {\rm QRPA} \rangle
 &=& - \hat{J}^{-1} \sum_{pn}
 \Bigg[  (v_{n} u_{p} X_{pn}^{JM}
 +  u_{n} v_{p} Y_{pn}^{JM})
 \langle
 j_{p} || O^{J} || j_{n} \rangle_{} \Bigg],
 \eeqn
 and for the isospin-lowering $T^-$ channel
 \beq
 \langle JM | O^{JM} | {\rm QRPA} \rangle =
 \hat{J}^{-1} \sum_{pn}    (-)^{j_p-j_{n}+J+1} 
 \Bigg[(
 u_{n} v_{p} X_{pn}^{JM}  +  v_{n} u_{p} Y_{pn}^{JM} ) \langle
 j_{n} || O^{J} || j_{p} \rangle_{}  \Bigg].
 \eeq

 For axially deformed nuclei,  the total angular momentum $J$ is not conserved and thus there is no definite $J$ for the excited states from the deformed QRPA calculation. Only the intrinsic $z$-projection ($K$) of the angular momentum and parity ($\pi$) are good quantum numbers. The excited states are thus labeled as $| K ^\pi \rangle$, and the two quasiparticle pairs satisfy the selection rule $ |\Omega_p - \Omega_n| = K $ and $\pi_p \pi_n = \pi$, where $\Omega$ is the projection of the single-particle angular momentum. In this case, one needs to transform the wave function from the intrinsic frame to the laboratory frame.  For the nuclei with permanent deformation, the GT transition strength by the deformed  QRPA calculation in the intrinsic frame turns out to be equivalent to that calculated in the laboratory frame~\cite{Krumlinde:1984NPA,Yousef:2009PRC}. However, for the nuclei with small deformation, the intrinsic states with different $K$ values are strongly mixed, in which case, the restoration of rotation symmetry is demanded. \jmy{It is worth mentioning that the solution of the deformed QRPA equation in a full single-particle basis is time-consuming. To speed up the  calculation for GT transition and $2\nu\beta\beta$ decay,  Hinohara~\cite{Hinohara:2019FAM} implemented the finite-amplitude method~\cite{Nakatsukasa:2007FAM} into the deformed QRPA for the corresponding nuclear matrix elements based on the linear response for nuclear density functional theory.} 
 

\subsection{The symmetry-projected Hartree-Fock-Bogoliubov (PHFB) theory}
\label{subsubsec:symmetry-projected-HFB}

The wave function $\ket{\Phi(\bm{q})}$ from the solution of HFB equation (\ref{eq:HFB_equation}) does not conserve particle number as the quasiparticle operators $(\beta^\dagger_k, \beta_k)$ defined in Eq.~(\ref{eq:Bogoliubov_Ttransformation}) mix particle creation and annihilation operators. In other words, the HFB wave function is not invariant under the U(1) rotation $\hat R(\varphi)=e^{i\varphi\hat N}$, where $\hat N=\sum_pc^\dagger_pc_p$ is particle-number operator and $\varphi$ the gauge angle. Even though the average particle numbers, like neutrons and protons, can be constrained to be correct in the HFB calculation by adding a term $\lambda_n(\hat N-N_0)+\lambda_p (\hat N-Z_0)$ with the Lagrangian multipliers $\lambda_n$ and $\lambda_p$, the contamination of configurations with particle numbers differing from that of the atomic nucleus of interest could spoil the quality of the description. In addition, the HFB calculation for open-shell nuclei yields a deformed wave function $\ket{\Phi(\bm{q})}$, where the SO(3) rotation symmetry is broken and the wave function is a mixture of eigenstates of angular momentum $\hat J^2$ with different eigenvalues $J$. The loss of angular momentum $J$ in the wave function makes it difficult to analyze nuclear spectroscopic data (including the energy spectrum and transitions).

 A way to restore broken symmetries in the mean-field wave function $\ket{\Phi(\bm{q})}$ is through the use of projection operators $\hat P$ (for details see Ref. \cite{Sheikh2021_JPG48-123001}). Its basic idea is constructing the wave function $\ket{\Psi}$ with a proper quantum number $\lambda$ as a linear combination of a set of rotated mean-field wave functions generated by the full rotation group  $\hat R(g)$\footnote{The following considerations apply strictly only for abelian symmetry groups, and not for the group SO(3) of rotations in three-dimensional space. This group will be discussed in detail further down.},
\beqn
\label{eq:PHFB4wf}
 \ket{\Psi^\lambda(\bm{q})}
    &=&\frac{d_{\lambda}}{n_{G}}\int dg  D^{\lambda\ast}(g) \hat R(g)\ket{\Phi(\bm{q})}
    \equiv \hat P^\lambda \ket{\Phi(\bm{q})},
\eeqn
 where the projection operator $\hat P^\lambda$ is defined as a sum of the full rotations weighted by $D^{\lambda\ast}(g)$ determined by the  irreducible representations (irreps) of the symmetry group $G=\{\hat R(g)\}$, labeled with the Greek letter $\lambda$. The symbol $g$ generates all group elements, $n_G$ is the order (the number of elements or the volume) of the group, $d_\lambda$ is the dimension of the irreps. The projection operator has the property $(\hat P^\lambda)^2=\hat P^\lambda$. 
 
The energy of the symmetry-projected HFB state (\ref{eq:PHFB4wf}) is given by the expectation value of the Hamiltonian with respect to this state. Because the Hamiltonian $H$ commutes with the projection operator $\hat P^\lambda$, one finds the energy of the symmetry-restored HFB state as follows,
\beqn
\label{eq:PHFB4E}
    E^\lambda (\bm{q},\bm{q})
    = \dfrac{\braket{\Psi^\lambda (\bm{q})| H |\Psi^\lambda(\bm{q})}}{\braket{\Psi^\lambda(\bm{q}) | \Psi^\lambda(\bm{q})}}
    =  \dfrac{\braket{\Phi (\bm{q})|   H  \hat P^\lambda |\Phi(\bm{q})}}{\braket{\Phi (\bm{q})|   \hat P^\lambda |\Phi(\bm{q})}} 
    =\int dg y^\lambda(g) \braket{\Phi(\bm{q}) | H  |\Phi(\bm{q},g)},
\eeqn
where the function $y^\lambda(g)$ is defined as
 \beqn
    y^\lambda(g)(\bm{q},\bm{q})
    = {\cal N}^{-1}_\lambda D^{\lambda\ast}(g)  \braket{\Phi(\bm{q}) |\hat R(g) |\Phi(\bm{q})},
\eeqn
and ${\cal N}_\lambda(\bm{q},\bm{q})=\braket{\Psi^\lambda(\bm{q}) | \Psi^{\lambda}(\bm{q})}$ is  the normalization factor. The rotated HFB wave function~\footnote{This definition can meet problems in the case of Egido poles \cite{Anguiano:2001NPA} when the overlap $\braket{\Phi(\bm{q}) |\hat R(g) |\Phi(\bm{q})}=0$, in which case special treatment is required and it will be discussed later. }
\begin{equation}
    \ket{\Phi(\bm{q},g)}
    \equiv \dfrac{\hat R(g) \ket{\Phi(\bm{q})}}{ \braket{\Phi(\bm{q}) |\hat R(g) |\Phi(\bm{q})}}
\end{equation}
can be defined as in Eq. (\ref{eq:HFB_wf}),
\beq
\label{eq:rotated_HFB_wf}
\ket{\Phi(\bm{q}, g)}
=\prod^M_{k=1} \beta_k(\bm{q}, g) \ket{0}
\eeq
using the rotated quasiparticle operator
 \begin{equation}
     \beta_k(\bm{q}, g) \ket{\Phi{(\bm{q},g)}} =0.
 \end{equation}
The two sets of quasiparticle operators ($\beta_k(\bm{q}), \beta^\dagger(\bm{q})$) and ($\beta_k(\bm{q}, g), \beta^\dagger(\bm{q}, g)$) are connected by the following transformation \cite{Ring:1980}
 \beq
 \label{Ttransformation}
 \begin{pmatrix}
    \beta(\bm{q})\\
    \beta^{\dagger}(\bm{q})
    \end{pmatrix}
    =
 \begin{pmatrix}
    \mathbb{U}^\dagger & \mathbb{V}^\dagger \\
    \mathbb{V}^T       & \mathbb{U}^T
 \end{pmatrix}
\begin{pmatrix}
    \beta(\bm{q},g)\\
    \beta^\dagger(\bm{q}, g)
 \end{pmatrix},
 \eeq
 where the matrices $\mathbb{U}$ and $\mathbb{V}$ are given by
 \bsub\begin{align}
 \label{eq:Thouless_U}
     \mathbb{U}(\bm{q};\bm{q},g)
     &\equiv U^\dagger(\bm{q}, g)U(\bm{q})
     +V^{\dagger}(\bm{q}, g)V(\bm{q}), \\
 \label{eq:Thouless_V}
      \mathbb{V}(\bm{q};\bm{q},g)
     &\equiv V^{T}(\bm{q}, g) U(\bm{q})
     +U^T(\bm{q}, g) V(\bm{q})
  \end{align}
   \esub
  with $U(\bm{q}, g)$ and $V(\bm{q}, g)$ defined by,
 \begin{equation}
 \label{eq:rotated_UV}
     U(\bm{q},g) = D(g) U(\bm{q}),\quad  V(\bm{q}, g) = D^\ast(g) V(\bm{q}).
 \end{equation}

 The Hamiltonian overlap in the projected energy (\ref{eq:PHFB4E}) can be determined with the generalized Wick theorem~\cite{Balian:1969},
\beqn
\label{eq:energy_overlap}
\braket{\Phi(\bm{q}) | H  |\Phi(\bm{q},g)}
&=&\sum_{pq} t^p_{q}\rho^p_q(\bm{q};\bm{q},g) \nonumber\\
&&+ \dfrac{1}{4}\sum_{pqrs}  v^{pr}_{qs}\Bigg(\rho^p_q(\bm{q};\bm{q},g){\cal \rho}^r_s(\bm{q};\bm{q},g) 
-\rho^p_s(\bm{q};\bm{q},g)\rho^r_q(\bm{q};\bm{q},g)
+\kappa^{pr}(\bm{q};\bm{q},g)\kappa_{qs}(\bm{q};\bm{q},g)\Bigg),
\eeqn
where the {\em mixed} normal and abnormal densities are defined as~\cite{Ring:1980,Yao:2009PRC}
\bsub\beqn
\label{eq:mix_dens}
\rho^p_q(\bm{q};\bm{q},g) &\equiv& \langle\Phi(\bm{q}) \vert c^\dagger_p c_q\vert\Phi(\bm{q},g)\rangle
=[V(\bm{q})\mathbb{U}^{-1}V^\dagger(\bm{q},g)]_{pq},\\
\label{eq:mix_pairing1}
\kappa^{pr}(\bm{q};\bm{q},g) &\equiv& \langle\Phi(\bm{q}) \vert c^\dagger_p c^\dagger_r\vert\Phi(\bm{q},g)\rangle
=[V(\bm{q})\mathbb{U}^{-1}U^\dagger(\bm{q},g)]_{pr},\\
\label{eq:mix_pairing2}
\kappa_{sq}(\bm{q};\bm{q},g) &\equiv&  \langle \Phi(\bm{q})\vert c_qc_s\vert\Phi(\bm{q},g)\rangle
=[U(\bm{q})\mathbb{U}^{-1}V^\dagger(\bm{q};\bm{q},g)]_{qs}.
\eeqn
\esub

In general, the HFB wave function $\ket{\Phi(\bm{q})}$ breaks the symmetries of parity, gauge U(1), or rotational SO(3). To restore these symmetries, one introduces the corresponding projection operators,
\begin{itemize}
    \item Parity $\pi$: For the mean-field wave function $\ket{\Phi(\bm{q})}$ with nonzero moments of odd multiplicity, the space-reversal symmetry is broken.  The wave function with the correct parity $\pi (=\pm)$ is constructed as
    \begin{equation}
        \ket{\Psi^\pi(\bm{q})}= \dfrac{1}{2}(\unity + \pi \hat P) \ket{\Phi(\bm{q})}
        \equiv  \hat P^\pi \ket{\Phi(\bm{q})},
    \end{equation}
    which defines the parity projection operator~\cite{Egido:1991}
    \begin{equation}
        \hat P^\pi\equiv \dfrac{1}{2}(\unity + \pi \hat P).
    \end{equation}
    where $\hat{P}=\exp(i\pi{\hat{N}_-})$ is the standard parity (inversion) operator. Here $\hat{N}_-=\sum_k^\prime c^\dag_k c^{}_k$ is a restricted summation over all the states $k$ with negative parity.

    \item Particle number $N_0$:   The wave function with the correct particle number $N_0$ can be constructed as
    \begin{equation}
    \centering
        \ket{\Psi^{N_0}(\bm{q})}\equiv  \hat P^{N_0} \ket{\Phi(\bm{q})},
    \end{equation}
     which defines a particle-number projection operator
     \begin{equation}
         \hat P^{N_0}\equiv  \dfrac{1}{2\pi} \int d\varphi   e^{-i\varphi  N_0}  \hat S(\varphi),\quad \hat S(\varphi)\equiv e^{i\varphi  \hat N}
     \end{equation}
     with the gauge angle $\varphi_N\in [0, 2\pi]$. In this case, the rotated quasiparticle operator $\beta^{\dagger}_k(\bm{q}, g)$ becomes $\beta^{\dagger}_k(\bm{q}, \varphi)$, which can be determined with the Baker–Campbell–Hausdorff(BCH) formula
     \footnote{The  single-particle operators under the rotation transformation are given by
     \beqn
     \begin{array}{l}
    e^{i \varphi \hat{N}} c_{p} e^{-i \varphi \hat{N}}=c_{p} e^{-i \varphi}, \quad e^{i \varphi \hat{N}} c_{\bar{p}} e^{-i \varphi \hat{N}}=c_{\bar{p}} e^{-i \varphi}, \\
    e^{i \varphi \hat{N}} c_{p}^{\dagger} e^{-i \varphi \hat{N}}=c_{p}^{\dagger} e^{i \varphi}, \quad e^{i \varphi \hat{N}} c_{\bar{p}}^{\dagger} e^{-i \varphi \hat{N}}=c_{\bar{p}}^{\dagger} e^{i \varphi}
   \end{array}
   \eeqn}
     \beqn
     \beta^{\dagger}_k(\bm{q}, \varphi)
     &\equiv&\hat S(\varphi) \beta^{\dagger}_k(\bm{q}) \hat S^\dagger(\varphi)\nonumber\\
     &=&\sum_{p} \Bigg[V_{pk}(\bm{q}) \hat S(\varphi) c_p \hat S^\dagger(\varphi) + U_{pk}(\bm{q})\hat S(\varphi)c^\dagger_p\hat S^\dagger(\varphi)\Bigg]\nonumber\\
     &\equiv& \sum_{p} \Bigg[V_{pk}(\bm{q}, \varphi) c_p   + U_{pk}(\bm{q}, \varphi)c^\dagger_p\Bigg],
     \eeqn
    where the wave function of the rotated HFB state $\ket{\Phi(\bm{q}, \varphi)}$ is determined by $V_{pk}(\bm{q}, \varphi)=V_{pk}(\bm{q})e^{-i \varphi}$ and $U_{pk}(\bm{q}, \varphi)=U_{pk}(\bm{q})e^{i \varphi}$. In other words, the representation of the U(1) operator in (\ref{eq:rotated_UV}) is $D(g)=e^{i\varphi}$.

     \item Angular momentum $J$: The corresponding group is the special orthogonal group  in three dimensions, i.e. SO(3), with a continuous rotation $\hat R(\phi,\theta,\psi)=e^{i\phi \hat J_z} e^{i\theta \hat J_y} e^{i\psi \hat J_z} $ in three-dimensional Euclidean space, where $\hat J_i$ is the component along the $i$-th axis of the angular momentum $\bm{\hat J}$, being generators of the Lie algebra of the SO(3) group, and the three Euler angles $\Omega=(\phi,\theta,\psi)$. The wave function with angular momentum $J$, $z$-projection $M$, and intrinsic $z$-projection $K$ is constructed as
    \begin{equation}
    \centering
        \ket{\Psi^{J}_{MK}(\bm{q})}
        \equiv  \hat P^{J}_{MK} \ket{\Phi(\bm{q})},
    \end{equation}
     where the operator $\hat P^{J}_{MK}$ is introduced as
     \begin{eqnarray}
     \label{eq:AMP_Operator}
     \hat P^{J}_{MK}
     &\equiv&  \dfrac{2J+1}{8\pi^2} \int^{2\pi}_0 d\phi
     \int^{\pi}_0  \sin\theta d\theta \int^{2\pi}_0 d\psi
     D^{J\ast}_{MK}(\phi,\theta,\psi) \hat R(\phi,\theta,\psi),
     \end{eqnarray}
     where $D^{J\ast}_{MK}(\Omega)$ is the complex conjugate of Wigner D-matrix of dimension $2J + 1$ in the spherical basis with elements
     \begin{eqnarray}
         D^{J}_{MK}(\phi,\theta,\psi)
        &\equiv& \langle JM | \hat R(\phi,\theta,\psi)| JK \rangle
        =e^{iM\phi} d^J_{MK}(\theta)e^{i K\psi}.
     \end{eqnarray}
   In other words, $\hat{P}_{MK}^J$ is an operator that projects a state onto components with well-defined angular momentum $J$, $z$-projection $M$, and intrinsic-$z$-projection $K$.  Since $K$ is not a good quantum number for a triaxially-deformed nucleus, components with all possible values of $K$ contribute to the GCM state (through ``$K$ mixing''). Therefore, the wave function with angular momentum $J$ is constructed as
    \begin{equation}
      \ket{\Psi^{J}_{M}(\bm{q})}= \sum^{K=J}_{K=-J} g^J_K \ket{\Psi^{J}_{MK}(\bm{q})}.
    \end{equation}
\end{itemize}

Considering a generally deformed HFB state $\ket{\Phi(\bm{q})}$ for atomic nucleus ($N_0, Z_0$), its wave function (\ref{eq:PHFB4wf}) projected onto quantum numbers ($N_0, Z_0, J, \pi$) can be constructed as
\begin{eqnarray}
\label{eq:general_HFB_wf_projection}
 \ket{\Psi^{N_0Z_0J\pi}(\bm{q})}
 &=&\sum_K g^{J\pi}_K \ket{N_0Z_0,J\pi MK,\bm{q}},
 \end{eqnarray}
 where the symmetry-conserving basis reads
 \begin{eqnarray}
 \ket{N_0Z_0,J\pi MK,\bm{q}}
 &\equiv &\hat P^{N_0}\hat P^{Z_0} \hat P^J_{MK}\hat P^\pi \ket{\Phi(\bm{q})}.
\end{eqnarray}
The weight coefficient $g^{J\pi}_K$ is determined from the variational principle, leading to the following generalized eigenvalue equation
\begin{equation}
\label{eq:HWG_K-mixing}
\sum_{K'} \big[ \mathcal{H}^{J\pi}_{KK'}(\bm{q},\bm{q}) - E^J
\mathcal{N}^{J\pi}_{KK'}(\bm{q},\bm{q}) \big] g^{J\pi}_{K'} = 0 \,,
\end{equation}
where the Hamiltonian and norm kernels $\mathcal{H}$ and $\mathcal{N}$ are defined as
\bsub\begin{eqnarray}
\label{eq:hamker}
\mathcal{H}^{J\pi}_{KK'}(\bm{q},\bm{q}) &=& \bra{N_0Z_0,J\pi MK,\bm{q}} \hat{H}
\ket{N_0Z_0,J\pi MK',\bm{q}},\\
\label{eq:ovker}
\mathcal{N}^{J\pi}_{KK'}(\bm{q},\bm{q}) &=& \langle N_0Z_0,J\pi MK,\bm{q}
|N_0Z_0,J\pi MK',\bm{q}\rangle,
\end{eqnarray}
\esub
and they are computed explicitly with the following expression
\beqn
\label{eq:kernels}
\mathcal{O}^{J\pi}_{KK'}(\bm{q},\bm{q})
   &=&  \dfrac{1}{(2\pi)^2} \int \int d\varphi_n d\varphi_p  \dfrac{2J+1}{8\pi^2} \int^{2\pi}_0 d\phi
     \int^{\pi}_0 \sin\theta d\theta \int^{2\pi}_0 d\psi  \nonumber\\
     &\times&
     e^{-i\varphi_n  N_0}
     e^{-i\varphi_p  Z_0}
     D^{J\ast}_{MK}(\phi,\theta,\psi)
     \bra{\Phi(\bm{q})} \hat O \hat R(\phi,\theta,\psi)\hat S_n(\varphi_n)\hat S_p(\varphi_p) \hat P^{\pi}\ket{\Phi(\bm{q})},
\eeqn
where the operator $\hat O$ in the overlap is $\hat H$ and $1$, respectively.

The computation of kernels (\ref{eq:kernels}) requires multidimensional integrals over rotation angles ($\varphi_n, \varphi_p, \phi, \theta, \psi$), which are transformed into summations of the overlaps evaluated in all discretized angles. The number of overlaps is scaling as $N^2_\varphi N_\phi N_\theta N_\phi \sim 10^{6}$. An efficient way to carry out the overlap integrals is demanded. In most cases, the Fomenko expansion \cite{Fomenko:1970JPA} is used for the integration over gauge angles ($\varphi_n, \varphi_p$)
\beq
\hat{P}^{N_0}
=\frac{1}{L} \sum_{n=1}^{L} e^{i(\hat{N}-N_0) \varphi_{n}}, \quad \varphi_{n}=\frac{\pi}{L} n
\eeq
with $L$ points in the expansion. The trapezoidal rule is applied for the integration over the angles $(\phi, \psi)$, and Gaussian-Legendre quadrature is used for integration over $\theta$. Recently, Johnson et al.~\cite{Johnson:2018JPG,Johnson:2017PRC} proposed an alternative competing method to carry out the quadrature with linear algebra.
Note that the integrals over Euler angles can be simplified significantly if the axial symmetry is preserved in the HFB calculation for $\ket{\Phi(\bm{q})}$, in which case,  two of the three integrals over the Euler angles in the angular momentum projection operator (
\ref{eq:AMP_Operator}) can be carried out analytically and the computation cost is reduced by at least two orders of magnitude.

 Note that the solution of (\ref{eq:HWG_K-mixing}) only guarantees that the energy
 \beq
 E^{N_0Z_0J\pi}(\bm{q},\bm{q}) = \dfrac{\sum_{KK'}g^{J\pi \ast}_Kg^{J\pi}_{K'}\mathcal{H}^{J\pi}_{KK'}(\bm{q},\bm{q})}{\sum_{KK'}g^{J\pi}_Kg^{J\pi}_{K'}\mathcal{N}^{J\pi}_{KK'}(\bm{q},\bm{q})}
 \eeq
 is minimized with respect to the unknown weight $g^{J\pi}_K$, not to the HFB wave function $\ket{\Phi(\bm{q})}$, the variation of which according to Thouless theorem can be written as~\cite{Thouless:1960NPA,Ring:1980}
 \beqn
 \ket{\Phi(\bm{q}+\delta \bm{q}) }
 =\exp\Bigg(\frac{1}{2}
 \sum_{kk'} \mathbb{Z}^{(\bm{q})}_{kk'} \beta^\dagger_{k}  \beta^\dagger_{k'}\Bigg)|\bm{q}\rangle,
 \eeqn
 where the unknown parameter is the skew-symmetric matrix $\mathbb{Z}^{(\bm{q})}$. The symmetry projected-energy gradient with respect to $\mathbb{Z}^{(\bm{q})}_{kk'}$ is defined as a matrix $G_{kk'}(\bm{q})$, which has the following form~\cite{Schmid:1984PRC}
 \beqn
 \begin{aligned}
G^{N_0Z_0J\pi}_{kk'}(\bm{q})
\equiv \frac{\partial E^{N_0Z_0J\pi} }{\partial \mathbb{Z}^{(\bm{q})\ast}_{kk'}}
=\sum_{K K^{\prime}}g^{J\pi \ast}_Kg^{J\pi}_{K'}
 \bra{\Phi(\bm{q})}  \beta_{k'}(\bm{q})\beta_k(\bm{q})
  \left(H-E^{N_0Z_0J\pi}\right)  \hat P^{N_0}\hat P^{Z_0} \hat P_{KK^{\prime} }^{J}\hat P^\pi \ket{\Phi(\bm{q})}.
\end{aligned}
\eeqn
In the case of  variation after particle-number projection (PNVAP), the matrix element $G_{kk'}$ is simplified as
 \beqn
 \label{eq:VAP_H20}
 \begin{aligned}
 G^{N_0Z_0}_{kk'}(\bm{q})
 =
 \bra{\Phi(\bm{q})}
 & \beta_{k'}(\bm{q})\beta_k(\bm{q})
  \left(H-E^{N_0Z_0}\right)  \hat P^{N_0}\hat P^{Z_0}  \ket{\Phi(\bm{q})}.
\end{aligned}
\eeqn
The variational principle requires the matrix elements $G^{N_0Z_0}_{kk'}(\bm{q})$ to vanish. It can be achieved with the help of the gradient descent method that has been employed to solve the standard HFB equation \cite{Mang:1976,Ring:1980}.  More details on the implementation of the  gradient descent method  into the PNVAP are introduced in Refs.~\cite{Anguiano:2001NPA,Bally:2021EPJA}. Alternatively, as demonstrated by Sheikh and Ring \cite{Sheikh:2000}, an equation of HFB-type can be derived from the variation of the symmetry-projected energy expectation (\ref{eq:PHFB4E}) with respect to the densities $\rho$ and $\kappa$, where the corresponding pairing field $\Delta$ and the mean-field $\Gamma$ are replaced by quantities depending on the projected quantum numbers. 

In Sec.~\ref{subsubsec:Selection of basis states} we will discuss that the VAP method provides a way to determine the relevant degrees of freedom for the energy of the state of interest. For simplicity, in most of the applications, symmetry-projected HFB calculations were carried out in the framework of projection after variation (PAV), where first, an HFB wave function $\ket{\Phi}$ is determined by the variation of the unprojected energy, and afterwards a projection is carried out. In the following, we will call this method PHFB. Note that VAP is  rarely applied, mostly for the particle-number projection. However, in many cases, the projected wave function (\ref{eq:general_HFB_wf_projection}) in either VAP or PHFB is not yet a good approximation of the true wave function. To improve the approximation, one needs to extend in such cases the model space by including either multi-collectively-correlated configurations within the framework of  symmetry-projected generator coordinate method~\cite{Ring:1980} or non-collectively-correlated configurations as in the projected shell model (PSM)~\cite{Hara:1995}, sometimes also in a stochastic way with a systematically improved truncation scheme~\cite{Honma:1995}.

\subsection{Configuration mixing with symmetry-projected  nonorthogonal basis}

There are many excellent review papers on the BMF studies of performing configuration mixing with symmetry-projected  nonorthogonal basis, see for instance, Refs.~\cite{Schmid:2004PPNP,Niksic:2011PPNP,Egido:2016PS,Robledo:2018JPG,Sheikh2021_JPG48-123001}. Here, we will provide a schematic review of the recent developments in both techniques and algorithms for BMF approaches with an emphasis on the recent achievements towards a precise description of nuclear $0\nu\beta\beta$ decay.

\subsubsection{The symmetry-projected generator coordinate method (PGCM)}

 The GCM is based on the assumption that the trial wave function  $\ket{\Psi_\alpha} $ of a nuclear state labeled with $\alpha$ is written as a continuous superposition of the basis functions $\ket{\Phi (\bm{q})}$, with 
 a continuous parameter set $\bm{q}$,
 \begin{equation}
 \label{GCM-state}
\ket{\Psi_\alpha} = \int d\bm{q}  f^\alpha_{\bm{q}}\ket{\Phi (\bm{q})},
\end{equation}
where the parameters $\bm{q}$ denote a set of collective variables--the so-called generator coordinates and they do not appear in the nuclear state wave function $\ket{\Psi_\alpha}$.  The weight $f(q)$ also called  "generator function" is folded into the basis functions to produce the wave function $\ket{\Psi_\alpha}$.  The internal degrees of freedom are supposed to be taken into account by the functions $\ket{\Phi (\bm{q})}$. In the Hill-Wheeler theory~\cite{Hill:1953}, it is assumed that the set $\ket{\Phi (\bm{q})}$ is predetermined, often by constraint HFB calculations based on the operator $\hat{H}-\bm{q}\hat{\bm{Q}}$~\cite{Ring:1980}. Only the weight function $ f(\bm{q})$ is unknown. As long as we are interested in the low-lying excited states, this approximation can be justified by arguing that the collective motions are much slower than the internal motions.

 \begin{figure}[tb]
\centering
\includegraphics[width=10cm]{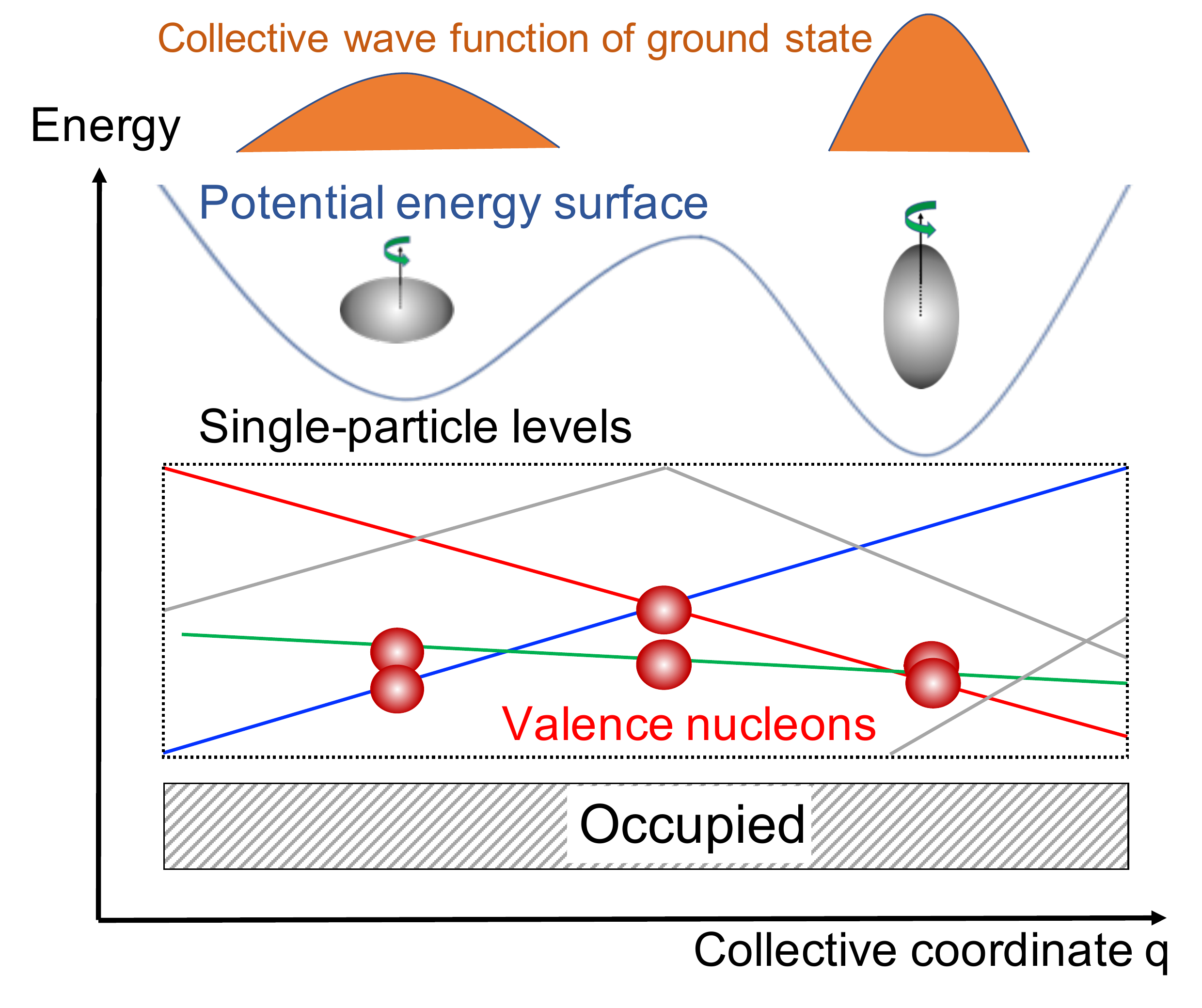}
\caption{\label{fig:GCM_Cartoon_collective} A schematic picture for illustrating the configuration-mixing calculation in the framework of PGCM based on non-orthogonal collective states labeled with the collective coordinate $\bm{q}$ (such as nuclear quadrupole deformation). }
\end{figure}

The GCM is a very flexible tool so that the basis functions $\ket{\Phi (\bm{q})}$ can be chosen to be any many-body wave functions, and the generator coordinate $\bm{q}$ can be a complex number. In many realistic applications to nuclear physics, the basis functions are usually chosen as a set of mean-field states, which are generated by collective coordinates $\bm{q}$ usually associated with mass multipole moments that characterize nuclear shapes \cite{Rodriguez:2010,Song:2014,Yao:2015}, nuclear size, or parameters characterizing nuclear clustering structure \cite{Zhou:2019} or pairing amplitudes \cite{Vaquero:2013,Hinohara:2014} depending on whether or not they are relevant for the physics of interest. The choice of $\bm{q}$ defines a model space of many-body configurations whose dimension is usually much smaller than that of full configuration interaction (CI) calculations~\cite{Caurier:2005RMP}, since many types of (collective) correlations are already built into the basis functions and many high-lying configurations can be neglected. In particular, the GCM turns out to be a powerful method for nuclei with strong shape mixing (the  wave function of one nuclear state, such as the ground state, is an admixture of prolate and oblate deformed shapes as illustrated in Fig.~\ref{fig:GCM_Cartoon_collective}) or shape coexistence~\cite{Heyde:2011RMP}. The latter is a nuclear phenomenon that the low-lying states of an atomic nucleus consist of two or more states which have well-defined and distinct properties and can be interpreted in terms of different intrinsic shapes. 

 In practical applications, the generator coordinate $\bm{q}$ in (\ref{GCM-state}) is discretized. The continuous integral over $\bm{q}$ becomes a sum of discretized states.  In the PGCM, the trial wave function is constructed as
 \begin{equation}
 \label{eq:BMF_GCM_wf}
 \ket{\Psi^{JNZ}_\alpha}= \sum^J_{K=-J}\sum^{N_q}_{n=1} f^{J\alpha}_{K,\bm{q}_n} \ket{JMK,\bm{q}_n} \,,
\end{equation}
where  $\ket{JMK,\bm{q}_n}$ are symmetry-projected quasiparticle vacua,
\begin{equation}
\ket{JMK,\bm{q}_n}=\hat{P}^J_{MK} \hat{P}^N \hat{P}^Z \ket{\Phi(\bm{q}_n)}.
\end{equation}
 The basis functions $\ket{\Phi(\bm{q})}$ labeled by the generator-coordinate parameters $\bm{q}(=q_1, q_2, \cdots, q_{N_q}$) are a set of quasiparticle vacua determined from HFB or (VAP-HFB) calculations with  constraints on the quantities $\bm{q}$.  The projection operators produce basis states that are not orthonormal to each other. Variation of the energy with respect to the weight function $f_{K,q}^{J\alpha}$ leads to the Hill-Wheeler-Griffin (HWG) equations \cite{Hill:1953,Griffin:1957},
\begin{equation}
\sum_{K',\bm{q}'} \big[ \mathcal{H}^J_{KK'}(\bm{q},\bm{q}') - E^J_\alpha
\mathcal{N}^J_{KK'}(\bm{q},\bm{q}') \big] f^{J\alpha}_{K',\bm{q}'} = 0 \,,
\end{equation}
where the Hamiltonian and norm kernels $\mathcal{H}$ and $\mathcal{N}$ are given by the expressions
\begin{eqnarray}
\mathcal{H}^J_{KK'}(\bm{q},\bm{q}') &=& \bra{JMK,\bm{q}} \hat{H}
\ket{JMK',\bm{q}'}\\ \label{eq:hamker_qq}
\mathcal{N}^J_{KK'}(\bm{q},\bm{q}') &=& \langle JMK,\bm{q}
|JMK',\bm{q}'\rangle\,,
\label{eq:ovker_qq}
\end{eqnarray}
and $E^J_\alpha$ is the energy of the state with angular momentum $J$. The HWG equation is solved in the standard way \cite{Ring:1980}, by diagonalizing the norm kernel to obtain a basis of ``natural states'' and then diagonalizing the Hamiltonian $H$ in that basis.  Because of the overcompleteness of the GCM-basis, the second diagonalization can be numerically unstable. This problem is taken care of by truncating the natural basis to include only states with norm eigenvalues larger than a reasonable value.

 The contribution of each collective state $\ket{\Phi(\bm{q})}$ to the GCM state $\ket{\Psi^{JNZ}}$ can be indicated from the distribution of the squared collective wave function $|g_{\alpha}^{J}|^2$, as schematically shown in Fig.~\ref{fig:GCM_Cartoon_collective}, which is defined via the weight function $f^{J}_{K,\bm{q}}$ by the following relation~\cite{Toledo:1977},
\beq
\label{eq:collective_wfs}
g_{\alpha}^{J}(K,\bm{q})
=\sum_{K',\bm{q}'}\Bigg[\mathcal{N}^J_{KK'}(\bm{q},\bm{q}')\Bigg]^{1/2} f^{J\alpha}_{K',\bm{q}'}.
\eeq

\subsubsection{Configuration mixing with noncollective basis states}

Another way to extend the model space is through the inclusion of noncollective states on top of the collectively correlated HFB states $\ket{\Phi(\bm{q)}}$. This extension  can be performed in a few different ways.

\begin{itemize}
    \item Including quasiparticle excitations. It follows the basic idea of the shell model in which the Hamiltonian is formulated in a set of orthonormal Slater determinants related via particle-hole excitations. In contrast, a set of non-orthonormal multi-quasiparticle states on top of the collectively correlated HFB state
    \begin{equation}
        \ket{\Phi_\kappa(\bm{q})}
        =\Bigg\{\ket{\Phi(\bm{q})},\quad \beta^\dagger_k\beta^\dagger_l\ket{\Phi(\bm{q})}, \quad \beta^\dagger_k\beta^\dagger_l\beta^\dagger_m\beta^\dagger_n\ket{\Phi(\bm{q})}, \ldots\Bigg\},
    \end{equation}
    are mixed for even-mass nuclei, or
        \begin{equation}
        \ket{\Phi_\kappa(\bm{q})}
        =\Bigg\{\beta^\dagger_k\ket{\Phi(\bm{q})}, \quad \beta^\dagger_k\beta^\dagger_l\beta^\dagger_m\ket{\Phi(\bm{q})}, \ldots\Bigg\},
    \end{equation}
    for odd-mass nuclei, where $\beta^\dagger_{k,l,\ldots}$ is a set of quasiparticle creation operators.  The basis functions constructed in this way for a fixed collective coordinate $\bm{q}$ was used in the projected shell model (PSM) \cite{Hara:1995} and configuration-interaction projected density functional theory (CI-PDFT) \cite{Zhao:2016PRC}. This type of basis has also been implemented in a  symmetry-projected GCM calculation with a pairing-plus-quadrupole residual interaction allowing for mixing of different $\bm{q}$~\cite{Chen:2017PRC}.  For the sake of simplicity, the above basis of quasiparticle  excitations is usually truncated up to two-quasiparticle states. Even with this approximation, the GCM calculation based on a general one-plus-two-body-interaction Hamiltonian is still a computational challenge. To select out the two quasiparticle states relevant for nuclear low-lying states, Jiao  et al.~\cite{Jiao:2019PRC} extended the model space of GCM  by the inclusion of collective superpositions of  two-quasiparticle excitations  based on the Thouless theorem
     \beqn
     \label{eq:QTDA_wf}
      \ket{\Phi_\kappa(\bm{q})}
     &=&\exp \left\{\frac{1}{2} \sum_{kk^{\prime}} Z_{kk^{\prime}}^{\kappa} \beta_{k}^{\dagger}(0) \beta_{k^{\prime}}^{\dagger}(0)\right\}\left|\Phi(\bm{q})\right\rangle.
  \eeqn
  The coefficients $Z_{kk^{\prime}}^{\kappa}$ are determined by quasiparticle  Tamm-Dancoff approximation:
   \beq
   \sum_{ll^{\prime}} A_{ kk^{\prime},ll^{\prime}} Z_{ll^{\prime}}^{r}=E_{\kappa}  Z_{kk^{\prime}}^{\kappa}
   \eeq
   where  the matrix elements $A_{kk',ll'}$  of the Hamiltonian in a basis of two quasiparticle excited states are defined as
   \beq
    A_{k k^{\prime} l l^{\prime}}
    =\left\langle\Phi(\bm{q})\left|\left[\beta_{k^{\prime}} \beta_{k},\left[H, \beta_{l}^{+} \beta_{l^{\prime}}^{+}\right]\right]\right| \Phi(\bm{q})\right\rangle.
   \eeq
    With a shell-model Hamiltonian, it has been found that this extension of the model space improves the PGCM description of the NME of $0\nu\beta\beta$ decay \cite{Jiao:2019PRC}.
    
   In the framework of PSM,  Wang et al. have successfully extended the configuration space to include higher-order quasiparticle states in order to describe the back-bending phenomenon in high-spin states  \cite{Wang:2014PRC} and the Gamow-Teller transitions in electron capture and $\beta$ decays \cite{Tan:2020PLB,Wang:2021PRL}. This type of calculations becomes feasible in the PSM thanks to the use of the Pfaffian algorithm for computing the overlap of two general HFB states, first introduced into nuclear physics by Robledo \cite{Robledo:2009PRC}.

     In the DFT-rooted no-core configuration-interaction (DFT-NCCI) model \cite{Satula:2016PRC}, nuclear wave function is constructed as an admixture of  a set of low-lying  deformed Slater determinants projected onto the correct isospin and angular momentum. The  Slater determinants are (multi)particle-(multi)hole excitations on top of the HF state determined from a self-consistent Skyrme-Hartree-Fock calculation.  Note that the model space of DFT-NCCI is very large even in the case that the configurations are restricted to only 1p-1h excitations. The DFT-NCCI has been successfully applied to study single-$\beta$ decay of a set of light nuclei \cite{Konieczka:2016PRC,Konieczka:2018PRC}. There, the convergence behavior for the observables of interest is investigated with respect to the number of 1p-1h configurations selected by energy ordering and symmetry requirements.

\item Including multicranked configurations. In the 1970s-1980s, three-dimensional angular momentum projection has already been successfully carried out to restore the angular momentum of cranking model wave functions in Refs. \cite{Bengtsson:1978ZPA,Islam:1979NPA,Baye:1984PRC}, even though only one intrinsic state with a definite cranking frequency was used. It was recommended later in Ref. \cite{Zdunczuk:PRC2007} that the state with angular momentum $J^c$ is better to be projected from the cranked state with the frequency that gives $\bra{\Phi(\bm{q}, \omega^{\rm rot})} \hat  J_y\ket{\Phi(\bm{q}, \omega^{\rm rot})}=J^c$. Here the quantity $\omega^{\rm rot}$ is the rotational (or cranking) frequency and $\hat J_y$ is the angular momentum operator around the rotation-axis (here chosen to be the y-axis). This can be understood by the fact that the cranking model wave function for angular momentum $J$ provides an approximate solution of the VAP method for this $J$~\cite{Beck:1970,Hara:1982NPA}. This scheme also allows for correcting the cranking moment of inertia at low spins. However, it has been pointed out in Ref. \cite{Shimada:2015PTEP} that this procedure is too complicated in practical applications because the angular momentum projection needs to be performed for each spin state from an intrinsic state with different cranking frequency. A general way to proceed is mixing multicranked states at different frequencies $\omega^{\rm rot}_i$, as originally proposed by Peierls and Thouless in Ref. \cite{Peierls:1962NP}, where the velocity has been used as a generator coordinate. The basis functions in this case are a set of cranked states
   \begin{equation}
        \ket{\Phi_\kappa(\bm{q})}
        =\Bigg\{\ket{\Phi(\bm{q}_1, \omega^{\rm rot}_1)}, \quad\ket{\Phi(\bm{q}_2, \omega^{\rm rot}_2)},\quad\cdots\Bigg\},
    \end{equation}
where each cranked state  $\ket{\Phi(\bm{q}_i, \omega^{\rm rot}_i)}$ is obtained from a constrained HFB calculation with a cranking term, namely the Hamiltonian transformed into the uniformly rotating frame (the so-called Routhian),
\begin{equation}
\label{eq:Routhian}
    H'_i = H -  \omega^{\rm rot}_i \hat J_y,
\end{equation}
is considered instead of the original Hamiltonian $H$.  This procedure has been used in recent studies for high-spin physics \cite{Shimada:2015PTEP,Shimada:2016PRC,Ushitani:2019PRC,Shimizu:2021PRC}. Alternatively, one can include a set of cranked states $\ket{\Phi(\bm{q}, J^c_i)}$ which are obtained from the constraint calculation based on the Routhian (\ref{eq:Routhian}) with the Lagrangian multiplier  $\omega^{J^c}_i$ ensuring the condition  $\bra{\Phi(\bm{q}_i, J^c_i)} \hat J_y\ket{\Phi(\bm{q}_i, J^c_i)}=\sqrt{J^c_i(J^c_i+1)}$. This procedure was used in the PGCM for nuclear low-lying states based on the Gogny force \cite{Borrajo:2015,Egido:2016PRL}. It turns out that adding  time-reversal symmetry breaking states with $J^c\neq 0$ squeezes notably the spectra, providing moments of inertia very close to the experimental ones. It is the result of the procedure that the excited states with a larger value of $J$ gain more energies than the ground state (with a lower value of $J$ which is zero for even-even nuclei) by including the $J^c\neq 0$ states.

\end{itemize}


\subsubsection{Selection of relevant basis states}
\label{subsubsec:Selection of basis states}

 The number of basis states $\ket{\Phi_\kappa(\bm{q})}$ grows quickly with the number of collective degrees of freedom  $\bm{q}=\{q_1, q_2, \cdots, q_{N_q}\}$, and in particular with that of noncollective ones. As a result, the computational cost increases significantly. Notice that some of these basis states, however, may closely resemble others or have little contribution to low-lying wave functions, and can therefore be omitted. In other words, the actual number of required basis states, if they are well selected, can be much smaller. The calculation with these selected basis states can be much cheaper with respect to computer resources.  Finding a proper way to select out the relevant basis states becomes important.  There are several algorithms that have been  implemented to select the relevant basis states in different BMF approaches:

 \begin{itemize}

    \item The VAP algorithm with spin-dependent basis functions. The advantage of using VAP is the possibility of including $mp$-$mh$ excitations in a rather cheap way. Schmid et al. proposed an angular momentum projected particle-hole theory on top of the HF solution (AMP-PH)  for excited states of deformed nuclei~\cite{Schmid:1976ZPA}, where angular momentum projection was carried out before the variation. In the practical application for $\nuclide[20]{Ne}$, the particle-hole configurations were truncated up to $1p$-$1h$ excitations. Note that even with this truncation, $mp$-$mh$ excitations are partially included through the VAP as both the HF-field and the $1p$-$1h$ mixing are spin-dependent.  This AMP-PH model was later extended to incorporate pairing correlations. The extended model was named MONSTER  (model for handling large numbers of number- and spin-projected two quasiparticle excitations with realistic interactions and model spaces) \cite{Schmid:1984PRC}, which approximates the nuclear wave functions by linear combinations of the angular momentum and number projected HFB vacuum and the corresponding equally spin and number projected two quasiparticle excitation. There are two versions of MONSTER approaches: MONSTER(HFB) and MONSTER(VAMPIR). For the former, the HFB wave function  is obtained before projection and is fixed for all the spin states. In contrast, the MONSTER(VAMPIR) approach belongs to the VAP class, where the HFB wave function is optimized for each projected state and is thus spin-dependent. More details have been introduced in the review papers~\cite{Schmid:1987RPP,Schmid:2004PPNP}.

    Later on, the VAP algorithm has been extensively employed in the BMF calculations based on antisymmetrized molecular dynamics (AMD) \cite{Kanada-Enyo:1998PRL},  HF \cite{Ohta:2004PRC,Gao:2015PRC,Gao:2021PLB}, and fermionic molecular dynamics (FMD)~\cite{Chernykh:2007PRL} or HFB \cite{Shimizu:2021PRC_VAP} solutions. In the VAP-AMD calculation~\cite{Kanada-Enyo:1998PRL}, the basis function used to construct the wave function of nuclear ground state is optimized by minimizing the energy of the angular momentum and parity projected state. The basis functions for the excited states are obtained using the orthogonality condition. In the VAP-HF by Ohta et al. \cite{Ohta:2004PRC},  a self-consistent HF calculation with variation after parity projection was carried out for low-lying states of light nuclei with a Skyrme EDF.  In the VAP-HF by Gao et al.~\cite{Gao:2015PRC,Gao:2021PLB}, a set of basis functions are optimized  by minimizing the sum of the energies of nuclear low-lying states formed by the basis functions. The optimization is guaranteed by the Cauchy's interlacing theorem or the Poincar\'e separation theorem~\cite{Bellman:1997Book,Magnus:2019Book}.
    A proof-of-principle calculation was carried out with a shell-model interaction and it turns out that the VAP can simplify significantly the BMF calculations of high-spin states~\cite{Gao:2021PLB}.    In VAP-FMD~\cite{Chernykh:2007PRL}, the nuclear wave function is spanned by Slater determinants built on single-particle wave packets of Gaussian shape. The single-particle wave functions of determinants are obtained in a variation after the angular momentum projection procedure.  In the VAP-HFB or also called quasiparticle vacua shell model (QVSM) by  Shimizu et al.~\cite{Shimizu:2021PRC_VAP},  the nuclear wave function is a superposition of the angular-momentum, parity, and particle-number projected quasiparticle vacua
    \beq
    \label{eq:QVSM}
    \ket{\Psi^{JNZ}_{N_b}}
    =\sum_{n=1}^{N_b} \sum_{K=-J}^{J} f_{JKn}^{N_b} P_{M K}^{J} \hat P^\pi P^{Z} P^{N}\ket{\Phi_n}.
    \eeq
     The HFB wave functions $\ket{\Phi_n}$ are determined at every $N_b$ so that the energy expectation value after the projections and superposition, $E^{JNZ}_{N_b} =\bra{\Psi^{JNZ}_{N_b}}H\ket{\Psi^{JNZ}_{N_b}}$ is minimized.  Shimizu et al.~\cite{Shimizu:2021PRC_VAP} applied both the Monte-Carlo shell-model (MCSM) and VAP-HFB to the NME of $0\nu\beta\beta$ decay for \nuclide[76]{Ge} and \nuclide[150]{Nd}. It was shown that the NME in the MCSM converges very slowly against the number of the basis states. In contrast, even though the VAP-HFB is more complex than the MCSM,  the energy converges faster with respect to $N_b$. The NME for $\nuclide[76]{Ge}$ also converges rapidly to the exact solution~\cite{Shimizu:2021PRC_VAP}.

     \item Stochastic sampling of Slater determinants. This algorithm was used in the MCSM calculation \cite{Otsuka:2001PPNP}, and a VAP-HF calculation \cite{Shinohara:2006PRC}. The nuclear wave function is still a linear combination of angular-momentum and parity projected HF states
    \beq
    \label{eq:MCSM}
    \ket{\Psi^{JNZ}_{N_{b}}}=\sum_{n=1}^{N_{b}} \sum_{K=-J}^{J} f_{JKn}^{\left(N_{b}\right)} \hat P_{M K}^{J} \hat P^\pi \ket{\Phi_n}
    \eeq
     but with the HF wave functions $\ket{\Phi_n}$ determined by stochastic sampling. The first HF state starts from the one with the lowest energy. The $m$-th ($m\ge2$) HF state is selected based on the criteria that this state provides the most amount of energy contribution to the ground state among a group  of HF states (let us say the number is $M$ which is also referred to as the number of iterations)  generated randomly with Monte-Carlo techniques. Once the $m$-th HF state is selected, the $(m+1)$-th state will be selected in the same way, i.e., selecting the basis function among another group of randomly generated basis functions. This procedure is repeated until the energy is not changing significantly when one more HF state is added. In the VAP-HF calculation by Shinohara  et al.~\cite{Shinohara:2006PRC},  the basis functions are selected as the intermediate configurations during the imaginary-time iteration when the rate of energy decrease becomes relatively slow.   Very recently, Ichikawa and Itagaki~\cite{Ichikawa:2021} proposed a replica-exchange Monte-Carlo (RXMC) method to sample important Slater determinants obtained by the Markov-chain Monte Carlo (MCMC) method  following the Boltzmann distribution on  multi-dimensional energy surface  under a given model space. This method turns out to be successful in the study of  clustering structures in the excited states of $\nuclide[12]{C}$.

    \item The energy-transition-orthogonality procedure  (ENTROP)  \cite{Romero:2021PRC}. It is worth mentioning that most algorithms including the previously discussed two are validated based on energy criteria. It is not guaranteed that other observables like transition strength are converging  at the same pace. In contrast, the ENTROP algorithm was proposed recently for computing the NME of $0\nu\beta\beta$ decay with GCM  by considering the energy, transition matrix element, and orthogonality altogether~\cite{Romero:2021PRC}. It is based on the following three observations: (1) Basis states with lower expectation values for the Hamiltonian are in general more important than those with higher expectation values.  (2) The largest contributions to NMEs often come from transitions between basis states of two different nuclei with the same values for the collective coordinates $\bm{q}$.  (3) Basis states that can nearly be represented as a linear combination of states in the selected subset do not need to be included themselves in the subset.  A proof-of-principle study with a shell-model interaction for $\nuclide[76]{Ge}$ and $\nuclide[76]{Se}$ showed that the NME of $0\nu\beta\beta$ decay converges quickly, reducing significantly the number of basis states needed for an accurate calculation~\cite{Romero:2021PRC}. There, this algorithm was also implemented into the {\em ab initio}  in-medium GCM calculation starting from a chiral nuclear force.

 \end{itemize}

\subsubsection{Overlap between basis states}
\label{subsubsec:overlap}

One of the most important ingredients in the PGCM calculations is the norm overlap $n(\bm{q};\bm{q}',g)$ which has the following general form
\begin{equation}
  n(\bm{q};\bm{q}',g)  \equiv \bra{\Phi(\bm{q})}\hat R(g)\ket{\Phi(\bm{q'})},
\end{equation}
 which, if the two wave functions are not orthogonal to each other, can be computed with the formula proposed by Onishi and Yoshida \cite{Onishi:1966NP, Ring:1980},
 \begin{equation}
 \label{eq:Onishi}
     n(\bm{q};\bm{q}',g)
     =\pm \sqrt{\det\mathbb{U}(\bm{q};\bm{q}',g)},
 \end{equation}
 where the matrix $\mathbb{U}$ is defined in Eq.(\ref{eq:Thouless_U}).
 It is well known that the Onishi formula (\ref{eq:Onishi}) suffers a sign problem. Hara, Hayashi and Ring were the first to use of the continuity and differentiability of the norm overlap to determine the sign~\cite{Hara:1982NPA}. However, this method becomes inefficient in the case of cranked HFB wave functions \cite{Oi:2004PLB}. A set of very dense mesh points is needed to identify correctly the phase, increasing significantly the computational cost.  It was pointed out later by Neergard and W\"{u}st  that the sign problem can be avoided by rewriting the norm overlap into the following form~\cite{Neergard:1983NPA},
 \begin{equation}
     n(\bm{q};\bm{q}',g)
     =\prod^{M/2}_{k>0} \left(u^{(\bm{q})}_k u^{(\bm{q}')}_{k}\right)  \prod^{M/2}_{l>0} (1+c_l),
 \end{equation}
 where $u^{(\bm{q})}_k$, $u^{(\bm{q}')}_{k}$ are the Bogoliubov-Valatin transformations coefficients (\ref{eq:uv-coefficients}) of the HFB wave functions $\ket{\Phi(\bm{q})}$ and  $\ket{\Phi(\bm{q}', g)}$. The  product $\prod^{M/2}_{l>0}$ runs over the pairwise degenerate eigenvalues $c_l$ of the $M\times M$ matrix $\mathbb{C}$,
 \begin{equation}
     \mathbb{C}(\bm{q};\bm{q}',g) = Z^{(\bm{q}')}_g Z^{(\bm{q})\dagger},
 \end{equation}
 where $Z^{(\bm{q})}=V(\bm{q})U^{-1}(\bm{q})$ and $Z^{(\bm{q}')}_g=V(\bm{q}',g )U^{-1}(\bm{q}',g)$ defined in Eq.(\ref{eq:rotated_UV}).
 The use of existing pairwise generated eigenvalues was also made recently by Mizusaki et al. \cite{Mizusaki:2018PLB} to trace the origin of the sign problem in the Onishi formula.
 Robledo made the first connection of the norm overlap  to the Pfaffian of a skew-symmetric matrix~\cite{Robledo:2009PRC} by expressing the HFB state with Grassmann numbers. This  work inspires a great interest in the derivation of varieties of expressions for the overlap of two general quasiparticle vacua \cite{Robledo:2011PRC,Bertsch:2011PRL,Gao:2014PLB,Carlsson:2021PRL}, and the formulas for the matrix elements $\bra{\Phi_\kappa(\bm{q})} \hat O\ket{\Phi_{\kappa'}(\bm{q}')}$ of many-body operators $\hat O$ between multi-quasiparticle excitation configurations for either even or odd-mass nuclei \cite{Oi:2011PLB,Avez:2012PRC,Hu:2014PLB}.  The exponential growth of the number of combinations of the contractions with the number of quasiparticle operators in the multi-quasiparticle excitation configurations can be handled  with the Pfaffian method. Here we overview some typical formulas which were proposed  for the norm overlap based on the Pfaffian method.

 \begin{itemize}
 \item The original formula proposed by Robledo~\cite{Robledo:2009PRC}
 \begin{equation}
 \label{eq:Robledo}
     n(\bm{q};\bm{q}',g)  =(-1)^{M(M+1)/2} \cdot {\rm Pf}(\mathbb{X}_g),
  \end{equation}
where the $\mathbb{X}_g$ is a $2M\times 2M$ matrix defined as
 \beqn
 \label{X-R}
 \mathbb{X}_g = \begin{pmatrix}
 Z^{(\bm{q}')}_g & -\mathbb{I} \\
 \mathbb{I}  & -Z^{(\bm{q})\ast}
 \end{pmatrix}.
 \eeqn
 Here $\mathbb{I}$ is a unity matrix $\mathbb{I}_{ij}=\delta_{ij}$.
 In the case of particle-number projection, the rotation operator $\hat R(\varphi_N)=e^{i\varphi_N \hat N}$, then  $Z^{(\bm{q}')}_{\varphi_N}$ in the canonical basis has a similar form as $Z^{(\bm{q}')}$, but replacing $u^{(\bm{q})}_k$ with $u^{(\bm{q}')}_ke^{i\varphi_N}$ and $v^{(\bm{q}')}_k$ with $v^{(\bm{q}')}_ke^{-i\varphi_N}$, respectively. 
 
 Later on, Mizusaki and Oi derived an equivalent formula~\cite{Mizusaki:2012Norm}
 \begin{equation}
 \label{eq:Mizisaki}
     n(\bm{q};\bm{q}',g)  =(-1)^{M(M+2)/2} \cdot {\rm Pf}(\tilde{\mathbb{X}}_g),
  \end{equation}
 for the norm overlap in the same way but with a different convention in the ordering of the Grassmann numbers. The matrix $\tilde{\mathbb{X}}_g$ is defined as
 \beqn
\tilde{\mathbb{X}}_g=\left(\begin{array}{cc}
Z^{(\bm{q}')}_g & -\tilde{\mathbb{I}}\\
\tilde{\mathbb{I}} & -\tilde{\mathbb{I}}Z^{(\bm{q})\ast} \tilde{\mathbb{I}}
\end{array}\right),
\eeqn
where $\tilde{\mathbb{I}}$ is defined as $\tilde{\mathbb{I}}_{i j}=\delta_{i+j, M+1}$.

     \item The formula by Bertsch and Robledo \cite{Bertsch:2011PRL}, who derived the norm overlap by writing the two HFB states in terms of quasiparticle operators,
     \beqn
     \label{eq:norm_Bertsch}
     n(\bm{q};\bm{q}',g)
     &=& \frac{\operatorname{det} C^{*}_{(\bm{q})} \operatorname{det} C_{(\bm{q}')}}{\prod^{M/2}_{k,k^{\prime}} v^{(\bm{q})}_{k} v_{k^{\prime}}^{(\bm{q}')}}
     \bra{0}\beta_{M}^\dagger(\bm{q}) \ldots \beta_{1}^\dagger(\bm{q}) \beta_{1}^\dagger(\bm{q}^\prime,g) \ldots \beta_{M}^\dagger(\bm{q}^\prime,g)\ket{0}
     \nonumber\\
     &=&
     (-1)^{M/2} \frac{\operatorname{det} C^{*}_{(\bm{q})}  \operatorname{det} C_{(\bm{q}')}}{\prod^{M/2}_{k,k^{\prime}} v^{(\bm{q})}_{k} v_{k^{\prime}}^{(\bm{q}')}} 
     \operatorname{pf}
     \left(\begin{array}{cc} V^{T}(\bm{q}) U(\bm{q}) & V^{T}(\bm{q}) V^{\ast}(\bm{q}',g) \\
     -V^{\dagger}(\bm{q}',g) V(\bm{q}) & U^{\dagger}(\bm{q}',g) V^{*}(\bm{q}', g)
     \end{array}\right)_{2M\times 2M},
     \eeqn
     where the matrix $C$ is defined in (\ref{eq:Bloch-Messiah}).
     As illustrated in Ref. \cite{Bertsch:2011PRL}, the norm overlap of two HFB states with odd number parity can be derived similarly.

 \end{itemize}

Note that special treatment is required while using the formula (\ref{eq:Robledo}) when one of the two states is an HF state~\cite{Robledo:2011PRC}. One may also encounter some numerical problems while using the formula (\ref{eq:norm_Bertsch}) when many of $v_{k}$s are small numbers. It can cause either a singularity problem in the denominator or a tiny number beyond the double-precision data type. These numerical problems can be avoided by evaluating the norm overlap using only occupied single-particle states with $v^2_k$ larger than a cutoff value \cite{Bonche:1990NPA,Valor:1999NPA,Yao:2009PRC}, or by introducing an {\em ad hoc} tiny numerical parameter $\epsilon$ to replace the small values of $v_k$ \cite{Gao:2014PLB}, or by artificially adding a tiny value to the pairing field $\Delta$ to prevent pairing collapse, or by scaling all the $v_k$ with a factor so that their products are within the precise of double-precision data type \cite{Yao:2020PRL}. Recently, Carlsson and Rotureau reshaped the formula (\ref{eq:norm_Bertsch}) by moving the product of $v_k$s in the denominator into the Pfaffian of the matrix \cite{Carlsson:2021PRL},
 \beq
\label{eq:norm_Carlsson}
 \begin{array}{l}
n(\bm{q};\bm{q}',g)
=(-1)^{M / 2} \mathrm{pf}\left(\begin{array}{cc}
-\bar{U}(\bm{q}) \sigma & \Lambda(\bm{q}) D^{\dagger}(\bm{q}) D^{\prime}(\bm{q}',g) \Lambda^{\prime}(\bm{q}',g) \\
-\Lambda^{\prime}(\bm{q}',g) D^{\prime T}(\bm{q}',g) D^{*}(\bm{q}) \Lambda(\bm{q}) & \sigma \bar{U}^{\prime}(\bm{q}',g)
\end{array}\right)_{2M\times 2M},
\end{array}
 \eeq
 which allows for a stable numerical computation of overlaps independently of how tiny the occupation number $v_k$ might be. In Eq.(\ref{eq:norm_Carlsson}),  $\sigma$ is the $M\times M$ tridiagonal skew-symmetric matrix with elements 1 and $-1$
\beqn
\sigma=\left(\begin{array}{ccccc}
0 & 1 & 0 & 0 & 0 \\
-1 & 0 & 0 & 0 & 0 \\
0 & 0 & 0 & 1 & 0 \\
0 & 0 & -1 & 0 & 0 \\
0 & 0 & 0 & 0 & \ddots
\end{array}\right),
\eeqn
and the $\Lambda$ matrix is a $M\times M$ diagonal matrix defined in terms of the occupation probabilities,
\beqn
\Lambda=\left(\begin{array}{ccccc}
\sqrt{v_{1}} & & & & \\
& \sqrt{v_{1}} & & & \\
& & \sqrt{v_{2}} & & \\
& & & \ddots & \\
& & & & \sqrt{v_{M/2}}
\end{array}\right).
\eeqn
This idea can also be implemented in the following simpler way~\cite{Frosini:2021b},
     \beqn
     \label{eq:norm_Frosini}
     n(\bm{q};\bm{q}',g)
     &=&
     (-1)^{M/2} \operatorname{det} C^\ast_{(\bm{q})}  \operatorname{det} C_{(\bm{q}')} 
     \operatorname{pf}
     \left[\frac{1}{
     \Bigg(\prod^{M/2}_{k,k^{\prime}} v^{(\bm{q})}_{k} v_{k^{\prime}}^{(\bm{q}')}\Bigg)^{1/M}}
     \left(\begin{array}{cc} V^{T}(\bm{q}) U(\bm{q}) & V^{T}(\bm{q}) V^{\ast}(\bm{q}',g) \\
     -V^{\dagger}(\bm{q}',g) V(\bm{q}) & U^{\dagger}(\bm{q}',g) V^{*}(\bm{q}', g)
     \end{array}\right)_{2M\times 2M}\right].
     \eeqn

\jmy{Besides, there are alternative methods to compute the norm overlap, such as the one proposed  recently by Bally and Duguet~\cite{Bally:2018PRC}.}

\subsubsection{Connection to  ab initio frameworks}

In the recent decade, remarkable progress has been achieved in the development of nuclear {\em ab initio} methods starting from nuclear interactions derived from chiral EFT. Thanks to the use of the similarity renormalization group (SRG) technique which is employed to integrate out the non-trivial coupling between low and high momentum states, the {\em ab initio} methods with some controllable truncations  have been extended to study atomic nuclei with  the mass number up to $A=100$ \cite{Hergert:2020}, from doubly-closed spherical nuclei to open-shell deformed nuclei, even though the latter ones are still challenging. To capture the collective correlations associated with nuclear deformation and pairing correlations, multi-particle multi-hole configurations need to be included in the model space and these configurations are becoming extremely difficult to include in {\em ab initio} methods for medium-mass or heavy nuclei.

As discussed in the previous section, collective correlations are relatively easy to be taken into account by introducing the mechanism of symmetry breaking in the mean fields. Therefore, a symmetry-breaking reference state has been introduced and the BMF techniques are implemented into  {\em ab initio} methods for strongly correlated systems. This idea gives birth to a new generation of {\em ab initio} methods, including the no-core Monte Carlo shell model~\cite{Liu:2012PRC,Abe:2012,Abe:2021}, multi-reference in-medium SRG (MR-IMSRG) for nuclei with pairing correlations~\cite{Hergert:2014} and deformation~\cite{Yao:2018wq,Yao:2020PRL},  tensor-optimized high-momentum antisymmetrized molecular dynamics \cite{Lyu:2018PRC}, Bogoliubov many-body perturbation theory (MBPT) \cite{Tichai:2018PLB}, and coupled-cluster theory for nuclei with pairing \cite{Qiu:2019} and deformation \cite{Novario:2020PRC}, and multi-reference MBPT~\cite{Frosini:2021a,Frosini:2021b,Frosini:2021c}. In these {\em ab initio} methods, the model space grows only polynomially with nuclear size, even though one still needs to restore the broken symmetries at some point which may bring extra computational costs.

%% file: 2theory4dbd.tex
 The theory of $\znubb$  decay was first formulated by Furry \cite{Furry:1939} and was further developed by Primkoff and Rosen \cite{Primakoff:1959}, Doi~\cite{Doi:1981,Doi:1985},  Haxton and Stephenson \cite{Haxton1984PPNP}, Tomoda \cite{Tomoda:1991}, Bilenky \cite{Bilenky:1987,Bilenky:2015}, Simkovic~\cite{Simkovic99}, and many others. Recently, a master formula for $\znubb$  decay was reformulated by Cirigliano et al. \cite{Cirigliano:2018JHEP} in chiral EFT. For completeness, we present the formulas for the $\znubb$  decay, including the half-life, transition operator, and leptonic phase space factor. Different treatments of the short-range correlation between nucleons are also discussed.

 \subsection{The transition operator  in the standard mechanism}
  
  \subsubsection{The long-range effective interaction}

The most general Lorentz invariant effective   Hamiltonian contributing to the $\znubb$ decay at low-energy scales can be parametrized in the following form \cite{Pas:1999PLB}
 \begin{eqnarray}
    \label{eq:general_weak_Hamiltonian}
 {\cal H}_{w} = \dfrac{G_\beta}{\sqrt{2}}
 \left[j^\mu_L J^\dagger_{L,\mu}+\sum_{\alpha,\beta} \epsilon^\beta_\alpha   j_\alpha  J^\dagger_{\beta}\right]+ {\rm h.c},
 \end{eqnarray}
where the first term in (\ref{eq:general_weak_Hamiltonian}) is the {\em standard} V-A  (vector minus axial vector)  weak interaction. The corresponding LEC is the Fermi coupling constant $G_\beta=G_F\cos\Theta_C$ with $G_F=1.166378\times 10^{-5}$ GeV$^{-2}$ \cite{PDG:2018}, where $\Theta_C$ is the Cabibbo angle with $\cos\Theta_C=0.974$. The second term in (\ref{eq:general_weak_Hamiltonian}) stands for all other exotic interactions  depending on specific particle models.  The dimensionless LECs $\epsilon^\beta_\alpha$ stand for different types of effective couplings (except for $\alpha=\beta=L$).  The indices $\alpha$ and $\beta$ in the hadronic current  
$J^\dagger_\alpha=\bar u{\cal O}_\alpha d$
and leptonic current 
$j_\beta=\bar e{\cal O}_\beta \nu$
run over $L/R$, $S\mp P$ and $T_{L/R}$, with the operators ${\cal O}_\alpha$ and ${\cal O}_\beta$ defined as $\gamma^\mu(1\mp\gamma_5)$, $(1\mp\gamma_5)$, and $\sigma_{\mu\nu}(1\pm\gamma_5)$, respectively, where $\sigma_{\mu\nu}=\frac{i}{2} [\gamma_\mu,\gamma_\nu]$ and $\gamma_i$ being a four-component Dirac matrix. The $L/R$-handed currents are defined as   \footnote{The left-handed neutrino is usually introduced as $\nu_{eL}\equiv\hat P_L \nu_e$, where the projection operator is defined as $\hat P_L\equiv(1-\gamma_5)/2$ and $\hat P^2_L=\hat P_L$. Therefore, the left-handed current is also written as $j^\mu_{L}=2\bar e\gamma^\mu\nu_{eL}= \bar e\gamma^\mu(1 -\gamma_5)\nu_{e}$.}
  
 \bsub
 \beqn
 \label{eq:leptonic_current}
 j^\mu_{L/R} &=& \bar e\gamma^\mu(1 \mp\gamma_5)\nu_{e},\\
 \label{eq:hadronic_current_quark}
 J^{\mu\dagger}_{L/R} 
 &=&\bar u\gamma^\mu \left(1 \mp  \gamma_5 \right)d.
 \eeqn
 \esub

\begin{figure}[t]
\centering  
\includegraphics[width=12cm]{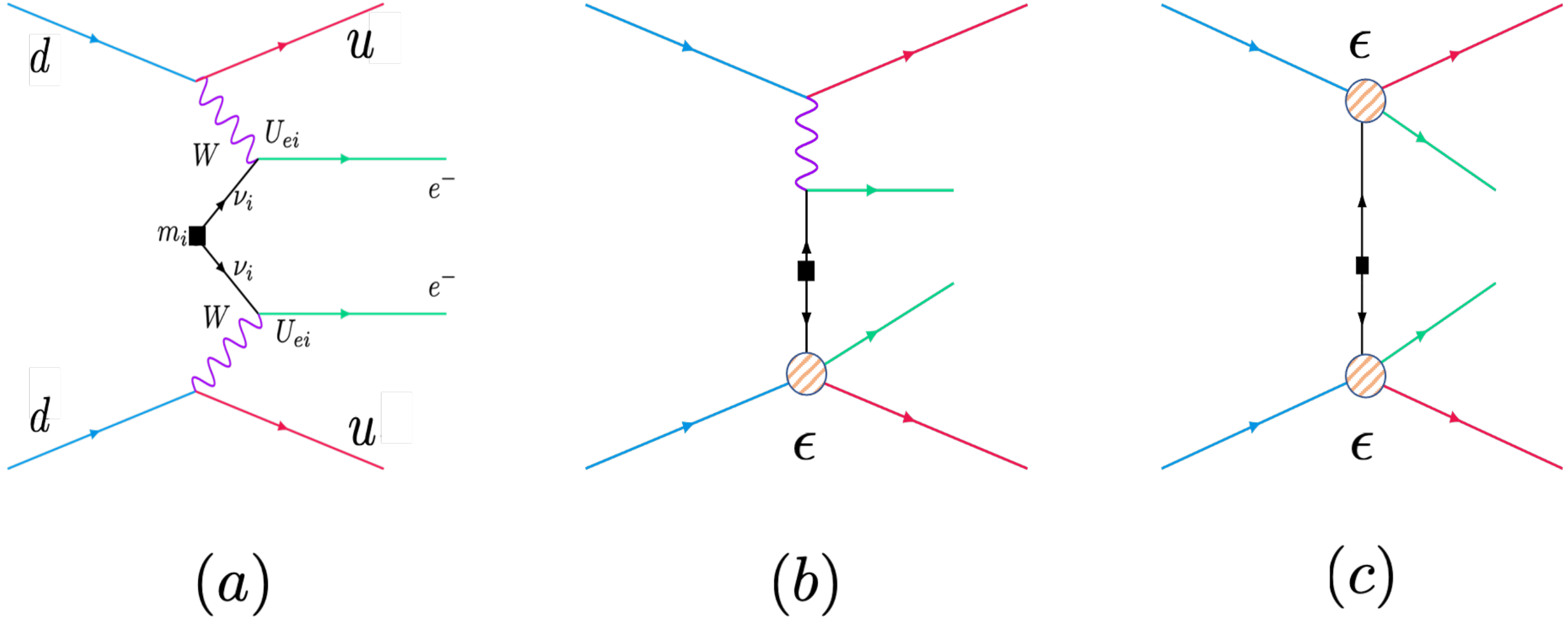} 
\caption{\label{fig:dd2uuee_cartoon_nldbd} Feynman diagrams of the  long-range operators  (\ref{eq:general_neutrino_potential})  to  $0\nu\beta\beta$ decay. The shaded circle stands for the vertices of non-standard weak interactions with the LECs $\epsilon^\beta_\alpha G_\beta/\sqrt{2}$.
(a) corresponds to the standard mechanism of $V-A$ type weak interaction; (b) is a product of an ordinary $V-A$ interaction and a non-standard one; (c) is a product of two non-standard interactions. }
\end{figure}

 The $\znubb$ decay of $\ket{i} \to \ket{f}$ is a second-order weak process with its transition amplitude ${\cal A}^{0\nu}$ proportional to the time-ordered product of the effective Hamiltonian, 
  \begin{eqnarray}
    \label{eq:general_neutrino_potential}
  {\cal A}^{0\nu}_{i\to f}
  &\propto& \bra{f}T\left[{\cal H}_{w}(x_1){\cal H}_{w}(x_2)\right] \ket{i}  \nonumber\\
  &\propto&  \bra{f} T\left( j^\mu_L J^\dagger_{L,\mu}j^\nu_L J^\dagger_{L,\nu}\right)
  +\sum_{\alpha,\beta} \epsilon^\beta_\alpha  T\left(j^\mu_L J^\dagger_{L,\mu} j_\alpha  J^\dagger_{\beta}\right) 
  +\sum_{\alpha,\beta,\gamma,\delta} \epsilon^\beta_\alpha  \epsilon^\delta_\gamma T\left( j_\alpha  J^\dagger_{\beta} j_{\gamma} J^\dagger_{\delta}\right)\ket{i}.  
 \end{eqnarray}
 The contribution of the above three terms is schematically depicted in Fig.\ref{fig:dd2uuee_cartoon_nldbd}(a), (b), and (c), respectively. The third term is quadratic in $\epsilon$ and is thus negligible. As noticed in Ref.~\cite{Pas:1999PLB}, the second term, corresponding to Fig.\ref{fig:dd2uuee_cartoon_nldbd}(b) contains two general cases:
 \begin{itemize}
     \item $j_\alpha$ is of either $S-P$ or $T_L$ type, the neutrino potential becomes
     \beq
     \label{eq:LL_case}
      \hat P_{L} \frac{q^{\mu} \gamma_{\mu}+m_{\nu}}{q_{\mu} q^{\mu}-m_{v}^{2}} \hat P_{L}=\frac{m_{\nu}}{q_{\mu} q^{\mu}-m_{\nu}^{2}}.
     \eeq
     \item $j_\alpha$ is of either $V+A, S+P$ or $T_R$ type, the neutrino potential becomes
     \beq
     \label{eq:LR_case}
      \hat P_{L} \frac{q^{\mu} \gamma_{\mu}+m_{\nu}}{q_{\mu} q^{\mu}-m_{v}^{2}} \hat P_{R}=\frac{q^{\mu} \gamma_{\mu}}{q_{\mu} q^{\mu}-m_{\nu}^{2}}.
     \eeq
 \end{itemize}

  In the left-right symmetric model \cite{Mohapatra:1975PRD}, only the $\alpha, \beta=L/R$ types of operators are considered for the $\znubb$ decay \footnote{In this model, the vector bosons $W^{\pm}_{L/R}$ are responsible for the left- and right-handed weak interaction and are integrated out in low-energy effective weak Hamiltonian.}.  The general weak Hamiltonian (\ref{eq:general_weak_Hamiltonian}) is simplified into the following form~\cite{Doi:1985,Hirsch:1996PLB}
 \beq
 \label{eq:LR_weak_Hamiltonian}
 {\cal H}^{LR}_{w} = \dfrac{G_\beta}{\sqrt{2}}\left[j^\mu_L(J^\dagger_{L,\mu}+\chi J^\dagger_{R,\mu})
 +j^\mu_R(\eta J^\dagger_{L,\mu}+\lambda J^\dagger_{R,\mu})\right] + {\rm h.c},
 \eeq
 where the LECs $\epsilon^\beta_\alpha$ are replaced with parameters $\chi, \eta, \lambda$ normalized to the first term. Their values reflect the importance of different parts of the right-handed weak interaction.  In this model, the $\znubb$ decay transition amplitude  becomes, 
  \begin{eqnarray}
    \label{eq:LR_neutrino_potential}
  {\cal A}^{0\nu}_{i\to f}
  &\propto&  \bra{f} T\left( j^\mu_Lj^\nu_L  J^\dagger_{L,\mu} J^\dagger_{L,\nu}\right) +  T\left(\chi j^\mu_L j^\nu_L J^\dagger_{L,\mu}  J^\dagger_{R,\nu} 
  + \eta j^\mu_L j^\nu_R J^\dagger_{L,\mu}  J^\dagger_{L,\nu} 
  + \lambda j^\mu_L j^\nu_R J^\dagger_{L,\mu}  J^\dagger_{R,\nu}\right) +\ldots \ket{i},
 \end{eqnarray}
 where "$\ldots$" stands for the negligible higher-order terms ${\cal O}(\epsilon^2)$. As shown in (\ref{eq:LR_case}),  the $\eta, \lambda$ terms in (\ref{eq:LR_neutrino_potential}) are not sensitive to neutrino mass $m_\nu$, as $q_\mu q^\mu >> m^2_\nu$.

 In most practical calculations for the NMEs of $\znubb$ decay, only  the first term in (\ref{eq:general_weak_Hamiltonian}) is considered
 \begin{eqnarray}
    \label{eq:LL_weak_Hamiltonian}
 {\cal H}^{LL}_{w} = \dfrac{G_\beta}{\sqrt{2}} j^\mu_L J^\dagger_{L,\mu} + {\rm h.c}.
 \end{eqnarray}
 For the sake of simplicity, we will present the formulas based on the above simplest weak Hamiltonian subsequently. Of course, all the following formulas can be generalized for the general weak Hamiltonian (\ref{eq:general_weak_Hamiltonian}).  
 With the simple weak Hamiltonian (\ref{eq:LL_weak_Hamiltonian}), only the Fig.\ref{fig:dd2uuee_cartoon_nldbd}(a) diagram exists. It corresponds to the so-called {\em standard} mechanism of light Majorana neutrino exchange.

 \jmy{In atomic nucleus, the so-called impulse approximation is usually employed. In this approximation,  the nucleus is described as a collection of free nucleons. At the nucleon level,  the quark current operator $J^\dagger_{L,\mu}$ in (\ref{eq:hadronic_current_quark}) is replaced with an effective single-nucleon current operator $\mathcal {J}_{L,\mu}^\dagger$, the matrix element of which between  nucleon states with momentum $p$ and $p'$ respectively is given by~\cite{Walecka:1975} 
  \beq
  \langle \psi(p')| \mathcal{J}_{L}^{\mu\dagger}(0) | \psi(p)\rangle
  =\int d^4xe^{-iqx}\langle \psi(p')| \mathcal{J}_{L}^{\mu\dagger}(x) | \psi(p)\rangle,
 \eeq
 where $q=p-p'$ is the transferred four-momentum from one nucleon state to another.} The effective one-body current operator takes the following form~\cite{Tomoda:1991},
 \beqn
 \label{eq:effect_nucleon_current}
  \mathcal J^{\mu\dagger}_{L}  
  &=&\bar \psi_N \gamma^\mu \Biggl[g_V(\vec{q}^2) - g_A(\vec{q}^2)  \gamma_5\Biggr]\tau^+\psi_N 
  + \bar \psi_N \Biggl[
   g_P (\vec{q}^2)q^\mu \gamma_5
   -  ig_W(\vec{q}^2) \sigma^{\mu\nu} q_\nu\Biggr]\tau^+\psi_N,
\eeqn
where $\psi_N$ is the nucleon field operator, $\tau^+$ is the isospin raising operator with nonzero matrix element $\braket{p|\tau^+|n}=1$. The operators in the second term of (\ref{eq:effect_nucleon_current}) are  the induced pseudoscalar and weak-magnetism currents which could contribute to the NME of  $\znubb$ decay more than 25\% \cite{Simkovic99}.   The finite-nucleon-size effects~\footnote{The  Fourier transformation of the dipole form factor reads 
  \begin{equation} 
   \int \frac{d^3q}{(2 \pi)^{3}} \mathrm{e}^{i\bm{q}\cdot\left(\boldsymbol{x}-\boldsymbol{r}_{n}\right)}\left(\frac{\Lambda^{2}}{\Lambda^{2}+\bm{q}^{2}}\right)^{2} \\ =\frac{\Lambda^{3}}{8 \pi} \mathrm{e}^{-\Lambda\left|\bm{x}-\bm{r}_{n}\right|},
   \end{equation}
   where $\bm{r}_{n}$ is the coordinate of the $n$-th nucleon. It is seen that the inverse of the cutoff value $\Lambda^{-1}$ represents the finite extension of the nucleon. } are taken into account using the momentum-dependent dipole form factors $g_V(\vec{q}^2)$, $g_A(\vec{q}^2)$, $g_P(\vec{q}^2)$, and $g_W(\vec{q}^2)$, which in the zero-momentum-transfer limit are the vector, axial-vector, induced pseudoscalar and weak-magnetism coupling constants, respectively. The dipole form factors $g_\alpha(\vec{q}^2)$ with $\alpha=(V, A, W, P)$ are usually parametrized as follows~\cite{Simkovic99} 
\bsub\begin{align}
\label{eq:dipole_form_factors_gV}
 g_V(\vec{q}^2) &= g_V(0)\left(1+ \vec{q}^2/\Lambda^2_V\right)^{-2},\\
\label{eq:dipole_form_factors_gA}
 g_A(\vec{q}^2) &= g_A(0)\left(1+ \vec{q}^2/\Lambda^2_A\right)^{-2},\\
\label{eq:dipole_form_factors_gP}
 g_P(\vec{q}^2) &= g_A(\vec{q}^2) \left(\dfrac{2m_p}{ \vec{q}^2+m^2_\pi}\right),\\
\label{eq:dipole_form_factors_gM}
 g_W(\vec{q}^2) &= g_V(\vec{q}^2) \frac{\kappa_1}{2m_p},
\end{align}
\esub
where $g_V(0)=1$, $g_A(0)=1.27$, and the cutoff values are $\Lambda_V=\SI{0.85}{\GeV}$ and 
$\Lambda_A=\SI{1.09}{\GeV}$.  According to the conserved vector current (CVC) hypothesis, $g_W(0)=\kappa_1/2m_p$ with $\kappa_1$ being the anomalous nucleon iso\-vector magnetic moment $\kappa_1=\mu^{(a)}_n-\mu^{(a)}_p\simeq 3.7$.

 \subsubsection{The half-life of $\znubb$ decay}
  The half-life of  the $\znubb$ decay $\ket{^AZ} \to \ket{^A(Z+2)} + 2e^-$ is defined in terms of decay width,
   \begin{equation}
     \label{eq:half-life-definition} 
    [T^{0\nu\beta\beta}_{1/2}]^{-1} 
    = \dfrac{1}{\ln 2}\Gamma^{0\nu},
  \end{equation}
 where  the decay width $ \Gamma^{0\nu}$   is given by~\cite{Fukugita:2003} 
 \begin{equation}
 \label{eq:decaywidth}  
      \Gamma^{0\nu}
     = \dfrac{1}{2} \int \dfrac{d^3\bm{k}_1}{(2\pi)^32\epsilon_1} 
      \int \dfrac{d^3\bm{k}_2}{(2\pi)^32\epsilon_2} 
     (2\pi)\delta(E_I-E_F-\epsilon_1-\epsilon_2) 
     \vert {\cal M}_{fi}\vert^2.  \nonumber\\
 \end{equation}
 Here $\epsilon_{1,2}=k^0_{1,2}$ and $\bm{k}_{1,2}$ are the energies and momenta of the two emitted electrons, respectively. The $E_{I, F}$ are the energies of initial and final nuclei, respectively. The full matrix element ${\cal M}_{fi}$  is determined by the $S$ matrix in the following way
\begin{eqnarray}
 \label{eq:S-matrix}
   \langle f|S^{(2)}|i\rangle
    &=& i(2\pi)\delta(E_I-E_F-\epsilon_1-\epsilon_2){\cal M}_{fi},
\end{eqnarray}  
where the $S$ matrix is determined by the second-order interaction (\ref{eq:LL_weak_Hamiltonian}),
\begin{eqnarray}
  \label{S-matrix1}
	\langle f|S^{(2)}|i\rangle
	&=& 4\frac{(-i)^2}{2!}\left( \frac{G_\beta}{\sqrt2} \right)^2 
	\int  d^4 x_1 \int  d^4 x_2
	\bar u(k_1,s_1)\te^{\ti k_1x_1} \gamma^\mu\langle 0| T\left(\nu_{eL}(x_1)\nu^T_{eL}(x_2) \right)|0\rangle \gamma^{\nu T}\bar u^T(k_2,s_2)\te^{\ti k_2x_2}\notag\\
	&&\times \langle \Psi_F | T\left( \mathcal J_{L,\mu}^\dagger(x_1)\mathcal J_{L,\nu}^\dagger(x_2)\right) |\Psi_I\rangle -(k_1\leftrightarrow k_2).
\end{eqnarray}
Here $k_ix_i=\epsilon_i t_i -\bm{k}_i\cdot\bm{r}_i$ and $u(k,s)$ is a Dirac spinor for the electron.  The $\ket{\Psi_I}$ and $\ket{\Psi_F}$ are wave functions of the initial and final nuclei.  The second term $(k_1\leftrightarrow k_2)$ arises due to the exchange between the two outgoing electrons. 

The hadronic current ${\cal J}_{L,\mu}^\dagger(x)$ in Eq.(\ref{eq:effect_nucleon_current}) is defined in the Heisenberg representation, i.e., ${\cal J}_{L,\mu}^\dagger(x)=e^{i H t} {\cal J}_{L,\mu}^\dagger(\bm{r})e^{-i H t}$ with $H$ being the Hamiltonian of the strong interaction. By inserting the identity, $\sum_{N} \ket{N}\bra{N} =1$ with $\ket{N}$ being nuclear intermediate states,  one has
\begin{itemize}
    \item case $t_1 > t_2$:
    \begin{eqnarray}
       \langle \Psi_F | T\left( \mathcal J_{L,\mu}^\dagger(x_1)\mathcal J_{L,\nu}^\dagger(x_2)\right) |\Psi_I\rangle 
    &=&  \sum_{N} e^{i(E_F-E_N)t_1}e^{i(E_N-E_I)t_2}
    \bra{\Psi_F} {\mathcal J}_{L,\mu}^\dagger(\bm{r}_1)
     \ket{N}\bra{N}
    {\mathcal J}_{L,\nu}^\dagger(\bm{r}_2)  \ket{\Psi_I}.
    \end{eqnarray}
     
     \item case $t_1 < t_2$:
    \begin{eqnarray}
       \langle \Psi_F | T\left( \mathcal J_{L,\mu}^\dagger(x_1)\mathcal J_{L,\nu}^\dagger(x_2)\right) |\Psi_I\rangle 
    &=&  \sum_{N} e^{i(E_F-E_N)t_2}e^{i(E_N-E_I)t_1}\langle \Psi_F | \mathcal \mathcal J_{L,\nu}^\dagger(\bm{r}_2) 
     \ket{N}\bra{N} 
     J_{L,\mu}^\dagger(\bm{r}_1) |\Psi_I\rangle.
    \end{eqnarray}
\end{itemize}

For the field operator of the electron neutrino, one has 
\begin{equation}
 \nu_{e, L}(x) = \sum^3_{j=1} U^L_{ej}  \nu_{j, L}(x),\quad C\bar\nu^T_{j,L}(x)=\xi_j\nu_{j,L}(x),
\end{equation} 
where $\nu_{j,L}(x)$ is the field of a Majorana neutrino with mass
$m_j$. The combination of phase factor $\xi_j$ and matrix element $U^L_{ej}$ yields the relation $(U^L_{ej})^2\xi_j=U_{ej}^2$~\cite{Bilenky:1987}. The charge-conjugation matrix $C$ has the following properties,
\begin{equation}
C \gamma_{\alpha}^{T} C^{-1}=-\gamma_{\alpha}, \quad C^\dagger C=1, \quad C^{T}=-C.
\end{equation} 
With the above relations, one obtains the contraction of electron neutrino fields
\begin{eqnarray}
 \langle 0| T\left(\nu_{eL}(x_1)\nu^T_{eL}(x_2) \right)|0\rangle =-i\sum_j U_{ej}^2m_j  \int\frac{d^4 q }{(2\pi)^4} \frac{e^{-i q(x_1-x_2)}}{q^2-m_j^2} P_LC
\end{eqnarray}
where  the additional relations $C\gamma_5C^{-1}=\gamma_5$ and $P_L\gamma^\mu q_\mu P_L=0$ are used.
Note that the second term denoted by $(k_1\leftrightarrow k_2)$ just contributes equally to the first term.  With the above considerations and the following integral formula
\beq
\int_{-\infty}^{a}dt e^{iEt} = \lim _{\epsilon \rightarrow 0} \int_{-\infty}^{a}dt e^{i(E-i \epsilon) t}=\lim _{\epsilon \rightarrow 0} \frac{-ie^{i(E-i \epsilon) a}}{E-i \epsilon},
\eeq
the $S$ matrix in (\ref{S-matrix1}) can be transformed into the following form after the integration over $t_{1,2}$,
\begin{eqnarray}
\label{eq:Smatrix_approximations}
    \langle f|S^{(2)}|i\rangle 
	&=& i4\pi \delta(\epsilon_1+\epsilon_2 +E_F  -E_I) \left( \frac{G_\beta}{\sqrt2} \right)^2 
		\int d^3 r_1d^3 r_2 \bar{e}_{\bm{k}_1,s_1}(\bm{r}_1)\gamma^\mu P_LC\gamma^{\nu T}  e^c_{\bm{k}_2,s_2}(\bm{r}_2)\notag\\
	&& \times  \sum_j U^2_{ej}  m_j\dfrac{1}{q^0_j}
\frac{ 1}{(2\pi)^3} \int d^3q  e^{i \bm{q}\cdot \bm{r}_{12} }  \notag\\
	&& \times 
     \sum_{N} \left[\dfrac{\bra{\Psi_F} {\mathcal J}_{L,\mu}^\dagger(\bm{r}_1)
     \ket{N}\bra{N}
    {\mathcal J}_{L,\nu}^\dagger(\bm{r}_2)  \ket{\Psi_I}}{\epsilon_2+q^0_j+E_N-E_I}
    +\dfrac{\bra{\Psi_F} {\mathcal J}_{L,\nu}^\dagger(\bm{r}_2)
    \ket{N}\bra{N}
    {\mathcal J}_{L,\mu}^\dagger(\bm{r}_1)  \ket{\Psi_I}}{\epsilon_1+q^0_j+E_N-E_I} \right] \nonumber\\
  &\simeq&  i  (2\pi) \delta(\epsilon_1+\epsilon_2 +E_F  -E_I) 
    \dfrac{g^2_A(0)G^2_\beta}{ 2\pi R_0} 	\int d^3 r_1d^3 r_2   \bar{e}_{\bm{k}_1,s_1}(\bm{r}_1)\gamma^\mu P_LC\gamma^{\nu T}  \bar{e}^T_{\bm{k}_2,s_2}(\bm{r}_2) \notag \\
	&& \times   \sum_j U^2_{ej}  m_j \dfrac{4\pi R_0}{g^2_A(0)}
  \int \frac{d^3q }{(2\pi)^3}  e^{i \bm{q}\cdot \bm{r}_{12}}  
	 \sum_N\dfrac{ \bra{\Psi_F} {\mathcal J}_{L,\mu}^\dagger(\bm{r}_1)  \ket{N}\bra{N}
    {\mathcal J}_{L}^{\mu\dagger}(\bm{r}_2) \ket{\Psi_I} }
    {q\left[q+E_N -(E_I+E_F)/2\right]},
\end{eqnarray}
where we define $\bar{e}_{\bm{k},s}(\bm{r})  \equiv \bar u(k,s)e^{-i \bm{k}\cdot\bm{r}}$ and $\bm{r}_{12}=\bm{r}_1-\bm{r}_2$. $E_N$ is the energy of the intermediate state $\ket{N}$.   In the above derivation, the relations $C\gamma^{\nu T}C^{-1}=-\gamma^\nu$, $\gamma_5\gamma^\nu=-\gamma^\nu\gamma_5$, 
 $\gamma^\mu\gamma^\nu=g^{\mu\nu}+\frac12 \left(\gamma^\mu\gamma^\nu-\gamma^\nu\gamma^\mu\right)$ were used. The energy of neutrino is approximated as $q^0_j=\sqrt{q^2+m^2_j}\simeq  q$ considering the fact that the average momentum carried by the virtual neutrino $q=|\bm{q}|\sim 1/\vert \bm{r}_{12}\vert \sim 100~\mathrm{MeV}$, much larger than the  masses of light neutrinos. Besides, $\epsilon_1\simeq \epsilon_2=(\epsilon_1+\epsilon_2)/2=(E_I-E_F)/2$. 

 For the transition with two electrons in the $S_{1/2}$ state, only the $L=0$ component in the plane-wave expansion is considered, $e^{-i\bm{k}\cdot \bm{r}} \to j_0(kr)\simeq 1$ as $kr\to 0$. In this case, the finite de Broglie wave length correction (FBWC) is not taken into account~\cite{Doi:1985}. This approximation is reasonable considering the fact that $\vert \bm{k}_{1,2}\vert\lesssim 1~\mathrm{MeV}$ and $\vert \bm{r}_{1,2}\vert <R_0$ ($R_0\simeq 1.2A^{1/3}~\mathrm{fm}$).
 
 From (\ref{eq:S-matrix}) and (\ref{eq:Smatrix_approximations}), one finds that the transition matrix element $M_{fi}$, in this case, can be written as
\begin{equation}
   {\cal M}_{fi}=   
   \dfrac{g^2_A(0)G^2_\beta}{2\pi R_0}    
   \langle m_{\beta\beta} \rangle   \bar u(k_1,s_1)  P_R C\bar u^T(k_2,s_2)M^{0\nu},
\end{equation}
where the effective electron neutrino mass $\langle m_{\beta\beta}\rangle$ has been defined in (\ref{eq:effective-mass}), and the NME of $0\nu\beta\beta$ decay
\begin{eqnarray}
\label{eq:NME-light}
	 M^{0\nu}
	 &\equiv&\dfrac{4\pi R_0}{g^2_A(0)}
	\int d^3 r_1d^3 r_2  \int \frac{d^3q }{(2\pi)^3}  e^{i \bm{q}\cdot \bm{r}_{12}}  
	 \sum_N\dfrac{ \bra{\Psi_F} {\mathcal J}_{L,\mu}^\dagger(\bm{r}_1) \ket{N}\bra{N} 
    {\mathcal J}_{L}^{\mu\dagger}(\bm{r}_2) \ket{\Psi_I} }
    {q\left[q+E_N -(E_I+E_F)/2\right]}.
\end{eqnarray}
Here, $R_0$ is introduced to make the NME $M^{0\nu}$ dimensionless. Substituting the transition matrix element ${\cal M}_{fi}$ into (\ref{eq:half-life-definition}), one finds the expression for the $0\nu\beta\beta$-decay half-life 
\beqn
\label{eq:0nubb-half-life}
 [T^{0\nu\beta\beta}_{1/2}]^{-1} 
  &= & G_{0\nu} g_A^4(0)\left|\frac{\langle m_{\beta\beta}\rangle}{m_e}\right|^2\left|M^{0\nu}\right|^2,
\end{eqnarray}
where the leptonic phase-space factor $G_{0\nu}$ is defined as 
\begin{eqnarray}
\label{eq:G0nu_general}
G_{0\nu} 
	&\equiv&  \dfrac{1}{(\ln 2)(2\pi)^5 }
  \frac{G^4_\beta m^2_e}{R^2_0}  \dfrac{1}{4}\int \int  
 \sum_{\rm spins}\Big\vert \bar u(k_1,s_1)  P_R C\bar u^T(k_2,s_2) \Big\vert^2 
k_1k_2d\epsilon_1  d\cos\theta_{12},
\end{eqnarray}

 \subsubsection{The neutrino potentials from  phenomenological nucleon currents}
 
\begin{figure}[tb]
\centering 
\includegraphics[width=7cm]{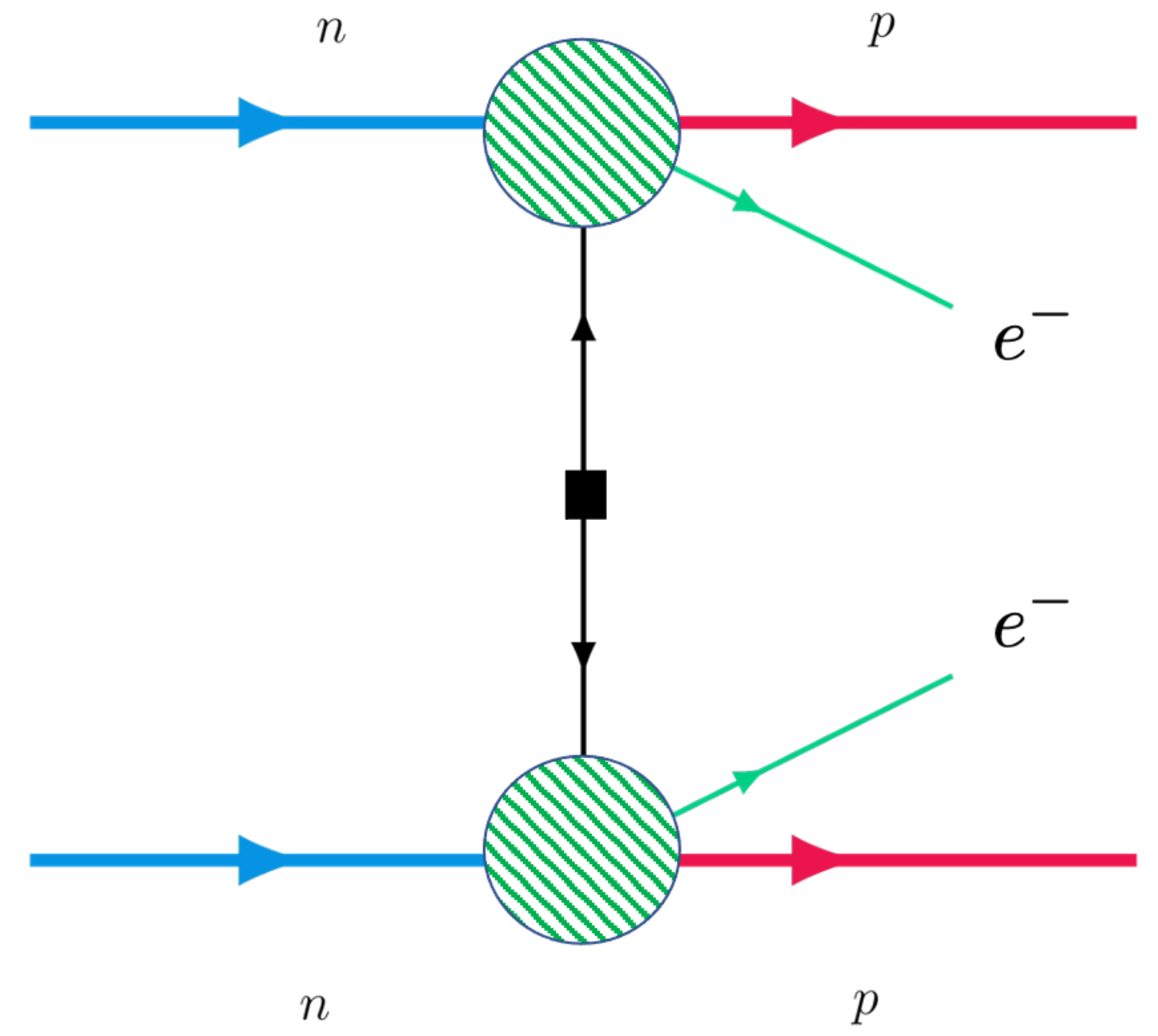} 
\caption{\label{fig:cartoon_dbd_LR_nnpp} 
The long-range contribution to the neutrino potentials of the  $0\nu\beta\beta$ decay at nucleon level. The shaded area stands for the standard model vertex with higher-order corrections to the one-body nucleon weak current, cf. Eq. (\ref{eq:effect_nucleon_current}).}
\end{figure}

 From the definition of the NME  in Eq.(\ref{eq:NME-light}), one can introduce a transition operator  $O^{0\nu}$  as follows
  \beq
  M^{0\nu} = \bra{\Psi_F}  O^{0\nu} \ket{\Psi_I},
  \eeq
  where the operator  $O^{0\nu}$ has the following form,
 \beqn
 \label{eq:dbd_operator_long}
  O^{0\nu} 
   &=&\frac{4\pi R_0}{g_A^2(0)} \iint \dd^3 r_1 \dd^3 r_2 \int\frac{\dd^3 q}{(2\pi)^3}\frac{e^{i \vec{q}\cdot \vec{r}_{12}}}{q}
   \sum_N\dfrac{  {\mathcal J}_{L,\mu}^\dagger(\bm{r}_1) \ket{N}\bra{N} 
    {\mathcal J}_{L}^{\mu\dagger}(\bm{r}_2)  }
    {q+E_N -(E_I+E_F)/2 }.
\eeqn
 The r.h.s. of the above equation requires nuclear intermediate states $\ket{N}$ the description of which is a challenge for many nuclear models. Therefore, in many applications the so-called closure approximation is adopted, namely, the excitation energies of all possible intermediate states $\ket{\Psi_N}$ are replaced by an estimate ``average" value $E_d=\langle E_N\rangle-(E_I+E_F)/2$ and the summation over these states can be eliminated by making use of the relation $\sum_N \ket{N}\bra{N}=1$. As a consequence, the summation of products of one-body matrix elements is approximated as one simple two-body matrix element.
 An empirical formula value $E_d=1.12A^{1/2}$ MeV proposed in Ref.~\cite{Haxton1984PPNP} is adopted for the average excitation energy. In this case, the transition operator is simplified as
  \beqn
 \label{eq:dbd_operator_long_closure}
  O^{0\nu} 
   &\simeq&\frac{4\pi R_0}{g_A^2(0)} \iint \dd^3 r_1 \dd^3 r_2 \int\frac{\dd^3 q}{(2\pi)^3}   \dfrac{  e^{i \vec{q}\cdot \vec{r}_{12}}}
    {q(q+E_d )} {\mathcal J}_{L,\mu}^\dagger(\bm{r}_1) 
    {\mathcal J}_{L}^{\mu\dagger}(\bm{r}_2). 
\eeqn
 As discussed before, the average value of $q\sim 100$ MeV, which is much larger than the average nuclear excitation energy ($10-20$ MeV) for long-ranged two nucleon processes, the closure approximation is accurate at the $10\%-20\%$ level, as demonstrated  in Refs.~\cite{Pantis:1990,Faessler:1991,Senkov2013}.  (see Fig.~\ref{fig:cartoon_dbd_LR_nnpp})
 
The product of hadronic current operators $\mathcal {J}_{L,\mu}^\dagger {\mathcal {J}^{\mu\dagger}_L}$ in (\ref{eq:dbd_operator_long_closure}) is composed of five terms labelled with VV, AA, PP, AP and MM respectively based on their vertex structures,
     \begin{eqnarray}  
     \label{eq:current_product_relativistic}
            \begin{array}{cc}
                g^2_V(\vec{q}^2)(\bar\psi\gamma_\mu\tau^+\psi)^{(1)}(\bar\psi\gamma^\mu\tau^+\psi)^{(2)}, & {\rm VV}  \\ 
                \\
                g^2_A(\vec{q}^2)(\bar\psi\gamma_\mu\gamma_5\tau^+\psi)^{(1)}(\bar\psi\gamma^\mu\gamma_5\tau^+\psi)^{(2)}, & {\rm AA}  \\ 
                \\
                g^2_P(\vec{q}^2)(q_\mu\gamma_5\tau^+\psi)^{(1)}(\bar\psi q^\mu\gamma_5\tau^+\psi)^{(2)}, & {\rm PP}  \\
                \\
                g_A(\vec{q}^2)g_P(\vec{q}^2)(\bm{\gamma}\gamma_5\tau^+\psi)^{(1)}(\bar\psi \bm{q}\gamma_5\tau^+\psi)^{(2)}, & {\rm AP}  \\
                \\
                g^2_W(\vec{q}^2)( \sigma_{\mu\nu} q_\mu\tau^+\psi)^{(1)}(\bar\psi  \sigma^{\mu\nu}  q^\mu\tau^+\psi)^{(2)}, & {\rm MM}  \\
            \end{array} 
     \end{eqnarray}

\begin{table} 
\centering
\tabcolsep=10pt
\caption{The non-relativistic reduction of the nuclear currents  arranged in the order of $(1/m_p)^k$~\cite{Tomoda:1991}, where $m_p$ is the nucleon mass, $q^\mu=p^{\mu}-p^{\prime \mu}$ and $\bm{Q}=\bm{p}+\bm{p}'$  with $p^\mu$ and $p^{\mu'}$ being the four-momenta of the initial and final nucleon states, respectively. } 
\begin{tabular}{CCCC}
\hline  \hline
& k=0 & k=1 & k=2 \\
\hline
V^{(k) 0} & 1 & 0 & -\frac{1}{8 m^{2}_p} \bm{q} \cdot(\bm{q}-\mathrm{i} \bm{\sigma} \times \bm{Q}) \\
W^{(k) 0} & 0 & -\frac{1}{2 m_p} \bm{q} \cdot(\bm{q}-\mathrm{i} \bm{\sigma} \times \bm{Q}) & 0 \\
A^{(k) 0} & 0 & \frac{1}{2 m_p}  \bm{\sigma}\cdot \bm{Q} & 0 \\
p^{(k) 0} & 0 & -\frac{1}{2 m_p} q^{0}  \bm{\sigma} \cdot \bm{q} & 0 \\
\bm{V}^{(k)}   & 0 & \frac{1}{2 m_p}(\bm{Q}-\mathrm{i} \bm{\sigma} \times \bm{q}) & 0 \\
\bm{W}^{(k)}   & -\mathrm{i} \bm{\sigma} \times \bm{q} & -\frac{1}{2 m_p} q^{0}(\bm{q}-\mathrm{i} \bm{\sigma} \times \bm{Q}) & \frac{1}{8 m^{2}_p}\left[\mathrm{i} \bm{q}^{2} (\bm{\sigma} \times \bm{q}) +\mathrm{i} (\bm{\sigma} \cdot \bm{Q}) (\bm{Q} \times \bm{q}) - \bm{q}^{2} \bm{Q} + (\bm{q} \cdot \bm{Q}) \bm{q}\right] \\
\bm{A}^{(k)} & \bm{\sigma} & 0 & -\frac{1}{8 m^{2}_p}\left[ \bm{\sigma} Q^{2}- \bm{Q} ( \bm{\sigma}\cdot\bm{Q})  + \bm{q}(\bm{\sigma}\cdot \bm{q})  -\mathrm{i} \bm{q} \times \bm{Q}\right] \\
\bm{P}^{(k)} & 0 & -\frac{1}{2m_p} \bm{q} (\bm{\sigma}\cdot \bm{q}) & 0 \\
\hline \hline
\end{tabular}
\label{tab:Foldy-Wouthuysen}
\end{table}

It has been found in the BMF covariant density functional theory (CDFT) studies for $\nuclide[150]{Nd}$ that relativistic corrections to the operators of light Majorana neutrino exchange are generally within 5\% \cite{Song:2014,Yao:2015}. Therefore, the non-relativistic transition operators are usually adopted in most studies and they can be derived from Eq.(\ref{eq:current_product_relativistic}) by using a Foldy-Wouthuysen (FW) transformation \cite{Foldy:1950}, which transforms the operator into the form that the odd operators connecting large and small components of the Dirac spin are suppressed successively. By setting $\beta=1$ in the transformed form \cite{Rose:1954,Friar:1966,Blin:1966}, one can obtain the non-relativistic form of the current operator which in first quantization reads~\cite{Tomoda:1991}
 \begin{eqnarray}
 \mathcal {J}^{\mu\dagger}_L(\bm{r})
 &\equiv&  \Bigg(J^{0\dagger}_L,\quad \bm{J}^{\dagger}_L \Bigg)\nonumber\\
 &=&\sum^A_{n=1} \tau^+_n\delta(\bm{r}-\bm{r_n})\sum_{k=0,1,\cdots}\left[g_V V^{(k)\mu}
 -g_A A^{(k)\mu}
 -g_P P^{(k)\mu}  
 +g_W W^{(k)\mu} \right]_n,
 \end{eqnarray}
 where $\bm{r_n}$ is the spatial coordinate of the $n$-th nucleon, and $k$ denotes the order of the suppression factor $(1/m_p)$. The expressions of all the terms up to $k=2$ are given in Table~\ref{tab:Foldy-Wouthuysen}. 
 \begin{itemize}
     \item The $k=0$ terms: 
     \bsub\begin{eqnarray}
        J^{0\dagger}_L (\bm{r})
        &=&\sum^A_{n=1} \tau^+_n\delta(\bm{r}-\bm{r_n})
            g_V,\\
        \bm{J}^{\dagger}_L(\bm{r}) 
        &=&\sum^A_{n=1} \tau^+_n\delta(\bm{r}-\bm{r_n})
            \Bigg[ -g_A \bm{\sigma}
                    +g_W(-i)\bm{\sigma}\times \bm{q}
            \Bigg]_n
        \end{eqnarray}
    \esub
    
 \item The $k=1$ terms:
       \bsub\begin{eqnarray}
        J^{0\dagger}_L (\bm{r})
        &=&\dfrac{1}{2m_p}\sum^A_{n=1} \tau^+_n\delta(\bm{r}-\bm{r_n})
           \Bigg[-g_A \bm{\sigma}\cdot\bm{Q}
            -g_W\bm{q}\cdot(\bm{q}-i\bm{\sigma}\cdot\bm{Q})
            +g_P q^0 \bm{\sigma}\cdot\bm{q} \Bigg]_n,\\
        \bm{J}^{\dagger}_L (\bm{r})
        &=&\dfrac{1}{2m_p}\sum^A_{n=1} \tau^+_n\delta(\bm{r}-\bm{r_n}) \Bigg[ g_V(\bm{Q}-i\bm{\sigma}\times \bm{q})   +g_P\bm{q}\bm{\sigma}\cdot \bm{q}  
                    -g_Wq^0(\bm{q}-i\bm{\sigma}\times \bm{Q})
            \Bigg]_n
        \end{eqnarray}
    \esub
 \end{itemize}
 Neglecting the nucleon recoil terms (depending on $Q$) and the energy transfer $q^0$ between the nucleons, the hadronic current  truncated up to the $1/m_p$ terms is simplified as 
 \bsub\beqn
 J^{0\dagger}_L(\bm{r})
 &=&\sum^A_{n=1} \tau^+_n\delta(\bm{r}-\bm{r_n})
 \Bigg(g_V-\dfrac{g_W}{2m_p} \bm{q}^2\Bigg)_n,\\
 \bm{J}^{\dagger}_L(\bm{r})
 &=&-\sum^A_{n=1} \tau^+_n\delta(\bm{r}-\bm{r_n})
 \Bigg(g_A \bm{\sigma}
     +ig_W\bm{\sigma}\times \bm{q} 
     +ig_V\dfrac{(\bm{\sigma}\times \bm{q})}{2m_p}
     - g_P\dfrac{\bm{q}\bm{\sigma}\cdot \bm{q}}{2m_p} \Bigg)_n.
 \eeqn
 \esub
The second term in the time-like component of the current is actually in the $(1/m_p)^2$ order and it is  usually dropped out as well. Combining the second and third terms in the spatial-like component by defining 
\beq 
\frac{g_M}{2m_p} \equiv g_W + \frac{g_V}{2m_p},
\eeq
or $g_M= g_V(1+\kappa_1)$, one rewrites the spatial-like component in terms $g_M$ as below
 \beq
 \label{eq:NR_currents}
  \bm{J}^{\dagger}_L(\bm{r})
  =-\sum^A_{n=1} \tau^+_n\delta(\bm{r}-\bm{r_n})
 \Bigg( 
      g_A \bm{\sigma}
     +ig_M\dfrac{\bm{\sigma}\times \bm{q} }{2m_p} 
     - g_P\dfrac{\bm{\sigma}\cdot \bm{q}}{2m_p}\bm{q} \Bigg)_n.
 \eeq
 \jmyr{It is noted that the momentum $\bm{q}$  has an opposite sign for $\bm{J}^{\dagger}_L(\bm{r}_1)$ and $\bm{J}^{\dagger}_L(\bm{r}_2)$ which affects the sign of the second term (proportional to $g_M$).}  Therefore, one obtains the expression for the product of the current operator \footnote{The following relation $
 (\bm{\sigma}_m\times\bm{q})\cdot (\bm{\sigma}_n\times\bm{q})
 =(\bm{\sigma}_m\cdot\bm{\sigma}_n)\bm{q}^2-(\bm{\sigma}_m\cdot\bm{q})(\bm{\sigma}_n\cdot\bm{q}) $ is used in the derivation.}
{\small \begin{eqnarray}
 \label{eq:NR_products}
   J^\dagger_{L,\mu}(\bm{r}_1)J^{\mu\dagger}_L (\bm{r}_2)
  &=&\sum^A_{m\neq n=1} \tau^+_m \tau^+_n\delta(\bm{r}_1-\bm{r_m})\delta(\bm{r}_2-\bm{r_n})
  \Bigg[g^2_V-g^2_A \bm{\sigma}_m\cdot\bm{\sigma}_n
  -g^2_P\dfrac{\bm{q}^2(\bm{\sigma}_m\cdot \bm{q})(\bm{\sigma}_n\cdot \bm{q})}{4m^2_p}\nonumber\\
  &&
   +g^2_M (-1)\dfrac{(\bm{\sigma}_m\times\bm{q})(\bm{\sigma}_n\times\bm{q})}{4m^2_p}  
   +2g_Ag_P \dfrac{(\bm{\sigma}_m\cdot\bm{q})(\bm{\sigma}_n\cdot\bm{q})}{2m_p}\Bigg]\nonumber\\
    &\equiv& -\sum^A_{m\neq n=1} \tau^+_m \tau^+_n\delta(\bm{r}_1-\bm{r_m})\delta(\bm{r}_2-\bm{r_n}) 
   \Bigg( h_F(\bm{q}) + h_{GT} \bm{\sigma}_m\cdot\bm{\sigma}_n
   -h_{T}S^{\bm{q}}_{mn}  \Bigg),
 \end{eqnarray}}
 where the spin-tensor operator in momentum space is introduced
 \beq
 \label{eq:spin_tensor_operator}
 S^{\bm{q}}_{mn}=  3(\bm{\sigma}_m\cdot\bm{q}) (\bm{\sigma}_n\cdot\bm{q})/\bm{q}^2 -\bm{\sigma}_m\cdot\bm{\sigma}_n,
 \eeq
 and
 \bsub
 \begin{eqnarray}
    h_F(\bm{q}^2) &=& -g^2_V+ g_Vg_W\frac{\bm{q}^2}{m_p}-g^2_W \frac{\bm{q}^4}{4m^2_p},\\
    h_{GT}(\bm{q}^2) &=& g^2_A   
   - g_Ag_P \dfrac{\bm{q}^2}{3m_p} 
    +g^2_P \dfrac{\bm{q}^4}{12m^2_p} 
   + g^2_M  \dfrac{\bm{q}^2}{6m^2_p},\\
   h_T(\bm{q}^2) 
   &=& g_Ag_P \dfrac{\bm{q}^2}{3m_p}
   -g^2_P \dfrac{\bm{q}^4}{12m^2_p} 
   +g^2_M  \dfrac{\bm{q}^2}{12m^2_p}. 
 \end{eqnarray}
 \esub
 The second and the third terms in $h_F(\bm{q}^2)$ are much smaller than the first term and are thus neglected in most studies. With the above expression, the transition operator $O^{0\nu}$ becomes
 \begin{align}
\label{eq:nldbd_operator}
  O^{0\nu} 
   &=\sum^A_{m\neq n=1} \tau^+_m \tau^+_n \iint \dd^3 r_1 \dd^3 r_2 \delta(\bm{r}_1-\bm{r_m})\delta(\bm{r}_2-\bm{r_n})\notag\\
  &\hphantom{{}={}}\times 
   \frac{4\pi R_0}{g_A^2(0)} \int\frac{\dd^3 q}{(2\pi)^3}\frac{e^{i \vec{q}\cdot \vec{r}_{12}}}{q(q+E_d)}  \Bigg[ h_F(\bm{q}^2) + h_{GT}(\bm{q}^2) \bm{\sigma}_m\cdot\bm{\sigma}_n
    -h_{T}(\bm{q}^2)S^{\bm{q}}_{mn}  \Bigg]\nonumber\\
     &\equiv\sum^A_{m\neq n=1}  
   \frac{4\pi R_0}{g_A^2(0)} \int\frac{\dd^3 q}{(2\pi)^3} e^{i \vec{q}\cdot \vec{r}_{mn}} V^{0\nu}_{mn}(q,E_d),
  \end{align}
  where the neutrino potential $V^{0\nu}_{mn}(q,E_d)$ is defined as
  \beqn
  \label{eq:neutrino_potential_q_Ed}
     V^{0\nu}_{mn}(q,E_d)   \equiv   \dfrac{\tau^+_m \tau^+_n}{q(q+E_d)}
  \Bigg[ h_F(\bm{q}^2) + h_{GT}(\bm{q}^2) \bm{\sigma}_m\cdot\bm{\sigma}_n
    -h_{T}(\bm{q}^2)S^{\bm{q}}_{mn} \Bigg]
  \eeqn
  The transition operator can be written as a summation of Gamow-Teller (GT), Fermi, and tensor parts,
  \beq
  O^{0\nu}=\sum_{\alpha} O^{0\nu}_{\alpha} 
  \eeq
  where the GT, Fermi, and tensor parts are
  \bsub\beqn
  O^{0\nu}_{\rm F} 
   &=& \sum^A_{m\neq n=1}  
   \frac{4\pi R_0}{g_A^2(0)} \int\frac{\dd^3 q}{(2\pi)^3} e^{i \vec{q}\cdot \vec{r}_{mn}} \dfrac{\tau^+_m \tau^+_n}{q(q+E_d)} h_F(\bm{q}^2),\\
   O^{0\nu}_{\rm GT} 
   &=& \sum^A_{m\neq n=1}  
   \frac{4\pi R_0}{g_A^2(0)} \int\frac{\dd^3 q}{(2\pi)^3} e^{i \vec{q}\cdot \vec{r}_{mn}} \dfrac{\tau^+_m \tau^+_n}{q(q+E_d)} h_{GT}(\bm{q}^2)\bm{\sigma}_m\cdot\bm{\sigma}_n,\\ 
   O^{0\nu}_{\rm T} 
   &=& -\sum^A_{m\neq n=1}  
   \frac{4\pi R_0}{g_A^2(0)} \int\frac{\dd^3 q}{(2\pi)^3} e^{i \vec{q}\cdot \vec{r}_{mn}} \dfrac{\tau^+_m \tau^+_n}{q(q+E_d)} h_T(\bm{q}^2) )S^{\bm{q}}_{mn}.
  \eeqn
  \esub
  With the plane wave expansion
  \beq
  e^{i \vec{q}\cdot \vec{r}_{mn}}
  =4\pi\sum_{LM}i^L j_L(qr_{mn})Y^\ast_{LM}(\hat{\bm{q}})Y_{LM}(\hat{\bm{r}}_{mn}),
  \eeq
  and the orthogonality relation
  \beq
  \int d\hat{\bm{q}} Y_{LM}(\hat{\bm{q}})Y^\ast_{L'M'}(\hat{\bm{q}})
  = \delta_{LL'}\delta_{MM'},\quad
  Y_{00}=\dfrac{1}{\sqrt{4\pi}},
  \eeq 
  one finds the expression for the Fermi part
  \begin{eqnarray}
  \label{eq:FermiOperator}
    O^{0\nu}_{F}   
      &=&\frac{4\pi R_0}{g_A^2(0)} \int\frac{d q}{(2\pi)^3}\frac{q^2}{q(q+E_d)}4\pi\sum_{LM}i^L j_L(qr_{mn}) \int d\hat{\bm{q}}  \sqrt{4\pi}\delta_{L0}\delta_{M0}Y_{LM}(\hat{\bm{r}}_{mn})  h_F(\bm{q}^2)\tau^+_m \tau^+_n
     \nonumber\\
      &=&\frac{2R_0}{\pi g_A^2(0)} \int dqq^2 \frac{h_F(\bm{q}^2)}{q(q+E_d)}   j_0(qr_{mn})\tau^+_m \tau^+_n.
  \end{eqnarray}
  Similarly, one can derive an expression for the GT part.  The tensor part reads,
   \begin{eqnarray}
     O^{0\nu}_{T} 
   &=&  -\frac{4\pi R_0}{g_A^2(0)} \int\frac{\dd^3 q}{(2\pi)^3}\frac{h_{T}(\bm{q}^2)}{q(q+E_d)}
   4\pi\sum_{LM}i^L j_L(qr_{mn})Y^\ast_{LM}(\hat{\bm{q}})Y_{LM}(\hat{\bm{r}}_{mn})S^{\bm{q}}_{mn}.\nonumber\\ 
  \end{eqnarray}
   
    By rewriting the spin-tensor operator in Eq.(\ref{eq:spin_tensor_operator}) into the following coupled form~\cite{Varshalovich:1988},
  \beq
  S^{\bm{q}}_{mn} =3\{ \bm{\sigma}_1\otimes  \bm{\sigma}_2\}_2\cdot  \{ \hat{\bm{q}}\otimes \hat{\bm{q}}\}_2.
  \eeq
  one can simplify the tensor operator further
   \begin{eqnarray}
    O^{0\nu}_T  
   &=&  -3 \frac{4\pi R_0}{g_A^2(0)} \int \frac{q^2dq}{(2\pi)^3}\dfrac{h_{T}(\bm{q}^2)}{q(q+E_d)}
   4\pi\sum_{LM}\sum_{\mu}(-1)^{\mu}i^L j_L(qr_{mn})Y_{LM}(\hat{\bm{r}}_{mn})   \{\bm{\sigma}_m\otimes  \bm{\sigma}_n\}_{2\mu}
   \int d\hat{\bm{q}} Y^\ast_{LM}(\hat{\bm{q}})
   \{\hat{\bm{q}}\otimes \hat{\bm{q}}\}_{2-\mu} \nonumber\\ 
     &=& -3 \frac{4\pi R_0}{g_A^2(0)} \int \frac{q^2dq}{(2\pi)^3}\dfrac{h_{T}(\bm{q}^2)}{q(q+E_d)}
   4\pi \sum_{\mu}(-1)^{\mu}i^2 j_2(qr_{mn})   \{\bm{\sigma}_m\otimes  \bm{\sigma}_n\}_{2\mu}
   \{\hat{\bm{r}}_{mn}\otimes \hat{\bm{r}}_{mn}\}_{2-\mu} \nonumber\\
    &=&   \frac{2 R_0}{\pi g_A^2(0)} \int q^2 dq \dfrac{h_{T}(\bm{q}^2)}{q(q+E_d)}
      j_2(qr_{mn})  S^{\bm{r}}_{mn},
  \end{eqnarray}
  where the minus sign is canceled by $i^2=-1$. Moreover, we have introduced the spin-tensor operator in coordinate space 
  \beqn
  S^{\bm{r}}_{mn}
  \equiv 3\{\bm{\sigma}_m\otimes  \bm{\sigma}_n\}_{2}\cdot
   \{\hat{\bm{r}}_{mn}\otimes \hat{\bm{r}}_{mn}\}_2
  =  3(\bm{\sigma}_m\cdot\hat{\bm{r}}_{mn}) (\bm{\sigma}_n\cdot\hat{\bm{r}}_{mn}) -\bm{\sigma}_m\cdot\bm{\sigma}_n
  \eeqn
  and used the relation \cite{Varshalovich:1988}
  \beq
  \{\hat{\bm{q}}\otimes \hat{\bm{q}}\}_{2-\mu} 
  =\sqrt{\dfrac{4\pi}{5}}\langle 1010|20\rangle Y_{2-\mu}(\hat{\bm{q}}),\quad
  \{\hat{\bm{r}}\otimes \hat{\bm{r}}\}_{2-\mu} 
  =\sqrt{\dfrac{4\pi}{5}}\langle 1010|20\rangle Y_{2-\mu}(\hat{\bm{r}}).
  \eeq
  Adding all the three terms together, one obtains the transition operator in coordinate space
  \beqn
  \label{eq:neutrino_potential_operator}
  O^{0\nu} 
  &=&    \sum^A_{m\neq n=1}\tau^+_m\tau^+_n\Bigg[H^{0\nu}_{F,0}(r_{mn}, E_d)
  +  H^{0\nu}_{GT,0 }(r_{mn}, E_d) \vec{\sigma}_m\cdot\vec{\sigma}_n 
 +H^{0\nu}_{T,2}(r_{mn}, E_d)S^{\bm{r}}_{mn}\Bigg],
\eeqn
where the neutrino potentials in coordinate space are defined as
\begin{align}
\label{eq:neutrino_potential_r}
 H^{0\nu}_{\alpha,L}(r_{mn}, E_d)&=
\dfrac{2R_0}{\pi g^2_A(0)} \int^\infty_0 \dd q \, q^2 \dfrac{ h_\alpha (\bm q^2)}{q(q+E_d)} j_L(qr_{mn}), 
\end{align}
with $\alpha$ being the index for F, GT, or T, respectively. The function $j_L(qr_{12})$ is the spherical Bessel function of rank $L$ with $L=0$ for the Fermi and GT terms, and $L=2$ for tensor term.

 \subsubsection{The neutrino potentials from chiral EFT}
 
The disadvantage of deriving the transition operators based on the Lorentz symmetry of the vertices in  (\ref{eq:general_weak_Hamiltonian}) and the phenomenological currents in (\ref{eq:effect_nucleon_current}) is that the error introduced by the transition operators cannot be estimated systematically. The EFT provides a model-independent way to construct  both strong and weak currents  consistently for low-energy nuclear physics based on symmetry consideration and power counting. In the chiral EFT, only $\pi$ mesons, nucleons, and neutrinos are the relevant degrees of freedom, and all possible interactions or operators fulfilling the symmetry requirements are expanded in the power of $(Q, m_\pi)/\Lambda_\chi$, where $Q$ is the typical momentum scale in the process.  Each term comes with a LEC absorbing the nonperturbative effect of QCD. For the nuclear strong interaction, one can see the recent review papers \cite{Epelbaum:2009RMP,Machleidt:2011PR,Hebeler:2021PR}. With the chiral EFT, Cirigliano et al. derived the neutrino potentials of $\znubb$ decay  based on light-neutrino exchange mechanism up to the N$^2$LO terms~\cite{Cirigliano:2018PRC}.

 \begin{figure}[tb]
\centering 
\includegraphics[width=12cm]{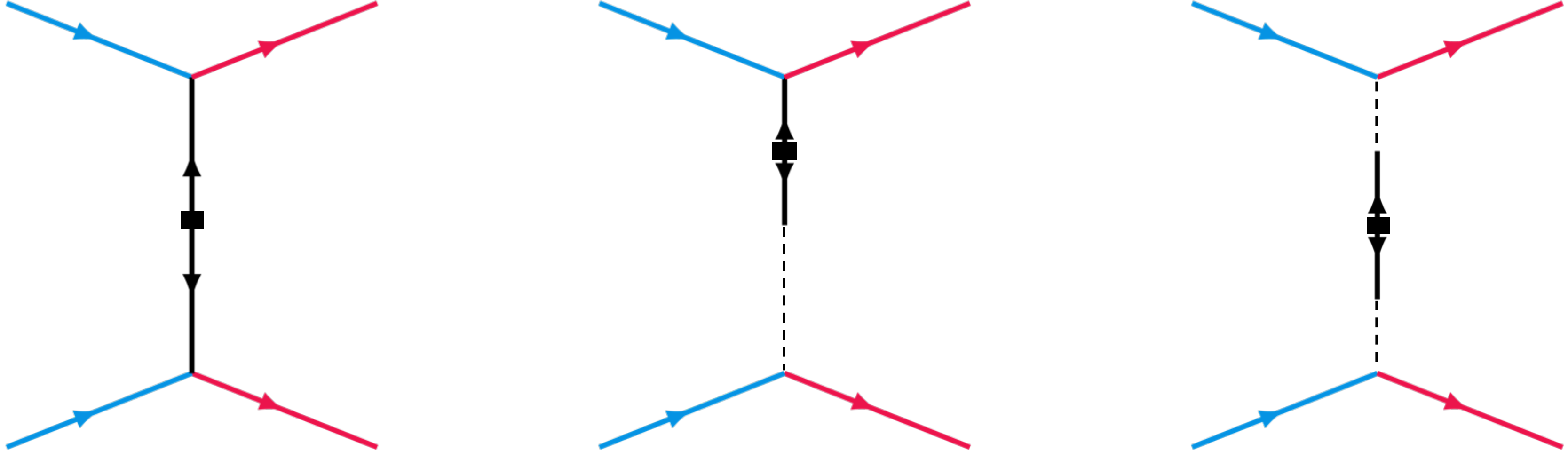}  \caption{\label{fig:cartoon_dbd_LR_LO} The leading-order long-range contributions to the neutrino potentials of   $0\nu\beta\beta$ decay via a direct neutrino exchange between neutrons or via intermediate $\pi$. The square represents an insertion of the neutrino mass $m_{\beta\beta}$ term. The electron lines are omitted.}
\end{figure}

At the LO level, the $\znubb$ transition is simply given by the tree-level neutrino exchange between two nucleons, mediated by the single-nucleon vector and axial currents, as shown in Fig.~\ref{fig:cartoon_dbd_LR_LO}. The corresponding neutrino potential with the pion field integrated out reads \cite{Cirigliano:2018PRC}
 \beqn
 \label{eq:LO_neutrino_potenital_chiral_EFT}
V^{\rm LO}_{mn,  {\rm LR}}(q)
&=& \frac{\tau_{m}^+ \tau_{n}^+}{\boldsymbol{q}^{2}}\left[1-g_{A}^{2} \boldsymbol{\sigma}_m \cdot \boldsymbol{\sigma}_n
+g_{A}^{2} \left(\boldsymbol{\sigma}_{m} \cdot \boldsymbol{q}\right) \left(\boldsymbol{\sigma}_{n} \cdot \boldsymbol{q}\right) \frac{2 m_{\pi}^{2}+\boldsymbol{q}^{2}}{\left(\boldsymbol{q}^{2}+m_{\pi}^{2}\right)^{2}}\right]\nonumber\\
&=& \frac{\tau_{m}^+ \tau_{n}^+}{\boldsymbol{q}^{2}}\left\{1-\frac{2 g_{A}^{2}}{3} \boldsymbol{\sigma}_m \cdot \boldsymbol{\sigma}_n\left[1+\frac{m_{\pi}^{4}}{2\left(\boldsymbol{q}^{2}+m_{\pi}^{2}\right)^{2}}\right] 
-\frac{g_{A}^{2}}{3} S^{\boldsymbol{q}}_{mn}\left[1-\frac{m_{\pi}^{4}}{\left(\boldsymbol{q}^{2}+m_{\pi}^{2}\right)^{2}}\right]\right\},
\eeqn
where the spin tensor operator $S^{\boldsymbol{q}}_{mn}$ is defined  in Eq.(\ref{eq:spin_tensor_operator}).  One can see that the $V^{\rm LO}_{mn, {\rm LR}}(q)$ is consistent with the {\em standard} neutrino potential $V^{0\nu}_{mn}(q,E_d=0)$ in (\ref{eq:neutrino_potential_q_Ed}) if the induced pseudoscalar and weak-magnetism terms are included.

 The transition operator (\ref{eq:LO_neutrino_potenital_chiral_EFT}) of the form $1/\bm{q^2}$ has a Coulomb-like behavior at large $\boldsymbol{q}^2$, which has a singularity at the short-distance with $\vert \bm{r}_{mn}\vert \to 0$. It causes ultraviolet divergences in the transition amplitude of the process $nn\to ppe^-e^-$.  Cirigliano et al. \cite{Cirigliano2018PRL,Cirigliano:2019PRC} studied the  transition amplitude ${\cal A}^{0\nu}$ for this process with both the two neutrons in the initial state and the two protons in the final state being in the $^1$S$_0$ channel. The LO nuclear strong interaction and LO neutrino potentials (\ref{eq:LO_neutrino_potenital_chiral_EFT}) were used. They have shown that a contact operator  
\begin{equation}
\label{eq:contact_term}
    V^{\rm LO}_{mn, {\rm CT}}(q,0) 
    =-2g^{NN}_{\nu} \tau_m^+ \tau_n^+
\end{equation}
has to be introduced  at  LO  to remove the divergence and the regulator dependence of the transition amplitude ${\cal A}^{0\nu}$.  It needs to be clarified that this LO contact operator (\ref{eq:contact_term}) is different from those contributed from the non-standard mechanism of dimension-9 weak interactions~\cite{Pas:2001PLB,Deppisch:2012JPG,Graesser:2016,Deppisch:2018PRD,Graf:2018,Cirigliano:2018JHEP,Liao:2020JHEP}. Instead, this contact operator is still within the standard mechanism. The finite part contributed from the LO contact operator encodes the contribution from the exchange of light neutrino with momentum larger than the nuclear energy scale. \jmyr{The  LEC $g^{NN}_{\nu}$ of this contact operator is {\em scale} and {\em scheme} dependent, where the {\em scheme} refers to the type of nuclear forces (phenomenological or derived from chiral EFT), and the {\em scale} refers to the regulation cutoff. Therefore, the LEC has to be determined by matching the related observable (like transition amplitude) by chiral EFT to the data if exists or to the value determined by a more fundamental theory (like lattice QCD).  A framework that enables this matching was developed recently~\cite{Davoudi:2020gxs}. In particular, Cirigliano et al. \cite{Cirigliano:2021PRL,Cirigliano:2021qko} developed a method to estimate the LEC in analogy to the Cottingham formula~\cite{Cottingham:1963,Harari:1966}.  They provide the value of the full transition amplitude at a given kinematic point. This amplitude
is (in principle) observable, and can be used as a synthetic datum to constrain the contact term in other regularization schemes of nuclear chiral forces employed in nuclear-structure calculations.} With their synthetic datum, Wirth et al.~\cite{Wirth:2021} evaluated the contribution of  this LO contact operator to the NMEs of light nuclei \nuclide[6,8]{He} and the candidate nucleus \nuclide[48]{Ca} starting from a chiral two-plus-three nucleon interaction EM1.8/2.0~\cite{Hebeler:2011}. It has been found that the contact term enhances the NME by 43(7) \% in $\nuclide[48]{Ca}$ in the IM-GCM calculation, where the uncertainty is propagated from the synthetic datum. The large impact of the short-range operator was also found in a recent study by Jokiniemi et al.~\cite{Jokiniemi:2021} who evaluated its contribution to the NME of heavy candidate nuclei with conventional shell model and QRPA using phenomenological nuclear forces.   

The NLO terms are given by the diagrams involving the first derivative of the pion field, which do not contribute to the NME of $0\nu\beta\beta$ decay ($0^+ \to 0^+$) due to parity conservation. The N$^2$LO terms in the Weinberg power counting consist of two types of contributions \cite{Cirigliano:2018PRC}
\begin{itemize}
    \item corrections to the single-nucleon current, which are often included via momentum-dependent dipole form factors $g_i(\bm{q}^2)$, cf. (\ref{eq:dipole_form_factors_gA})-(\ref{eq:dipole_form_factors_gM}),
     \item genuine N$^2$LO two-body effects from loop corrections, which induce the short-range neutrino potential that cannot be absorbed by the one-body weak currents (\ref{eq:effect_nucleon_current}),  and have seldom been considered.
     Determination of  these loop corrections completely is challenging as it requires the information of intermediate states to ensure renormalization conditions \cite{Cirigliano:2018PRC}. Parts of these corrections were estimated in  a recent {\em ab initio} calculation \cite{Pastore:2018} and they turn out to be varying from sub-percent level to  ${\mathcal O}$(10\%) of the LO term.
\end{itemize}

\subsubsection{The chiral two-body nucleon currents}
\label{subsubsec:two-body-currents}

 \begin{figure}[t]
\centering 
\includegraphics[width=8cm]{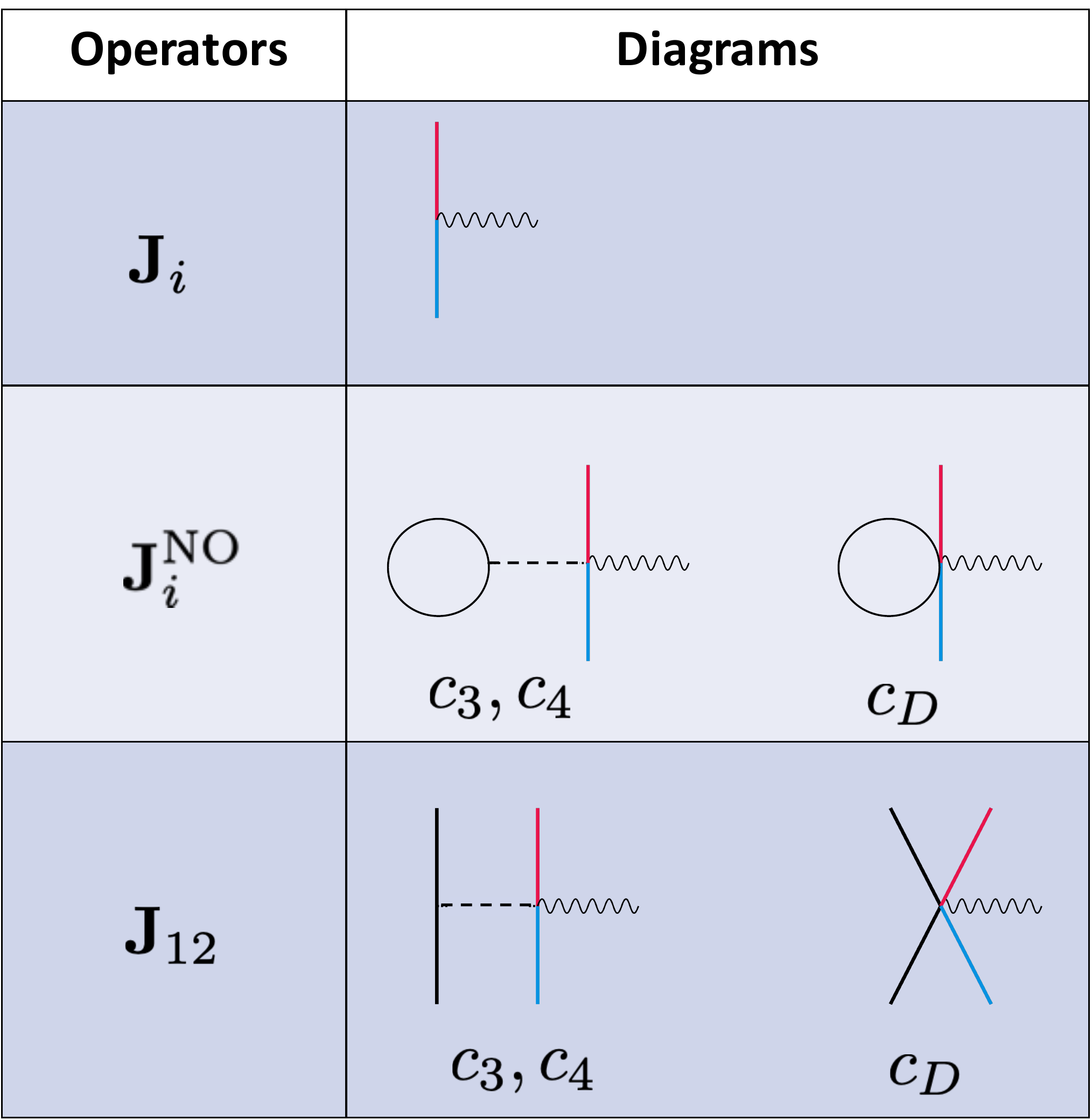}  
\caption{\label{fig:cartoon_two_body_currents} Schematic pictures for the one  and two-body  weak current operators $\bm{J}_i$ and $\bm{J}_{12}$.   By choosing a reference state which defines the density $\rho$, one can normal-order the two-body current into an effective one-body current operator $\bm{J}^{\rm NO}_i$.  The wiggly line stands for the weak field (current) and the dashed line for the  $\pi$. See text for details.}
\end{figure}

The two-body nucleon currents which induce three-nucleon interactions appear at N$^3$LO in the chiral EFT. The two-body currents contain the contributions from one-pion-exchange term depending on the LECs $c_{3}, c_{4}$  and short-range coupling term depending on the LECs $d_{1}, d_{2}$, where only the combination  $d_{1}+2 d_{2}=c_{D} /\left(g_{A} \Lambda_{\chi}\right)$ with  $\Lambda_{\chi}\simeq 700$ MeV is a free parameter due to the Pauli principle~\cite{Park:2003,Menendez:2011}, as shown in Fig.~\ref{fig:cartoon_two_body_currents}. Its expression has been derived by several groups~\cite{Park:2003,Krebs:2017AP,Krebs:2020EPJA,Baroni:2021}. The two-body currents can be generally written as
\begin{equation}
\label{eq:tbc_total}
\bm{J}_{2B}
= \sum_{1<2}^{A} \bm{J}_{12},
\end{equation}
where $\bm{J}_{12}$ in momentum space has the following form~\cite{Park:2003,Menendez:2011}
\beqn
\label{eq:tbc_momentum}
\bm{J}_{12}
&=&-\frac{g_{A}}{2 f_{\pi}^{2}} \frac{1}{m_{\pi}^{2}+\bm{k}^{2}}
\left[-\frac{i}{2 m_p} \bm{k} \cdot\left(\bm{\sigma}_{1}-\bm{\sigma}_{2}\right) \bm{p} \tau_{\times}^{+}
+4 c_{3} \bm{k} \cdot\left(\bm{\sigma}_{1} \tau_{1}^{+}+\bm{\sigma}_{2} \tau_{2}^{+}\right) \bm{k}
+\left(c_{4}+\frac{1}{4 m_p}\right) \bm{k} \times\left(\bm{\sigma}_{\times} \times \bm{k}\right) \tau_{\times}^{+}
\right]\nonumber\\
&& 
+\frac{g_{A}}{4f_{\pi}^{2}}\left[2 d_{1}\left(\bm{\sigma}_{1} \tau_{1}^{+}+\bm{\sigma}_{2} \tau_{2}^{+}\right)+d_{2} \bm{\sigma}_{\times} \bm{\tau}_{\times}^{+}\right], 
\eeqn
where $f_{\pi}=92.4$ MeV  is the pion decay constant,  and $m_p$ is the nucleon mass. The compound spin and isospin operators are constructed as
\bsub\beqn
 \bm{\sigma}_{\times}&=&\bm{\sigma}_{1} \times \bm{\sigma}_{2},\\ 
 \bm{\tau}_{\times}^{+} &=& \left(\bm{\tau}_{1} \times \bm{\tau}_{2}\right)^{+}.
\eeqn 
\esub
and the momenta
\bsub
\beqn
 \bm{k}&=&\frac{1}{2} (\bm{p}_{2}^{\prime}-\bm{p}_{2}-\bm{p}_{1}^{\prime}+\bm{p}_{1}),\\  \bm{p}&=&\frac{1}{4}(\bm{p}_{1}+\bm{p}_{1}^{\prime}-\bm{p}_{2}-\bm{p}_{2}^{\prime}).
\eeqn
\esub

After Fourier transformation, the two-body currents can be written in coordinate space~\cite{Park:2003,Wang2018}
\beqn
\label{eq:tbc_coordinate}
\bm{J}_{12}(\boldsymbol{x})
&=& \frac{2 c_{3} g_{A}}{m_pf_{\pi}^{2}}
\Bigg\{m_{\pi}^{2}\left[\left(\frac{\bm{\sigma}_{2}}{3}-\bm{\sigma}_{2} \cdot \hat{\bm{r}} \hat{\bm{r}}\right) Y_{2}(r)
-\frac{\bm{\sigma}_{2}}{3} Y_{0}(r)\right]+\frac{\bm{\sigma}_{2}}{3} \delta(\bm{r})\Bigg\} \tau_{2}^{+} \delta\left(\bm{x}-\bm{r}_{2}\right)+(1 \leftrightarrow 2) \nonumber\\
&&+\left(c_{4}+\frac{1}{4}\right) \frac{g_{A}}{2 m_{p} f_{\pi}^{2}}\left\{m_{\pi}^{2}\left[\left(\frac{\bm{\sigma}_{\times}}{3}-\bm{\sigma}_{1} \times \hat{\bm{r}} \sigma_{2} \cdot \hat{\bm{r}}\right) Y_{2}(r)-\frac{\bm{\sigma}_{\times}}{3} Y_{0}(r)\right]+\frac{\bm{\sigma}_{\times}}{3} \delta(\bm{r})\right\} \tau_{\times}^{-} \delta\left(\bm{x}-\bm{r}_{1}\right)+(1 \leftrightarrow 2) \nonumber\\
&&-\frac{g_{A}}{4 m_{p} f_{\pi}^{2}}\left[2 d_{1}\left(\bm{\sigma}_{1} \tau_{1}^{+}+\bm{\sigma}_{2} \tau_{2}^{+}\right)+d_{2} \bm{\sigma}_{\times} \tau_{\times}^{+}\right] \delta(\bm{r}) \delta\left(\bm{x}-\bm{r}_{1}\right),
\eeqn
 where the Yukawa functions are  defined as
\begin{equation}
    Y_{0}(r)=\frac{e^{-m_{\pi} r}}{4 \pi r},\quad Y_{2}(r)=\frac{1}{m_{\pi}^{2}} r \frac{\partial}{\partial r} \frac{1}{r} \frac{\partial}{\partial r} Y_{0}(r),
\end{equation} 
with $\hat{\bm{r}} \equiv \bm{r}/r$ and $\bm{r}=\bm{r}_{1}-\bm{r}_{2}$. It has been pointed out in Ref.~\cite{Park:2003} that the two-body currents $\hat{\bm{J}}_{2B}$ are valid only up to a certain cutoff in momentum space and one needs to regularize the $\delta$-function in (\ref{eq:tbc_coordinate}) with for instance a Gaussian function.

 \begin{figure}[t]
\centering 
\includegraphics[width=10cm]{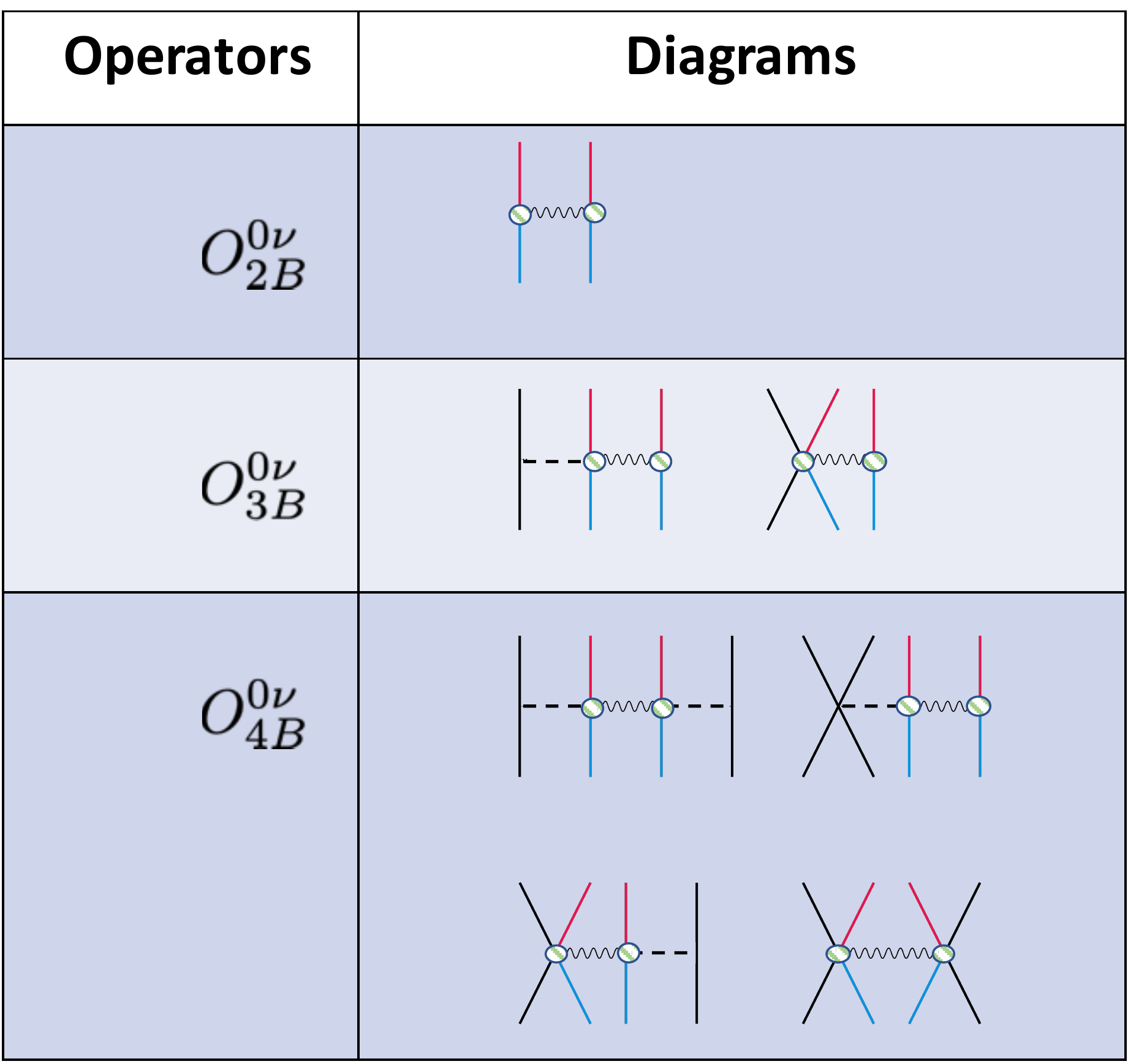}  
\caption{\label{fig:cartoon_many_body_transition_operators} The leading contribution from two-body, three-body and four-body transition operators generated by products of the one-body and two-body weak currents in Fig.~\ref{fig:cartoon_two_body_currents}. The wiggly line stands for the neutrino potential and the dashed line for the  $\pi$. All the Goldstone diagrams discussed in Ref.~\cite{Wang2018} are not given here.}
\end{figure}

 The presence of the two-body currents, together with the one-body current operator (\ref{eq:effect_nucleon_current}), generates additional three-body and four-body transition operators contributing to $\znubb$ decay, as diagrammatically described in Fig.~\ref{fig:cartoon_many_body_transition_operators}. The calculation of the NME using higher-body transition operators is computationally challenging. For simplicity,  Men\'endez et al.~\cite{Menendez:2011} derived an effective density-dependent one-body axial-vector current $\bm{J}_{i, A}^{\rm NO}$  by normal-ordering the chiral two-body currents in momentum space with respect to the non-interacting ground state of spin- and isospin-symmetric nuclear matter (S.N.M.) \footnote{It was suggested in Ref.~\cite{Gazit:2019PRL_Erratum} that a factor of $-1/4$ is multiplied to the LEC $c_D$.}
\beqn
\label{eq:effective_1b_current}
\bm{J}_{i, A}^{\rm NO}({\rm S.N.M.})
&=&g_{A} \bm{\sigma}_{i} \bm{\tau}_{i}^{+} \frac{\rho}{f_{\pi}^{2}}\Bigg[\jmyr{-\frac{c_{D}}{4}}\frac{1}{g_{A} \Lambda_{\chi}}
+\frac{2c_{3}}{3}  \frac{q^{2}}{4 m_{\pi}^{2}+q^{2}}
+\frac{I(\rho, Q)}{3}\left(2 c_{4}-c_{3}+\frac{1}{2m_p}\right)\Bigg],
\eeqn
where  $\rho=2 k_{F}^{3} /\left(3 \pi^{2}\right)$  is the density of the reference state with  $k_{F}$  the corresponding Fermi momentum, $\bm{q}=\bm{p}_i-\bm{p}'_i$, $\bm{Q}=\bm{p}_i+\bm{p}'_i$. The  $I(\rho, Q)$ is due to the summation in the exchange term~\cite{Menendez:2011,Menendez:2012PRD},
\beqn
\label{eq:I-factor}
I(\rho, Q) 
&=& 1-\frac{3 m_{\pi}^{2}}{k_{F}^{2}}+\frac{3 m_{\pi}^{3}}{2k_{F}^{3}} \operatorname{cot}^{-1}\left[\frac{m_{\pi}^{2}+Q^{2} / 4-k_{F}^{2}}{2 m_{\pi} k_{F}}\right] +\frac{3 m_{\pi}^{2}}{4 k_{F}^{3} Q}\left(k_{F}^{2}+m_{\pi}^{2}-\frac{Q^{2}}{4}\right) \ln \left[\frac{m_{\pi}^{2}+\left(k_{F}-Q / 2\right)^{2}}{m_{\pi}^{2}+\left(k_{F}+Q/ 2\right)^{2}}\right].
\eeqn

We note that even in nuclear {\em ab initio} studies the impact of the two-body currents on the $\znubb$-decay NME is actually   {\em  scale} and {\em scheme} dependent. Realistic nuclear forces are usually formulated at a high-energy  scale.   The technique of SRG was usually introduced to soften the realistic nuclear forces~\cite{Bogner:2010}. The evolved nuclear force with a lower-energy  scale is prone to speed up the convergence rate of many-body approaches. In this framework, the $\znubb$-decay operator has to be evolved accordingly.  However, the evolution of one-body weak current operator down to a lower-energy scale will induce higher-body current operators~\cite{Bogner:2010}, bringing an additional contribution to the bare two-body weak currents.  In the studies with conventional nuclear models, the situation is more complicated as  phenomenological nuclear interactions with parameters fitted directly to the properties of finite nuclei are usually adopted. In this framework,  the contributions from the bare and induced two-body currents are messed up in which case an effective two-body current operator is usually introduced with the parameters adjusted directly to corresponding data. For the $\znubb$ decay, this idea does not work as there is no related data.  Therefore, to draw a more solid conclusion on the impact of the two-body currents on the NME of $\znubb$ decay, one needs to adopt the nuclear interaction and currents derived consistently from the same framework such as the chiral EFT.  In a consistent chiral EFT, however, the contribution of the two-body currents will be divergent and it must be renormalized by introducing another contact operator~\cite{Wang2018}. The corresponding LEC has to be determined by a more fundamental non-perturbative theory, such as lattice QCD.

\subsubsection{The  value of axial-vector coupling strength $g_A$}
\label{subsubsec:gA_problem}

The $0\nu\beta\beta$ decay is dominated by the GT-type transition, the matrix element of which is proportional to the square of the axial-vector coupling strength  $g_A$. Thus, the value of $g_A$ plays a decisive role in the half-life of $0\nu\beta\beta$ decay, as shown in Eq.(\ref{eq:0nubb-half-life}),  and it has a direct impact on the deduced upper limit of the effective neutrino mass $|\langle m_{\beta\beta}\rangle|$, c.f. Eq.(\ref{eq:effective-mass}).  The coupling strengths $g_i$ appear in Eq.(\ref{eq:effect_nucleon_current}) when the quark currents are expressed in terms of nucleons. If nature respected chiral symmetry exactly, the $g_V$ and  $g_A$ would be exactly one. However, chiral symmetry is explicitly broken by nonzero quark masses and is spontaneously broken by the QCD vacuum. The conservation of vector current  hypothesis~\cite{Feynman:1958} and partially conserved axial-vector current~\cite{Nambu:1960PRL,Gell-Mann:1960} yield the free-nucleon $g_V=1.0$ and  $g_A=1.27$, consistent with the measured value $g_{A}^{\rm exp}=1.2723(23)$~\cite{PDG:2016} and the value $g_{A}^{\rm LQCD}=1.271 (13)$ from a lattice QCD calculation~\cite{Chang:2018nature} for free-space neutron $\beta$ decays.

In atomic nuclei, many-body currents may modify the coupling strengths. Because of charge conservation, the $g_V$ remains unchanged. In contrast,  many-body currents bring an additional contribution to the GT-type operator, renormalizing the free-space $g_A$ if only a one-body axial-vector current is employed. Therefore, an effective axial-vector coupling strength $g^{\rm eff}_A$ must be introduced to take into account this contribution. As discussed in Sec.\ref{subsubsec:two-body-currents}, the contribution of two-body currents is scale and scheme-dependent, and so is the $g^{\rm eff}_A$. Besides, a variety of truncations are employed in modeling atomic nuclei. The contribution from the outside of the model space should be included as well in the  effective one-body current and this contribution will renormalize further the $g^{\rm eff}_A$. In other words, the $g^{\rm eff}_A$ value to be adopted is expected to vary from the employed nuclear models with different truncations. 

For single-$\beta$ decay and $2\nu\beta\beta$ decay, only the $g_A$ at zero momentum transfer matters. The effective value $g^{\rm eff}_A$ is determined by fitting the results of each nuclear model to corresponding data. It is defined in terms of a quenching factor $\mathcal{Q}_A$~\cite{Wilkinson:1973NPA},
\beq
\mathcal{Q}_A \equiv g^{\rm eff}_A/g_A  = 1-\delta \mathcal{Q}_A
\eeq
In the studies for nuclear single-$\beta$ decays with valence-space shell models,  the quenching factor $\delta\mathcal{Q}_A$ turns out to be increasing systematically with nuclear mass number $A$, i.e. $\delta\mathcal{Q}_A(0)\simeq19\%$ for the $p$ and $sd$-shell nuclei~\cite{Brown:1985,Chou:1993,Siiskonen:2001PRC} and 23\% in the $fp$-shell nuclei~\cite{Pinedo:1996PRC,Siiskonen:2001PRC}, 44\% and 59\% for  heavier nuclei such as \nuclide[56]{Ni} and \nuclide[100]{Sn}~\cite{Siiskonen:2001PRC}.  The value of $g^{\rm eff}_A$  was also extracted from the ISM~\cite{Caurier:2007IJMPE}, QRPA~\cite{Suhonen:2013PLB} and IBM~\cite{Barea:2013,Barea:2015} calculations for the NMEs of $2\nu\beta\beta$ decay. To reproduce available data, the optimized $g^{\rm eff}_A$ values for different nuclei were parametrized as below,
 \bsub\beqn
 g^{\rm eff}_A&=&1.269 A^{-0.12},\quad {\rm ISM},\\
 g^{\rm eff}_A&=&1.269 A^{-0.18},\quad {\rm IBM}.
 \eeqn
 \esub
 A detailed introduction to the determination of the $g^{\rm eff}_A$ value from the data of single-$\beta$ decay and $2\nu\beta\beta$ decay can be found for instance in the Review~\cite{Suhonen:2017FP}.

Different from the single-$\beta$ decay and $2\nu\beta\beta$ decay, the $0\nu\beta\beta$ decay is dominated by the contribution at high-momentum transfer ($q\sim 100-200$ MeV~\cite{Menendez:2011,Engel:2014}). Therefore, the $g_A$ is expected to be renormalized differently. Since the half-life of $0\nu\beta\beta$ decay is proportional to the fourth power of $g_A$, different choice of the $g^{\rm eff}_A$ value leads to significant uncertainty in the determined neutrino mass. Because of no data on $0\nu\beta\beta$ decay, it is essential to understand all the possible sources that may renormalize the $g_A$ in each nuclear model and finally to determine the $g^{\rm eff}_A$ of the employed nuclear model for $0\nu\beta\beta$ decays. This has been a long-standing issue in nuclear physics. See for instance the review papers~\cite{Towner:1987PR,Suhonen:2017FP,Engel:2017,Ejiri:2019PR} and references therein. 
 
  The need of a quenched $g^{\rm eff}_A$ for nuclear single-$\beta$ decay, $2\nu\beta\beta$ decay and other related processes has been attributed to the missing of contributions from nuclear many-body correlations~\cite{Towner:1987PR,Bertsch:1982,Brown:1988,Siiskonen:2001PRC,Holt:2013} and many-body currents~\cite{Towner:1987PR,Park:1997PLB,Ney:2021}. However, it is still not clear which one dominates the observed quenching effect.  We note the recent studies by Coraggio et al.~\cite{Coraggio:2017PRC,Coraggio:2019PRC,Coraggio:2020} with the realistic shell model starting from the CD-Bonn NN potential. It has been demonstrated that the empirical quenching of the $g_A$ can be obtained from the renormalization procedure used to derive the effective shell-model Hamiltonian and decay operators without the introduction of meson-exchange currents~\cite{Coraggio:2019PRC}.  In particular,  the renormalization effect of the  $0\nu\beta\beta$-decay operator is found to be less relevant than that for single-$\beta$ and $2\nu\beta\beta$ decays~\cite{Coraggio:2020}.  Of course, it should be pointed out that most of the previous discussions were carried out within the conventional nuclear models based on either phenomenological nuclear interactions and/or currents, where systematic uncertainty is difficult to assess.  As discussed before, the chiral EFT provides a framework to derive consistently nuclear interactions and electroweak currents up to different order. With the contribution of the chiral two-body currents (\ref{eq:tbc_total}), the one-body axial-vector current will be renormalized 
  \beqn
  \bm{J}_{1b, A}^{\rm eff}(q, Q, \rho) 
  = \bm{J}_{1b, A}+\bm{J}_{1b, A}^{\rm NO} 
  \equiv   \sum_i \bm{J}_{i,A}^{\rm eff}(q, Q, \rho),
  \eeqn
  where the $\bm{J}_{1b, A}^{\rm NO}$ is the contribution from two-body currents obtained with the normal-ordering approximation, see Eq.(\ref{eq:effective_1b_current}).  The above effective one-body current $\bm{J}_{i,A}^{\rm eff}$ defines an effective coupling strength $g^{\rm eff}_A$ depending on the quantities $(q, Q, \rho)$ in the following way 
 \beqn
\bm{J}_{i,A}^{\rm eff}(q, Q, \rho)
&\equiv & - g^{\rm eff}_{A}(q, Q, \rho) \bm{\sigma}_i \bm{\tau}^{+}_i \nonumber\\
&=& -g_{A}  \bm{\sigma}_i \bm{\tau}^{+}_i  \left\{1-\frac{\rho}{f_{\pi}^{2}}\Bigg[-\frac{c_{D}}{4}\frac{1}{g_{A} \Lambda_{\chi}}
+\frac{2c_{3}}{3}  \frac{q^{2}}{4 m_{\pi}^{2}+q^{2}}
+\frac{I(\rho, Q)}{3}\left(2 c_{4}-c_{3}+\frac{1}{2m_p}\right)\Bigg]\right\}.
\eeqn
For the special case of zero center-of-mass momentum $Q=0$, the effective axial-vector coupling strength $g_{A}^{\rm eff}(q, Q, \rho)$ is simplified as~\cite{Engel:2014}
 \beqn
 \label{eq:effective_gA_q}
g^{\rm eff}_A(q, 0, \rho) 
\equiv g_{A} \left\{1-\frac{\rho}{f_{\pi}^{2}}\Bigg[-\frac{c_{D}}{4}\frac{1}{g_{A} \Lambda_{\chi}}
+\frac{2c_{3}}{3}  \frac{q^{2}}{4 m_{\pi}^{2}+q^{2}}
+\frac{I(\rho, 0)}{3}\left(2 c_{4}-c_{3}+\frac{1}{2m_p}\right)\Bigg]\right\}.
\eeqn
The $I(\rho, 0)$ defined in Eq.(\ref{eq:I-factor}) becomes~\cite{Menendez:2012PRD,Klos:2013PRD}
\beqn
I(\rho, 0) &=& 1-\frac{3 m_{\pi}^{2}}{k_{F}^{2}}
+\frac{3 m_{\pi}^{3}}{ 2k_{F}^{3}} \operatorname{tan}^{-1}\left[\frac{ 2k_F/m_{\pi}}{1-(k_F/m_{\pi})^2}\right]\nonumber\\
&=&1-\frac{3 m_{\pi}^{2}}{k_{F}^{2}}+\frac{3 m_{\pi}^{3}}{ k_{F}^{3}} \operatorname{tan}^{-1}\left(\frac{ k_F}{m_{\pi}}\right),
\eeqn
where the following relation is used,
\beq
\operatorname{tan}(2x)=2\frac{\operatorname{tan}(x)}{1-\operatorname{tan}^2(x)},\quad x = \operatorname{tan}^{-1}\left(\frac{ k_F}{m_{\pi}}\right).
\eeq

From Eq.(\ref{eq:effective_gA_q}), one immediately finds the expression for the quenching factor $\mathcal{Q}_A(q, \rho)$,
\beqn
\label{eq:Q-factor-chiral}
\mathcal{Q}_A(q, \rho) \equiv g^{\rm eff}_A(q, 0, \rho) /g_A 
= 1 + A[q]\rho + B\rho^{1/3} + C,
\eeqn
where the coefficients $A, B$ and $C$ are defined as
\bsub
\begin{align}
 A[q] &=   \frac{c_{D}}{4f^2_\pi}\frac{1}{g_{A} \Lambda_{\chi}}  -
 \frac{1}{3f^2_\pi}  \left[ \left(2 c_{4}-c_{3}+\frac{1}{2m_p}\right) 
 + 2c_3  \frac{q^{2}}{4 m_{\pi}^{2}+q^{2}}\right],\\
 B & = \frac{m^2_\pi}{f^2_\pi} \left(\frac{2}{3\pi^2}\right)^{2/3} \left(2 c_{4}-c_{3}+\frac{1}{2m_p}\right),\\
 C & = -\frac{2m^3_\pi}{3\pi^2f^2_\pi}\left(2 c_{4}-c_{3}+\frac{1}{2m_p}\right)  \operatorname{tan}^{-1}\left(\frac{ k_F}{m_{\pi}}\right).
\end{align}
\esub
Note that the coefficient $C$ also depends on the density $\rho$ in terms of $k_F=(\frac{3\pi^2}{2})^{1/3}\rho^{1/3}$ which changes from $1.06$ fm$^{-1}$ to $1.33$ fm$^{-1}$ for the density $\rho$ increasing from $0.08$ fm$^{-3}$ to $0.16$ fm$^{-3}$. For the case of zero momentum transfer $q=0$, the effective axial-vector coupling strength only depends on the density $\rho$ of the reference state with respect to which the two-body current is normal-ordered, and the coefficient $A[0]$ is   
 \beq
 A[0] =   \frac{c_{D}}{4f^2_\pi}\frac{1}{g_{A} \Lambda_{\chi}}  -
 \frac{1}{3f^2_\pi}  \left(2 c_{4}-c_{3}+\frac{1}{2m_p}\right),
 \eeq
 where the values of the LECs ($c_3, c_4, c_D$) vary with different  chiral nuclear forces. For  commonly used chiral nuclear forces, the $c_3\simeq -c_4 \simeq -4$ GeV$^{-1}$~\cite{Entem:2003,Epelbaum:2005NPA}, and  $c_D\in [-4,4]$~\cite{Gazit:2019PRL_Erratum,Huther:2020PLB}.

 \begin{figure}[]
\centering 
\includegraphics[width=8cm]{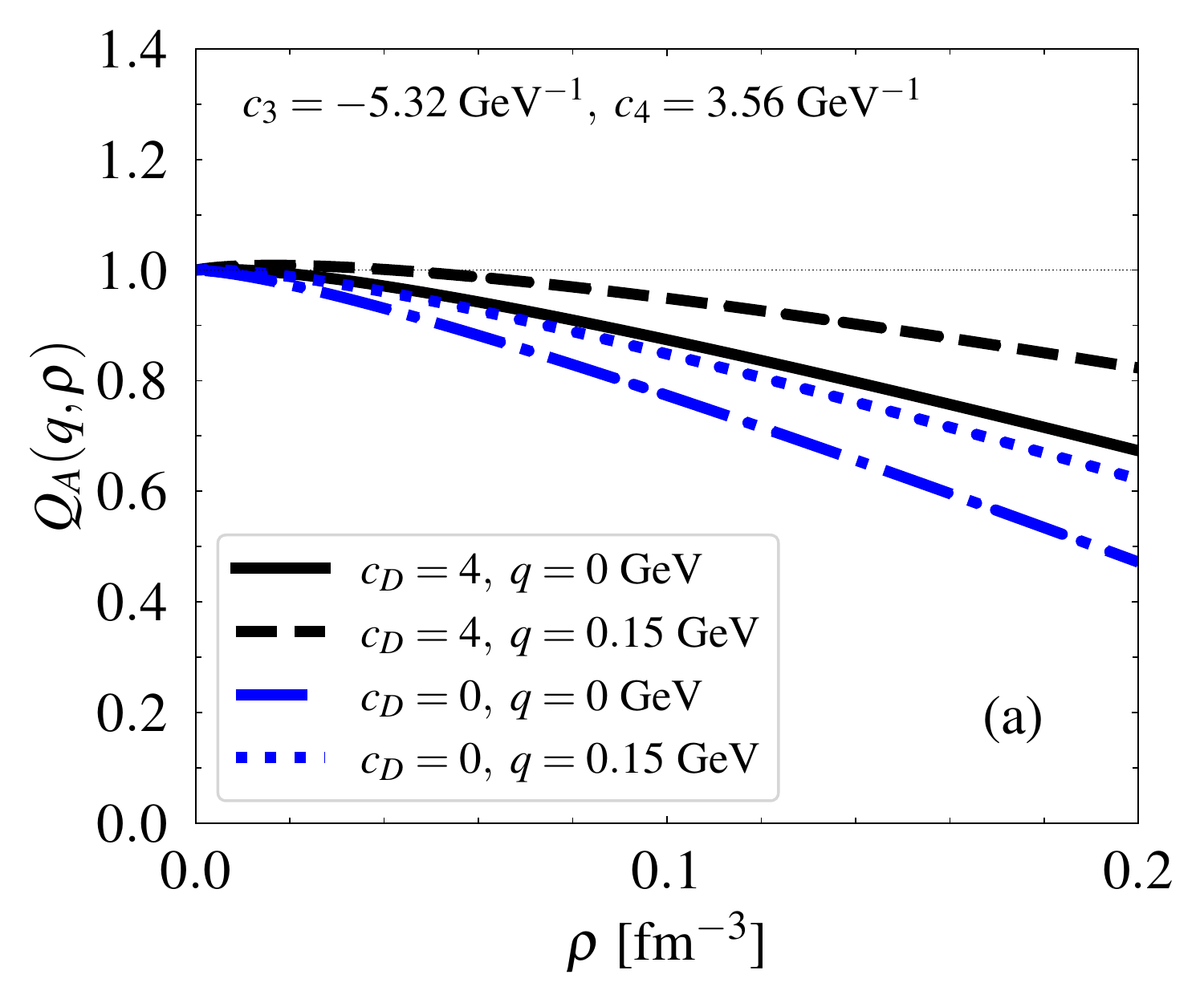}
\includegraphics[width=8cm]{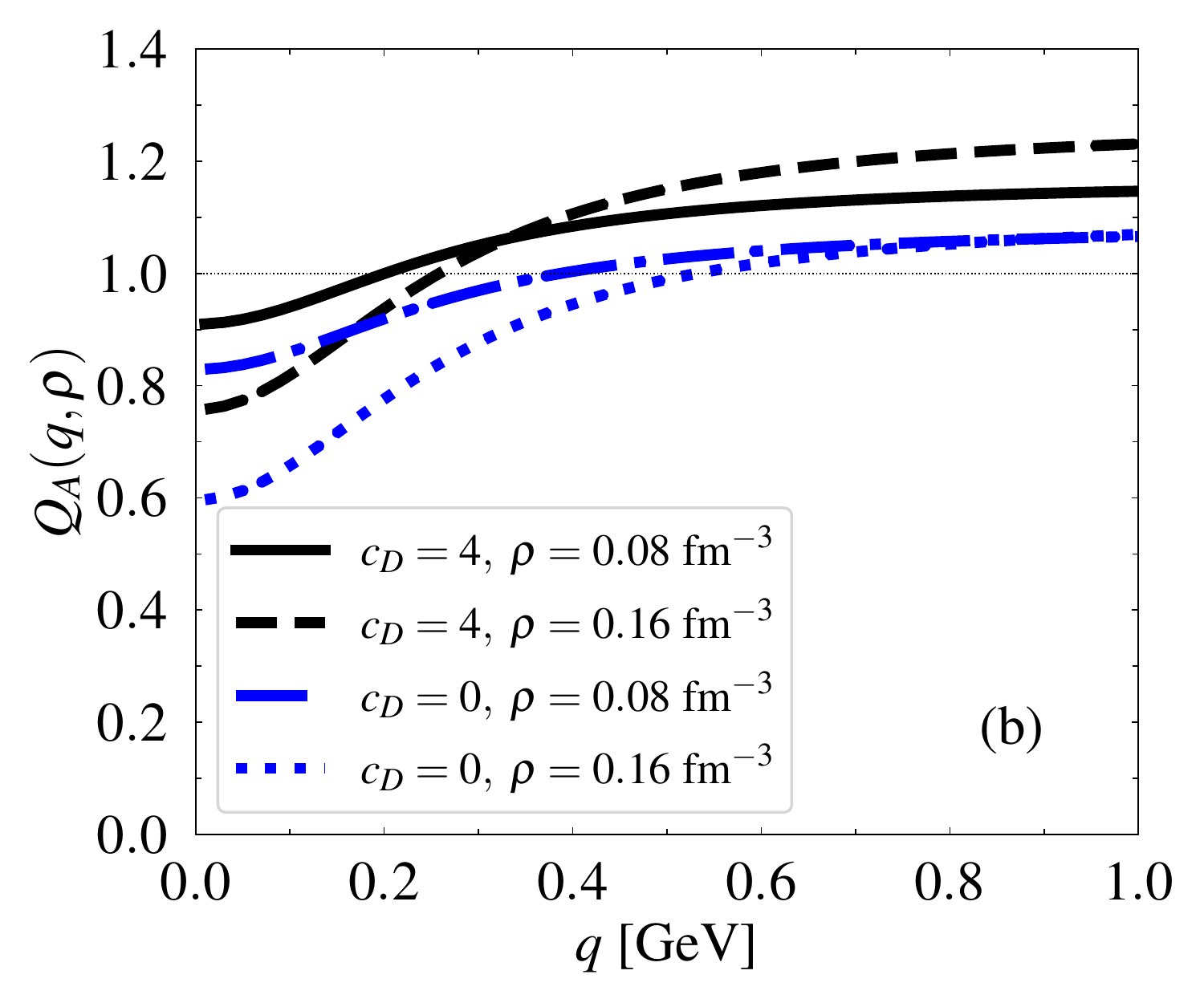}  
\caption{\label{fig:gA} The quenching factor $Q_A$ in (\ref{eq:Q-factor-chiral}) as a function of (a) density $\rho$ and (b) momentum transfer $q$. The LECs $c_D, c_3, c_4$ are chosen from the new chiral NN+3N (N$^3$LO) interaction~\cite{Huther:2020PLB} for illustration, where the $c_D=0$ case is also plotted for comparison. }
\end{figure}

It has been discussed in Refs.~\cite{Menendez:2011,Gysbers:2019df} and can be seen from Eq.(\ref{eq:Q-factor-chiral}) and Fig.~\ref{fig:gA} that 
\begin{itemize}
    \item For the case of $q=0$: the coefficient $A[0]$ is generally negative.
    Therefore, the $\mathcal{Q}_A$ is decreasing with the density $\rho$, consistent with the findings of phenomenological studies that the $\mathcal{Q}_A$ decreases with nuclear mass number $A$~\cite{Caurier:2007IJMPE,Barea:2013,Barea:2015}. 
    \item For the case of $q\neq0$: especially $q\simeq m_\pi$, a typical value for $\znubb$ decay, the $A[q]$ is still dominated by the second term, i.e. the long-range one-pion exchange current depending on the LECs $c_3, c_4$. However, the nonzero third ($q$-dependent) term reduces significantly the $A[q]$ value, and thus weakens the quenching effect, as shown in Fig.~\ref{fig:gA}(a). Besides, the term $B\rho^{1/3}$ which enhances the factor $Q_A$ may become important, as illustrated in  Fig.~\ref{fig:gA}(b) and also Fig.3 in Ref.~\cite{Menendez:2011}. This enhancement may even bring the quenching factor larger than one in the high-$q$ region ($q>300-500$ MeV) depending on the $c_D$ value.   
    As the $c_D$ increases from $0$ to $4$, the absolute value of $A[q]$ is decreasing. As a result, the quenching effect from the chiral two-body current is decreasing. See also Fig. 16 in the appendix of Ref.~\cite{Gysbers:2019df}. 
    \end{itemize} 
 In short, one can see that the $g_A$ quenching effect is mainly from the one-pion exchange current. However, uncertainty in the $c_D$ of the contact term could lead to a significant uncertainty in the quenching effect.  The $c_D$ which enters into both the  three-nucleon force and two-body current is usually determined by the properties of triton and other observables~\cite{Gazit:2009PRL}.  The study by H\"{u}ther et al.~\cite{Huther:2020PLB} shows that the choice of $c_D\simeq 4$ in the new chiral EMN (NN+3N) interaction can reproduce nuclear binding energies of nuclei in different mass regions. It indicates that the short-range contact ($c_D$) term lessens the quenching effect.

  Men\'endez et al.~\cite{Menendez:2011} carried out a valence-space shell model calculation using the chiral two-body currents with the normal-ordering approximation. It has been found there that the two-body currents quench the NNE of $0 \nu \beta \beta$ decay in the mass range $A=48-136$ by a factor in between $-35 \%$ and $10 \%$ depending on the values of the LECs.  Engel et al.~\cite{Engel:2014} carried out a similar study in the framework of QRPA, where a slightly less quenching effect of about $20\%$ was obtained.  Recently, Wang et al.~\cite{Wang2018} reexamined the two-body currents on the NME in a PGCM calculation by evaluating explicitly the two- and three-body transition operators generated from the product of the one- and two-body currents.  They found that the NME is quenched only by about 10\%, less than those suggested by the prior works~\cite{Menendez:2011,Engel:2014}, which neglected portions of the operators due to the normal-ordering approximation. It provides us a hint that the renormalization effect of two-body currents on the $g_A$ for $0 \nu \beta \beta$ decay is mild. Again, it should be pointed out that the Hamiltonians employed in these works are not consistent with the chiral nuclear currents. Besides, the renormalization effect from the outside of model space was difficult to consider in these conventional nuclear models. 
       
 Ekstr\"om et al.~\cite{Ekstrom:2014PRL} carried out the first full consistent {\em ab initio} study on the impact of the chiral two-body currents on the Gamow-Teller transitions in $^{14} \mathrm{C}$ and ${ }^{22,24} \mathrm{O}$ with a coupled-cluster method, where the three-nucleon interaction together with consistent two-body currents from chiral EFT were employed. It was shown that the two-body currents reduce the Ikeda sum rule, corresponding to a quenching factor $Q^{2}_A \approx 0.84-0.92$~\cite{Ekstrom:2014PRL}. A recent {\em ab initio} study with NCSM, coupled-cluster method, and VS-IMSRG by Gysbers et al.~\cite{Gysbers:2019df} has shown that this effect, together with many-body correlations, is important for  accurate calculation of nuclear single-$\beta$ decay rates. 
 In other words, it is not necessary to introduce a quenching factor on the $g_A$ in nuclear {\em ab initio} methods for nuclear single-$\beta$ decay when the two-body currents are included in the calculations.  The quenching factor from the chiral two-body current in the coupled-cluster calculation turns out to be around 0.81 for $^{48}$Ca~\cite{Gysbers:2019df}. With this quenching factor,  the NME of $2\nu\beta\beta$ decay in $^{48}$Ca can be reproduced~\cite{Novario:2021PRL}. Of course, it is still an open question whether or not one should use this quenching factor in the {\em ab intio} calculation of $0\nu\beta\beta$ decay. To answer this question requires one to perform {\em ab initio}  calculations of $0\nu\beta\beta$ decay with the two-body currents included explicitly and consistently with the nuclear force.

 \subsection{Modeling the short-range correlation between two nucleons}

Different nuclear model calculations using different interactions produce wave functions that substantially differ at short distances~\cite{Cruz-Torres:2021_nature_physics}. As a result, the impact of short-range correlation (SRC) on the NMEs is also {\em  scale} and {\em scheme} dependent, which is similar to the two-body current effect.  In most of the BMF approaches, effective nuclear interactions with a low energy resolution scale are used and thus the SRC between nucleons is missing in the nuclear wave functions from BMF calculations. Here we make a comparison among several ways employed in literature to consider  the SRC effect on the NME of $\znubb$ decay. 
\begin{itemize}
    \item The Jastrow-like correlation function \cite{Jastrow:1955PR} provides an analytical way to consider the SRC effect via modifying the two-body wave function in the short distance. In this method, a radial two-body correlation function $f_{\mathrm{J}}(r)$ is multiplied to the neutrino potential (\ref{eq:neutrino_potential_r}), \footnote{Hereafter, the relative distance  $r_{mn}$ between two nucleons is abbreviated with $r$ for simplicity.} 
\begin{equation}
     H^{0\nu}_{\alpha,L}(r, E_d) \to  H^{0\nu}_{\alpha,L}(r, E_d)f^2_{\mathrm{J}}(r),
\end{equation}
where  the function $f(r)$ is chosen in a form that  should vanish as the distance of two nucleons $r\to 0$ and approach unity for $r\to \infty$ \cite{Jastrow:1955PR}.  To be consistent with the correlation function for nuclear matter, Miller and Spencer (MS) proposed the following form for the two-body correlation function $f_{\mathrm{J}}(r)$~\cite{Miller:1976AP},
\jmyr{
\beq
\label{eq:Jastrow_MS_function}
f_{\mathrm{J}}(r)=1 -  e^{-a r^2}\left(1-b r^{2}\right),
\eeq
which was extended  by \v{S}imkovic et al. \cite{Simkovic:2009PRC} to the following form
\beq
\label{eq:Jastrow_Simkovic_function}
f_{\mathrm{J}}(r)=1 -  ce^{-a r^2}\left(1-b r^{2}\right),
\eeq
to reproduce the SRC from the coupled-cluster method (CCM) calculation using either the charge-dependent Bonn potential (CD-Bonn)~\cite{Machleidt:2001PRC} or the Argonne V18 potential~\cite{Wiringa:1995PRC}. The  parameter $c$ was found to be below unity for these two potentials.  As a result, the central depletion is reduced and the maxima of the correlation functions are shifted from 1.5 fm to 1.0 fm, leading to a quenched SRC effect~\cite{Simkovic:2009PRC} .}
Cruz-Torres et al.~\cite{Cruz:2018PLB} adopted the following form for the squared correlation $F(r)\equiv f^2_{\mathrm{J}}(r)$,
\beq
\label{eq:Jastrow_Cruz_function}
F(r)=1-e^{-\alpha r^2}\left(\gamma + \sum^3_{i=1} \beta_i r^{i+1}\right).  
\eeq
The parameters $(\alpha=3.17~{\rm fm^{-2}}$, $\gamma=0.995$, $\beta_1=1.81~{\rm fm^{-2}}$, $\beta_2=5.90~{\rm fm^{-3}}$, $\beta_3=-9.87~{\rm fm^{-4}})$ were obtained by fitting to the results  from cluster variational Monte Carlo calculations for the proton-proton/neutron-neutron correlation functions. 

Figure~\ref{fig:Jastrow_function_MS} displays the distribution of the MS type of correlation function (\ref{eq:Jastrow_MS_function}) for a different set of parameters $(a, b, c )$, where $c$ is set to be one. It is seen that the function suppresses the contribution at the short distance with $r<1.0$ fm and enhances the contribution at the region with $r \simeq 1.5$ fm, depending on the choice of parameters. The smaller values of $a$ and $b$, the larger SRC effect will be generated by the correlation function. The parameter set $a=1.1~\mathrm{fm}^{-2}$ and $b=0.68~\mathrm{fm}^{-2}$  is usually used in the literature.  

Compared to the MS parameterization, the correlation function by (\ref{eq:Jastrow_Cruz_function}) is shifted inward with a  peak located around $r \simeq 1.2$ fm. In contrast, the correlation function by CCM enhances the two-nucleon wave function  significantly around  $r \simeq 1.0$ fm, which compensates partially for the suppression effect at the short distance. Therefore, it is expected that the SRC effect by the CCM is the smallest among all the three parameterizations. This observation is indeed confirmed in the numerical calculations and will be discussed later.

 \begin{figure}[tb] 
\centering
\includegraphics[width=10cm]{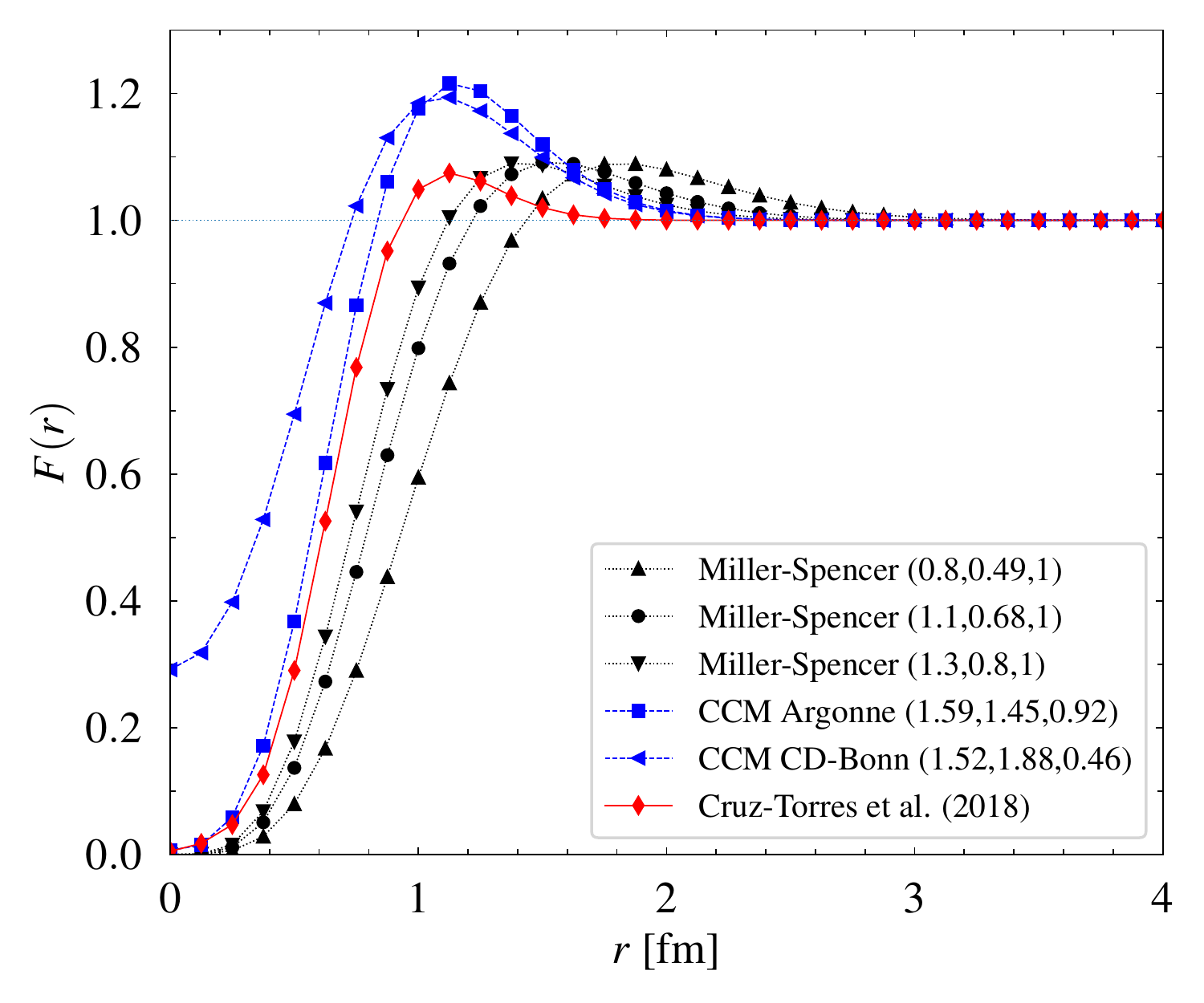}  
\caption{\label{fig:Jastrow_function_MS}  The squared $F(r)\equiv f^2_J(r)$ Jastrow-like two-body correlation function (\ref{eq:Jastrow_MS_function}) as a function of the relative distance between two nucleons. Results are shown for the  Miller-Spencer parameterization \cite{Miller:1976AP}  and those obtained by \v{S}imkovic et al. \cite{Simkovic:2009PRC} to reproduce the short-range correlation from the coupled-cluster method (CCM) calculation with either AV18~\cite{Wiringa:1995PRC} and CD-Bonn nuclear force. The result by Cruz-Torres et al. \cite{Cruz:2018PLB} is given by Eq.(\ref{eq:Jastrow_Cruz_function}). }
\end{figure}

Since the Jastrow-like function $f_{\mathrm{J}}(r)$ changes the coordinate behavior of the relative wave function of the two nucleons, in which case the norm of the relative wave functions is altered. A renormalization procedure is required in this case, and the NME should be determined by
\beqn
    M^{0\nu} &=& \dfrac{\langle 0^+_F\vert f_J\hat O^{0\nu}  f_J\vert 0^+_I \rangle}{\sqrt{\bra{0^+_F}f_J\ket{0^+_F}\bra{0^+_I}f_J\ket{0^+_I}}}.
 \eeqn

    \item \jmy{The correlation function} from the unitary correlation operator method (UCOM) \cite{Feldmeier:1998NPA,Roth:2005PRC}. In this method,  the SRC is considered with a correlated nuclear wave function $\ket{\overline{0^+_{I/F}}}$ which are related to the uncorrelated one $\ket{0^+_{I/F}}$~\footnote{Here the correlated wave function refers to the two-nucleon wave function with short-range correlation.}
    \beq
    \ket{\overline{0^+_{I/F}}}
    = {\cal C} \ket{0^+_{I/F}},
    \eeq
    where ${\cal C}$ is a unitary correlation operator, which is a product of two unitary operators: ${\cal C} = {\cal C}_\Omega {\cal C}_r$, where ${\cal C}_\Omega$ describes short-range tensor correlation and ${\cal C}_r$ central correlation which shifts a pair of two nucleons in the radial direction away from each other so that they get out of the range of the repulsive core. The expressions for the ${\cal C}$ are given in Ref.~\cite{Feldmeier:1998NPA,Roth:2005PRC}, which actually depend on the choice of nuclear forces. For the Bonn-A and AV18 potentials, the UCOM parameters can be found in Ref.~\cite{Neff:2003NPA}. The NME given by the correlated wave function reads
    \beqn
    M^{0\nu}=\left\langle\overline{0^+_F}|\hat O^{0\nu}| \overline{0^+_I}\right\rangle
    =\left\langle 0^+_F|\overline{\hat O^{0\nu}}| 0^+_I\right\rangle,
    \eeqn
    where the transformed operator is defined as
    \beq
    \overline{\hat O^{0\nu}}
    ={\cal C}^{\dagger} \hat O^{0\nu} {\cal C}.
    \eeq

    \item \jmy{The spin- and isospin-dependent correlation function $f_{ij}$, the operator structure of which reflects the structure of the underlying $NN$ potential, has the form~\cite{Benhar:2014PRC}
 \beqn
f_{mn}(r)=\sum_{\alpha=1}^{6} f^{(\alpha)}(r_{mn}) O_{mn}^{(\alpha)},
\eeqn
with
\beq
O_{mn}^{(\alpha)}=\left[1,\bm{\sigma}_{m} \cdot \bm{\sigma}_{n}, S^r_{mn}\right] \otimes\left[1,\bm{\tau}_{m} \cdot \bm{\tau}_{n} \right].
\eeq 
The implementation of the above correlation function into the $\znubb$-decay operator mixes the contributions from the GT and Fermi parts. In Ref.~\cite{Benhar:2014PRC}, the $f_{ij}$ is truncated up to $\alpha=4$ and the radial dependence of the functions $ f^{(\alpha)}(r_{mn})$ was obtained from a realistic nuclear Hamiltonian including the Argonne $v'_6$ $NN$ potential. It has been found that this type of correlation function quenches the NME of \nuclide[48]{Ca} by $\sim20\%$ with respect to the result of shell model calculation.}

\end{itemize}

As mentioned before that how the short-range correlation modifies the distribution of the two-nucleon wave function depends on the employed nuclear interactions~\cite{Knoll:2019_Master}. In other words, one should choose a correlation function or UCOM parameters consistent with the nuclear interaction, which is however impossible for the BMF approaches starting from an effective nuclear interaction or EDF. Therefore, in all the available phenomenological shell-model and EDF-based calculations, the SRC effect was not considered in a fully consistent way. Nevertheless, it has been found by Kortelainen et al.~\cite{Kortelainen:2007PLB}  with nuclear shell model that  the use of MS type of Jastrow functions mainly cuts off significantly the contributions from high values of angular momentum of the intermediate states and finally reduce the NME of $\znubb$ decay by 30\%-40\%, which tends to overestimate the SRC effect~\cite{Engel:2011PRC}. In contrast,  the UCOM quenches the NME only by $7\%-16\%$. This finding is consistent with the studies with QRPA \cite{Kortelainen:2007PRC,Simkovic:2008}. Of particular interest is the finding that the SRC effect by the CCM quenches the NME  only $\sim 5\%$ when the finite-nucleon-size (FNS) effect is already considered via momentum-dependent dipole form factors \cite{Simkovic:2009PRC,Song:2017}. Recently, more consistent 
treatment of both nuclear potential and neutrino potential was carried out  with a shell model \cite{Coraggio:2019}, in which both the CD-Bonn NN potential and neutrino potentials were renormalized with the same $V_{\rm low-k}$ procedure~\cite{Bogner:2010}. The renormalized nuclear potential was subsequently used to derive an effective Hamiltonian for a given valence space using many-body perturbation theory. It was shown that the renormalization effect on the NME of $\znubb$ decay is only about 2\%. The above-mentioned studies indicate that the SRC effect is comparable to or even smaller than the N$^2$LO corrections if the $0\nu\beta\beta$ decay is driven by the standard mechanism. In contrast, if the  $0\nu\beta\beta$ is governed by  the exchange of heavy neutrinos which is beyond the scope of this Review, the SRC effect may alter significantly the corresponding NME~\cite{Simkovic99,Vergados:2012RPP,Barea:2013,Hyvarinen:2015PRC,Song:2017}.   


 \subsection{The leptonic phase-space factor}
   
 \subsubsection{The plane wave approximation}
The leptonic phase-space factor (\ref{eq:G0nu_general}) can be simplified if the  electron wave functions  are approximated as plane waves. In this case, one has the following relations,
\bsub
\begin{align}
   \sum_{s_1}  u(k_1,s_1)\bar u(k_1,s_1) &= \gamma_\mu k^\mu_1 + m_e,\\
   \sum_{s_2} u^c(k_2,s_2) \bar u^c(k_2,s_2)  &= \gamma_\mu k^\mu_2 - m_e,
\end{align}
\esub
and ${\rm Tr}[\gamma_5]={\rm Tr}[\gamma_\mu]={\rm Tr}[\gamma_\mu\gamma_5]={\rm Tr}[\gamma_\mu\gamma_\nu\gamma_5]=0$, ${\rm Tr}[\gamma^\mu\gamma^\nu]=4g^{\mu\nu}$, $\gamma_5^2=1$, from   which  one finds
\beqn
\sum_{s_1,s_2} \big\vert \bar u(k_1,s_1)P_R C \bar u^T(k_2,s_2)\big\vert^2 
 &=&\dfrac{1}{4} {\rm Tr}\left[(1+\gamma_5)(\gamma_\mu k^\mu_2 + m_e)(1-\gamma_5)(\gamma_\nu k^\nu_1-m_e)\right]\nonumber\\
&=&2(\epsilon_1\epsilon_2 - \bm{k}_1\cdot \bm{k}_2).
\eeqn
With the above relations, the phase-space factor is simplified as below  
\begin{eqnarray}
\label{eq:G0nu_general_plan_wave}
G_{0\nu} 
	&=&   
  \frac{G^4_\beta m^2_e}{2\ln(2)(2\pi)^5R^2_0}  \int \int  
(\epsilon_1\epsilon_2 - k_1k_2\cos\theta_{12}) 
k_1k_2d\epsilon_1  d\cos\theta_{12}. 
\end{eqnarray}
where $\theta_{12}$ is the angle between two electron momenta $\bm{k}_1$ and $\bm{k}_2$, and $\epsilon_2=E_I-E_F-\epsilon_1$.  In the above derivation, we have applied the relation $k^2_idk_i = k_i \epsilon_id\epsilon_i$.  It is noted that the $G_{0\nu}$ has been defined differently in different literature. \footnote{The phase-space factor by Tomoda {\it et al.} \cite{Tomoda:1991} was given in units of  yr$^{-1}$ fm$^2$. As noted in Ref. \cite{Cowell:2006}, a scale of $R_0=r_0 A^{1/3}$ was usually multiplied onto the NME in subsequent studies to make the NME dimensionless. Besides, as pointed out in Ref. \cite{Kotila:2012}, an additional factor of $4$ is necessary to be introduced in the formalism of Ref. \cite{Tomoda:1991} to make the calculation consistent with Ref. \cite{Boehm:1992}.
The $G_{0\nu}$ by Doi et al  in Ref. \cite{Doi:1985}  includes the factor $g^4_A\approx 2.6$ and the Fermi-Primakoff-Rosen approximation was used. } The  phase-space factor $G_{0\nu}$ in (\ref{eq:G0nu_general_plan_wave}) is consistent with the definition of Kotila and Iachello \cite{Kotila:2012}. 

For the transition of $0^+ \to 0^+$ with two outgoing electrons in the $S_{1/2}$ state, the phase-space factor $G_{0\nu}$ becomes 
\beqn
\label{eq:G0nu_free}
G^{(00)}_{0\nu} 
&=& \dfrac{G^4_\beta m^2_e} {(2\pi)^5 R^2_0}\dfrac{1}{\ln(2)}\int^{m_e+Q_{\beta\beta}}_{m_e}  k_1k_2 \epsilon_1\epsilon_2 d\epsilon_1.
\eeqn

\subsubsection{Distorted electron wave functions}

The  wave functions of the outgoing electrons are essential ingredients for computing the phase-space factor $G_{0\nu}$. Due to the presence of Coulomb interaction generated by the protons inside the nucleus, the wave functions of electrons are distorted from plane wave functions.  The Coulomb correction ${\cal F}(Z,\epsilon)$ to the electron current, defined as the square of the ratio of the scattering solution for electrons to a plane wave, has been determined either in non-relativistic case or relativistic case.  Two approaches are often employed to take into account this effect.

\begin{itemize}
    \item The Fermi-Primakoff-Rosen (FPR) approximation: 
    
    In this approximation, the Coulomb correction $F(Z_F,\epsilon)$ can be derived analytically under the approximation that the two electrons scatter off a point charge $Z_F$ in non-relativistic kinematics \cite{Haxton1984PPNP},
 \beq
 \label{eq:Fermi}
 {\cal F}^{\rm NR}(Z_F,\epsilon) =e^{\pi\eta}\mid\Gamma(1+i\eta)\mid^2
 = \dfrac{2\pi \eta}{1-\exp(-2\pi \eta)},
 \eeq
 where the Sommerfeld parameter $\eta = \alpha Z_F \epsilon/k$ with $k = |\bm{k}|$ and $\epsilon$ being the momentum and energy of the electron, respectively. $\alpha\simeq1/137$ is the fine structure constant. We note that ${\cal F}^{NR}(Z_F,\epsilon)\to 1$ if $2\pi \eta<<1$. Primakoff and Rosen (PR) \cite{Primakoff:1959} made a further approximation $\epsilon/k\approx 1$ in which case the Coulomb correction becomes
    \beqn
   {\cal F}^{\rm NR}(Z_F,\epsilon) 
   &\approx& {\cal F}^{\rm PR}(Z_F,\epsilon)
    = \left(\dfrac{\epsilon}{k}\right) F(Z_F),
    \eeqn
where $F(Z_F)\equiv\dfrac{2\pi\alpha Z_F}{1-\exp(-2\pi\alpha Z_F)}$ is an energy-independent Fermi function which can be taken out from the integral. With the correction under the FPR approximation, the  phase-space factor (\ref{eq:G0nu_free}) is replaced with
\begin{eqnarray}
\label{eq:phase_FPR}
 G^{(00)}_{0\nu}  ({\rm FPR}) 
&=&  \dfrac{G^4_\beta m^2_e} {(2\pi)^5 R^2_0}\dfrac{1}{\ln(2)}\int^{m_e+Q_{\beta\beta}}_{m_e}  k_1k_2 \epsilon_1\epsilon_2 {\cal F}^{\rm PR}(Z_F,\epsilon_1){\cal F}^{\rm PR}(Z_F,\epsilon_2) d\epsilon_1\nonumber\\
&\simeq&   \dfrac{F^2(Z_F)}{R^2_0} \dfrac{G^4_\beta m^2_e} {(2\pi)^5 }\dfrac{1}{\ln(2)}\int_0^{Q_{\beta\beta}} d T_1  \epsilon^2_1\epsilon^2_2,
\end{eqnarray}
 where we have applied the PR approximation $k_1k_2\simeq \epsilon_1\epsilon_2$ again. Neglecting the recoil energy of atomic nucleus, the kinetic energy of electron is $T_1=\epsilon_1-m_e$. The energy  of one electron $\epsilon_2$ is related to that of another electron $\epsilon_1$ by $\epsilon_2=Q_{\beta\beta}+2m_e-\epsilon_1$. The total released kinetic energy ($Q_{\beta\beta}$ value)  is fixed by the energies $E_{I/F}$ of the initial $(A, Z_I)$ and final $(A, Z_F)$ nuclei in a given decay
\beqn
Q_{\beta\beta} 
= E_I-E_F-2m_e 
=B(A,Z+2) - B(A,Z) + 2\Delta_{npe},
\eeqn
with $B(A,Z)$ standing for the binding energy of nucleus ($A, Z$) and $\Delta_{npe}=m_n-m_p-m_e$. The  integral in (\ref{eq:phase_FPR}) can be carried out  analytically  
\beqn 
G^{(00)}_{0\nu} ({\rm FPR})  
&=& \dfrac{F^2(Z_F)}{R^2_0}\dfrac{G^4_\beta m^2_e} {(2\pi)^5 }\dfrac{1}{\ln(2)} \int^{m_e+Q_{\beta\beta}}_{m_e} (Q_{\beta\beta}+2m_e-\epsilon_1)^2 \epsilon^2_1 d\epsilon_1\nonumber\\
&=& \dfrac{F^2(Z_F)}{R^2_0} \dfrac{G^4_\beta m^7_e} {(2\pi)^5 }\dfrac{1}{\ln(2)}
\left(\dfrac{\tilde T^5_0}{30}
      +\dfrac{\tilde T^4_0}{3}
      +\dfrac{4\tilde T^3_0}{3} 
      +2\tilde T^2_0 +\tilde T_0,
\right)
\eeqn
where $\tilde T_0=Q_{\beta\beta}/m_e$  following the notations of Ref. \cite{Haxton1984PPNP}  \footnote{In Ref \cite{Bilenky:1987}, the symbol $\epsilon_0$, instead of $\tilde T_0$, was used.}. After adding back $\hbar$ and $c$, the phase-space factor with the Coulomb correction to the electron wave functions given by the FPR approximation finally becomes
\beqn
\label{eq:G0nu-PR}
G^{(00)}_{0\nu} ({\rm FPR})  
&=&g_{0\nu}\dfrac{F^2(Z_F)}{A^{2/3}}   
\left(\dfrac{\tilde T^5_0}{30}
      +\dfrac{\tilde T^4_0}{3}
      +\dfrac{4\tilde T^3_0}{3} 
      +2\tilde T^2_0 +\tilde T_0
\right)
\eeqn
with the quantity $g_{0\nu}$ defined as
\beq
 g_{0\nu}\equiv  \dfrac{G^4_\beta (m_ec^2)^7 (\hbar c)^2}{(1.2)^2\ln(2)(2\pi)^5}  
 = 2.895\times 10^{-17} {\rm yr}^{-1}.
\eeq
It is seen from Eq.(\ref{eq:G0nu-PR}) that the phase-space factor $G^{(00)}_{0\nu}$ is approximately proportional to $Q^5_{\beta\beta}$ and $Z^2_F/A^{2/3}$. In other words,  the heavier isotope with a larger $Q_{\beta\beta}$ value has a larger phase-space factor for $0\nu\beta\beta$ decay.

\item Solution of a Dirac equation for the electron wave functions:

In recent studies \cite{Kotila:2012,Mirea:2015nsl,Stefanik:2015,Stoica:2019}, the phase-space factor was computed using the {\em exact} electron wave functions obtained from the solution of a Dirac equation for the electron in a Coulomb potential  corrected with finite nuclear size and screening effects.

The  scattering wave function $e_{\bm{k},s}(\bm{r})$ of an electron  with asymptotic momentum $\bm{k}$ and spin projection $s$ can be generally expanded in term of different spherical partial waves~\cite{Doi:1985,Tomoda:1991,Graf:2018},
\beq
\label{eq:wf4electron}
e_{\bm{k},s}(\bm{r}) 
=\sum_{\kappa,\mu} \begin{pmatrix}
		g_{\kappa}(\epsilon, r) \chi_{\kappa\mu}\\ 
		if_{\kappa}(\epsilon, r) \chi_{-\kappa\mu}\\ 
	\end{pmatrix}, 
\eeq
where the spin-angular part is
\beq
	\chi_{\kappa\mu}
	=\sum_s \langle \ell_\kappa \mu-s 1/2 s \vert j \mu\rangle Y_{\ell_\kappa \mu-s}(\bm{\hat r})\chi_{1/2,s}.
\eeq
The $\kappa$ is the eigenvalue of the operator $\hat K\equiv\beta(\bm{\Sigma}\cdot \bm{L}+1)$ \footnote{Here, $\beta, \bm{\Sigma}$ are four-component Dirac matrices and the operator $\hat K$ is given by
\beq
\hat K = \begin{pmatrix}
    \bm{\sigma}\cdot \bm{\ell}+1 & 0 \\
    0 & -(\bm{\sigma}\cdot \bm{\ell}+1)
\end{pmatrix},\quad
\hat K \psi_{\kappa\mu} = -\kappa \psi_{\kappa\mu}.
\eeq}, taking positive and negative integer numbers $\kappa=\pm(j+1/2)=\pm1, \pm2, \pm3, \ldots$. The label $\mu$ is for the projection of total angular momentum $j_\kappa=\vert \kappa\vert -1/2$. The  orbital angular momentum $\ell_k=\kappa$ if $\kappa>0$ or  $\ell_k=\vert\kappa\vert-1$ if $\kappa<0$.  The  radial wave functions of large and small components  $g_{\kappa}(\epsilon, r)$ and $f_{\kappa}(\epsilon, r)$ in (\ref{eq:wf4electron}) are   determined by the following spherical Dirac equation
\beq
\label{eq:Dirac4electron}
 \begin{pmatrix}
		m-\epsilon+V_C(r) & \bm{\sigma}\cdot\bm{k}\\
		 \bm{\sigma}\cdot\bm{k} & -m-\epsilon+V_C(r)\\ 
	\end{pmatrix} 
 \begin{pmatrix}
		g_{\kappa}(\epsilon, r) \chi_{\kappa\mu}\\ 
		if_{\kappa}(\epsilon, r) \chi_{-\kappa\mu}\\ 
	\end{pmatrix}
	=0,
\eeq
where the electron energy $\epsilon=\sqrt{m^2_e+\bm{k}^2}$. The Coulomb potential $V_C(r)$ is induced by a charged sphere with the total $Z_F$ charge uniformly distributed inside a sphere of radius $R_0$,
\beqn
\label{eq:Vc}
V_C(r)=\left\{\begin{array}{ll}
-\dfrac{\alpha Z_F}{2R_0} \left(3-(r/R_0)^2\right), & r<R_0,\\
&\\
-\dfrac{\alpha Z_F}{r}, & r\ge R_0. 
\end{array}\right.
\eeqn
The radial wave functions satisfy the asymptotic boundary condition  at large values of $kr$
 \begin{eqnarray}
 \begin{pmatrix}
 g_{\kappa}(\epsilon,r)\\
 f_{\kappa}(\epsilon,r)
 \end{pmatrix}
 \xrightarrow{kr\to\infty} e^{-i\delta_\kappa}\dfrac{1}{kr}
  \begin{pmatrix}
 \sqrt{\dfrac{\epsilon+m_e}{2\epsilon}}\sin\phi_\kappa(kr)\\
 \sqrt{\dfrac{\epsilon-m_e}{2\epsilon}}\cos\phi_\kappa(kr)
 \end{pmatrix}
 \end{eqnarray}
 where $\delta_\kappa$ is the phase shift, and the angle $\phi_\kappa(kr)$ is defined as
 \beq
 \phi_\kappa(kr) = kr+\eta\ln(2kr)-\dfrac{1}{2}\pi\ell_\kappa+\delta_\kappa.
 \eeq

The screening effect can be taken into account by multiplying the Coulomb potential $V_C(r)$ with the Thomas-Fermi function $\varphi(x)$, which is a solution of the following equation
\beq
\label{eq:TF_function}
\dfrac{d^2\varphi(x)}{dx^2}=\dfrac{\varphi^{3/2}(x)}{\sqrt{x}},\quad x=r/b
\eeq
with 
\beq
b = \dfrac{1}{2} (\dfrac{3\pi}{4})^{2/3}\dfrac{\hbar^2}{m_ee^2}Z^{-1/3}_F
\approx 0.8853a_0Z^{-1/3}_F,
 \eeq
 where $a_0=\hbar c/(\alpha m_e)=5.3\times10^4$ fm is the Bohr radius. There are different ways to solve  Eq. (\ref{eq:TF_function}), see for instance Refs. \cite{Esposito:2001,Liao:2003} for details. \footnote{Notice the boundary conditions for the TF function:  $\varphi(0)=1, \quad  \varphi(\infty)=2/Z_F$, which take into account the fact that the final atom is a positive ion with charge $+2$.}
The screening effect is of order $\alpha^2 Z^{4/3}_F$, which is  about two orders of magnitude smaller than that of Coulomb effect. Thus, the impact of the screening effect on the phase-space factor is negligible~\cite{Stefanik:2015}.

For the $0\nu\beta\beta$ decay with the emitted electrons in
$S_{1/2}$ ($\kappa=-1$, $j_\kappa=1/2$, $\ell_\kappa=0$) states, the corresponding electron wave function is
\begin{align} \label{eq:Dirac_spinor}
	e^{S_{1/2}}_{\bm{k},s}(\bm{r})
	=
	\begin{pmatrix}
		g_{-1}(\epsilon, r) \chi_s\\ 
		   f_1(\epsilon, r)(\bm{\sigma}\cdot \hat{\bm{k}}) \chi_s 
	\end{pmatrix}, 
\end{align} 
 with which one finds~\cite{Graf:2018}
 \beqn
 \label{eq:electron_wavefunctions_sum}
\dfrac{1}{4}\sum_{s_1,s_2}  \big\vert   \bar{e}^{S_{1/2}}_{\bm{k}_1,s_1}(\bm{r}_1)  (1+\gamma_5)C  e^{c,S_{1/2}}_{\bm{k}_2,s_2}(\bm{r}_2) \big\vert^2
=
\dfrac{1}{2}\left(f_{11}^{(0)} + f_{11}^{(1)}\cos\theta_{12} \right).
\eeqn
Substituting the relation (\ref{eq:electron_wavefunctions_sum}) into the phase-space factor (\ref{eq:G0nu_general}), one finds
\begin{eqnarray}
\label{eq:G0nu_exact}
 G_{0\nu} 
 &=&\dfrac{G^4_\beta m^2_e} {(2\pi)^5 R^2_0}\dfrac{1}{\ln(2)}\dfrac{1}{2}\int  d\cos\theta_{12}\left(f_{11}^{(0)} + f_{11}^{(1)}\cos\theta_{12} \right) k_1k_2 d\epsilon_1 \nonumber\\
 &=&\dfrac{G^4_\beta m^2_e} {(2\pi)^5 R^2_0}\dfrac{1}{\ln(2)}\int^{m_e+Q_{\beta\beta}}_{m_e} f^{(0)}_{11}(\epsilon_1,\epsilon_2) k_1k_2 \epsilon_1\epsilon_2 d\epsilon_1,
\eeqn
 where the functions $f_{11}^{(0)}$ and $f^{(1)}_{11}$ are defined as \cite{Tomoda:1991}
\bsub\beqn
f_{11}^{(0)}
&=&|g_{-1}(\epsilon_1) g_{-1}(\epsilon_2)|^2
+|f_{1}(\epsilon_1) f_{1}(\epsilon_2)|^2 
+|g_{-1}(\epsilon_1) f_{1}(\epsilon_2)|^2+|f_{1}(\epsilon_1) g_{-1}(\epsilon_2)|^2,\\
f_{11}^{(1)}&=&-2 \operatorname{Re}\left[g_{-1}(\epsilon_1) g_{-1}(\epsilon_2)f^\ast_{1}(\epsilon_1) f^\ast_{1}(\epsilon_2) +g_{-1}(\epsilon_1) f_{1}(\epsilon_2) f^\ast_{1}(\epsilon_1) g^\ast_{-1}(\epsilon_2)\right],
\eeqn
\esub
with $g_{-1}(\epsilon)$ and $f_{1}(\epsilon)$ chosen as the radial wave functions at the nuclear surface, i.e. $r=R_0$,
 \begin{eqnarray}
g_{-1}(\epsilon) =g_{-1}(\epsilon,r=R_0),\quad
f_{1}(\epsilon)  =f_{1}(\epsilon,r=R_0). 
\end{eqnarray}
 
 \end{itemize}
 Different levels of approximations are usually taken to determine the  radial wave functions.
 \begin{enumerate}
     \item  An "exact" solution of the Dirac equation (\ref{eq:Dirac4electron}) is obtained numerically by using for example the package RADIAL \cite{Salvat:1995}.  
  
\item By expansion the radial wave functions $g_\kappa(\epsilon,r)$ and $f_\kappa(\epsilon,r)$ with $r<R$  in terms of a power series of $r$ \cite{Buhring:1963}. The lowest terms were adopted by Doi et al.~\cite{Doi:1985} and Tomoda \cite{Tomoda:1991}  in the calculations of the phase-space integrals of $\beta\beta$ decay processes. The difference between these two calculations is that the expansion is up to $r^2$ terms in the calculation by Doi  et al.~\cite{Doi:1985} for the $S_{1/2}$ state.
If only the leading terms in the expansion are considered, the radial wave functions of $S_{1/2}$ state take the form \cite{Stefanik:2015}
\begin{eqnarray}
 \left(
   \begin{array}{c}
      g_{-1}(\epsilon,r) \\
      f_{+1}(\epsilon,r)
   \end{array}
 \right)
&\approx& 
\sqrt{{\cal F}_0(Z_f,\epsilon)} 
\left(
   \begin{array}{c}
      \sqrt{\frac{\epsilon+m_e}{2\epsilon}} \\
      \sqrt{\frac{\epsilon-m_e}{2\epsilon}}
   \end{array}
 \right),
\end{eqnarray}
where ${\cal F}_{n-1}$ (for $n=1,2, \cdots$) is given by
\begin{eqnarray}
 {\cal F}_{n-1}(Z_f,\epsilon) 
 &=& \left[ \frac{\Gamma(2n+1)}{\Gamma(n)\Gamma(1+2\gamma_{n})}
\right]^2 (2 kr)^{2(\gamma_{n}-n)} e^{\pi \eta}
\mid\Gamma(\gamma_{n}+i\eta)  \mid^2,
\end{eqnarray}
where 
\begin{eqnarray}
\gamma_{n} =\sqrt{n^2-(\alpha Z_f)^2} ,\quad \eta = \alpha Z_f \frac{\epsilon}{k}.
\end{eqnarray}
For the $S_{1/2}$ state, $n=1$, and $\kappa=-1$,
\begin{eqnarray}
{\cal F}_{0}(Z_f,\epsilon) 
&=&4
 (2kr)^{2(\gamma_{1}-1)} e^{\pi \eta}
 \Big\vert \frac{\Gamma(\gamma_{1}+i\eta) }{ \Gamma(1+2\gamma_1)}\Big\vert^2,
\end{eqnarray}
 where $\gamma_1 =\sqrt{1-(\alpha Z_f)^2}$. One can see that for the case $\alpha Z_F<<1$, $\gamma_1 \to 1$ and the  ${\cal F}_0(Z_F,\epsilon)\to {\cal F}^{NR}(Z_F,\epsilon)$ in Eq.(\ref{eq:Fermi}).
In general, the relativistic leading-terms approximation gives a larger value for the phase-space factor than the non-relativistic FPR approximation, as discussed in Ref.\cite{Haxton1984PPNP} and demonstrated below.

\item From a point-like atomic nucleus approximation. In this approximation, the radial wave functions take the form~\cite{Stefanik:2015}
\begin{eqnarray}
g_{\kappa}(\epsilon,r) &=& \frac{\kappa}{n}
 \frac{1}{k r} \sqrt{\frac{\epsilon+m_e}{2\epsilon}}
       \frac{|\Gamma(1+\gamma_n+i\eta)|}{\Gamma(1+2\gamma_n)} 
       (2p r)^{\gamma_n} e^{\pi \eta/2}\nonumber\\
       && {\rm Im} \left\{ e^{i(k r+\xi)}~
        _1F_1(\gamma_n-i\eta,1+2\gamma_n,-2ik r) \right\}, \nonumber\\
f_{\kappa}(\epsilon,r) &=& \frac{\kappa}{n}
\frac{1}{k r} \sqrt{\frac{\epsilon-m_e}{2\epsilon}}
       \frac{|\Gamma(1+\gamma_n+i\eta)|}{\Gamma(1+2\gamma_n)} 
       (2k r)^{\gamma_n} e^{\pi \eta/2}\nonumber\\
       && {\rm Re} \left\{ e^{i(k r+\xi)}~
        _1F_1(\gamma_n-i\eta,1+2\gamma_n,-2ik r) \right\},     
\end{eqnarray}     
where $_1F_1(a,b,z)$ is the confluent hypergeometric function and
\begin{eqnarray}
e^{-2i\xi} = \frac{\gamma_n-i\eta}{\kappa-i\eta m_e/\epsilon}.
\end{eqnarray} 
The radial wave function for the $S_{1/2}$ state ($n=1$) is given by
\begin{eqnarray}
g_{-1}(\epsilon,r) &=&
 -\frac{1}{kr} \sqrt{\frac{\epsilon+m_e}{2\epsilon}}
       \frac{|\Gamma(1+\gamma_1+i\eta)|}{\Gamma(1+2\gamma_1)} 
       (2kr)^{\gamma_1} e^{\pi \eta/2}\nonumber\\
       && {\rm Im} \left\{ e^{i(kr+\xi)}~
        _1F_1(\gamma_1-i\eta,1+2\gamma_1,-2ikr) \right\}, \nonumber\\
f_{-1}(\epsilon,r) &=& 
-\frac{1}{kr} \sqrt{\frac{\epsilon-m_e}{2\epsilon}}
       \frac{|\Gamma(1+\gamma_1+i\eta)|}{\Gamma(1+2\gamma_1)} 
       (2kr)^{\gamma_1} e^{\pi \eta/2}\nonumber\\
       && {\rm Re} \left\{ e^{i(kr+\xi)}~
        _1F_1(\gamma_1-i\eta,1+2\gamma_1,-2ikr) \right\}.    
\end{eqnarray}  
It has been shown in Ref.~\cite{Stefanik:2015} that the electron wavefunction of $S_{1/2}$ state in this approximation agrees well with the exact solution. 

 \end{enumerate}

 \begin{figure}
\centering
\includegraphics[width=8cm]{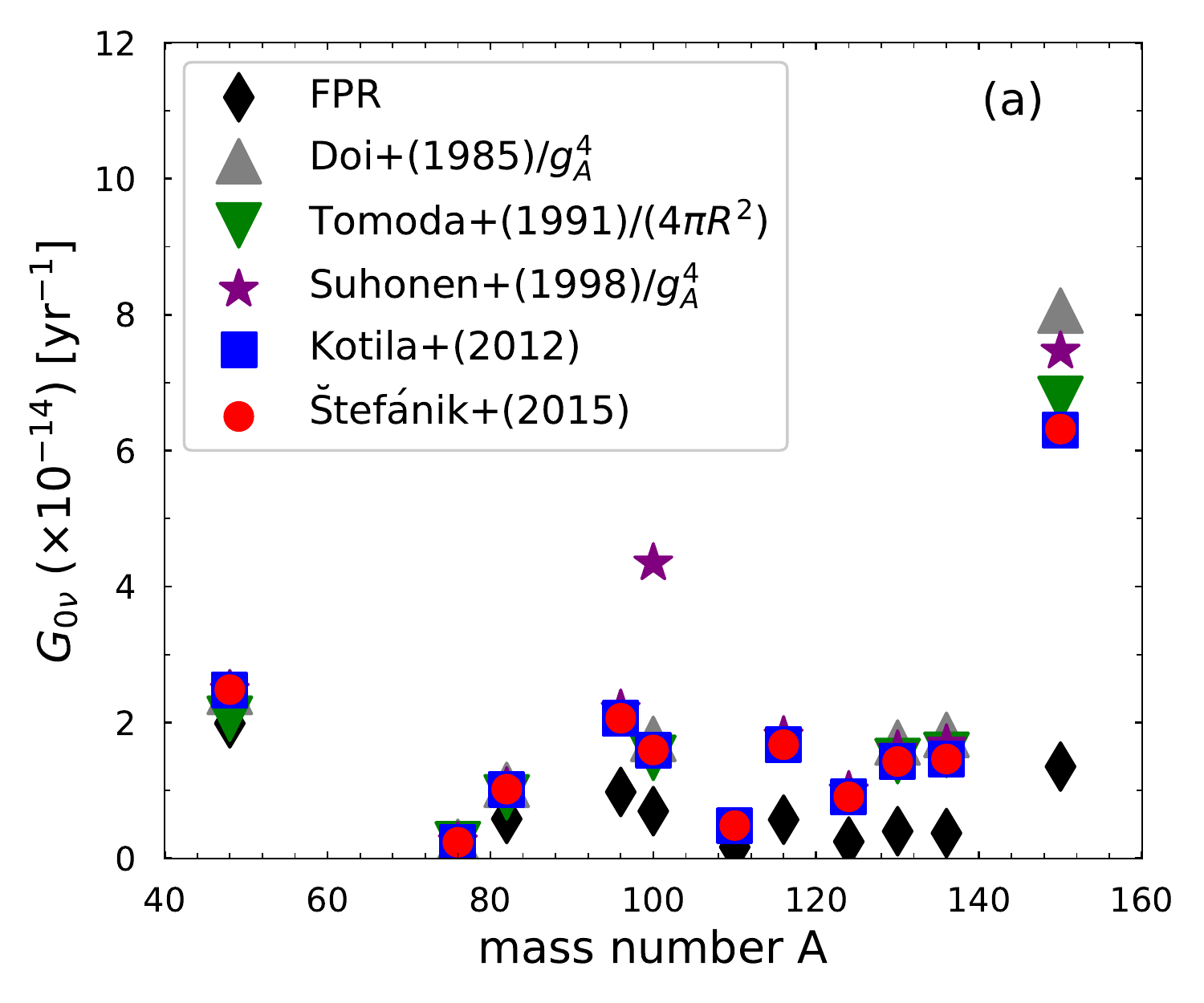}  
\includegraphics[width=8cm]{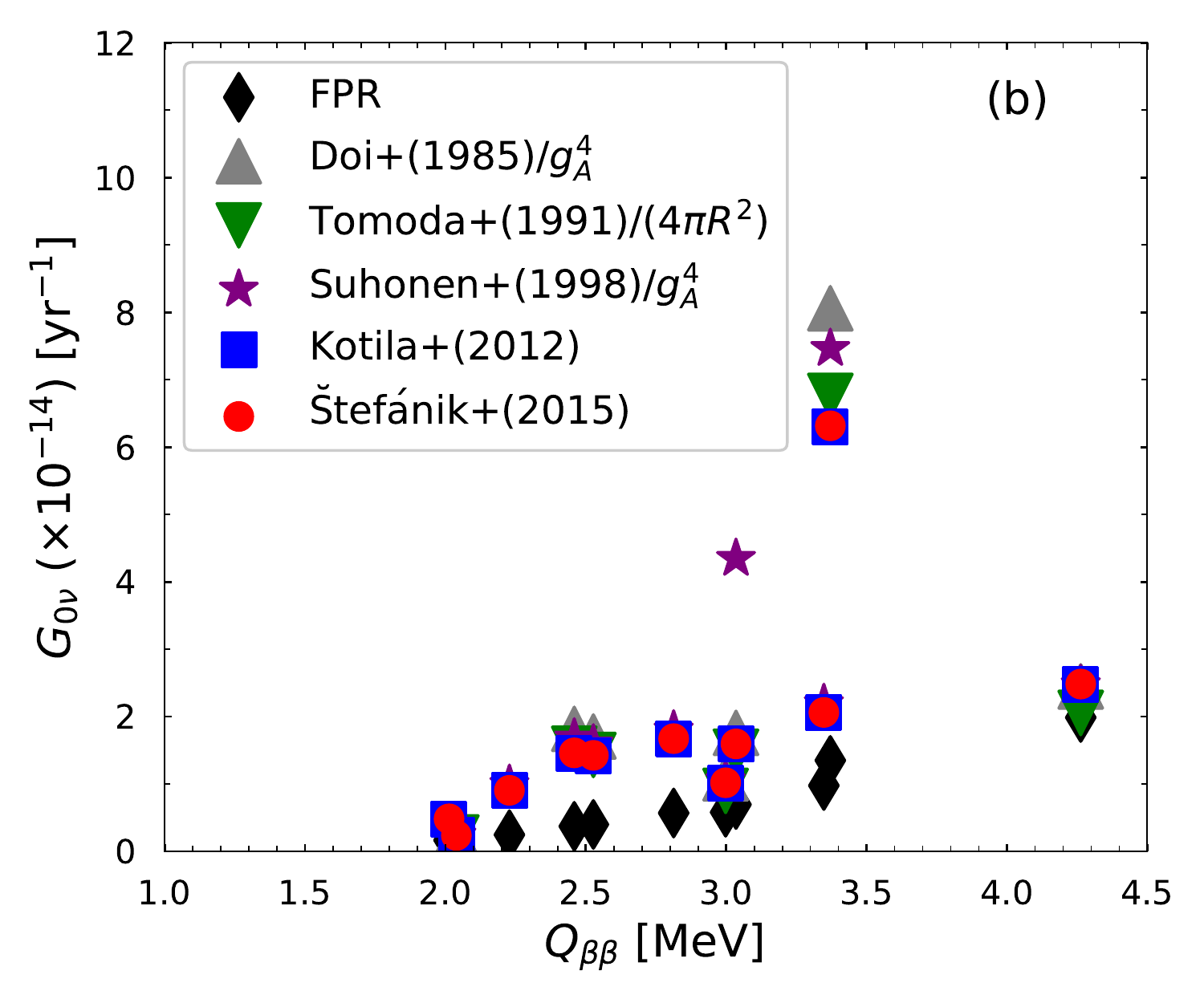}
\caption{\label{fig:G0nu} The phase-space factors $G_{0\nu}$ for candidate $0\nu\beta\beta$ decays calculated using the FPR approximation (\ref{eq:G0nu-PR}) in comparison with the calculations using electron wavefunctions by the relativistic leading-terms approximation \cite{Doi:1985,Tomoda:1991,Suhonen:1998} and by the numerically exact solution of a Dirac equation \cite{Kotila:2012,Stefanik:2015}   as a function of (a) mass number $A$ and (b) $Q_{\beta\beta}$ value, respectively.}
\end{figure}

\subsubsection{Comparison of results from different calculations}
Figure~\ref{fig:G0nu}(a) displays the phase-space factors $G_{0\nu}$ for candidate $0\nu\beta\beta$ decays from different calculations as a function of both the mass number $A$ and $Q_{\beta\beta}$ value. It is shown that the non-relativistic FPR approximation (\ref{eq:G0nu-PR}) systematically underestimates the $G_{0\nu}$, in particular for the heavier candidate nuclei, compared to the calculations by Doi et al. \cite{Doi:1985}, Suhonen  et al.~\cite{Suhonen:1998}, Kotila  et al.~\cite{Kotila:2012}  and \v{S}tef\'{a}nik  et al.~\cite{Stefanik:2015}  using electron wave functions determined from the  solution of the Dirac equation, the results of which generally agree with each other except for $\nuclide[100]{Mo}$. The $G_{0\nu}$ of candidate decays ranges between $10^{-15}$ yr$^{-1}$ and $10^{-14}$ yr$^{-1}$. Figure~\ref{fig:G0nu}(b) shows that as expected from (\ref{eq:G0nu-PR}), the $G_{0\nu}$  generally  increases with the $Q_{\beta\beta}$ value except for $\nuclide[150]{Nd}$, which lies much higher than the systematic trend. It indicates a significant distortion Coulomb effect on the electron wave functions in heavy nuclei, for which, the non-relativistic FPR approximation turns out to be worst.  For $\nuclide[150]{Nd}$, the relativistic leading-terms approximation overestimates the phase-space factor evidently.
It is worth mentioning that in Ref. \cite{Stoica:2013} the Coulomb potential (\ref{eq:Vc}) for electrons was constructed using a realistic proton density in the daughter nucleus. The latter was determined from the solution of a Woods-Saxon potential. This effect turns out to be less than 1\%. In short, the phase-space factors $G_{0\nu}$ for the $0^+\to 0^+$ $\znubb$ decay is rather well understood and can be described very precisely using the electron wave functions from the numerical solution of the Dirac equation with the presence of Coulomb potential generated by protons inside the atomic nuclei.

%% file: 3nuclear_structure_calculations.tex
 \subsection{The two decompositions of the NME}
 \jmyr{Expressing the one-body current operator in terms of second quantization form, one can rewrite the NME of the $\znubb$ decay of $0^+_I \to 0^+_F$ in (\ref{eq:NME-light}) formally  as below
    \beqn
    \label{eq:second_quantum_NME_general}
      M^{0\nu} 
      &=& \sum_{pp'nn'} \sum_N\langle pp'|O(E_N)|nn'\rangle  \langle 0^+_F\vert    c^\dagger_{p'}c_{n'}\ket{N}\bra{N}c^\dagger_p c_n\vert 0^+_I\rangle
      \eeqn 
 where the indices $p, p'$ are for proton states, and $n, n'$ for neutron states.  The $\langle pp'|O(E_N)|nn'\rangle$ is a matrix element depending on the energies $E_N$ of the intermediate states $\ket{N}$. Its expression can be found for instance in Ref.~\cite{Simkovic:2008} and will be discussed later.  In the angular-momentum coupled form, the above NME can be written as~\cite{Simkovic:2008,Senkov2013}
\beqn
\label{eq:total_NME_general}
M^{0\nu} &=& \sum_{I,J^\pi} \sum_{\kappa(pp^{\prime}nn^{\prime})}  (-1)^{j_{n}+j_{p^{\prime}}+J+I} \sqrt{2I+1}
\left\{\begin{array}{lll}
j_{p} & j_{n} &  J \\
j_{n^{\prime}} & j_{p^{\prime}} & I
\end{array}\right\} 
\left\langle pp'(I)\left\| O (E_N)  
\right\| nn' (I)\right\rangle \nonumber \\
& &\times\left\langle 0_{F}^{+}  \left\| \left[c_{p^{\prime}}^{+} \tilde{c}_{n^{\prime}}\right]^{J} \right\| N(J^\pi) \right\rangle
\left\langle  N(J^\pi) \left\|\left[c_{p}^{+} \tilde{c}_{n}\right]^{J}\right\| 0_{I}^{+}\right\rangle.
\eeqn
where $J^\pi$ is the spin-parity of  the intermediate state, while $I$ is the coupled angular momentum of proton pair $(p,p')$ or neutron pair $(n,n')$. The symbol $\kappa$ stands for the set of indices $(p, p^{\prime}, n, n^{\prime})$. The doubly barred matrix elements are reduced in the angular momentum space. The operator  $\tilde c_{jm}$ is defined as $\tilde c_{jm}=(-1)^{j+m}c_{j-m}$. As discussed in Refs.~\cite{Simkovic:2008,Senkov2013,Senkov:2016}, the total NME in Eq. (\ref{eq:total_NME_general}) can be decomposed 
\beq
\label{eq:k-decomposition}
 M^{0\nu}  =  \sum_{\kappa}  M^{0\nu}_{\kappa},
\eeq
where the $M^{0\nu}_{\kappa}$ can be further decomposed either into a summation over all the intermediate states (referred to as $J$-decomposition)
\beq
\label{eq:ph-decomposition}
M^{0\nu}_{\kappa} = \sum_{J} M^{0\nu}_{J,\kappa}({\rm ph}),
\eeq
or into a summation over all the two-particle states  (referred to as $I$-decomposition)
\beq
\label{eq:pp-decomposition}
M^{0\nu}_{\kappa} = \sum_{I} M^{0\nu}_{I,\kappa}({\rm pp}).
\eeq
It has been shown by \v{S}imkovic et al.~\cite{Simkovic:2008} that the two types of components can be related to each other by the following formula,
\beqn
\sum_{\kappa} M^{0\nu}_{J,\kappa}({\rm ph})
&\equiv & \sum_{\kappa,I,I^{\prime}}(-1)^{I+I^{\prime}} \hat{I} \hat{I}^{\prime} \hat{J}^{2}\left\{\begin{array}{lll}
j_{p} & j_{n} & J \\
j_{n^{\prime}} & j_{p^{\prime}} &  I^{\prime}
\end{array}\right\}  \left\{\begin{array}{lll}
j_{p} & j_{n} & J \\
j_{n^{\prime}} & j_{p^{\prime}} & I
\end{array}\right\} 
\frac{Z_{I^{\prime},\kappa}}{Z_{I,\kappa}} M^{0\nu}_{I,\kappa}({\rm pp}),
\eeqn
where the short-hand notation $Z_{I,\kappa}$ is introduced as $Z_{I,\kappa}\equiv\left\langle pp'(I)\left\| O (E_N)  
\right\| nn' (I)\right\rangle$. Note that the matrix element $Z_{I,\kappa}$ is an unsymmetrized matrix.
}
 
 \subsection{The NME in the closure approximation}
   
  \jmyr{If we approximate the energies $E_N$ of intermediate states  with the average value $\langle E_N\rangle$, the transition operator $O^{0\nu}$ in Eq.(\ref{eq:dbd_operator_long_closure}) and the matrix element $\langle pp'|O(E_N)|nn'\rangle$ will be independent on the intermediate states. Under this approximation, one can employ the  closure relation $ \sum_N\ket{N}\bra{N}=1$ to get rid of the intermediate states. In this case,  the NME of the transition $0^+_I \to 0^+_F$ in (\ref{eq:second_quantum_NME_general}) can be rewritten as below
    \beqn
    \label{eq:NME_M_scheme}
      M^{0\nu} 
      &=&  \sum_{pp'nn'} \langle pp'|O|nn'\rangle  \langle 0^+_F\vert    c^\dagger_{p'}c^\dagger_p c_nc_{n'}\vert 0^+_I\rangle\nonumber\\
      &\equiv&\dfrac{1}{4}\sum_{pp'nn'} O_{pp'nn'}  \langle 0^+_F\vert  c^\dagger_p c^\dagger_{p'}c_{n'}c_n\vert 0^+_I\rangle \nonumber\\
      &=& \sum_{p\leq p'; n\leq n'} \dfrac{1}{(1+\delta_{pp'})(1+\delta_{nn'})}O_{pp'nn'} \rho_{pp'nn'},
      \eeqn 
       where we introduce an antisymmetrized two-body matrix element $O_{pp'nn'}\equiv-(1-P_{pp'})(1-P_{nn'})\langle pp'|O|nn'\rangle$ with $P_{ij}$ being the exchange operator of the indices $i$ and $j$. }

       The  two-body transition density $\rho_{pp'nn'}$  is defined as
       \beq
       \rho_{pp'nn'} \equiv  \langle 0^+_F\vert  c^\dagger_p c^\dagger_{p'} c_{n'}c_n\vert  0^+_I\rangle.
       \eeq
   It is shown in Eq.(\ref{eq:NME_M_scheme}) that the NME becomes a sum of products of the two-body transition matrix elements with the two-body transition density. The former only depends on the transition operator and is thus universal for different candidate nuclei. It can be computed in advance and can be used as inputs for different nuclear models.  The latter is given by nuclear wave functions and is, therefore, nucleus- and model-dependent. In addition, it is convenient to rewrite the Eq.(\ref{eq:NME_M_scheme}) in angular-momentum coupled form,  
 \beqn
 \label{eq:SM-NME-I}
 M^{0\nu} &\equiv & \sum_I M^{0\nu}_{I} 
  =\sum_I\sum_{p\leq p'; n\leq n'}  \dfrac{(2I+1)}{(1+\delta_{pp'})(1+\delta_{nn'})}
 {\cal O}^I_{pp'nn'} \rho^I_{pp'nn'},
 \eeqn
 where the angular-momentum coupled unnormalized two-body transition matrix element is defined as
  \beq
 {\cal O}^I_{pp'nn'}
 \equiv \sqrt{(1+\delta_{pp'})(1+\delta_{nn'})}   \langle pp'(I)\vert O \vert nn'(I)\rangle
 \eeq
 with 
 \beqn  
 \langle pp'(I)\vert O \vert nn'(I)\rangle 
 &=& \sum_{m_pm_{p'}m_nm_{n'}}
    \langle j_pm_p j_{p'}m_{p'}\vert IM\rangle
   \langle j_nm_n j_{n'}m_{n'}\vert IM\rangle O_{pp'nn'},
\eeqn 
 and the angular-momentum coupled two-body transition density 
 \beqn
 \rho^I_{pp'nn'} 
 &\equiv& -\dfrac{1}{\sqrt{2I+1}}\langle 0^+_F\vert \left[[c^\dagger_{p}c^\dagger_{p'}]^I
 [\tilde c_{n}\tilde c_{n'}]^I\right]^0\vert 0^+_I\rangle\nonumber\\
 &=&\sum_{m_pm_{p'}m_nm_{n'}} \langle j_pm_{p'} j_n m_{n'}\vert IM\rangle \langle j_nm_n j_{n'}m_{n'}\vert IM\rangle
 \langle 0^+_F\vert c^\dagger_p  c^\dagger_{p'} c_{n'} c_n \vert 0^+_I\rangle.
 \eeqn

\jmyr{It has been pointed out in Ref.~\cite{Simkovic:2008} that after antisymmetrization it is not possible to recover the particle-hole decomposition of the total NME into the multipoles $J^\pi$ of the intermediate states as shown in Eq.(\ref{eq:ph-decomposition}). Therefore, it was suggested to use the unsymmetrized two-body matrix element $Z_{I,\kappa}$ in order to compare the results of QRPA calculations without the closure approximation. We note that once the closure approximation is employed in the calculations,  one can compute the $I$-decomposition $M^{0\nu}_{I}$ of the total NME directly with Eq.(\ref{eq:SM-NME-I})~\cite{Iwata:2016,Jiao:2017,Coraggio:2020,Yao:2020PRL}.}
    
 Subsequently, we will present the formulas for computing the two-body transition matrix element and transition density on a spherical HO basis, respectively. The calculation of transition density requires  wave functions of initial and final states, which are determined for instance with the PGCM.
 
 \subsubsection{The two-body transition matrix element}
 In this section, we present the two-body transition matrix element of $0\nu\beta\beta$-decay operator in {\em unnormalized} $jj$-coupling two-body spherical HO basis which is related to that in the $LS$ coupling form  via the $9$-j symbol \cite{Edmonds:1955}
  \beqn
  \label{eq:jj2LS} 
     &&\ket{n_1l_1j_1 t_1,  n_2 l_2 j_2 t_2; IM}\nonumber\\
     &=&\sum_{\lambda S} \hat j_1\hat j_2 \hat\lambda  \hat S
     \nj{l_1}{1/2}{j_1}{l_2}{1/2}{j_2}{\lambda }{S}{I} 
     \ket{n_1l_1n_2l_2 (l_1l_2)\lambda S;IM}\otimes \ket{t_1t_2},
 \eeqn
  where  $n$, $l$, $j$ and $t$ are principal, angular momentum and isospin quantum numbers, respectively. The $t_1, t_2$ are indices for isospin projections.
  The symbol $\hat j$ is defined as $\hat j\equiv \sqrt{2j+1}$. The orbital angular momenta $(l_1, l_2)$ of the two nucleons couples to $\lambda$ and their spin momenta $(1/2, 1/2)$  couples to $S$.  
  
   The two-body transition operator $O^{0\nu}_\alpha$  can be rewritten as follows,
\beqn
\label{eq4app:transition}
O^{0\nu}_\alpha 
 \equiv \sum_{1<2} O^{0\nu}_\alpha(1,2)
 =
 \dfrac{1}{2}\sum_{1,2} V_{\alpha,K}(r_{12}) \left(C^K_\alpha \cdot  S^K_\alpha\right) \tau^+_1\tau^+_2,
 \eeqn 
 where $\alpha$ stands for either F, GT, or T, respectively. The nonzero matrix element of the isospin raising operator is $\bra{pp}  \tau^+_1\tau^+_2\ket{nn}=1$. The $V_{\alpha,K}(r_{12})$ is just twice of the neutrino potential $H_{\alpha,K}(r_{12})$,
\begin{align}
 V_{\alpha,K}(r_{12}) 
 &=
\dfrac{4R_0}{\pi g^2_A(0)} \int^\infty_0 \dd q \, q^2 \dfrac{ h_\alpha (\bm q^2)}{q(q+E_d)} j_K(qr_{12}), 
\end{align}
where the function $j_K(qr_{12})$ is the spherical Bessel function of rank $K$ with $\hat{\vec{r}}_{12}=\vec{r}_{12}/\lvert\vec{r}_{12}\rvert$. 
For  the case of $\alpha={\rm F}$ and GT, $K=0$, and for $\alpha=$T, $K=2$. As the tensor operator $S^{\bm{r}}_{12}$ can be rewritten as~\cite{Horie:1961}
\beqn
S^{\bm{r}}_{12}
&\equiv & 3(\vec{\sigma}_1\cdot \hat{\bm{r}}_1)(\vec{\sigma}_2\cdot \hat{\bm{r}}_2)-\vec{\sigma}_1\cdot\vec{\sigma}_2 \nonumber\\
&=& 3\{\vec{\sigma}_1\otimes \vec{\sigma}_2\}^2\cdot \{\hat{\vec{r}}_1\otimes \hat{\vec{r}}_2\}^2
\nonumber\\
&=& 3 \{\vec{\sigma}_1\otimes \vec{\sigma}_2\}^2\cdot C^2_T,
\eeqn
with $C^2_T=\sqrt{4\pi/5}\langle 1010|20\rangle Y_2(\hat{\vec{r}}_{12})$,  the scalar product of two tensors  $\left(C^K_\alpha \cdot  S^K_\alpha\right)$ is thus given by
\bsub\begin{align}
C^0_{\rm F} &= 1, \quad S_{\rm F}^0=1, \\
C^0_{\rm GT}&=1,\quad S_{\rm GT}^0=\bm{\sigma}_1\cdot \bm{\sigma}_2, \\
C^2_{\rm T}&=\sqrt{\dfrac{24\pi}{5}}Y_2(\hat{\bm{r}}_{12}),\quad S_{\rm T}^2=\left[\vec{\sigma}_1\otimes \vec{\sigma}_2\right]^2.
\end{align}
\esub 

With the transition operator (\ref{eq4app:transition}),  the two-body transition matrix element in the {\em unnormalized} $jj$-coupling two-body basis (\ref{eq:jj2LS}) is given by \footnote{See also Refs.~\cite{Iwata:2016btn,Yoshinaga:2018}.}
\beqn  
&& \langle 1'2' (I)\vert O^{0\nu}_\alpha \vert 12 (I)\rangle \nonumber \\
&\equiv&\langle n'_1 l'_1 j'_1 t'_1, n'_2 l'_2 j'_2 t'_2; I|O_{\alpha}^{0 \nu}  | n_1 l_1 j_1 t_1, n_2 l_2 j_2 t_2; I  \rangle  \nonumber\\
 &=& \sum_{SS', \lambda\lambda'}  
 \langle n_1' l_1' n_2' l_2'; \lambda' | V_{\alpha,K}(r_{12})  |  n_1 l_1 n_2 l_2; \lambda \rangle   
 \langle (l_1' l_2') \lambda' S'; I|C^K_\alpha  \cdot S^K_{\alpha} |  (l_1 l_2) \lambda S; I \rangle\nonumber \\
 &&
 \hat {j_1'} \hat{j_2'}\hat {S'} \hat {\lambda'} 
  \hat {j_1} \hat{j_2}\hat {S} \hat {\lambda} 
  \left\{
\begin{array} {ccc}
l_1' & 1/2 & j_1' \\
l_2' & 1/2 & j_2' \\
\lambda' & S' & I 
\end{array}
\right\}  \left\{
\begin{array} {ccc}
l_1 & 1/2 & j_1 \\
l_2 & 1/2 & j_2 \\
\lambda & S & I 
\end{array}
\right\}\delta_{t_1',1/2}\delta_{t_2',1/2}\delta_{t_1,-1/2}\delta_{t_2,-1/2}.
\eeqn
 
 By using the Talmi-Moshinsky transformation which connects the product of two single-particle HO wavefunctions $\ket{n_1l_1n_2 l_2;\lambda\mu}$ to the HO wave function $\ket{n l NL; \lambda\mu}$ in the center-of-mass and relative coordinate 
 \begin{equation}
    \ket{n_1l_1n_2 l_2;\lambda\mu}
    =\sum_{nlNL} \bra{n lNL; \lambda} n_1l_1n_2 l_2; \lambda\rangle \cdot  \ket{n l NL; \lambda\mu},
\end{equation}
where 
\beqn
\ket{n_1l_1n_2 l_2;\lambda\mu}
 &\equiv&\sum_{m_1m_2}\langle l_1m_1l_2m_2\vert \lambda\mu\rangle 
 R_{n_1l_1}(r_1)Y_{l_1m_1}(\hat{\bm{r}}_1) R_{n_2l_2}(r_2)Y_{l_2m_2}(\hat{\bm{r}}_2),\\
\ket{n l NL; \lambda\mu}
&\equiv&\sum_{mM}\langle lmLM\vert \lambda\mu\rangle 
R_{nl}(\rho)Y_{lm}(\hat{\bm{\rho}}) R_{NL}(R)Y_{LM}(\hat{\bm{R}}),
\eeqn
where the relative coordinate is introduced as $\bm{\rho}=(\bm{r}_1-\bm{r}_2)/\sqrt{2}$, and  center-of-mass coordinate $\bm{R}=(\bm{r}_1+\bm{r}_2)/\sqrt{2}$,
one rewrites the matrix element  
\beqn
 && \langle n_1' l_1' n_2' l_2'; \lambda' | V_{\alpha,K}(r_{12}) |  n_1 l_1 n_2 l_2; \lambda \rangle  
 \langle (l_1' l_2') \lambda' S'; I| C^K_\alpha  \cdot S^K_{\alpha} |  (l_1 l_2) \lambda S; I \rangle\nonumber \\
  &=& \sum_{nlNLn'l'N'L'} 
 \bra{n lNL; \lambda} n_1l_1n_2l_2; \lambda \rangle
 \bra{n' l'N'L'; \lambda'} n'_1l'_1n'_2l'_2; \lambda' \rangle\nonumber\\
 &&\times
 \langle n' l' | V_{\alpha,K}(\sqrt{2}\rho)  | n l \rangle
  \langle N'L'\vert    NL\rangle  
 \langle (l' L') \lambda' S'; I| C^K_\alpha  \cdot S^K_{\alpha} |  (lL) \lambda S; I \rangle\nonumber\\
 &=& \sum_{nln'l'NL} 
 \bra{n lNL; \lambda} n_1l_1n_2l_2; \lambda \rangle
 \bra{n' l'N'L; \lambda'} n'_1l'_1n'_2l'_2; \lambda' \rangle\nonumber\\
 &&\times
 \langle n' l' | V_{\alpha,K}(\sqrt{2}\rho)  | n l \rangle  
 \langle (l' L) \lambda' S'; I| C^K_\alpha  \cdot S^K_{\alpha} |  (lL) \lambda S; I \rangle,
\eeqn
where the fact that the center-of-mass orbital angular momentum is conserved in the spin part of the operator, i.e., $L=L'$ is applied, and  $\langle N'L\vert NL\rangle=\delta_{NN'}$. 

The radial matrix element of the neutrino potential  
\beqn
\label{eq:radial}  
 \langle n' l'|  V_{\alpha,K}(\sqrt{2} \rho) | n l \rangle
 &=&\int d\rho \rho^2 R_{n'l'}(\rho)R_{nl}(\rho)H_{\alpha,K}(\sqrt{2} \rho),
\eeqn
 is responsible for the amplitude of each transition from a state with $n$, $l$ to another state with $n'$, $l'$. One way of choice to compute the matrix element (\ref{eq:radial}) can be found in Ref. \cite{Neacsu:2012}. 
  
  The spin-dependent part depends on the type of the transition operator. 
  \begin{itemize}
      \item For the Fermi type ($K=0$), 
  \beqn
  \langle (l' L') \lambda' S'; I| C^0_{\rm F}  \cdot S^0_{\rm F} |  (lL) \lambda S; I \rangle
  = 
  \delta_{SS'}\delta_{ll'}\delta_{\lambda\lambda'}. 
 \eeqn
 
 \item For the GT type ($K=0$),
   \beqn
  \langle (l' L') \lambda' S'; I| C^0_{\rm GT}  \cdot S^0_{\rm GT} |  (lL) \lambda S; I \rangle
  =
  \left[2S(S+1)-3\right]\delta_{SS'}\delta_{ll'}\delta_{\lambda\lambda'},
 \eeqn
 where the relation $\bm{\sigma}_1\cdot \bm{\sigma}_2=2(S^2-s^2_1-s^2_2)$ is used.
 
 \item For the tensor type ($K=2$),
    \beqn
  \langle (l' L) \lambda' S'; I| C^2_{\rm T}  \cdot S^2_{\rm T} |  (lL) \lambda S; I \rangle 
  &=&   \langle (l' L') \lambda' S'; I| 
  \sqrt{24\pi} \left[Y_2(\hat{\bm{r}}_{12}) [\bm{\sigma}_1\otimes \bm{\sigma}_2]^2\right]^0|  (lL) \lambda S; I \rangle\nonumber\\
  &=&(-1)^{L+I+1}\sqrt{120} \hat \lambda
  \hat \lambda'\hat l\hat l' \delta_{S1}\delta_{S'1}\nonumber\\
  &&\times \Gj{I}{1}{\lambda'}
  {2}{\lambda}{1}
  \Gj{l'}{\lambda'}{L}
  {\lambda}{l}{2}
  \tj{l'}{2}{l}{0}{0}{0},
 \eeqn 
 where $\langle 2020|00\rangle=1/\sqrt{5}$ and $\langle l' || Y_2||l\rangle=(-1)^{l}\sqrt{5/4\pi}\hat l\hat l'$.
  \end{itemize}
  
  \subsubsection{The two-body transition density}
    With the ground-state wave functions of initial and final nuclei defined in (\ref{eq:BMF_GCM_wf}) with $J=K=0$,  one derives the two-body transition density 
       \beqn
       \rho_{pp'nn'} 
       &=& \sum_{\kappa_I\kappa_F}   f_{\kappa_F} f_{\kappa_I}\langle\Phi_{\kappa_F}\vert    \hat P^{N_F}\hat P^{N_F} \hat P^{J=0}_{00} c^\dagger_p c^\dagger_{p'}c_{n'}c_n  \hat P^{N_I}\hat P^{Z_I} \hat P^{J=0}_{00}\vert \Phi_{\kappa_I}\rangle\nonumber\\
       &=& \sum_{\kappa_I\kappa_F}  f_{\kappa_F} f_{\kappa_I} 
       \int^{2\pi}_0 \dfrac{e^{-iN\varphi_N}}{2\pi}d\varphi_N
      \int^{2\pi}_0 \dfrac{e^{-iZ\varphi_Z}}{2\pi} d\varphi_Z \nonumber\\
      &&\times
      \dfrac{1}{8\pi^2}\int^{2\pi}_0 d\phi \int^{2\pi}_0d\psi \int^\pi_0 \sin\theta d\theta  d^{J=0}_{00}(\theta)
      \langle\Phi_{\kappa_F}\vert   c^\dagger_p c^\dagger_pc_{n'}c_n  e^{i\varphi_N\hat N}e^{i\varphi_Z\hat Z}e^{-i\phi\hat J_z} e^{-i\theta\hat J_y} e^{-i\psi\hat J_z}\vert \Phi_{\kappa_I}\rangle
     \eeqn
     In the case that the intrinsic configurations $\ket{\Phi_\kappa}$ are invariant with respect to the rotation $\hat R_z(\phi)=e^{-i\hat J_z\phi}$, i.e., $\hat R_z(\phi)\ket{\Phi_\kappa}=\ket{\Phi_\kappa}$, the integration over the $\phi$ and $\psi$ can be carried out analytically, and the two-body transition density is simplified as
     \beqn
     \rho_{pp'nn'} 
      &=& \sum_{\kappa_I\kappa_F}   f_{\kappa_F} f_{\kappa_I} 
      \int^{2\pi}_0 \dfrac{e^{-iN\varphi_N}}{2\pi}d\varphi_N
      \int^{2\pi}_0 \dfrac{e^{-iZ\varphi_Z}}{2\pi} d\varphi_Z\dfrac{1}{2}
      \int^1_{-1}  d\cos\theta\nonumber\\
      &&\times \langle\Phi_{\kappa_F}\vert   c^\dagger_p c^\dagger_{p'}c_{n'}c_n  e^{i\varphi_N\hat N}e^{i\varphi_Z\hat Z} e^{-i\theta\hat J_y}\vert \Phi_{\kappa_I}\rangle
    \eeqn
    
    \begin{itemize}
        \item If the overlap $\langle\Phi_{\kappa_F}\vert  e^{i\varphi_N\hat N}e^{i\varphi_Z\hat Z} e^{-i\theta\hat J_y}\vert \Phi_{\kappa_I}\rangle$ is nonzero, one can define 
        the rotated state 
          \beq
           \vert \Phi_{\kappa_I}(\varphi_N,\varphi_Z,\theta)\rangle
           \equiv\dfrac{e^{i\varphi_N\hat N}e^{i\varphi_Z\hat Z} e^{-i\theta\hat J_y}\vert \Phi_{\kappa_I}\rangle}{ \langle\Phi_{\kappa_F}\vert  e^{i\varphi_N\hat N}e^{i\varphi_Z\hat Z} e^{-i\theta\hat J_y}\vert \Phi_{\kappa_I}\rangle},
          \eeq 
 in terms of which the two-body transition density can be written as, 
   \beqn
     \rho_{pp'nn'}  
      &=& \sum_{\kappa_I\kappa_F}   f_{\kappa_F} f_{\kappa_I} 
      \int^{2\pi}_0 \dfrac{e^{-iN\varphi_N}}{2\pi}d\varphi_N
      \int^{2\pi}_0 \dfrac{e^{-iZ\varphi_Z}}{2\pi} d\varphi_Z\dfrac{1}{2}
      \int^1_{-1}  d\cos\theta\nonumber\\
      &&\times
         \langle\Phi_{\kappa_F}\vert   c^\dagger_p c^\dagger_{p'}c_{n'}c_n \vert \Phi_{\kappa_I}(\varphi_N,\varphi_Z,\theta)\rangle \cdot  \langle\Phi_{\kappa_F}\vert  e^{i\varphi_N\hat N}e^{i\varphi_Z\hat Z} e^{-i\theta\hat J_y}\vert \Phi_{\kappa_I}\rangle.
    \eeqn 
    
    The overlap in the above integral can be calculated with the generalized Wick theorem \cite{Balian:1969},
           \beqn
         && \langle\Phi_{\kappa_F}\vert    c^\dagger_p c^\dagger_{p'}c_{n'}c_n \vert \Phi_{\kappa_I}(g)\rangle\nonumber\\
          &=&\Bigg(\rho^p_n(\bm{q};\bm{q},g){\cal \rho}^{p'}_{n'}(\bm{q};\bm{q},g)  -\rho^p_{n'}(\bm{q};\bm{q},g)\rho^{p'}_n(\bm{q};\bm{q},g)
+\kappa^{pp'}(\bm{q};\bm{q},g)\kappa_{nn'}(\bm{q};\bm{q},g)\Bigg),
          \eeqn 
  where the label $g$ stands for $(\varphi_N,\varphi_Z,\theta)$ and the mixed densities are determined by the formulas (\ref{eq:mix_dens}), (\ref{eq:mix_pairing1}) and (\ref{eq:mix_pairing2}). 
    
  \item The overlap is zero if the basis states of both initial and final nuclei are HF states, in which case one cannot use the generalized Wick theorem. We label the HF states as $\ket{\Phi_{\kappa_I}(A,Z)}$ and  $\ket{\Phi_{\kappa_F}(A,Z+2)}$, respectively, which can be rewritten as
  \beqn
  \ket{\Phi_{\kappa_I}(A,Z)} &=& \prod^N_{n=1} b^\dagger_n \ket{0}\otimes\prod^Z_{p=1} b^\dagger_p \ket{0},\\
  \ket{\Phi_{\kappa_F}(A,Z+2)} &=& \prod^{N-2}_{n=1} a^\dagger_n \ket{0}\otimes\prod^{Z+2}_{p=1} a^\dagger_p \ket{0},
  \eeqn
  where $N+Z=A$. The  $a^\dagger, b^\dagger$ are introduced as the particle creation operators of the final and initial HF states, respectively. For neutrons, 
 \beqn
 a^\dagger_n = \sum_m \alpha^\ast_{mn} c^\dagger_m,\quad
 b^\dagger_n = \sum_m \beta^\ast_{mn} c^\dagger_m,
 \eeqn 
 and for protons
  \beqn
 a^\dagger_p = \sum_m \gamma^\ast_{mp} c^\dagger_m,\quad
 b^\dagger_p = \sum_m \delta^\ast_{mp} c^\dagger_m,
 \eeqn 
the operators conserve particle number.  The $c^\dagger_m$ is a particle creation operator in HO basis. For convenience, one introduces the rotated neutron creation operator 
  \beq
  \tilde b^\dagger_n(\theta)
  =e^{-i\theta\hat J_y} b^\dagger_n e^{i\theta\hat J_y}
  =\sum_m \beta^\ast_{mn} \langle m'\vert e^{-i\theta\hat J_y}\vert m\rangle^\ast c^\dagger_{m'}
  = \sum_{m'} \tilde\beta^\ast_{m'n}(\theta) c^\dagger_{m'},
  \eeq
  and proton creation operator
  \beq
  \tilde b^\dagger_p(\theta)
  =\sum_{m'} \tilde\delta^\ast_{m'p}(\theta) c^\dagger_{m'},
  \eeq
  where $\tilde\beta^\ast_{m'n}(\theta)\equiv\sum_m R_{m'm}(\theta)\beta^\ast_{mn}$ and $\tilde\delta^\ast_{m'p}(\theta)\equiv\sum_m R_{m'm}(\theta)\delta^\ast_{mp}$  with $R_{m'm}(\theta)=\langle n'l'j'm'| e^{-i\theta\hat J_y}|nljm\rangle$ being the small Wigner-D function. In terms of the rotated operators,  one finds the overlap for neutrons
  \beqn
  &&
  \bra{0} a_{n_{N-2}} \ldots  a_{n_2}a_{n_1} c_{n'}c_n   \tilde b^\dagger_{n_{1}}(\theta)\tilde b^\dagger_{n_{2}}(\theta)\ldots\tilde b^\dagger_{n_N}(\theta) \ket{0}\nonumber\\
  &=&
   (-1)^{(2N-4)(2N-4-1)/2})\bra{0} a_{n_1}a_{n_2}\ldots a_{n_{N-2}} c_{n'}c_n   \tilde b^\dagger_{n_{1}}(\theta)\tilde b^\dagger_{n_{2}}(\theta)\ldots\tilde b^\dagger_{n_N}(\theta) \ket{0}\nonumber\\ 
&=& \operatorname{det}({\cal A}^{(n'n)T} {\cal B}^\ast),
\eeqn
where the dimensions of both matrices ${\cal A}$ and ${\cal B}$ are $N\times N$.  The matrix ${\cal A}=\alpha$ except for the elements of the $(N-1)$-th and $N$-th columns ${\cal A}^{(n'n)}_{jN-1}=\delta_{jn'}$ and ${\cal A}^{(n'n)}_{jN}=\delta_{jn}$ with $j=1,\ldots, N$, and ${\cal B}=\tilde \beta(\theta)$. The candidate nuclei are even-even nuclei, therefore the phase $(-1)^{(N-2)(2N-4-1)/2})=1$. Similarly, one obtains the overlap for protons, 
\beqn
 &&\bra{0} a_{p_{Z+2}}\ldots a_{p_2}a_{p_1} c^\dagger_p c^\dagger_{p'} \tilde b^\dagger_{p_{1}}(\theta)\tilde b^\dagger_{p_{2}}(\theta) \tilde b^\dagger_{p_{Z}}(\theta)  \ket{0}\nonumber\\
 &=& (-1)^{(2Z+4)(2Z+4-1)/2})
 \bra{0} a_{p_1}a_{p_2}\ldots  a_{p_{Z+2}} c^\dagger_p c^\dagger_{p'} \tilde b^\dagger_{p_{1}}(\theta)\tilde b^\dagger_{p_{2}}(\theta) \tilde b^\dagger_{p_{Z}}(\theta)  \ket{0}\nonumber\\ 
&=& \operatorname{det}({\cal X}^T {\cal Y}^{(pp')\ast}),
\eeqn
where the dimensions of both matrices ${\cal X}$ and ${\cal Y}$ are $(Z+2)\times (Z+2)$. The matrix ${\cal X}=\gamma$, and ${\cal Y}^{(pp')}=\tilde \delta(\theta)$ except for the elements of the $(Z+1)$-th and $(Z+2)$-th columns ${\cal Y}^{(pp')}_{jZ+1}=\delta_{jp}$ and ${\cal Y}^{(pp')}_{jZ+2}=\delta_{jp'}$ with $j=1,\ldots,Z+2$. 

With the above relations,  two-body transition density is simplified into the following form 
  \beqn
     \rho_{pp'nn'}  
      &=&  \sum_{\kappa_I\kappa_F}   f_{\kappa_F} f_{\kappa_I}  
      \dfrac{1}{2}
      \int^1_{-1}  d\cos\theta \langle\Phi_{\kappa_F}\vert   c^\dagger_p c^\dagger_{p'}c_{n'}c_n    e^{-i\theta\hat J_y}\vert \Phi_{\kappa_I}\rangle\nonumber\\
      &=&
      \sum_{\kappa_I\kappa_F}   f_{\kappa_F} f_{\kappa_I} 
      \dfrac{1}{2}
      \int^1_{-1}  d\cos\theta 
         \operatorname{det}({\cal A}^{(n'n)T} {\cal B}^\ast)\operatorname{det}({\cal X}^T {\cal Y}^{(pp')\ast}).
    \eeqn  
     
     \end{itemize}

 \subsection{The NME without the closure approximation}
 
 In the QRPA and shell-model calculations, the NME of $0\nu\beta\beta$ can be evaluated without the closure approximation, i.e., the energies of intermediate states  are computed explicitly. \jmyr{According to Eq.(\ref{eq:FermiOperator}), the NME in Eq.~(\ref{eq:second_quantum_NME_general}) can be rewritten as
\beqn
 M^{0\nu}_\alpha 
 &=& \frac{8R_0}{g_A^2(0)} \sum_{N} \sum_{pnp'n'} \langle  N |  c^\dagger_n c_p |   0_F^+ \rangle     \langle  N |  c^\dagger_{p'} c_{n'} |   0_I^+ \rangle  K_{pnp'n'}^\alpha,
\eeqn
where the use of the following relation 
  \beq
  j_0( q r_{mn}) = 4\pi \sum_{L=0}^\infty j_L(qr_m) j_L(qr_n) \sum_{M=-L}^{L} Y^*_{LM}(\hat{\vec{r}}_m )  Y_{LM} (\hat{\vec{r}}_n),
  \eeq
  is made, and the $K$-matrix element is defined as
\beq 
 K_{pnp'n'}^\alpha =  \int dq q  \sum_{LM}   \frac{1} { q+E_N - (E_I+E_F)/2 } \langle  n | {\mathcal O}_\alpha^- | p  \rangle^* \langle  p' |  {\mathcal O}_\alpha^+ | n'  \rangle
\eeq
where $\alpha$ stands for either GT or F. The one-body  transition operator ${\mathcal O}^{\pm}_\alpha$ is defined as follows,
 \bsub
 \beqn
 \label{eq:QRPA-Fermi}
 {\mathcal O}^{\pm}_{\rm F} &=&    j_L(qr) Y_{LM}(\hat{\vec{r}}) \tau^{\pm},\\
  \label{eq:QRPA-GT}
 {\mathcal O}^{\pm}_{\rm GT} &=&     j_L(qr) Y_{LM}(\hat{\vec{r}}) \vec{\sigma}   \tau^{\pm}.
\eeqn
\esub
}

 \begin{figure}[t] 
\centering
\includegraphics[width=10cm]{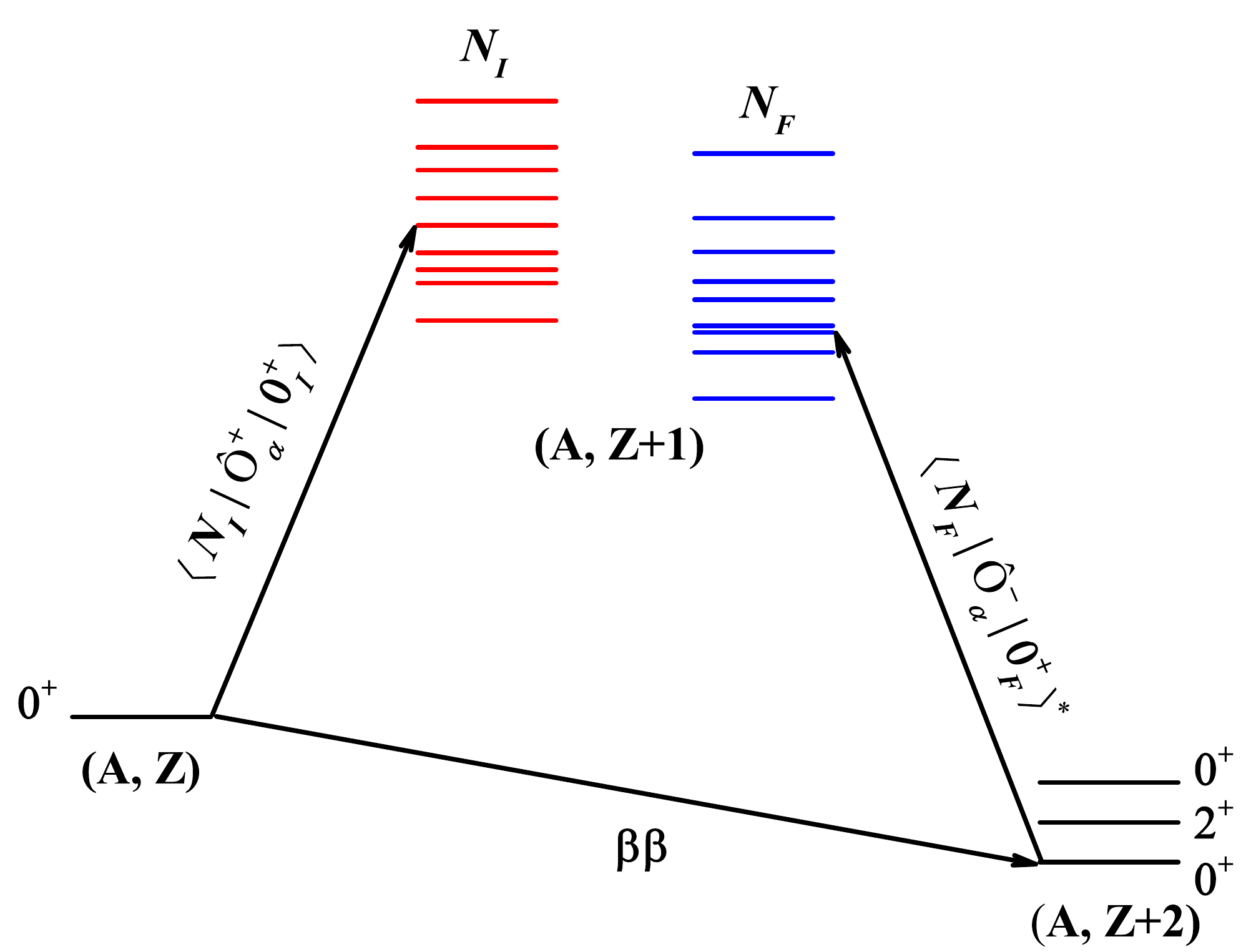} 
\caption{\label{fig:QRPA4DBD} A schematic picture illustrating the QRPA calculation of the NME of $\znubb$ decay, where the transition operators $\hat O^{\pm}_\alpha$ are defined as $\hat O^{+}_\alpha\equiv \sum_{pn}\langle  p | {\mathcal O}_\alpha^+ | n  \rangle c^\dagger_p c_n$ and $\hat O^{-}_\alpha\equiv \sum_{pn}\langle  n | {\mathcal O}_\alpha^- | p  \rangle c^\dagger_n c_p$, respectively. See text for details.}
\end{figure}

In the canonical basis representation of the QRPA, the one-body transition density has the following form 
 \bsub
 \beqn 
  \langle  N |  c^\dagger_n c_p |   0_F^+ \rangle  &=& X_{pn} u_n v_p + Y_{pn} v_n u_p, \\
  \langle  N |  c^\dagger_p c_n |   0_I^+ \rangle  &=& -(X_{pn} v_n u_p + Y_{pn} u_n v_p).
\eeqn
\esub

For the sake of convenience, we introduce $\ket{N_I}$ for the  intermediate states created by exciting the ground state  $\ket{0_I^+}$ of the initial nucleus and  $\ket{N_F}$ for those created by exciting the ground state  $\ket{0_F^+}$ of the final nucleus, c.f. Fig.~\ref{fig:QRPA4DBD}. Both the $\ket{N_I}$ and $\ket{N_F}$  are calculated with the QRPA. These two sets of intermediate states are based on different quasiparticle vacua $\ket{\Phi({\bm{q}_\alpha})}$, defined in Eq.(\ref{eq:HFB_wf}). Therefore,  the total NME (neglecting the tensor term) is written with the presence of the overlap factor
\beq
  M^{0\nu}  =  \frac{8R_0}{g_A^2(0)} \sum_{N_FN_I} \sum_{pnp'n'} \langle  N_F |  c^\dagger_n c_p |   0_F^+ \rangle \langle N_F | N_I \rangle     \langle  N_I |  c^\dagger_{p'} c_{n'} |   0_I^+ \rangle  (K_{pnp'n'}^{\rm F}  + K^{\rm GT} _{pnp'n'} ),
\eeq
where the overlap factor $\langle  N_F | N_I  \rangle $  of these two intermediate states is given by
\beq
 \langle N_F | N_I \rangle = a_{N_FN_I} \langle \Phi({\bm{q}_F}) | \Phi({\bm{q}_I}) \rangle.
\eeq
Here,
\beq
a_{N_FN_I}  = \sum_{kk'll' } C^{FI}_{k l} C^{FI}_{k' l'} (X^{N_F}_{k k'} X^{N_I}_{ll'} - Y^{N_F}_{kk'} Y^{N_I}_{ll'} ) \Bigg[u^{(\bm{q}_F)}_{k} u^{(\bm{q}_I)}_{l} + v^{(\bm{q}_F)}_{k} v^{(\bm{q}_I)}_{l}\Bigg]\Bigg[u^{(\bm{q}_F)}_{k'} u^{(\bm{q}_I)}_{l'} + v^{(\bm{q}_F)}_{k'} v^{(\bm{q}_I)}_{l'}\Bigg],
\eeq
where $C^{FI}_{kl}\equiv \braket{k^{(q_F)}|l^{(q_I)}}$  is the overlap of the $k$-th and $l$-th single-particle wave functions of the HFB states $\ket{\Phi({\bm{q}_F})}$ and $\ket{\Phi({\bm{q}_I})}$ in canonical basis, respectively. The norm overlap $\langle \Phi({\bm{q}_F}) | \Phi({\bm{q}_I}) \rangle$ of the two HFB states can be calculated using the formulas introduced in Sec.~\ref{subsubsec:overlap}.

\begin{itemize}
    \item In the spherical case, the proton and neutron pair can be coupled to a good angular momentum $J$. The above formulas can be written as $J$-coupled form. 
\beq
\label{eq:NME_QRPA}
  M^{0\nu}  =  \frac{8R_0}{g_A^2(0)} \sum_{N_FN_I} \sum_{J} \frac{1}{2J+1}\sum_{pnp'n'} \langle  N^J_F ||  [c^\dagger_{n} \tilde c_{p} ]^{J} ||   0_F^+ \rangle \langle N^J_F | N^{J}_I \rangle     \langle  N^{J}_I ||  [c^\dagger_{p'} \tilde c_{n'}]^J ||   0_I^+ \rangle  (K_{pnp'n'}^{J; \rm F}  + K^{J; \rm GT} _{pnp'n'} ),
\eeq
with 
\bsub\beqn 
 K_{pnp'n'}^{J;F} &=&  \int dq q     \frac{h_F(\vec{q}^2 )} { q+\bar{E}_N - (E_I+E_F)/2 } 
 \int r^2 dr u_n(r) j_J(qr) u^*_p(r)  
 \langle  j_n l_n || Y_J || j_p l_p  \rangle^* \nonumber\\
 && \int r^{'2} dr' u^*_{p'}(r') j_J(qr') u_{n'}(r') \langle  j_{p'}l_{p'} ||  Y_J || j_{n'} l_{n'}  \rangle    , \\
  K_{pnp'n'}^{J;GT} &=&  \int dq q   \sum_{L=J-1}^{J+1} \frac{h_{GT}(\vec{q}^2 )} { q+ \bar{E}_N - (E_I+E_F)/2}  \int r^2 dr u_n(r) j_L(qr) u^*_p(r)   \langle  j_n l_n ||  [Y_L\sigma]_J || j_p l_p  \rangle^* \nonumber\\
 && \int r^{'2} dr' u^*_{p'}(r') j_L(qr') u_{n'}(r') \langle  j_{p'} l_{p'} ||  [Y_L\sigma]_J  || j_{n'} l_{n'}  \rangle.
\eeqn
\esub
The energy of the intermediate state is chosen as $\bar{E}_N=(E_{N^J_I}+E_{N^J_F})/2$, where $E_{N^J_I}$ and $E_{N^J_F}$ are the energies of the intermediate states $\ket{N^J_I}$ and $\ket{N^J_F}$, respectively. The one-body transition density is 
\bsub\beqn
 \langle  N^J_F ||  [c^\dagger_{n} \tilde c_{p} ]^{J} ||   0^+_F \rangle 
 &=&
 \sum_{pn}   \sqrt{2J+1}   (-)^{j_p-j_{n}+J+1} \Bigg[
 u^{(\bm{q}_F)}_{n} v^{(\bm{q}_F)}_{p} X_{pn}^{N^J_F}  
 +  v^{(\bm{q}_F)}_{n} u^{(\bm{q}_F)}_{p} Y_{pn}^{N^J_F} \Bigg] , \\
  \langle  N^J_I ||  [c^\dagger_{p} \tilde c_{n} ]^{J} ||   0^+_I \rangle 
  &=& \sum_{pn}     \sqrt{2J+1}  
   \Bigg[ v^{(\bm{q}_I)}_{n} u^{(\bm{q}_I)}_{p} X_{pn}^{N^J_I}
 +  u^{(\bm{q}_I)}_{n} v^{(\bm{q}_I)}_{p} Y_{pn}^{N^J_I}\Bigg].
\eeqn
\esub

 \item In the deformed case with axial symmetry, the angular momenta of single-particle states in the transition density are not conserved, but their projections along the $z$-axis can still be used to label them. In this case, the NME   in (\ref{eq:NME_QRPA})  can be evaluated either in a set of spherical harmonic oscillator basis~\cite{Fang:2011} or in cylindrical coordinate space~\cite{Mustonen:2013}. 
\end{itemize}

\subsection{Status of calculations}
In this section, we will provide an overview of the status of calculations of the NMEs for $0\nu\beta\beta$ decays, where the transition operator is derived from the standard mechanism  and the wave functions are computed from the PGCM and QRPA based on different nuclear interactions or EDFs.  The recent studies starting from realistic nuclear forces will also be discussed.

\subsubsection{The valence-space shell-model Hamiltonian}   
 The shell-model interaction is usually parametrized as follows  
\beqn
H&=& \jmy{H_{\rm sp}} 
+  \frac{1}{4}\sum_{pqrs}\sum_{I M}  {\cal V}^J_{pqrs} A^{\dagger}(pq; I M) A(rs ; I M), 
\eeqn 
where the first term $\jmy{H_{\rm sp}}$ is given by the sum of single-particle energies $\varepsilon_p$,   
 \beq
 \jmy{H_{\rm sp}} = \sum_{p} \varepsilon_p c^\dagger_p c_p,
 \eeq
 and the second term for the two-body residual interaction. The angular-momentum-coupled two-particle creation and annihilation operators are defined as
\beqn
A^{\dagger}(pq ; I M)=\sum_{m_{p}, m_{q}}\langle j_{p} m_{p} j_{q} m_{q} \mid I M\rangle c_{j_{p} m_{p}}^{\dagger} c_{j_{q} m_{q}}^{\dagger}\\
A(rs ; I M)=\sum_{m_{r}, m_{s}}\langle j_{r} m_{r} j_{s} m_{s} \mid I M\rangle c_{j_{s} m_{s}} c_{j_{r} m_{r}}
\eeqn
  The single-particle energies $\varepsilon_p$ and the unnormalized two-body interaction matrix elements
 \beq
 {\cal V}^I_{pqrs}
 =\sqrt{(1+\delta_{pq})(1+\delta_{rs})} \langle pq(I)\vert V \vert rs(I)\rangle
 \eeq
 are free parameters fitted to nuclear low-lying states. Since the shell-model interaction is usually defined in a small valence space composed of a few orbits, for which the exact solution is available, it has been frequently implemented into  BMF approaches
 to examine corresponding truncation approximation~\cite{Lauber:2021}. This kind of benchmark is helpful to understand the discrepancy in the predicted NMEs of $0\nu\beta\beta$ decay.  Here we highlight some recent efforts along this direction. Wang et al. \cite{Wang:2018PRC} compared both yrast and non-yrast nuclear states in the $sd$ shell from the VAP and shell-model calculations using the USDB interaction \cite{Brown:2006PRC}. With the mixture of only two projected states determined by the VAP, the difference in the two models for the energies of low-lying states is reduced to be within 300 keV. Bally et al.~\cite{Bally:2019PRC} compared the PGCM and shell-model calculations for the ground-state and excitation energies for even-even and odd-even Calcium isotopes using the KB3G interaction~\cite{Poves:2000NPA} defined within the $fp$ shell. Shimizu et al., \cite{Shimizu:2021PRC}  extended this study from the $fp$ shell to those of \nuclide[132,133]{Ba} in the $50<N, Z<82$ model space. For higher spin states, a constraint  on the $x$-component of angular momentum $\langle \hat J_x\rangle$ was added.   Very recently, Sanchez-Fernandez et al.~\cite{Sanchez-Fernandez:2021nfg} compared the results of PNVAP and PGCM calculations against exact shell-model calculations for the energies and electromagnetic properties of $sd$-shell nuclei in a systematic way using the  USD interaction~\cite{Wildenthal:1984PPNP}. It was found that the best variational method in this study was the PGCM using the triaxial deformations as collective coordinates and intrinsic states with neutron-proton mixing.

 \begin{figure}[t] 
\centering
\includegraphics[width=5.4cm]{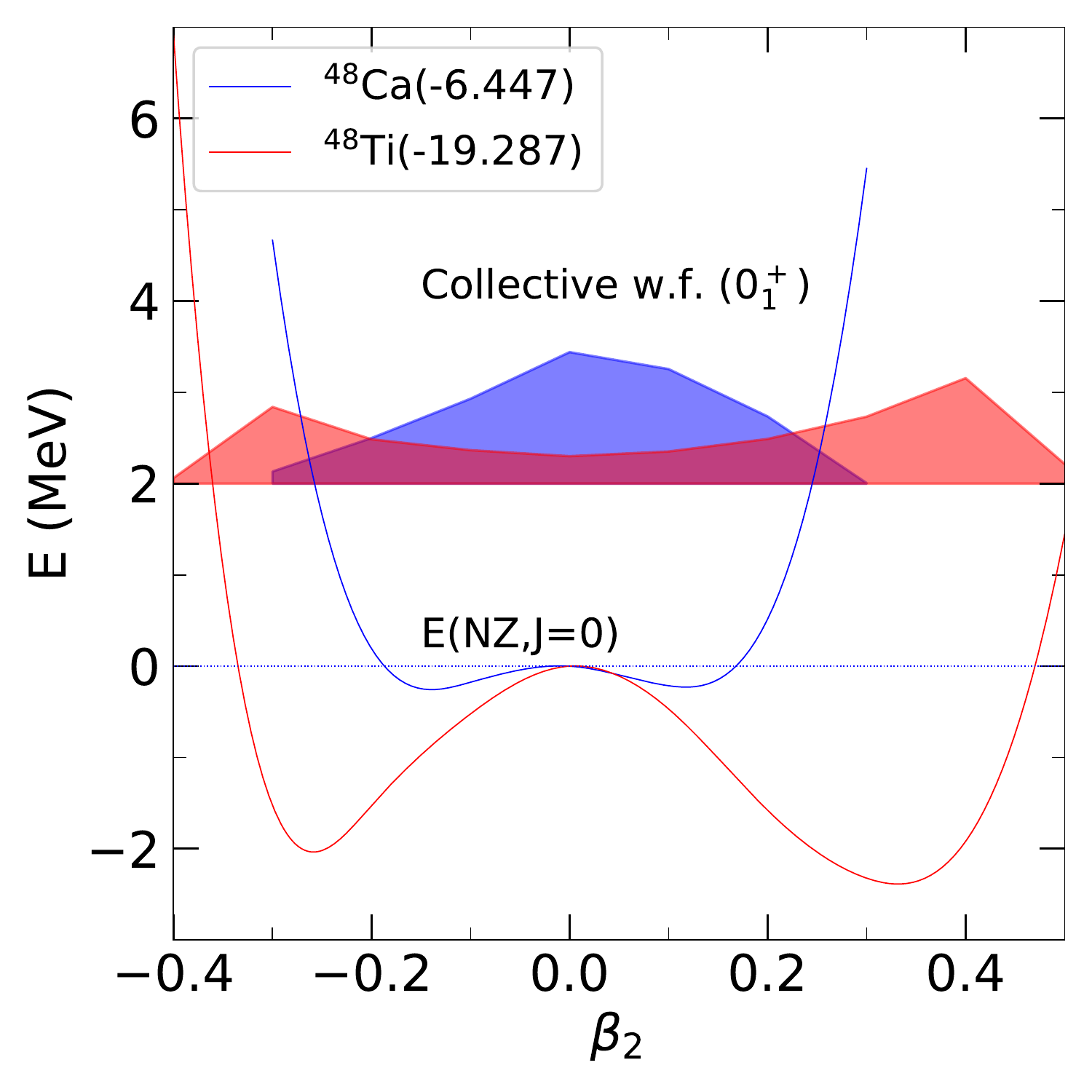}
\includegraphics[width=5.4cm]{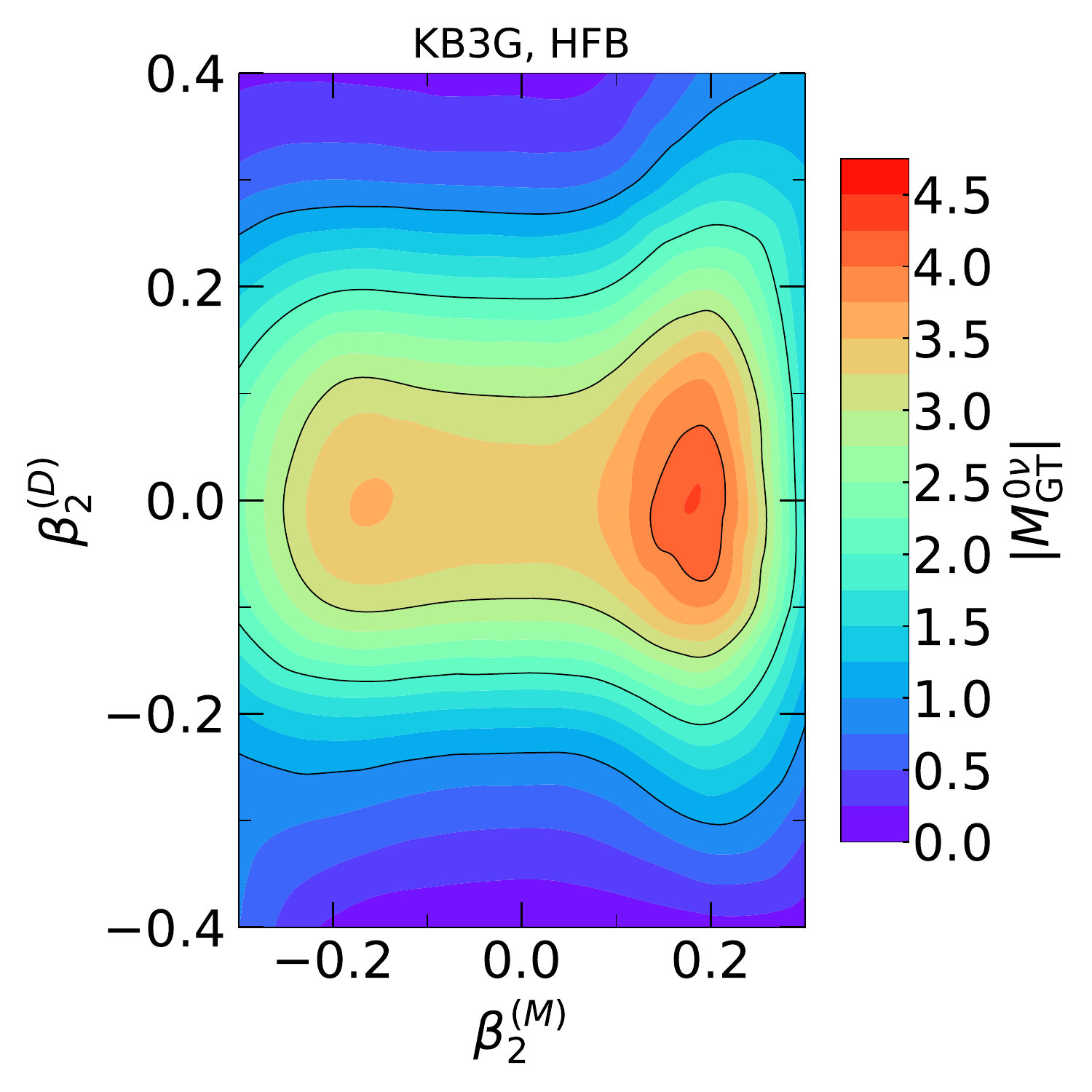}  
\includegraphics[width=5.4cm]{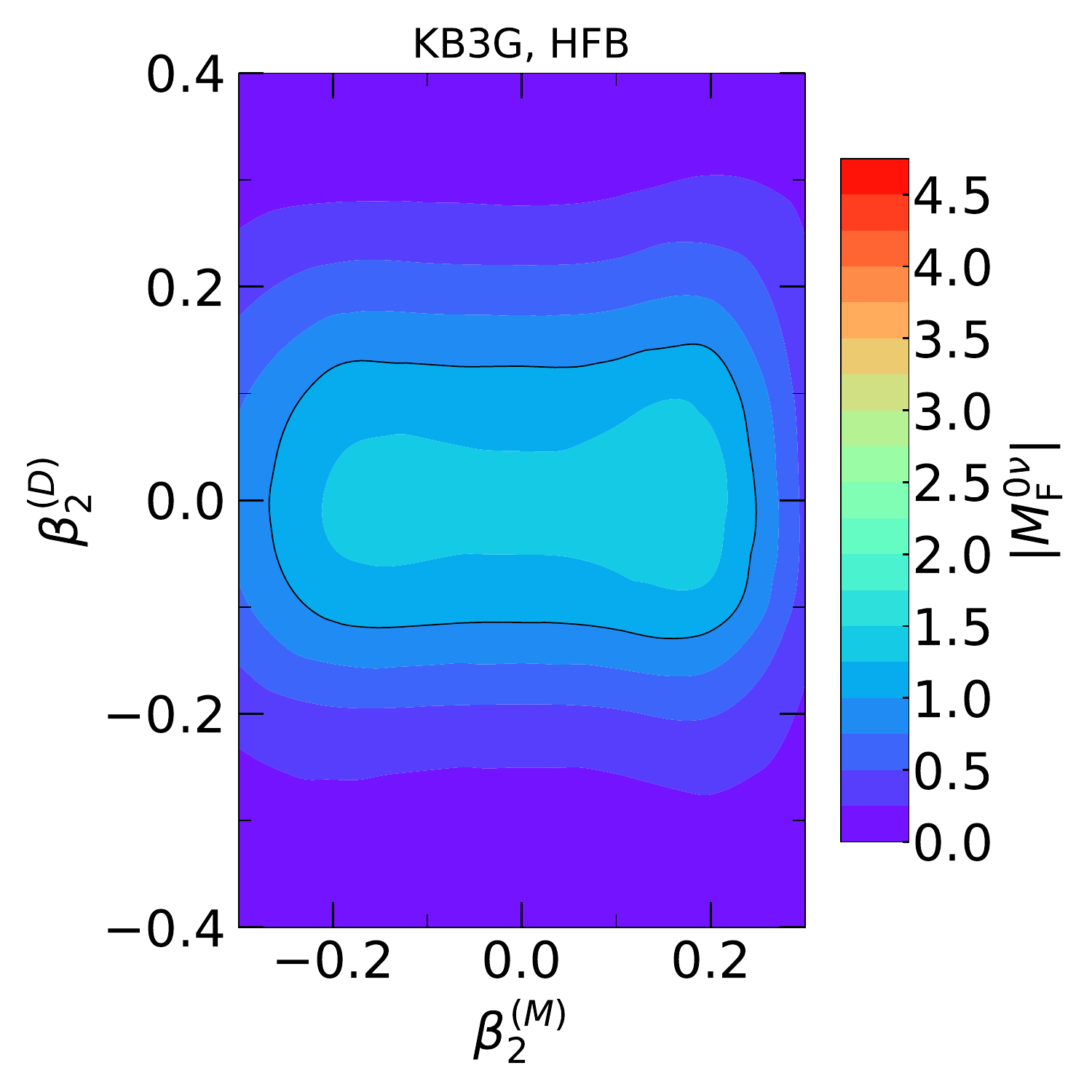}  
\caption{\label{fig:KB3G_CaTi_EHFB} (Left) The energies of  mother nucleus \nuclide[48]{Ca} and daughter nucleus \nuclide[48]{Ti} as a function of quadrupole deformation $\beta_2$ from the symmetry-projected HFB calculations with the KB3G interaction~\cite{Poves:2000NPA}. The projected energy curves are normalized to the spherical states. For \nuclide[48]{Ca} and \nuclide[48]{Ti}, the energy of the corresponding spherical state from the PHFB calculation is $-6.447$ MeV and $-19.287$ MeV, respectively. The squared collective wave functions $|g^J_\alpha(0^+_1)|^2$ of their ground states are also displayed.  In the middle and right panels are shown the absolute values of the GT  and Fermi  components of the normalized NME $\tilde {\cal M}^{0\nu}$ for the $0\nu\beta\beta$ decay of $\nuclide[48]{Ca}$ as a function of the quadrupole deformation parameters of HFB states for mother ($\beta^{(M)}_2$) and daughter ($\beta^{(D)}_2$) nuclei. The HFB states are projected onto good particle numbers and angular momentum $J=0$ when they are employed to compute the NME. }
\end{figure}

The NME of $0\nu\beta\beta$ decay provides a more sensitive probe to nuclear wave functions than the energies  as its value depends on the wave functions of two nuclei which differ from each other by replacing two neutrons with two protons and may possess different types of correlations. The validity of truncation schemes employed in the BMF approaches could be different, which would have a significant impact on the predicted NME  of $0\nu\beta\beta$ decay. Therefore, one should examine both the energies and the NME of $0\nu\beta\beta$ decay. 

The lightest candidate of $0\nu\beta\beta$ decay --$\nuclide[48]{Ca}$-- has been frequently selected for benchmarking nuclear models. The left panel of Fig.~\ref{fig:KB3G_CaTi_EHFB} displays the  energies of $\nuclide[48]{Ca}$ and $\nuclide[48]{Ti}$ as a function of the quadrupole deformation parameter $\beta_2$ for the projected HFB states with the KB3G interaction, where the deformation parameter $\beta_2$ is calculated from the expectation value of mass quadrupole moment operator
  \beq
  \beta_2 = \dfrac{4\pi}{3A R^2_0}\bra{\Phi(\bm{q})}  r^2 Y_{20}\ket{\Phi(\bm{q})}.
  \eeq
  After mixing the axially deformed states in the PGCM calculation, the ground-state energy of $\nuclide[48]{Ca}$ is $-7.12$ MeV,  while that of $\nuclide[48]{Ti}$ is $-22.18$ MeV.   Compared to the shell-model values $-7.57$ MeV and $-23.66$ MeV, respectively, the PGCM underestimates the ground-state energies of both nuclei by about 6\%.   The collective wave functions $|g^J_\alpha(0^+_1)|^2$ of their ground states defined in (\ref{eq:collective_wfs}) are also displayed, where one can see that  \nuclide[48]{Ca} is dominated by spherical state while \nuclide[48]{Ti} is dominated by deformed configurations. 

The middle and right panels of Fig.~\ref{fig:KB3G_CaTi_EHFB}  display the distribution of the GT and Fermi parts of the NME $\tilde {\cal M}^{0\nu}(\bm{q}_F, \bm{q}_I) $ as a function of the quadrupole deformation parameters of $\nuclide[48]{Ca}$ and $\nuclide[48]{Ti}$, respectively. The NME $\tilde {\cal M}^{0\nu}(\bm{q}_F, \bm{q}_I) $ is calculated by Eq.(\ref{eq:NME_M_scheme}) with the two-body transition density replaced by the configuration-dependent one  
    \beqn
    \label{eq:config_dependent_TD2B}
     \rho_{pp'nn'}(\bm{q}_F, \bm{q}_I)  
      &=& \dfrac{1}{\sqrt{{\cal N}^{N_FZ_FJ=0}_{q_F, q_F}}\sqrt{{\cal N}^{N_IZ_IJ=0}_{q_I, q_I}}} 
      \int^{2\pi}_0 \dfrac{e^{-iN\varphi_N}}{2\pi}d\varphi_N
      \int^{2\pi}_0 \dfrac{e^{-iZ\varphi_Z}}{2\pi} d\varphi_Z\nonumber\\
      &&\times \int^1_{0}  d\cos\theta
         \bra{\Phi(\bm{q}_{F})}   c^\dagger_p c^\dagger_{p'}c_{n'}c_n \ket{ \Phi(\bm{q}_I; \varphi_N,\varphi_Z,\theta)} \bra{\Phi(\bm{q}_F)}  e^{i\varphi_N\hat N}e^{i\varphi_Z\hat Z} e^{-i\theta\hat J_y} \ket{\Phi(\bm{q}_{I})},
    \eeqn 
    where ${\cal N}^{N_\alpha Z_\alpha J=0}_{\bm{q}_\alpha, \bm{q}_\alpha}$ is the normalization factor defined as
 \begin{equation}
{\cal N}^{N_\alpha Z_\alpha J=0}_{\bm{q}_\alpha, \bm{q}_\alpha}
=\langle \Phi (\bm{q}_\alpha) \vert \hat P^{N_\alpha}\hat P^{Z_\alpha} \hat P^{J=0} \vert \Phi(\bm{q}_\alpha)\rangle
 \end{equation}
 with $\alpha(=I/F)$ labels the initial (or mother) and final (or daughter) nuclei, respectively.

 \begin{figure}[t] 
\centering 
\includegraphics[width=9.2cm]{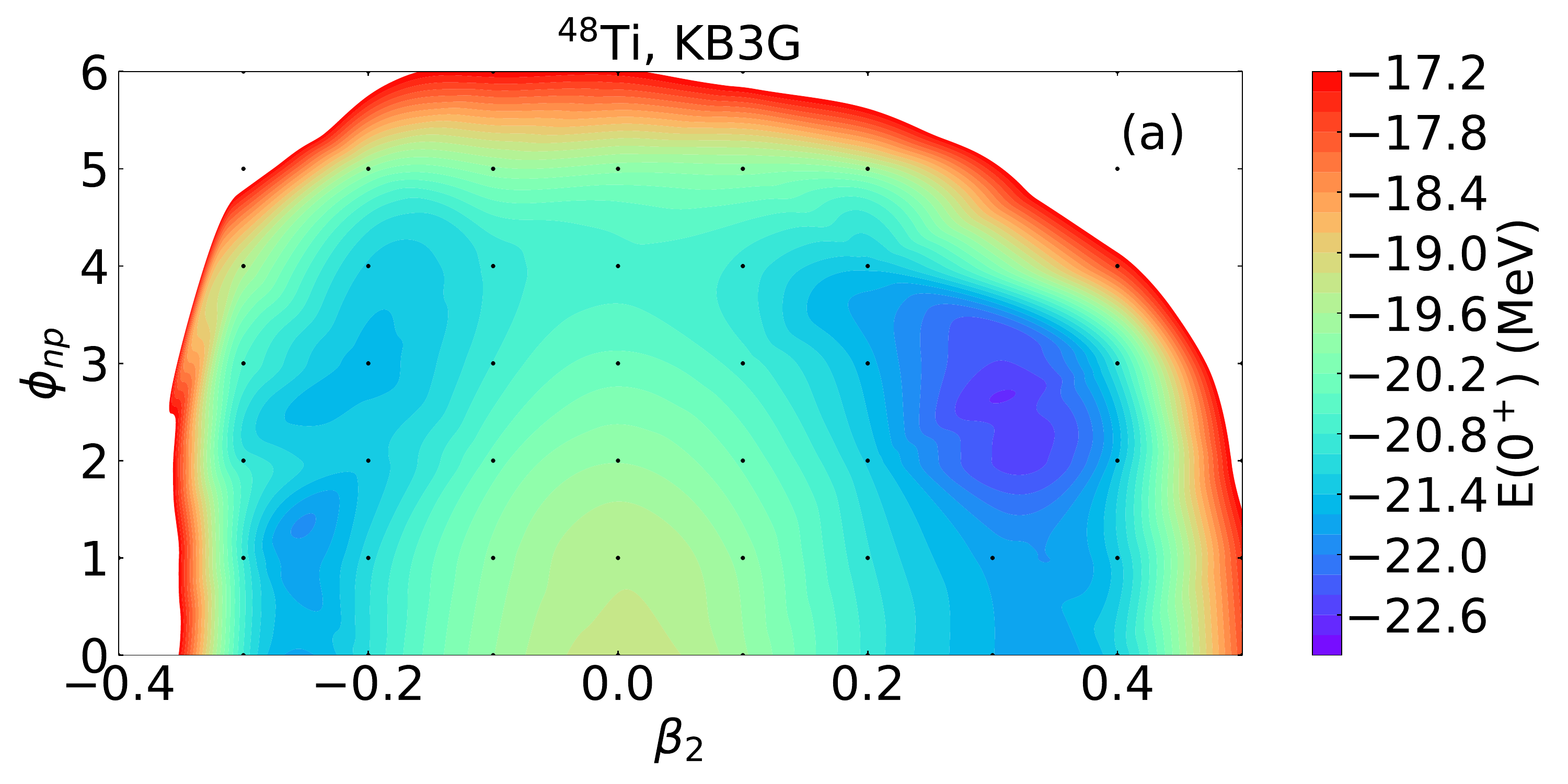}  
\includegraphics[width=9.2cm]{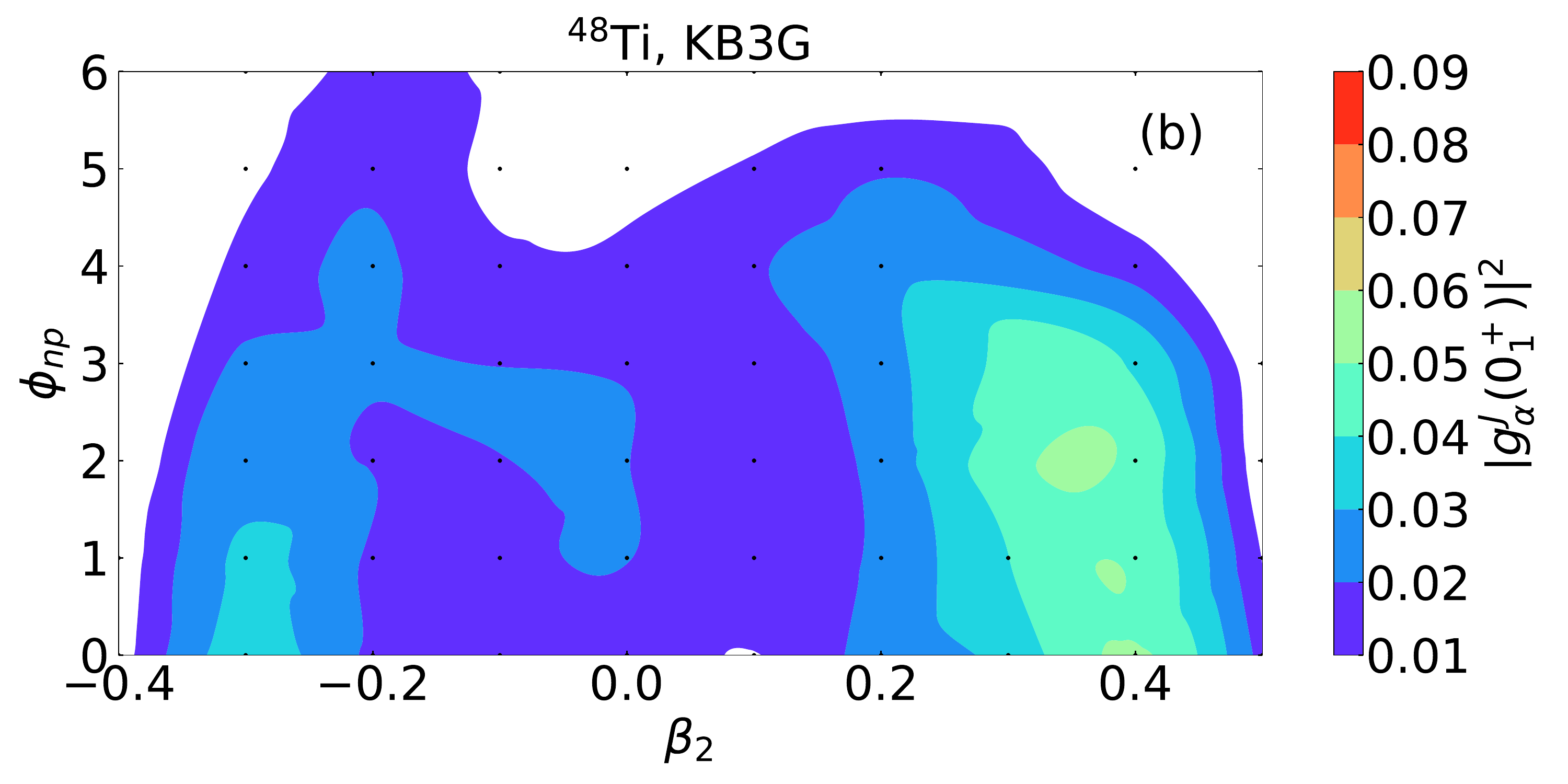}  
\caption{\label{fig:KB3G_Ti48_PES_np} (a) The energy of projected HFB state with quantum numbers $(NZ,J=0)$ and (b) the squared collective wave function $|g^J_\alpha(0^+_1)|^2$ for $\nuclide[48]{Ti}$ as a function of quadrupole deformation $\beta_2$ and neutron-proton pairing amplitude $\phi_{np}$ from the calculation using the KB3G interaction.}
\end{figure}

  It is shown in Fig.~\ref{fig:KB3G_CaTi_EHFB} that the NME is sensitive to the configuration of daughter nucleus $\nuclide[48]{Ti}$. In contrast, the NME is almost constant when varying the configuration of the mother nucleus $\nuclide[48]{Ca}$ around the spherical shape. If only the spherical configuration is considered for both nuclei, the GT part of the NME is $M^{0\nu}_{\rm GT}\simeq 3.37$. After mixing all the axially deformed configurations, it becomes $M^{0\nu}_{\rm GT}\simeq 1.78$. The deformation effect decreases significantly the value of Fermi part from 1.29 to 0.46. Fig.~\ref{fig:KB3G_Ti48_PES_np} shows the projected energy surface and distribution of the  collective wave function of  ground-state for \nuclide[48]{Ti} as a function of both quadrupole deformation $\beta_2$ and neutron-proton isoscalar pairing amplitude $\phi_{np}$, which is defined as $\phi_{np}=\langle\Phi(\bm{q})\vert (\hat P^\dag_{10}+\hat P_{10})\ket{\Phi(\bm{q})}$ with the operator 
  $\hat P^\dag_{10} = \frac{1}{\sqrt{2}} \sum_\ell  \hat{\ell} [c^\dag_\ell c^\dag_\ell]^{L=0,J=1,T=0}_{000}$. One can see that the energy surface around the minimal is rather soft towards the states with nonzero $\phi_{np}$. As a result, the collective wave function extends to the configurations with neutron-proton isoscalar pairing.  It has been illustrated  in Ref.~\cite{Jiao:2017} that the inclusion of neutron-proton isoscalar pairing fluctuation in the PGCM calculation decreases the NME further to $M^{0\nu}_{\rm GT}\simeq 0.92$, pretty close to the value $M^{0\nu}_{GT}\simeq 0.85$ from the shell-model calculation.

\begin{figure}[t]
\centering
\includegraphics[width=7.4cm]{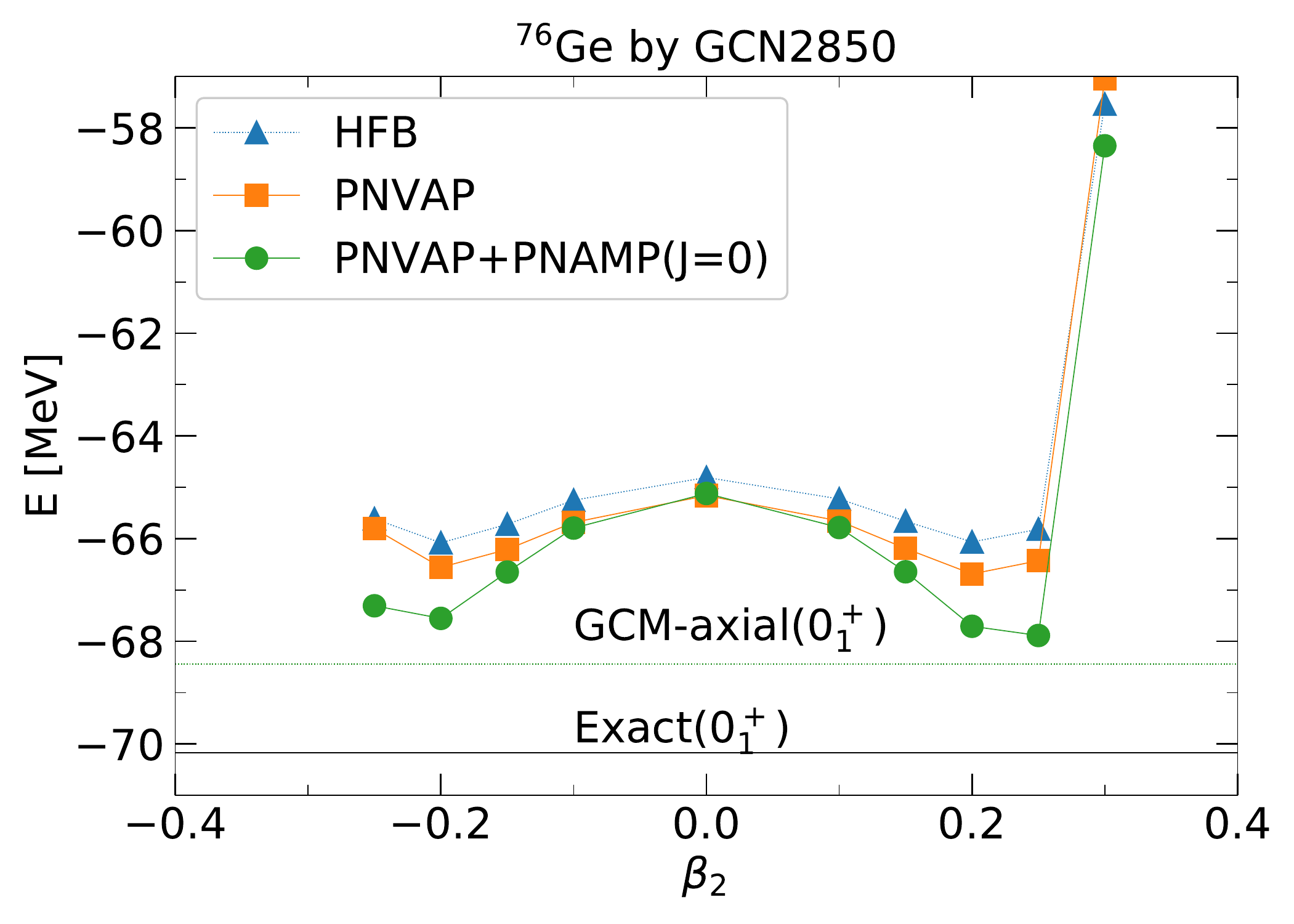} 
\includegraphics[width=7.4cm]{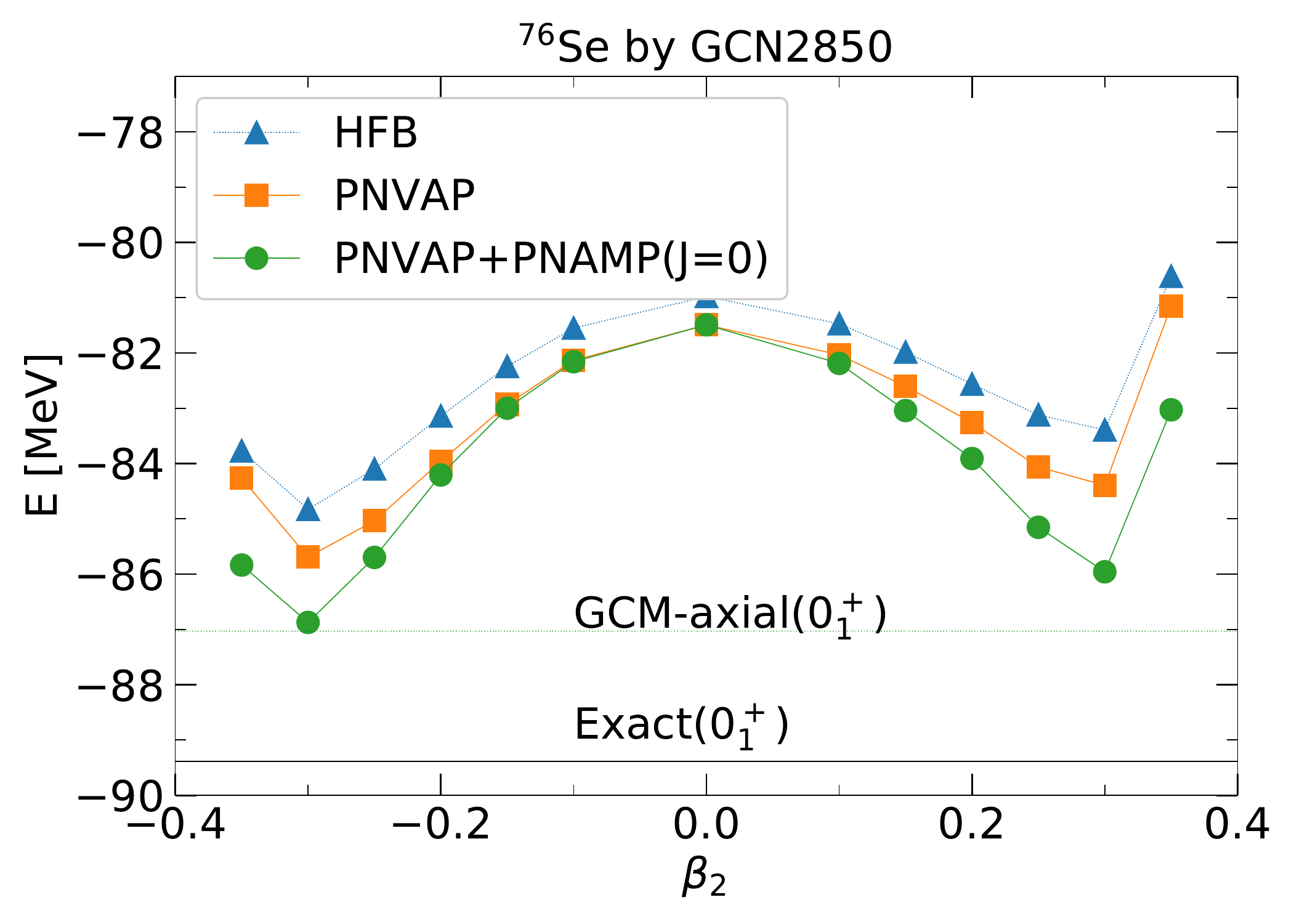} 
\caption{\label{fig:GCN2850_ENZJ_beta2}  The energies of $\nuclide[76]{Ge}$ and $\nuclide[76]{Se}$ as a function of the quadrupole deformation parameter $\beta_2$ from different calculation using the  shell-model GCN2850 interaction~\cite{Menendez:2009}. The model space spanned by the valence single-particle states limits the range of quadrupole deformation $\beta_2$.  }
\end{figure}

\begin{figure}  
\centering
\includegraphics[width=6.4cm]{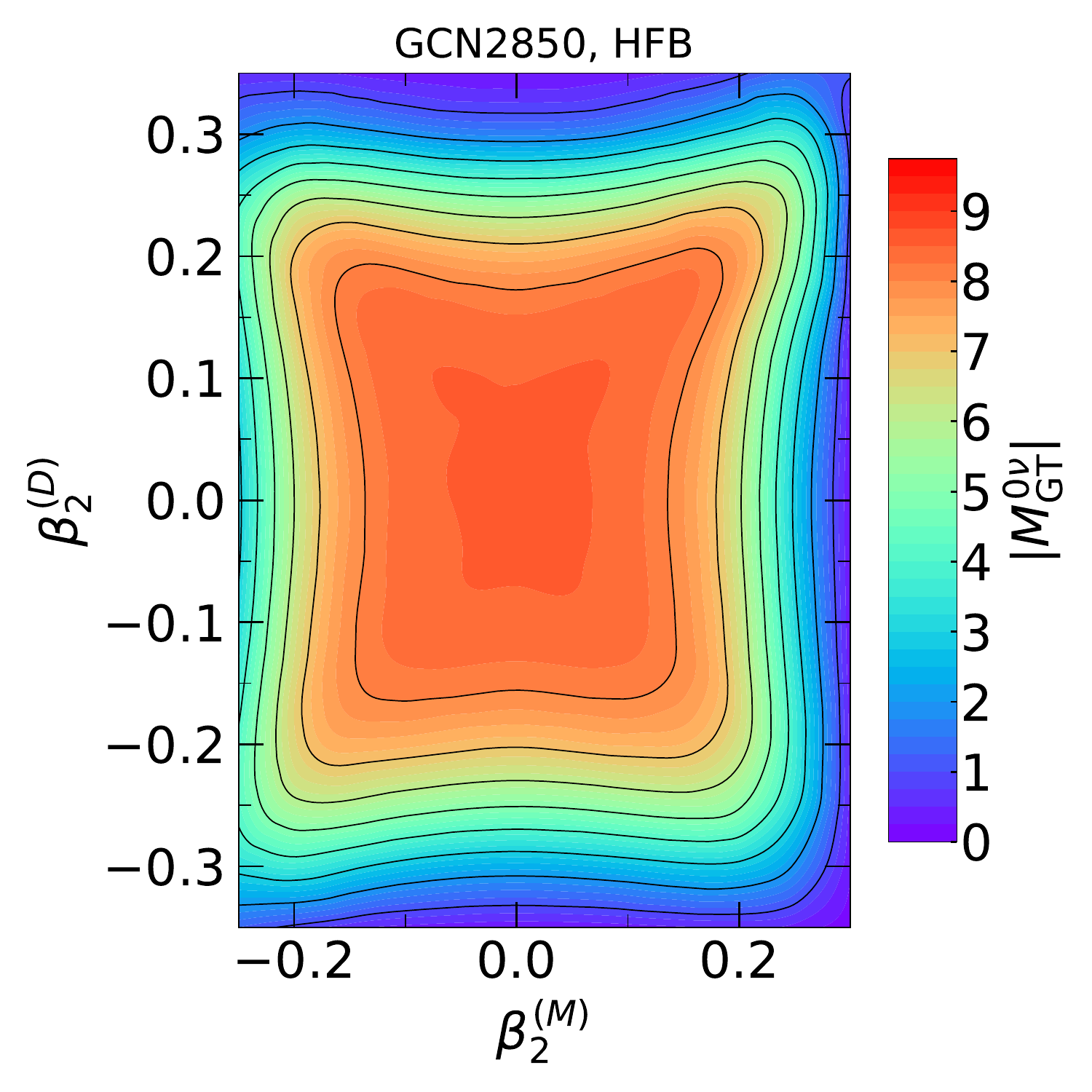} 
\includegraphics[width=6.4cm]{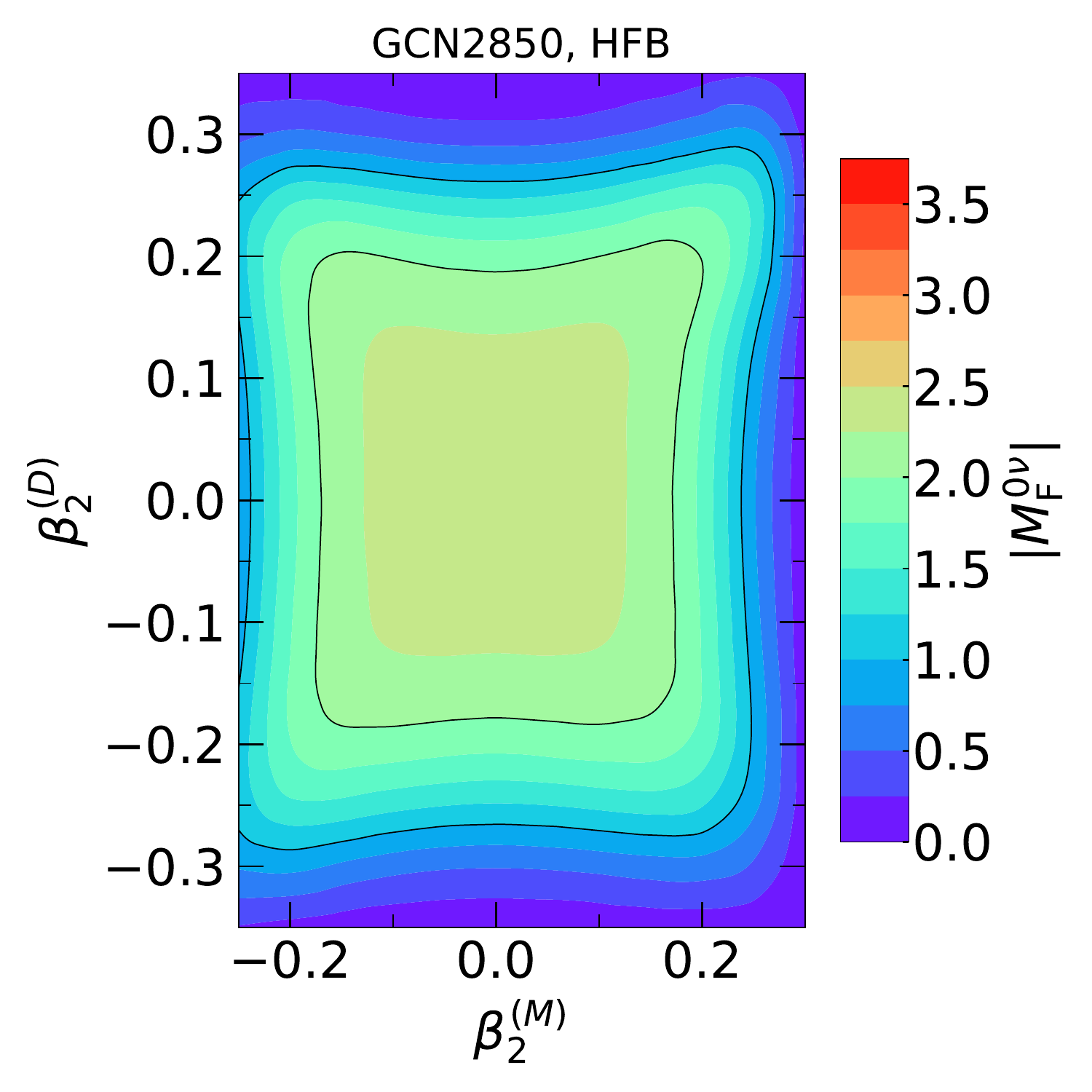}  
\caption{\label{fig:GCN2850_DBD_beta2} The same as Fig.~\ref{fig:KB3G_CaTi_EHFB}, but for $\nuclide[76]{Ge}$ from the HFB calculation with the GCN2850 interaction. }
\end{figure}

\begin{figure}
\centering
\includegraphics[width=8.4cm]{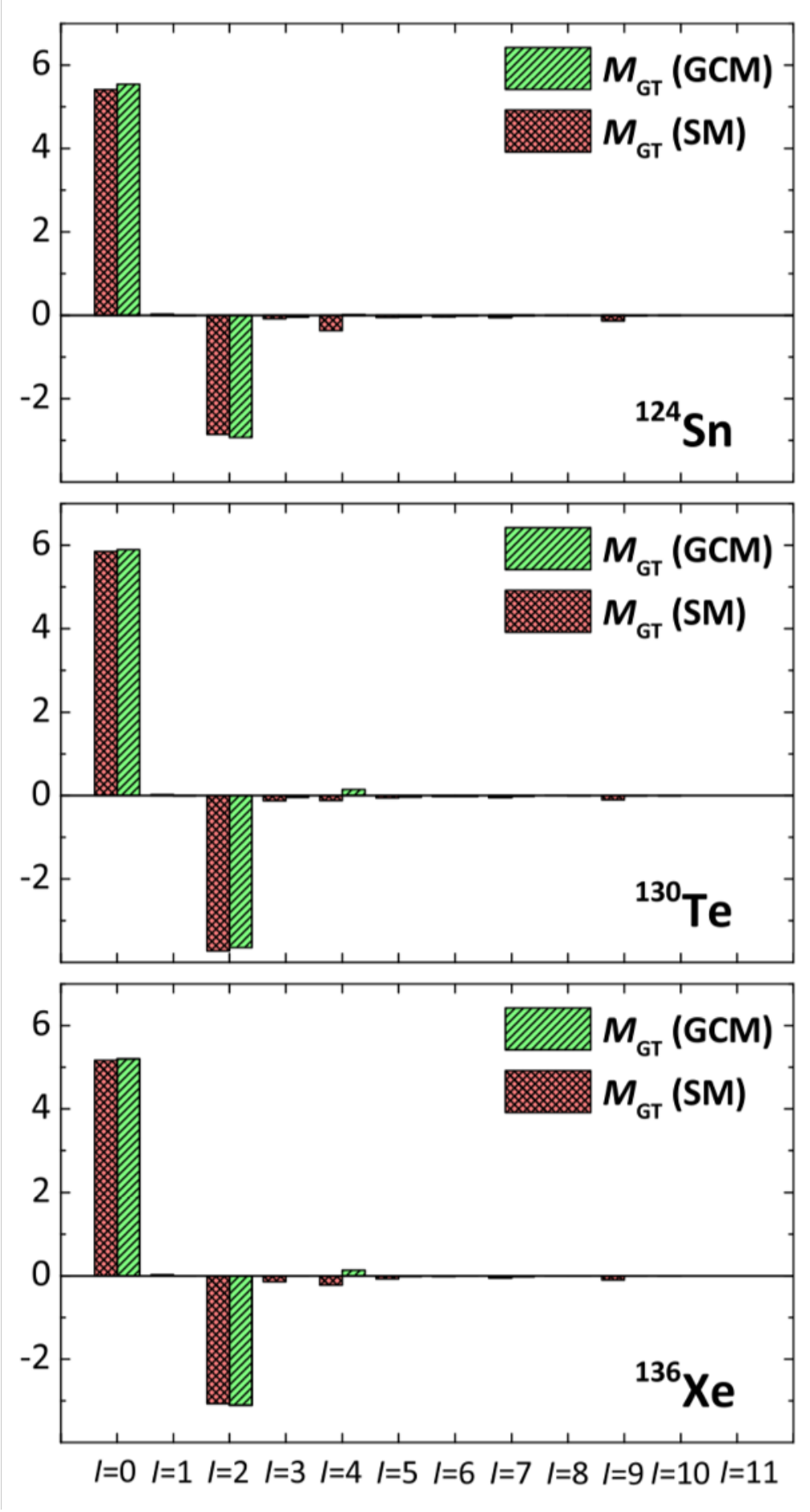}
\caption{\label{fig:Jiao_GCM_SM_NME_J}  Comparison of the $I$-decomposition $M^{0\nu}_I$ (\ref{eq:SM-NME-I}) of the GT matrix elements for the $0\nu\beta\beta$ decay of $^{124}$Sn, $^{130}$Te, and $^{136}$Xe from shell-model and PGCM calculations based on the same effective interactions.  See Ref.~\cite{Jiao:2018PRC} for details. Figure reprinted with permission from the American Physical Society. }
\end{figure}

 One of the  most popular candidates of $0\nu\beta\beta$ decay is $\nuclide[76]{Ge}$.
 Figure~\ref{fig:GCN2850_ENZJ_beta2} displays the potential energy surfaces of $\nuclide[76]{Ge}$ and $\nuclide[76]{Se}$ as a function of quadrupole deformation parameter $\beta_2$ from the calculations of pure HFB, the HFB with particle-number projection before variation (PNVAP-HFB) and that with additional particle-number and angular momentum projections after variation (PNVAP+PNAMP) based on the GCN2850 interaction~\cite{Menendez:2009}.  Because of the limited model space, the quadrupole deformation $\beta_2$ of the HFB state is confined into a finite range, in which two local energy minimal show up at both prolate and oblate deformed configurations with $\beta_2$ around $\pm0.2$ and $\pm0.3$ for $\nuclide[76]{Ge}$ and $\nuclide[76]{Se}$, respectively. The structure of the ground states of both nuclei is expected to be dominated by these two configurations. 
 Figure~\ref{fig:GCN2850_DBD_beta2} displays the distribution of the GT and Fermi parts of the NME $\tilde {\cal M}^{0\nu}$ as a function of the quadrupole deformation parameters of $\nuclide[76]{Ge}$ and $\nuclide[76]{Se}$. One can see that the GT and Fermi components share a similar pattern in the distribution, i.e., both components are peaked when the configurations of two nuclei are spherical and are quenched when those two have distinctly different shapes. After mixing axially deformed configurations in both mother and daughter nuclei with PGCM, the predicted NME for the GT, Fermi, and tensor parts is 2.60, 0.58, and -0.03, respectively. A comprehensive study of  $\nuclide[76]{Ge}$ with the inclusion of triaxiality and neutron-proton isoscalar pairing in the PGCM calculation predicts the total NME to be $M^{0\nu}=2.33$~\cite{Jiao:2017}, somewhat smaller than the value  $M^{0\nu}=2.81$ of shell-model calculation \cite{Menendez:2009}.  Figure~\ref{fig:Jiao_GCM_SM_NME_J} displays  the $I$-decomposition $M^{0\nu}_I$ (\ref{eq:SM-NME-I}) of the GT NMEs of heavier candidates $^{124}$Sn, $^{130}$Te, and $^{136}$Xe from both PGCM and shell-model calculations~\cite{Jiao:2017,Jiao:2018PRC}. It is seen that the NMEs are dominated by the cancellation of the $I=0$ and $I=2$ components. Again, the two predicted NMEs generally agree well with each other. The remaining discrepancy appears in the higher angular-momentum components, which is expected to be reproduced in the PGCM calculation with the inclusion of non-collective configurations~\cite{Jiao:2019PRC}.

\subsubsection{The separable multipole-multipole Hamiltonian}

In configuration-mixing calculations with interactions derived from {\em ab initio} methods one found~\cite{Zuker:1994,Dufour:1996} that the two-body interaction term in the shell-model Hamiltonian $H$  can be approximated by separable multipole-multipole interactions. The quadrupole part with multipolarity $\lambda=2$ is dominant in the ph-channel,
and in the pp-channel one has the monopole part with $\lambda=0$ introduced originally by Kerman in the seniority model~\cite{Kerman1961_APNY12-300}. This consideration leads to the old pairing-plus-quadrupole (PPQQ) model of Baranger and Kumar \cite{Baranger:1968,Kisslinger:1963RMP}. Sometimes a quadrupole pairing force with $\lambda=2$ is added in the pp-channel~\cite{Broglia1974_PLB50-213,Hara:1980NPA} and one ends with the separable Hamiltonian
\beq
  \label{eq:PPQQ}
  H= H_{\rm sp}  -\zeta_{QQ}\frac{1}{2} \chi \sum^2_{\mu=2} \hat{Q}_{2\mu}^{\dagger} \hat{Q}_{2\mu}-G_{M} \hat{P}^{\dagger} \hat{P}-G_{Q} \sum^2_{\mu=-2} \hat{P}_{2\mu}^{\dagger} \hat{P}_{2\mu},
  \eeq
 composed of the single-particle term $H_{\rm sp}$, a quadrupole-quadrupole interaction, as well as a monopole and a quadrupole pairing interaction,  
  \bsub \begin{align}
    \hat{Q}_{2\mu} &=\sum_{pq}\bra{p} r^2Y_{2\mu} \ket{q} c^\dagger_p c_q, \\
   \hat{P} &=\frac{1}{2} \sum_{p>0} c_pc_{\bar p}, \\
   \hat{P}_{2\mu} &=\frac{1}{2} \sum_{pq}\bra{p} r^2Y_{2\mu} \ket{q} c_pc_{\bar q}.
   \end{align} 
   \esub
  The interaction strengths are usually determined as follows~\cite{Hara:1995}: the quadrupole-quadrupole interaction strength $\chi$ is adjusted in such a way that the empirical value of quadrupole deformation $\beta_2$ is reproduced in the HFB calculation. In practical calculations, the values of $\chi$ and $G_M$ depend on nuclear mass number and model space under consideration. The PPQQ Hamiltonian is usually defined in three major HO shells around the Fermi surfaces of protons and neutrons. Here, a reduction factor $\zeta_{QQ}$ is introduced additionally in the quadrupole-quadrupole interaction term so that one is able to study the correlation between the NME of $0\nu\beta\beta$ decay and the strength of the quadrupole-quadrupole interaction term, which is responsible for the onset of quadrupole deformation in atomic nuclei.
  
 The above PPQQ Hamiltonian (\ref{eq:PPQQ}) has been frequently employed in the PHFB, PSM, and shell-model calculations with great success in the description of nuclear low-lying states which are dominated by the rotation and vibration collective excitations associated with quadrupole shapes. In particular, with the inclusion of configurations of multi-quasiparticle excitations, the PSM turns out to be a powerful tool of choice for nuclear high-spin states \cite{Hara:1995}.

 \begin{figure}[] 
\centering
\includegraphics[width=10cm]{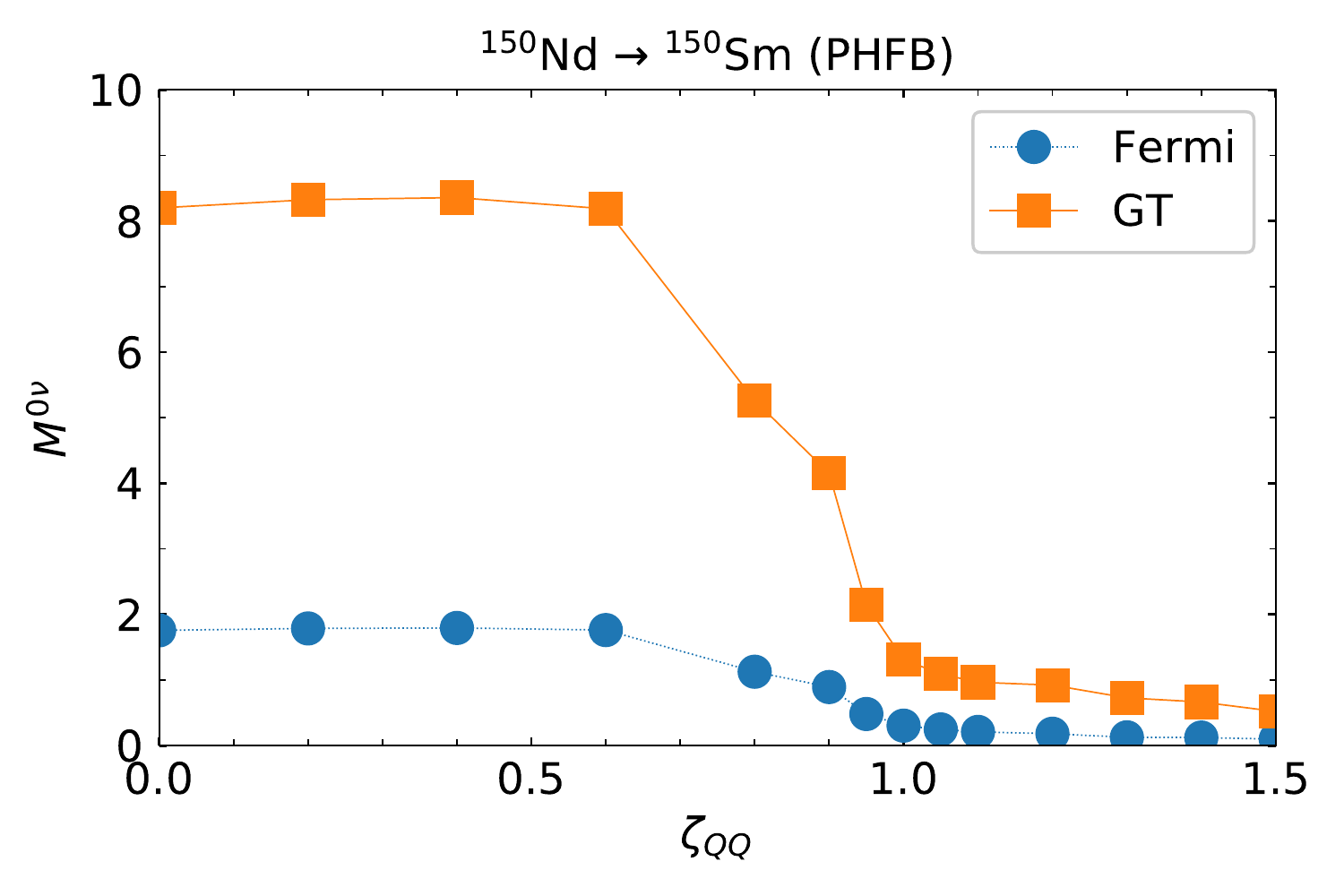}  
\caption{\label{fig:PHFB_Nd_chi}  The NME of $0\nu\beta\beta$ decay for $\nuclide[150]{Nd}$ as a function of the strength of the quadrupole-quadrupole coupling in (\ref{eq:PPQQ}) from the PHFB calculation. The results are taken from Ref.~\cite{Chaturvedi:2008}. }
\end{figure}

In the PHFB, the NME of $0\nu\beta\beta$ decay is determined by only one single configuration $\ket{\Phi(\bm{q}_{\alpha})}$ for each nucleus,
 \beqn
 M^{0\nu} 
 = \tilde {\cal M}^{0\nu} (\bm{q}_F, \bm{q}_I) 
 = \sum_{p\leq p'; n\leq n'} \dfrac{1}{(1+\delta_{pp'})(1+\delta_{nn'})}
 O^{0\nu}_{pp'nn'}  \rho_{pp'nn'}(\bm{q}_F, \bm{q}_I),
 \eeqn
where the configuration-dependent transition density $\rho_{pp'nn'}(\bm{q}_F, \bm{q}_I)$ has been defined in Eq.(\ref{eq:config_dependent_TD2B}).  Chandra et al.~\cite{Chaturvedi:2008,Chandra:2009} and Rath et al.~\cite{Rath:2013} employed the PPQQ interaction in the PHFB calculation, where the angular momentum projection was implemented to project the HFB wave function onto angular momentum $J=0$. By choosing the single-configuration $\ket{\Phi(q_{\alpha})}$ with different quadrupole deformation, they have shown  that the difference in the quadrupole deformation of initial and final nuclei has a strong quenching  effect on the NME. In particular, the impact of quadrupole-quadrupole interaction strength on the NME was investigated in detail by introducing the reduction factor $\zeta_{QQ}$ in (\ref{eq:PPQQ}). Their results for $\nuclide[150]{Nd}$ are displayed in Fig.~\ref{fig:PHFB_Nd_chi}, which shows clearly that the NME decreases significantly with the parameter $\zeta_{QQ}$ increasing from 0.6 to 1.0, consistent with the picture of shape transition from spherical to deformed shapes. See a recent review~\cite{Rath:2019} for the applications of the PHFB to $0\nu\beta\beta$ based on different mechanisms. It is worth noting that particle-number projection has not been implemented in the PHFB studies yet. According to the studies in Refs.~\cite{Yao:2015,Meng:2017}, restoration of particle numbers may increase significantly the NME of $0\nu\beta\beta$ decay in the case that pairing correlation collapses artificially in the mean-field solution for one of the two nuclei.  
 
\begin{figure}
\centering
\includegraphics[width=12cm]{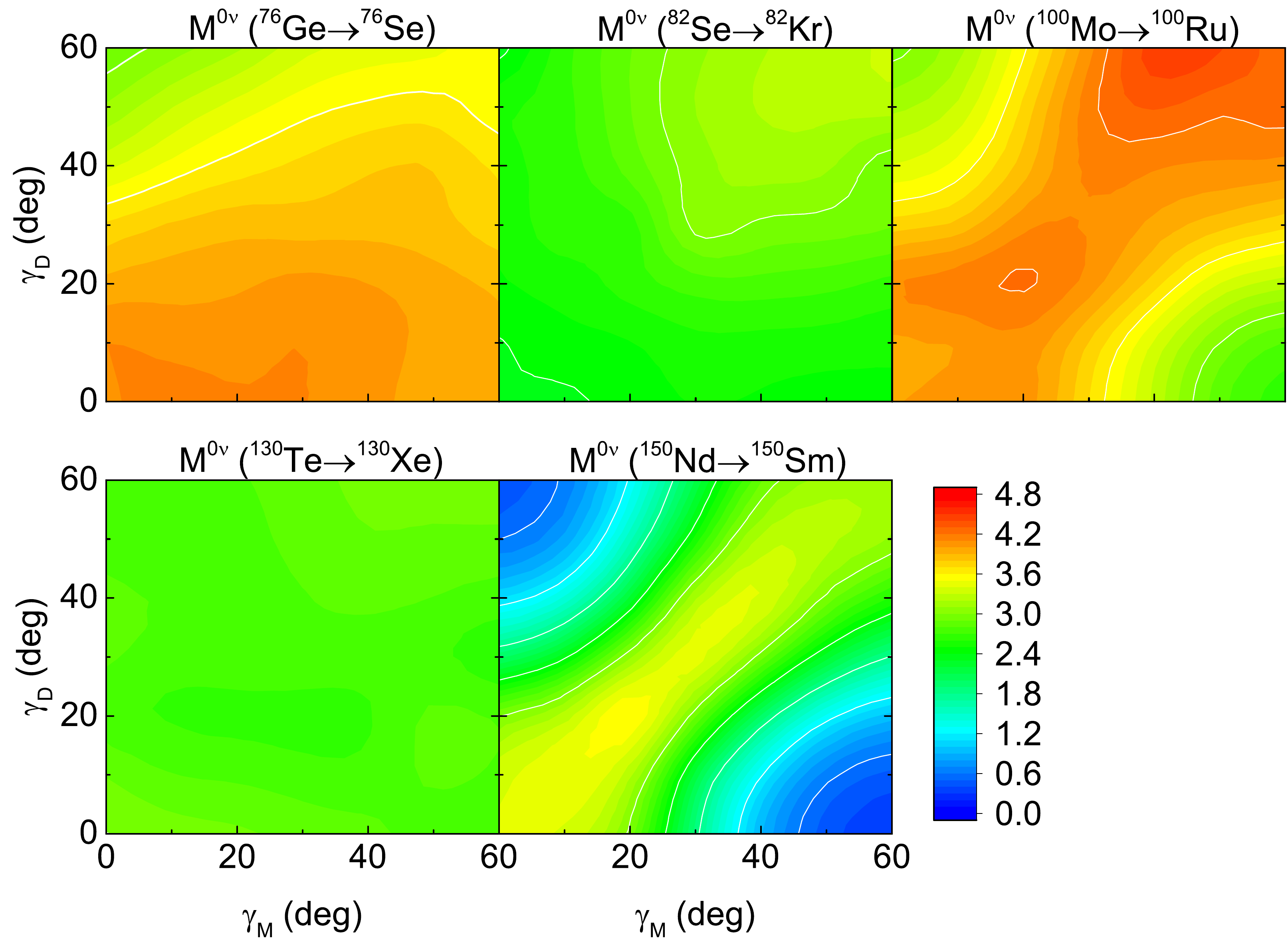}  
\caption{\label{fig:Wang2021_TPSM} The NMEs of $0\nu\beta\beta$ decay for several candidate nuclei as functions of the triaxial deformation parameters $\gamma_{M/D}$ for the mother and daughter nuclei, respectively. See Ref.~\cite{Wang:2021} for details. Figure reprinted with permission from the American Physical Society.   }
\end{figure}

 Recently, Wang et al.~\cite{Wang:2021} extended the PHFB to triaxial PSM by including triaxiality and  two-quasiparticle excitations. It was shown that the inclusion of the two quasiparticle configurations reduces the NMEs  of $0\nu\beta\beta$ decay moderately, by about 5\%. Fig.~\ref{fig:Wang2021_TPSM} displays how the NME changes as functions of the triaxiality parameters $\gamma_{M/D}$ of both mother and daughter nuclei, where the $\gamma$ is defined as
 \begin{equation}
 \gamma_\alpha =\tan ^{-1}\left(\sqrt{2} \frac{\bra{\Phi(\bm{q}_\alpha)}  r^2 Y_{22}\ket{\Phi(\bm{q}_\alpha)}}{\bra{\Phi(\bm{q}_\alpha)}  r^2 Y_{20}\ket{\Phi(\bm{q}_\alpha)}}\right).
 \end{equation} 
 As noticed in Ref.~\cite{Wang:2021} that the NME is generally sensitive to the $\gamma_{M/D}$ values with the exception that both nuclei are predicted to be $\gamma$-soft, such as $\nuclide[82]{Se}$ and $\nuclide[130]{Te}$.  By choosing the axially deformed configurations with $\gamma_M=0$ and $\gamma_D=60^\circ$, respectively,  the NME  for $\nuclide[76]{Ge}$ is $M^{0\nu}=3.17$. When the triaxiality $\gamma$ is considered, a triaxially deformed energy minimum is predicted for both nuclei. The use of the configurations corresponding to the triaxially deformed energy minimum enhances the NME to 3.72~\cite{Wang:2021}. This enhancement is probably because the overlap of the wave functions of the two nuclei becomes larger. This finding might be altered when shape fluctuation is allowed in the calculation. As demonstrated in the PGCM calculation~\cite{Jiao:2017}, the inclusion of triaxiality could reduce the NME.  It is worth mentioning that the NME in the PHFB/PSM calculation is very sensitive to the choice of the deformation parameters for the atomic nuclei and also the parameters of the PPQQ Hamiltonian and thus quantum fluctuations are better to be considered in the PHFB and PSM calculations. 
 
\begin{figure}[] 
\centering
\includegraphics[width=8.4cm]{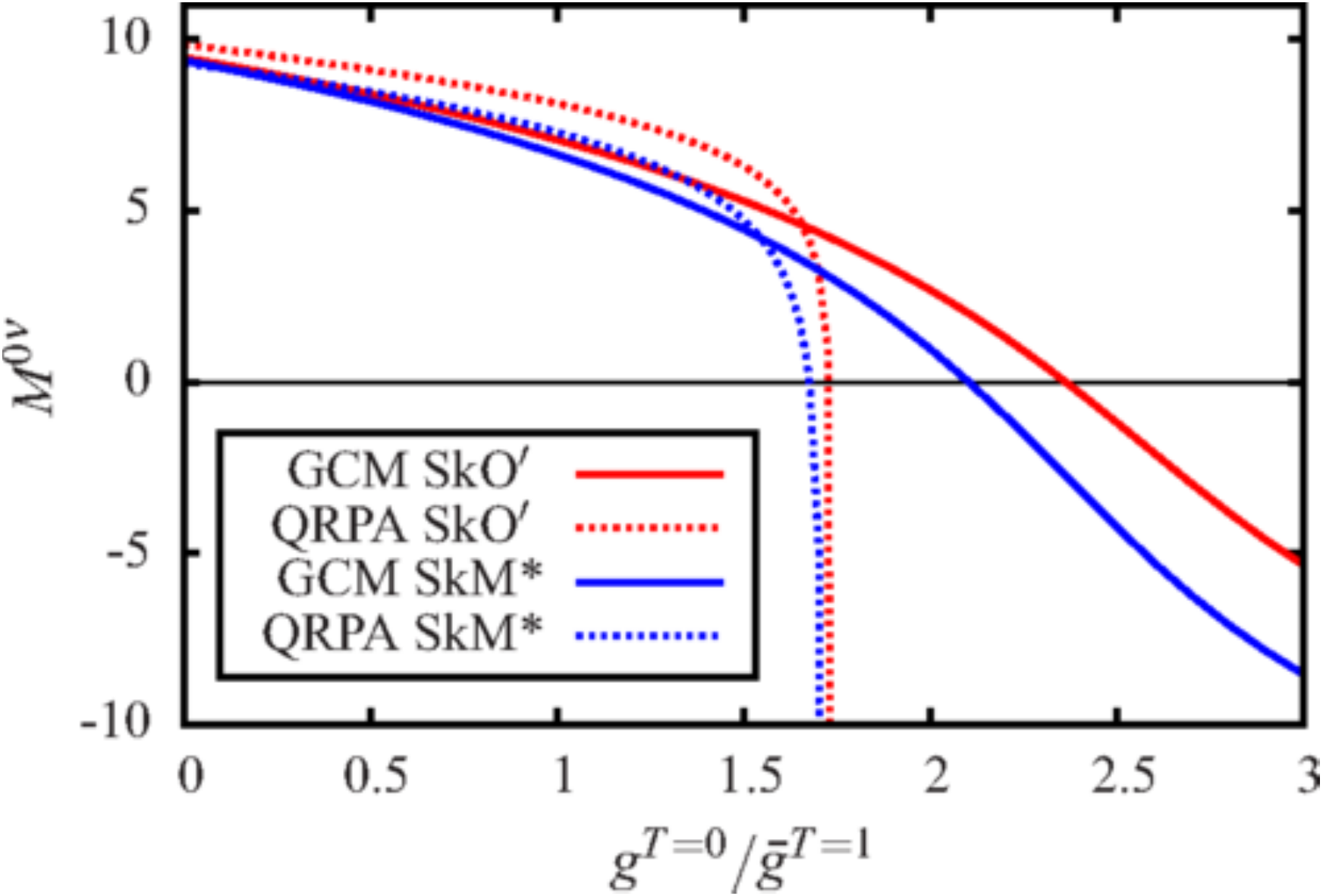}  
\caption{\label{fig:Hinohara2014PRC}  The NME of $0\nu\beta\beta$ decay in the PGCM (solid) and QRPA (dashed) calculation as a function of the ratio of isoscalar pairing strength $g^{T=0}$ to isovector pairing strength $\bar g^{T=1}=\dfrac{1}{2} (g^{T=1}_{-1}+g^{T=1}_1)$. The parameters of the Hamiltonian (\ref{eq:Hinohara2014}) are fitted to reproduce the results of HFB calculations based on either SkO' or SkM* parametrization  of Skyrme EDF. See Ref.~\cite{Hinohara:2014} for details. Figure reprinted with permission from the American Physical Society.   }
\end{figure}

  Hinohara and Engel \cite{Hinohara:2014} carried out a PGCM calculation with both particle-number and angular-momentum projections starting from a Hamiltonian -- an extended version of  the SO(8) \cite{Engel:1996PRC,Dussel:1986NPA} and the PPQQ Hamiltonian
\beqn
\label{eq:Hinohara2014}
H 
&=& H_{\rm sp}
-\dfrac{\chi}{2}\sum^2_{\mu=-2}  \hat Q^\dagger_{2\mu}\hat Q_{2\mu}
-\sum^1_{\mu=-1} g^{T=1}_\mu \hat S^\dagger_{1\mu} \hat S_{1\mu} 
-g^{T=0} \sum^1_{\mu=-1}  \hat P^\dagger_{1\mu} \hat P_{1\mu} 
+ g_{ph}\sum^1_{\mu,\nu=-1} \hat F^{\mu\dagger}_\nu \hat F^{\mu}_\nu
 \eeqn
 where $\hat S^\dag_{1\mu}$ creates a correlated isovector pair with orbital angular momentum $L=0$ and spin $S=0$ (and with $\mu$ labeling the isospin
component $T_z$), $\hat P^\dag_{1\mu}$ creates an isoscalar \textit{pn} pair with $L=0$
and $S=1$ ($S_z=\mu$), and the $\hat F^\mu_\nu$ are the components of the Gamow-Teller operator,
\bsub\begin{align} 
\hat S^\dag_{1\mu}  &= \frac{1}{\sqrt{2}} \sum_\ell \hat{\ell} [c^\dag_\ell c^\dag_\ell]^{001}_{00\mu},   \\ 
 \hat P^\dag_{1\mu} &= \frac{1}{\sqrt{2}} \sum_\ell  \hat{\ell} [c^\dag_\ell c^\dag_\ell]^{010}_{0\mu 0},   \\ 
\hat F^{\mu}_{\nu} &= \frac{1}{2} \sum_{ij}(\sigma^{\mu}  \tau^{\nu})_{ij}c^\dagger_i c_j,
\end{align}
\esub
where $\ell$ is the orbital angular momentum of a single-particle state. The basis states $\ket{\Phi(\bm{q})}$ are obtained from the constrained HFB calculation allowing for both isovector and isoscalar pairings, which are competing to each other in the solution.  The corresponding quasiparticle operator $\beta_k$ mixes the creation and annihilation operators of both neutrons and protons.  Figure~\ref{fig:Hinohara2014PRC} shows how the NME for $\nuclide[76]{Ge}$ changes as a function of the ratio of isoscalar pairing strength to isovector pairing strength from both QRPA and PGCM calculations, where the parameters in Hamiltonian (\ref{eq:Hinohara2014}) are determined by the results of HFB calculations with two different Skyrme forces. It is seen in Fig.~\ref{fig:Hinohara2014PRC} that in both QRPA and PGCM calculations, the NMEs are decreasing with the strength of isoscalar pairing strength $g^{T=0}$. It is because the increase of isoscalar pairing weakens the isovector pairing and thus the NME. Moreover, it is shown that the QRPA curves lie slightly above the PGCM ones until the ratio $g^{T=0}/\tilde{g}^{T=1}$ reaches a critical value slightly larger than 1.5, where a mean-field phase transition from an isovector pair condensates to an isoscalar condensate causes the collapse of QRPA solution. The collapse is spurious and it makes the QRPA unreliable near the critical point. Thus, in the weak pairing case,  the PGCM works better than the QRPA. The PGCM calculation has shown that the inclusion of neutron-proton pairing amplitude as generator coordinate almost halves the NME of $0\nu\beta\beta$ decay in $\nuclide[76]{Ge}$ \cite{Hinohara:2014}. The collapse of the QRPA is due to the  introduction of quasi-boson approximation which violates the Pauli-exclusion principle. \jmyr{This deficiency can be partially remedied in the renormalized QRPA~\cite{Toivanen:1995PRL,Toivanen:1997PRC}  which extends the occurring of collapse to a larger value for the neutron-proton interaction strength and thus lessens the sensitivity of the NME to the choice of the value. For details, see review papers~\cite{Faessler:1998JPG,Schuck:2021PR}.}
  
\begin{figure}[] 
\centering
\includegraphics[width=8.4cm]{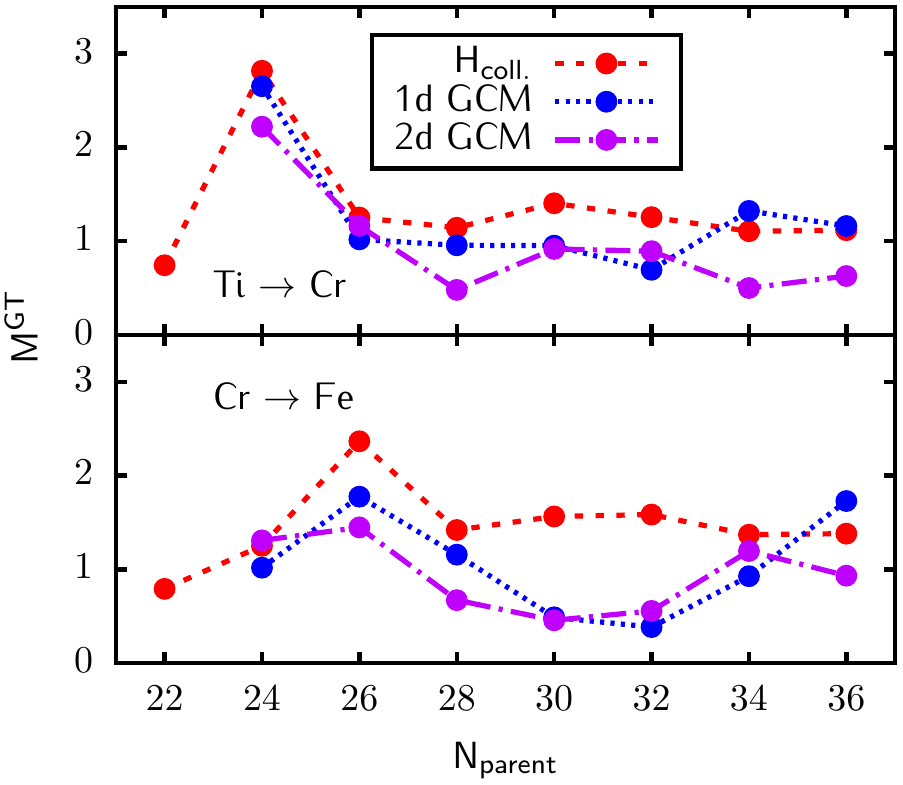}  
\caption{\label{fig:Menendez:2016PRC}  Gamow-Teller part of the NME for the $0\nu\beta\beta$ decay of Ti isotopes into Cr (top panel) and Cr isotopes into Fe (bottom), as a function of the neutron number in the parent nucleus. Results are shown for the shell-model calculation with the Hamiltonian (\ref{eq:Hinohara2014}) (red, dashed line), and the GCM calculation with the same $H$. For the latter, generator coordinate includes isoscalar pairing amplitude, with (2d) or without (1d) the axial quadrupole deformation parameter $\beta_2$.  See Ref.~\cite{Menendez:2016PRC} for details. Figure reprinted with permission from the American Physical Society.}
\end{figure}

  Starting from the same Hamiltonian (\ref{eq:Hinohara2014}), Menendez et al. carried out both shell-model and PGCM calculations for the NME of $0\nu\beta\beta$ decay from Ti  to Cr isotopes and from Cr  to Fe  isotopes~\cite{Menendez:2016PRC}. The main results are displayed in Fig.~\ref{fig:Menendez:2016PRC}. The NMEs by the PGCM generally agree with the shell-model predictions, except for the nuclei around neutron number $N=28-32$ with weak collectivity. Besides, one can see that the inclusion of quadrupole shape fluctuation only changes slightly the NME. It is attributed to the fact that their ground states are predicted to have moderate quadrupole deformation~\cite{Menendez:2016PRC}. These studies, together with those based on the shell-model Hamiltonians, demonstrate that the PGCM is able to provide a reliable prediction for the NME of $0\nu\beta\beta$ decay when the relevant degrees of freedom are taken into account.

\subsubsection{The multi-reference energy density functionals}
In the EDF approaches, atomic nuclei are modeled in terms of nuclear densities and their derivatives, where the nucleon degree of freedom is integrated out. From this point of view, it provides a lower-resolution and more economic way than the shell model to describe nuclear  global properties. On the other hand, the EDF does not have an exact solution to which one can use to examine model approximations. Therefore, the validity of these approaches is usually demonstrated in comparison with experimental data. In the case without data, one needs to examine the convergence of the results with respect to the employed model truncation. It is exactly the case for the NME of $0\nu\beta\beta$ decay, a precise description of which requires a careful check of all possible correlations.  

\begin{figure}[] 
\centering
\includegraphics[width=14cm]{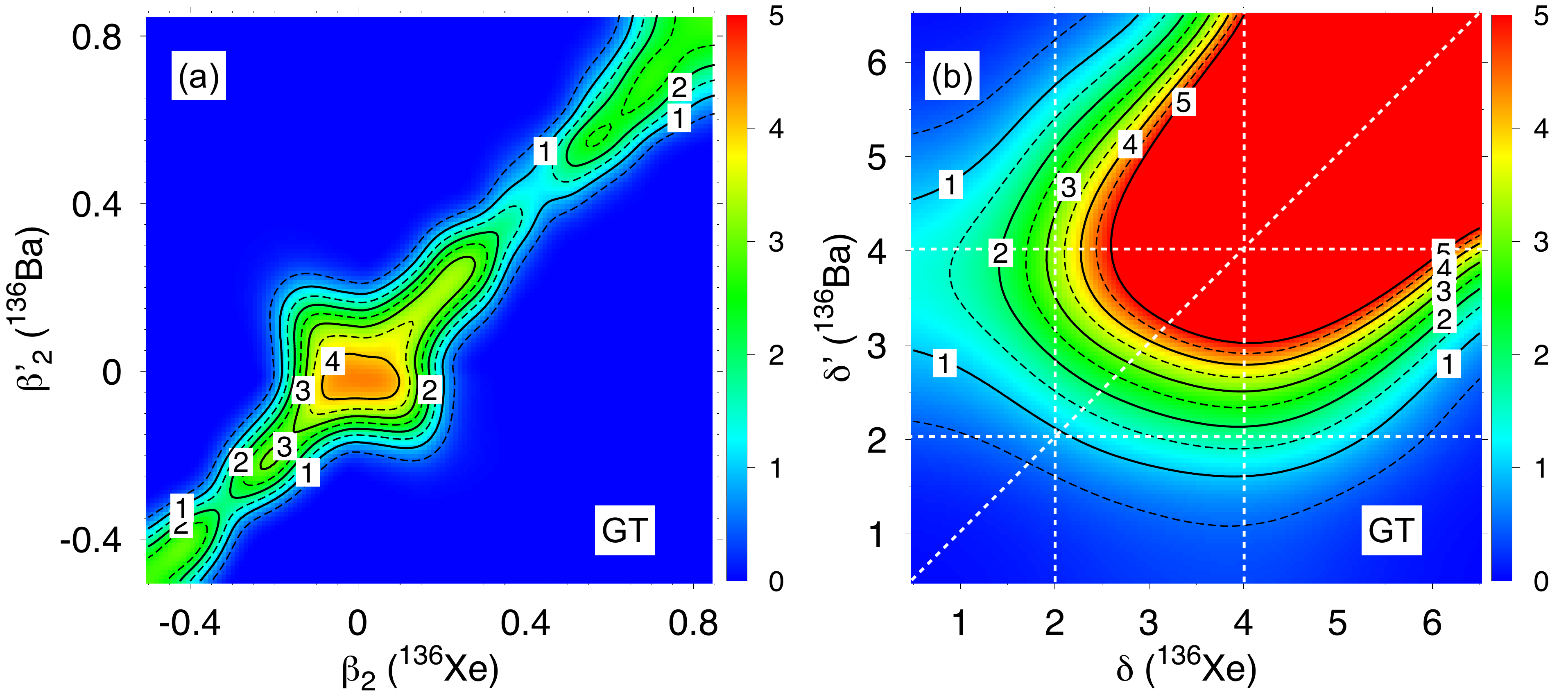}  
\caption{\label{Fig:Vaquero2013_PRL} The normalized  transition matrix element $\tilde{\cal M}^{0\nu}_{\mathrm GT}$ as a function of (a) the quadrupole deformation and (b) pairing of the initial   $^{136}$Xe  and final  $^{136}$Ba states by the Gogny D1S force. The horizontal and vertical dotted lines delimit the region where the wave functions of both nuclei take the largest values. Contour lines are separated by 0.5 units.  See Ref.~\cite{Vaquero:2013} for details. Figure reprinted with permission from the American Physical Society. }
\end{figure}

Rodriguez et al. carried out the first BMF study of the NMEs for  $0\nu\beta\beta$ decay based on the non-relativistic EDF of Gogny D1S force~\cite{Rodriguez:2010,Rodriguez:2010PPNP}. This MR-EDF framework has achieved great success in the studies of nuclear low-lying states~\cite{Egido:2016PS,Robledo:2018JPG}. The wave functions were obtained in the PGCM with particle number and angular momentum projections. Axial deformation $\beta_2$ was selected as the generator coordinate $\bm{q}$. It has been confirmed in the MR-EDF that the NME is strongly quenched by quadrupole deformation, especially when the mother and daughter nuclei are dominated by different shapes. Later on, Vaquero et al. extended this study by including additional neutron-neutron and proton-proton isovector pairing fluctuations~\cite{Vaquero:2013}. The normalized configuration-dependent NME $\tilde{\cal M}^{0\nu}_{\mathrm GT}$ is displayed in Fig.~\ref{Fig:Vaquero2013_PRL}. It is shown that the isovector pairing fluctuation can either increase or decrease the NME, depending on the distribution of nuclear collective wave functions. After convolution with the collective wave functions, the NME is generally  enhanced by 10\%–40\%~\cite{Vaquero:2013}.

\begin{figure}[] 
\centering
\includegraphics[width=10cm]{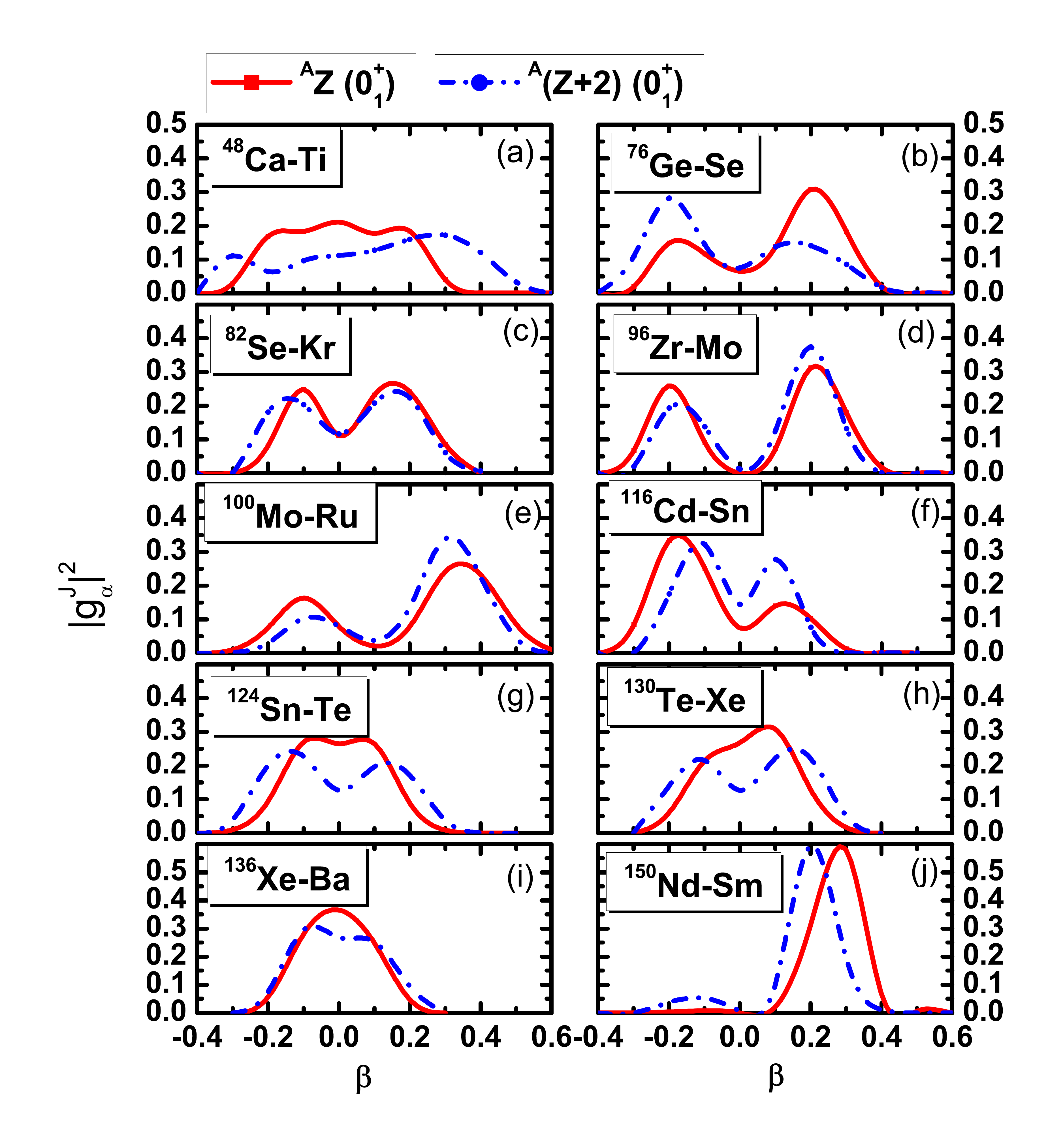}  
\caption{\label{fig:Yao2015_wfs_dist}   Distribution of collective wave functions $\left|g_{\alpha}^{J}(\beta_2)\right|^{2}$ as a function of nuclear mass quadrupole deformation parameter $\beta_2$ for the ground state of initial ${ }^{A} Z$ and final ${ }^{A}(Z+2)$ nuclei in the $0 \nu \beta \beta$ decay.  See Ref.~\cite{Yao:2015} for details. Figure reprinted with permission from the American Physical Society.  }
\end{figure}

Song et al.~\cite{Song:2014}  carried out the first full relativistic BMF description of the NME using the relativistic EDF PC-PK1~\cite{Zhao:2010PRC}. The wave functions were obtained from the PGCM calculation with the mixing of axially deformed states. The relativity effect turns out to be within $5\%$. Later on, Yao et al.~\cite{Yao:2015} carried out a systematic study of the NMEs of $\znubb$ decays for other candidate nuclei within this framework. Fig.~\ref{fig:Yao2015_wfs_dist} shows the distribution of the collective wave functions of ground states for both mother and daughter nuclei as a function of the deformation parameter $\beta_2$. It is seen that the ground states of mother and daughter nuclei in most cases are dominated by different quadrupole deformations. Moreover, shape fluctuation is shown to be significant in the light $0 \nu \beta \beta$ candidate nuclei.

 \begin{figure}[] 
\centering
\includegraphics[width=10cm]{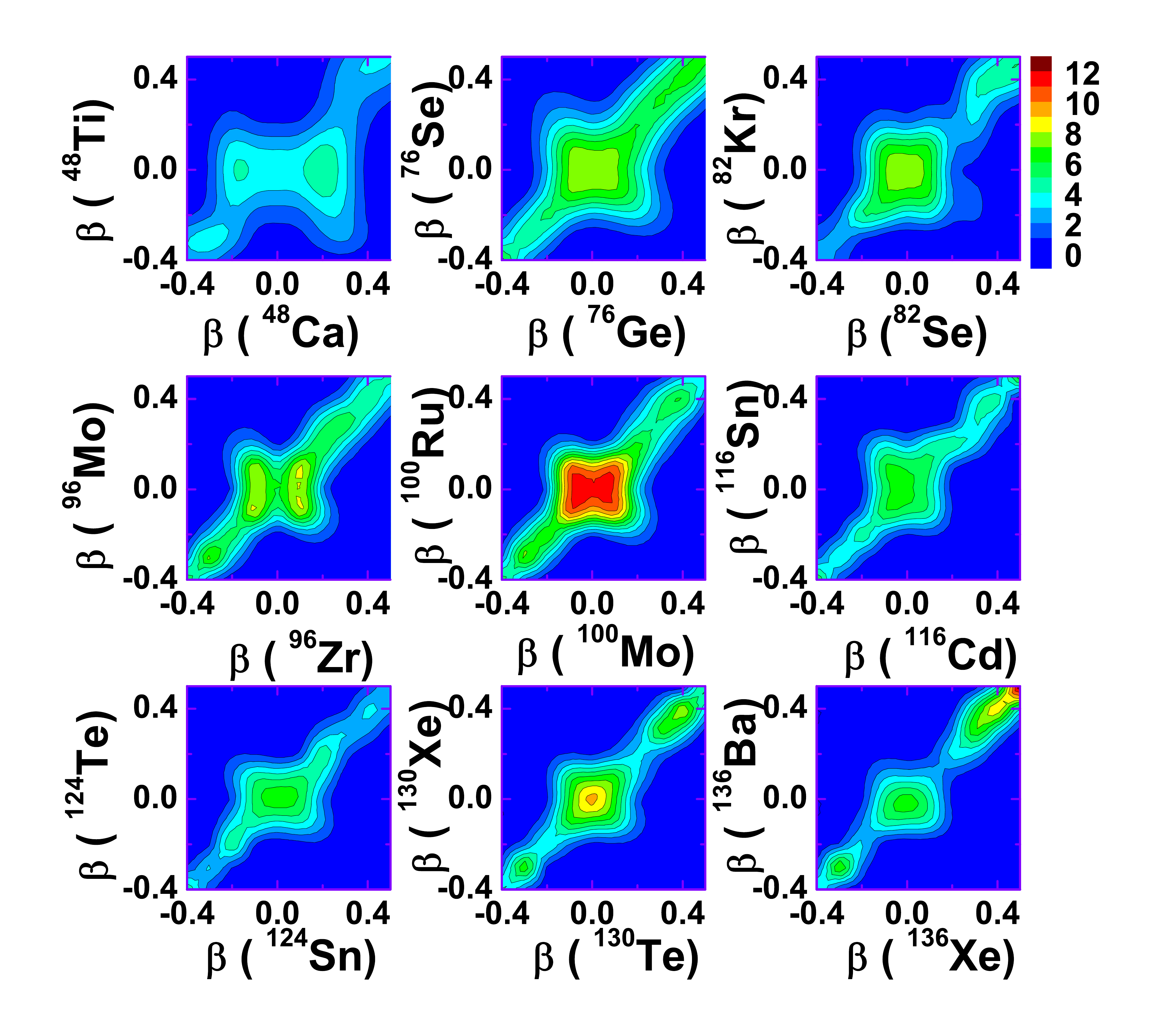}  
\caption{\label{fig:Yao2015_NME_dist} The normalized   transition matrix element ${\cal M}^{0\nu}_{\mathrm GT}$ as a function of the mass quadrupole deformation $\beta_2$ of the initial and final  states by the relativistic EDF PC-PK1. See Ref.~\cite{Yao:2015} for details. Figure reprinted with permission from the American Physical Society.  }
\end{figure}

Figure~\ref{fig:Yao2015_NME_dist} displays the NME $\tilde{\cal M}^{0 \nu}$ as a function of the intrinsic quadrupole deformations $\beta_{I}$ and $\beta_{F}$ of the mother and daughter nuclei, respectively. Similar to the behavior of the GT part shown in the MR-EDF calculation with the Gogny D1S, the NME is concentrated rather symmetrically along the diagonal line with $\beta_{I}=\beta_{F}$, implying again that the decay between nuclei with different deformation is strongly hindered.  After the convolution with the collective wave functions, the final NMEs predicted by the PC-PK1 are generally consistent with those by the non-relativistic D1S except for $\nuclide[150]{Nd}$~\cite{Yao:2015}. The impact of octupole deformation on the NME of $\nuclide[150]{Nd}$ was studied in Ref.~\cite{Yao:2016PRC}, where the octupole deformation is characterized with parameter $\beta_3$,
  \beq
  \beta_3 = \dfrac{4\pi}{3A R^3_0}\bra{\Phi(\bm{q})}  r^3 Y_{30}\ket{\Phi(\bm{q})}.
  \eeq
The total NME is shown to be reduced by about 7\% in the PGCM calculation with $(\beta_2, \beta_3)$ as the generator coordinates,  compared to the PGCM calculation with only $\beta_2$. In short, the static deformation quenches significantly the NME of $0 \nu \beta \beta$ decay, while the shape fluctuation effect generally moderates this quenching effect. We note that the advent of multi-dimensionally constrained CDFT \cite{Lu:2011PRC,Zhou:2016MD-CDFT,Meng:2019} and the CDFT in 3D lattices~\cite{Ren:2017PRC,Ren:ScienceChina2019,Ren:2020PLB} makes it feasible to include the configurations with higher-order deformations to examine their impact on the NME in the future.

\subsubsection{The QRPA calculations}
The QRPA studies of  $0\nu\beta\beta$ decay can be  classified into two types roughly according to the employed interactions. 
\begin{itemize}
    \item  The neutron-proton QRPA based on a Woods-Saxon mean-field potential, together with a realistic residual interaction, has usually been adopted to calculate the NMEs of $0\nu\beta\beta$ decays. The renormalized $ph$ strength $g_{\rm ph}$ and the $pp$ strength $g_{\rm pp}$ are determined by the position of GT resonance  and the NME of $2\nu\beta\beta$ decay. This framework originally restricted to spherical symmetry has been employed to compute the NME of  $0\nu\beta\beta$ decay~\cite{Rodin:2006,Simkovic:2013,Hyvarinen:2015PRC}. Later on, starting from an axially deformed QRPA developed by Yousef et al.~\cite{Yousef:2009PRC}, Fang et al.~\cite{Fang:2010,Fang:2011} studied the deformation effect on the NME of $\znubb$ decay. For \nuclide[150]{Nd}, the NME is suppressed by about 40\% by the deformation effect. After the restoration of isospin symmetry by imposing the Fermi part of $2\nu\beta\beta$ decay to be zero, the $M^{0\nu}_F$ of the $\znubb$ decay in \nuclide[150]{Nd} is reduced by about 20\%~\cite{Fang:2015}. Generally, the deformation quenches the NMEs of candidate nuclei by a factor ranging from 30\% to about 60\%~\cite{Fang:2018}, mainly due to the overlap between the quasiparticle vacua of initial and final nuclei.

 \item In recent years, the QRPA approaches based on EDFs have been applied to $0\nu\beta\beta$ decays. Mustonen and Engel~\cite{Mustonen:2013} carried out an axially deformed neutron-proton  QRPA calculation with the SkM* EDF for the NMEs of $0\nu\beta\beta$ decays of \nuclide[76]{Ge}, \nuclide[130]{Te}, \nuclide[136]{Xe}, and \nuclide[150]{Nd}. The obtained NMEs for  \nuclide[130]{Te} and \nuclide[136]{Xe} are significantly smaller than the values from spherical QRPA calculations based on realistic nuclear forces, while the values of the deformed nuclei \nuclide[76]{Ge} and \nuclide[150]{Nd} are not much changed. We note that this kind of comparisons is not much meaningful as totally different interactions were employed in these two calculations.  
  
 Terasaki  \cite{Terasaki:2015PRC} studied the NMEs of $0\nu\beta\beta$ decays with an axially deformed like-particle QRPA based on the SkM* EDF. Different from Ref.~\cite{Mustonen:2013} with the neutron-proton QRPA, the NME of $0\nu\beta\beta$ decay in this study is related to two-particle-transfer  transition matrix elements. The obtained NME for \nuclide[150]{Nd} is 3.604, not much different from the value 3.14 by the neutron-proton QRPA using the same EDF~\cite{Mustonen:2013}.   Recently, Terasaki and Iwata \cite{Terasaki:2021} compared the QRPA and shell-model calculations for the NME of \nuclide[48]{Ca}.  To compensate the drawbacks of each model, they proposed a phenomenological way to  modify each result by introducing an enhancing or quenching factor. With this factor, the predicted NMEs by the two models, which originally differ from each other by a factor of two, agree with each other. It would be very interesting to extend this kind of comparison to heavier candidates in the future.
 
 Recently, Gambacurta et al.~\cite{Gambacurta:2020} have shown that the inclusion of two-particle-two-hole (2p-2h) configurations in the QRPA calculation with Skyrme EDFs quenches the GT strengths in $\nuclide[48]{Ca}$ and $\nuclide[78]{Ni}$ significantly, which are in better agreement with data.  It is worthwhile to examine this effect on the NME of $\znubb$ decay in the near future.

Full 2p-2h calculations for heavy nuclei require a considerable configurations space and therefore there are only few applications in the literature. For the calculation of the width of giant resonances, where one would need, in principle, 2p-2h configurations, one has used a method to proceed in two steps: In a first step important correlations in the ph-space are taken into account by calculating RPA-phonons in the 1p-1h space. In a second step the low-lying collective phonons (surface vibrations) are coupled to single-particle states. According to Landau-Migdal theory of finite Fermi systems~\cite{Migdal1967} this leads to dressed Landau quasiparticles. Using such techniques of particle vibrational coupling (PVC) one ends up with an energy dependent Hamiltonian whose solutions are, in principle, the exact solutions of the 2p-2h problem. In practice only a few low-lying phonons are sufficient to include the most important correlations. This avoids the complicated diagonalization of the interaction in the full 2p-2h space. Such techniques of PVC have been applied successfully for the calculation of the fragmentation of single particle states due to coupling to surface vibrations~\cite{Litvinova2006_PRC73-044328,Colo2010_PRC82-064307} and of spectroscopic factors, to the description of the width of giant resonances~\cite{Litvinova2007_PRC75-064308,Roca-Maza:2017JPG}, but also for the calculation of GT-resonances~\cite{NIU-YF2012_PRC85-034314,Niu:2014PRC,Litvinova:2014PLB,Niu:2016PRC,NIU-YF2018_JPCS966-012046,Caroline:2019PRL} and single-$\beta$ decay~\cite{NIU-YF2015_PRL114-142501,Robin2016_EPJA52-1,Niu:2018PLB}. To our knowledge such methods have not been applied, so far, for $2\nu\beta\beta$ decay nor for $0\nu\beta\beta$ decay. Therefore this method, which definitely goes beyond mean field, remains to be used  also for the description of these problems in the future.

\end{itemize}
 
\subsubsection{The studies with realistic nuclear forces}

 Computing atomic nuclei from first principles have been the goal of nuclear physicists for a long time. Building nuclear models starting from realistic nuclear forces serves as a major step towards this goal. It is particularly important for the determination of the NME of $0\nu\beta\beta$ decay. There are many different ways to implement a realistic nuclear force into many-body calculations. Here we give a brief overview of them. 
\begin{itemize}
 \item Implemented directly into nuclear many-body solvers, like quantum Monte Carlo (QMC) methods~\cite{Carlson:2015}.  This framework is able to solve light nuclei {\em exactly}. The corresponding results are thus valuable to validate other many-body approaches.  Pastore et al.~\cite{Pastore:2018,Wang:2019,Cirigliano:2019PRC}  have calculated the NMEs of $^{6,8,10}$He and $^{10,12}$Be  with the QMC method based on the NN interaction AV18 and the 3N interaction IL7. The same QMC calculation was later carried out based  on the local Norfolk chiral NN+3N potentials \cite{Baroni:2018} for the NMEs of $\nuclide[6]{He}$ and $\nuclide[12]{Be}$ \cite{Cirigliano:2019PRC}.

 \item Preprocessed with the techniques of $V_{\rm low-k}$~\cite{Bogner:PR2003} or SRG \cite{Bogner:2010} and then implemented  into nuclear many-body solvers. In the  resultant effective low-momentum nuclear potential, the off-diagonal elements coupling low- and high-momentum states are suppressed in the SRG or the contribution from the high-momentum states is integrated into the $V_{\rm low-k}$. This feature is helpful for speeding up the convergence of many-body approaches, such as NCSM~\cite{Barrett:2013}, coupled-cluster method~\cite{Hagen:2014}, IMSRG~\cite{Hergert:2016} and MBPT~\cite{Tichai:2020}. Starting from a preprocessed chiral NN interaction, Basili et al.~\cite{Basili2020} carried out a benchmark calculation with both NCSM and IMSRG for the NME of $^{6}$He. Yao et al.~\cite{Yao:2021PRC} extended the benchmark calculation to the NMEs of isospin-changing $0\nu\beta\beta$ decays with $\Delta T=2$  in a set of light nuclei with the IMSRG and importance-truncated (IT)-NCSM  starting from a chiral NN+3N interaction EM1.8/2.0~\cite{Hebeler:2011}. \jmy{In the IT-NCSM~\cite{Roth:2009}, the dimension of the shell-model space is truncated by {\em a priori} selection of the  most relevant basis states based on an importance measure derived from multiconfigurational perturbation theory.}
 In the meantime, Novario et al.~\cite{Novario:2021PRL} performed a benchmark calculation with the coupled-cluster method and NCSM based on the same chiral NN+3N interaction. A reasonable agreement among all these  {\em ab initio}  calculations for both the ground-state energies and the NMEs of $0\nu\beta\beta$ decay was achieved.  In particular,  Yao et al. \cite{Yao:2020PRL} and Novario  et al. \cite{Novario:2021PRL} computed the NME for the lightest candidate nucleus \nuclide[48]{Ca} with the IM-GCM~\cite{Yao:2018wq} and symmetry-breaking coupled-cluster method, respectively.

 \item The preprocessed nuclear force with the SRG is further decoupled into a valence space with the techniques of MBPT~\cite{Hjorth-Jensen:1995PR,Coraggio:2008PPNP}, coupled-cluster method~\cite{Jansen:2014PRL,Sun:2021} or VS-IMSRG \cite{Bogner:2014PRL,Stroberg:2019}. The decoupling of the model space from the excluded space is usually accomplished at the two-body level, and the induced off-diagonal three-body terms are included perturbatively~\cite{Sun:2021}.   The interaction obtained in this way can be implemented into a conventional valence-space shell model which enables a systematic study of nuclei from helium to iron~\cite{Stroberg:2021PRL}.  Within this framework, Belley et al.~\cite{Belley2021PRL} computed the NMEs of \nuclide[48]{Ca}, \nuclide[76]{Ge}, and \nuclide[82]{Se} with a chiral NN+3N interaction. Coraggio et al. carried out a realistic shell-model (RSM) calculation starting from the CD-Bonn potential which was renormalized with the $V_{\rm low-k}$ procedure~\cite{Coraggio:2020}.

  \end{itemize}

 \begin{figure}
\centering  
\includegraphics[width=9cm]{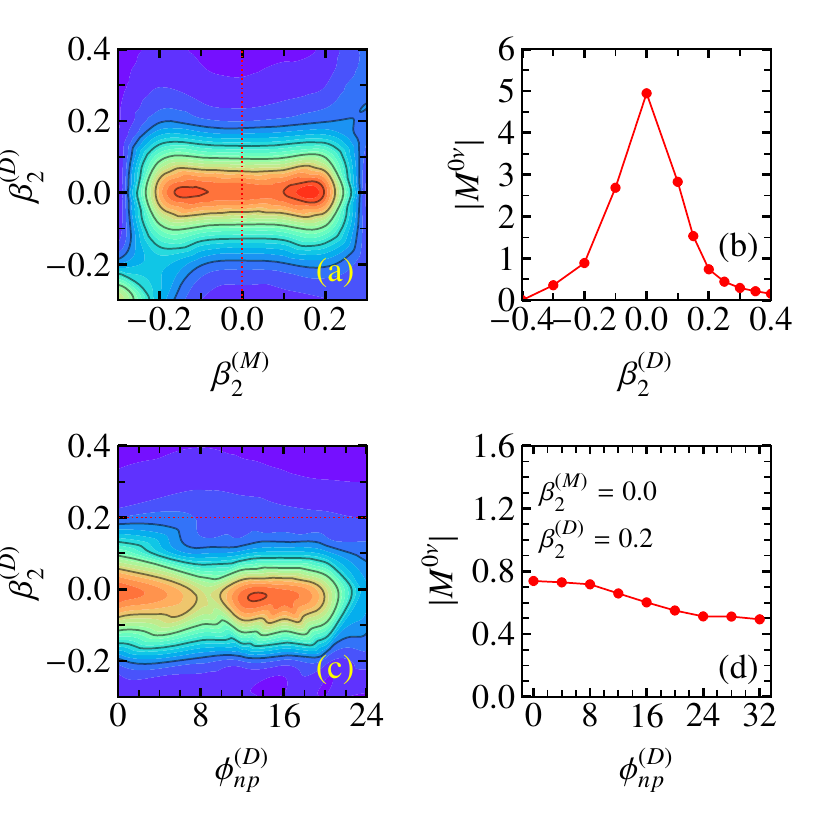}  
\caption{\label{fig:Yao2020PRL_NME}  (a) Contributions to the NME $\left|\tilde{\cal M}^{0\nu}\right|$ in the $\left(\beta^{(M)}_{2}, \beta^{(D)}_{2}\right)$ plane (see text). Neighboring contour lines here and in (c) are separated by $0.50 .$ (b) The NME $\left|M^{0 v}\right|$ as a function of the quadrupole deformation parameter $\beta^{(D)}_{2}$ in ${ }^{48}$Ti. (c) Contributions to $\left|M^{0 v}\right|$ in the $\left(\beta^{(D)}_{2}, \phi^{(D)}_{np}\right)$ plane. (d) The NME $\left|M^{0 v}\right|$ as a function of the proton-neutron pairing amplitude $\phi^{(D)}_{np}$ in $^{48}$Ti. See Ref.~\cite{Yao:2020PRL} for details. Figure reprinted with permission from the American Physical Society.    }
\end{figure}

 \begin{figure}
\centering  
\includegraphics[width=9cm]{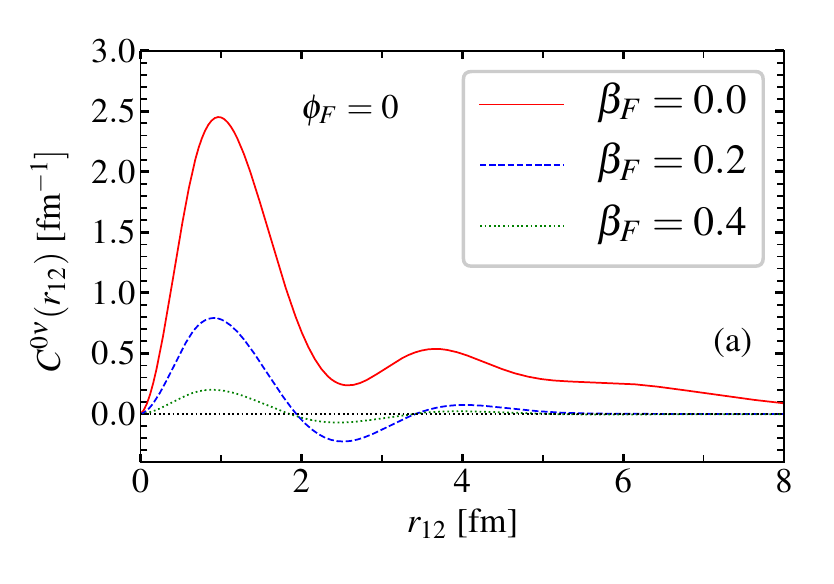}  
\includegraphics[width=9cm]{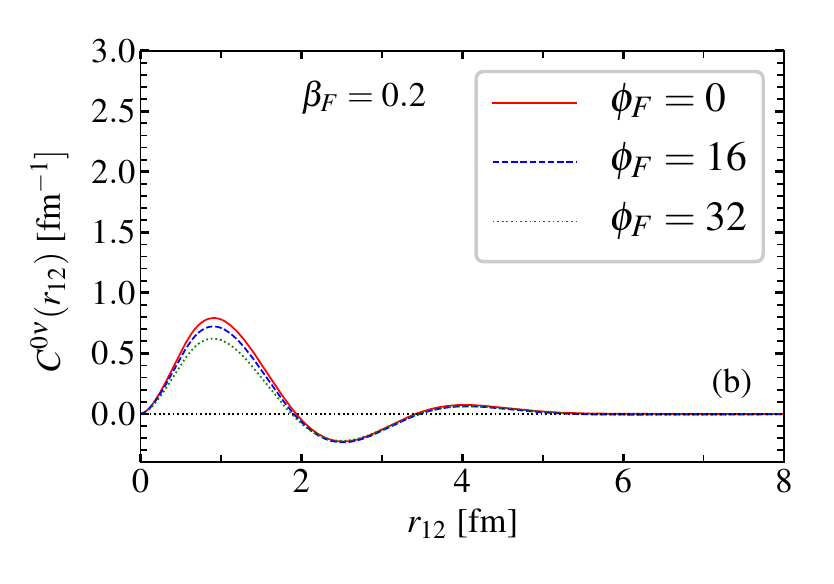}  
\caption{\label{fig:Yao2020PRL_TD} The distribution of $C^{0v}(r_{12})$ as a function of the relative coordinate $r_{12}$ corresponding to the transition from spherical $\nuclide[48]{Ca}$ to different configuration of $\nuclide[48]{Ti}$ by the chiral interaction EM1.8/2.0 distinguished with different values of quadrupole deformation parameter $\beta_{2}$(a) and different values of  neutron-proton isoscalar pairing amplitude $\phi_{F}$ of \nuclide[48]{Ti}. See Ref.~\cite{Yao:2020PRL} for details. Figure reprinted with permission from the American Physical Society.   }
\end{figure}

  Note that all the operators need to be preprocessed in the same way as that is applied for the nuclear force.   Since the $0\nu\beta\beta$-decay operator is essentially a long-range operator, the application of SRG to this operator in free space is not expected to induce a significant change. Therefore, in all the previously mentioned {\em ab intio} studies, the bare transition operator was used. However, the second decoupling in nuclear many-body space via either MBPT, shell-model coupled-cluster method, or IMSRG has a non-trivial impact on the operator and this effect has been taken into account.   
 
  The IM-GCM is a combination of IMSRG and PGCM. Here we introduce in rather detail the application of this method to the $0\nu\beta\beta$ decay of $\nuclide[48]{Ca}$. Fig.~\ref{fig:Yao2020PRL_NME} shows the NME $\left|\tilde{\cal M}^{0\nu}\right|$ as a function of quadrupole deformation parameters of both nuclei  and the neutron-proton isoscalar pairing amplitude $\phi_{np}$ in $\nuclide[48]{Ti}$, where the IMSRG-evolved transition operator was used in the calculation. \jmyr{One can see again the feature shown in Fig.~\ref{fig:KB3G_CaTi_EHFB} that the NME changes rapidly with the quadrupole deformation of $^{48}$Ti, but does not change much when the deformation of $^{48}$Ca varies from $\beta_2=-0.2$ to $\beta_2=0.2$.} Besides,  the NME is decreasing smoothly with the isoscalar pairing. However, after convolution with a nuclear collective wave function, its overall effect on the NME is mild.
  
  Figure~\ref{fig:Yao2020PRL_TD} displays the distribution $C^{0\nu}(r_{12})$ of the NME for $\nuclide[48]{Ca}$, defined as~\cite{Simkovic:2008}
  \beq
  M^{0\nu}=\int_{0}^{\infty} \mathrm{d} r_{12} C^{0\nu}(r_{12}).
  \eeq
  as a function of  the relative distance $r_{12}$  between the two neutrons that are transformed into protons.  One can see clearly that the quadrupole correlation quenches the NME in both the long-ranged and short-ranged regions. In contrast, the isoscalar pairing quenches the NME mainly in the short-ranged region, as demonstrated in Ref.~\cite{Yao:2020PRL}. Note that the distributions with deformed final states are qualitatively similar to those obtained in the phenomenological shell-model calculation of Ref.~\cite{Menendez:2009}, featuring the appearance of the main peak at $r_{12}\approx1.0$ fm. In other words, the typical momentum of the neutrino is around $q\approx 100$ MeV.

\subsection{Model comparison and discussion}

 \begin{figure}
\centering
\includegraphics[width=14cm]{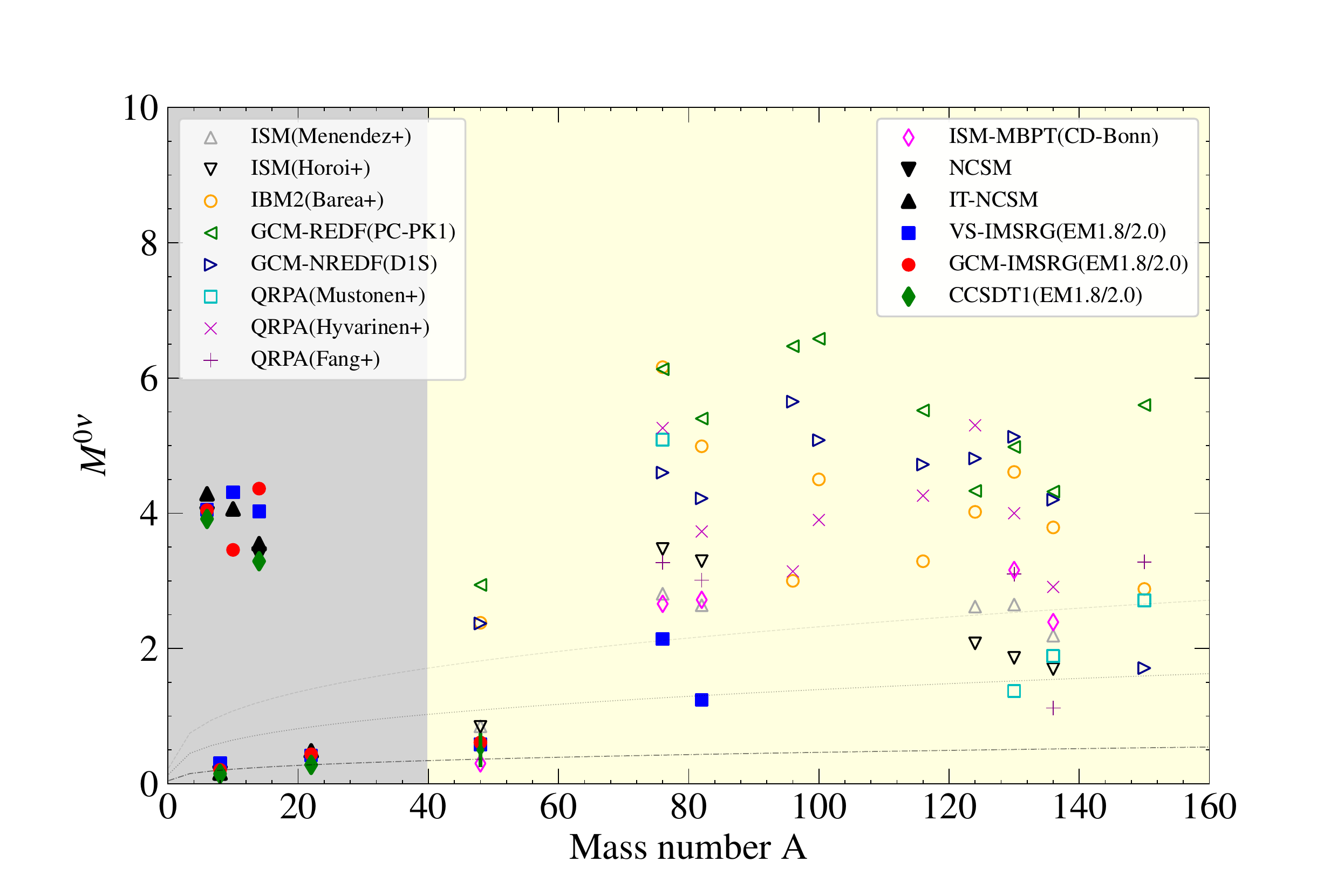}  
\caption{\label{fig:NME_com_all}  The NMEs of $0\nu\beta\beta$ decays in light nuclei (gray area, no experimental interest) and candidate nuclei from various model calculations using the same transition operator derived from the standard mechanism. The three curves of $\alpha A^{1/3}$ with $\alpha=0.1, 0.3$ and $0.5$, respectively, are added as guidelines. }
\end{figure}

 \begin{table*}[t]
 \centering
 \tabcolsep=3pt
 \caption{The NMEs $M^{0\nu}$ of the candidate $0\nu\beta\beta$ decays from various phenomenological model calculation using the unquenched value of $g_A$. }
 \begin{tabular}{ccccccccc|c}
  \hline\hline
 Models &  REDF  & NREDF   & QRPA(G)   &QRPA(SkM*)  & TPSM  & PHFB  & ISM  & IBM2  & [Min, Max]\\
 REF &   \cite{Yao:2015} &  \cite{Rodriguez:2010} &  \cite{Fang:2018} & \cite{Mustonen:2013}   & \cite{Wang:2021} &  \cite{Rath:2019} &  \cite{Menendez:2009} &  \cite{Barea:2013} & \\ 
 \hline
 $^{48}$Ca  $\to ^{48}$Ti  & 2.94   & 2.37   &     -      &  -    &     -  &     -    &  0.85   & 2.38 & [0.85, 2.94] \\
 $^{76}$Ge  $\to ^{76}$Se  & 6.13   & 4.60    & 3.12     &  5.09 & 3.17  &   -      &  2.81   & 6.16 & [2.81, 6.16] \\
 $^{82}$Se  $\to ^{82}$Kr  & 5.40   & 4.22    &  2.86     & -     &  2.59 &  -      &  2.64   &  4.99 & [2.59, 5.40] \\
 $^{96}$Zr  $\to ^{96}$Mo  & 6.47   & 5.65    &   -       & -     & -     &  2.86(26)   &   -   &  3.00  & [2.86, 6.47]  \\
 $^{100}$Mo $\to ^{100}$Ru & 6.58   & 5.08   &     -      & -     & 3.92  &   6.25(64)   &  -   &  4.50  & [3.92, 6.58]\\
 $^{116}$Cd $\to ^{116}$Sn & 5.52   & 4.72   &  -         & -     &   -   &  -     &    -     &  3.29 & [3.29, 5.52] \\
 $^{124}$Sn $\to ^{124}$Te & 4.33   & 4.81   &  -        & -      & -     &    -     &  2.62   &  4.02  & [2.62, 4.81]  \\
 $^{130}$Te $\to ^{130}$Xe &4.98    & 5.13   &   2.90     & 1.37 &  2.92    &  4.05(50)   &   2.65  &  4.61  & [1.37, 5.13] \\
 $^{136}$Xe $\to ^{136}$Ba &4.32    & 4.20   &  1.11     & 1.55  &    &    -        &   2.19  &  3.79 & [1.11, 4.32] \\
 $^{150}$Nd $\to ^{150}$Sm & 5.60   & 1.71    & 3.01     & 2.71  &  3.29   & 2.83(43)   &   -      &  2.88 & [1.71, 5.60] \\
 \hline \hline
\end{tabular}
 \label{tab:NMES_comparisons_pheno_models}
\end{table*}

\begin{table}[t]
\centering
\tabcolsep=8pt
\caption{The NMEs $M^{0\nu}$ of the candidate $0\nu\beta\beta$ decays from different calculations starting from realistic nuclear forces. The same chiral NN+3N interaction EM1.8/2.0 \cite{Hebeler:2011} was employed in the IM-GCM, VS-IMSRG and CCSD-T1, while the $V_{\rm low-k}$ pre-processed CD-Bonn $NN$ potential \cite{Machleidt:2001PRC} was used in the RSM calculation. } 
\begin{tabular}{lccc|c}
\hline \hline Decay  & CCSD-T1~\cite{Novario:2021PRL}  & IM-GCM~\cite{Yao:2020PRL}  & VS-IMSRG~\cite{Belley2021PRL}  & RSM~\cite{Coraggio:2020}  \\
\hline${ }^{48} \mathrm{Ca} \rightarrow{ }^{48} \mathrm{Ti}$ 
& 0.25-0.75 & 0.61 &  0.58  & $0.30$    \\
${ }^{76} \mathrm{Ge} \rightarrow{ }^{76} \mathrm{Se}$ 
&&  &  2.14   &  $2.66$  \\
${ }^{82} \mathrm{Se} \rightarrow{ }^{82} \mathrm{Kr}$ 
&&  & 1.24    &  $2.72$   \\
${ }^{130} \mathrm{Te} \rightarrow{ }^{130} \mathrm{Xe}$ 
&&  &   &    $3.16$   \\
${ }^{136} \mathrm{Xe} \rightarrow{ }^{136} \mathrm{Ba}$ 
&  &   &  &  $2.39$   \\
\hline \hline
\end{tabular}
\label{tab:NME_comparison_ab-initio}
\end{table}

 The NMEs of $0\nu\beta\beta$ decay from different model calculations assuming the standard mechanism are summarized in Fig.~\ref{fig:NME_com_all}, including the NMEs for light nuclei without experimental interest from {\em ab initio} calculations~\cite{Yao:2021PRC}. For the light nuclei, the NME is either around 4.0 for isospin-conserving transitions or less than 0.5 for isospin-changing transitions. For the candidate nuclei, the NMEs by different models disagree by a factor of about 3. The values from the model calculations starting from phenomenological interactions/EDFs are given in Tab.~\ref{tab:NMES_comparisons_pheno_models}, and those from the calculations starting from realistic nuclear forces are tabulated in Tab.~\ref{tab:NME_comparison_ab-initio}. The results from spherical QRPA~\cite{Hyvarinen:2015PRC}, GCM based on shell-model Hamiltonian~\cite{Jiao:2019PRC}, EFT~\cite{Brase:2021} based on the correlation from the shell-model calculation~\cite{Shimizu:2018PRL} are not included for comparison. As the study of the $0\nu\beta\beta$ decay based on nuclear chiral forces is still in the early stage, only the NME for the lightest candidate has been computed with different {\em ab initio} methods. The value is shown to be consistently smaller than those of phenomenological models, but larger than the value by the RSM starting from the CD-Bonn NN interaction. The computation demanding of VS-IMSRG(2) and RSM is similar to each other in the sense that both rely on the techniques of diagonalization of effective Hamiltonian in a valence space. Therefore, these two methods can be extended to the NMEs of rather heavy candidate nuclei without much computation challenge. However, one can see that the NMEs for $\nuclide[48]{Ca}$, $\nuclide[76]{Ge}$ and $\nuclide[82]{Se}$ do not agree with each other. This discrepancy could be due to the employment of different nuclear interactions. Either the extension of IM-GCM or coupled-cluster (CCSD-T1) with singles, doubles and leading triples  to heavier candidate nuclei or the implementation of the chiral interaction into the RSM will provide an insight into the origin of the discrepancy. 
  
It is relatively more difficult to resolve the discrepancy among phenomenological models. Nevertheless, remarkable progress has been achieved. In Ref.~\cite{Menendez:2014}, Menendez et al. compared the NMEs from the spherical EDF and seniority-zero shell-model calculations, which turns out to agree with each other. It is in strong contrast with the full NME calculations that the shell-model NMEs are about half of the values by the EDF, regardless of the employed effective interaction or EDF. It implies that the  correlations associated with collective deformation in the EDF ~\cite{Rodriguez:2013PLB} and the  correlations associated with high-seniority components in the shell model~\cite{Caurier:2008PRL} quench the NMEs in different amounts, leading to the  discrepancy of a factor of two. On one hand, it indicates that some kinds of correlations that could quench the NME further are missing in the current EDF calculations.  Indeed, it has been shown by Hinohara et al.~\cite{Hinohara:2014} that the inclusion of isoscalar pairing correlations can quench the NME significantly. The inclusion of this correlation in the EDF calculations is expected to reduce the discrepancy.  On the other hand, Iwata et al.~\cite{Iwata:2016} have shown that the NME for $\nuclide[48]{Ca}$  in the valence-space shell model calculation is  enhanced by about 30\% when the valence space is enlarged from $fp$ shell to include $sd$ shell. For the NMEs of $\nuclide[76]{Ge}$ and other heavier candidates, some spin-orbit partners are not included in the model spaces of shell-model calculations. It could be one of the reasons that the shell model gives the small predicted NMEs. Enlarging the model space in the shell model calculations to include spin-orbit partners is demanded. However, it was also found in the benchmark study of QRPA and shell-model by Brown et al. ~\cite{Brown:2015compare} that the corrections from the particle-particle and particle-hole parts due to the inclusion of spin-orbit partners approximately cancel each other, resulting in only a small change in the NME. From the above analysis, one may expect that the NMEs are generally small, as predicted by the shell models and available  {\em ab initio}  studies.

\subsection{Constraints with correlation relations}

Exploration of possible correlation relations between the $0\nu\beta\beta$-decay NMEs with nuclear-structure observable is of much interest as they may provide a model-independent constraint on the NMEs\cite{Freeman:2012JPG}.  Rodriguez et al. illustrated the correlation between nuclear isovector pairing energy and the NME in Refs.~\cite{Rodriguez:2010,Rodriguez:2013PLB}. It has been shown that large NMEs are obtained if the pairing correlations in the initial and final states are considered and the difference in the deformations of the two nuclei is small. This correlation is helpful for understanding the discrepancy in different mean-field-based approaches. However, pairing energy is not observable, this correlation cannot provide a model-independent way to constraint the NME directly. 

In recent years, Shimizu et al. \cite{Shimizu:2018PRL} and Romeo et al. \cite{Romeo:2021} demonstrated that the NME $M^{0\nu}$ of $0\nu\beta\beta$ decay is linearly correlated to the NME $M^{\rm DGT}$ of the ground-state to ground-state DGT transition  
\begin{align}  
\label{eq:DGT}
M^{\rm DGT} &=\left\langle 0^+_f\left|\sum_{1,2}\left[\boldsymbol{\sigma}_{1}  \otimes \boldsymbol{\sigma}_{2}  \right]^{0}\tau^+_1\tau^+_2\right| 0^+_i\right\rangle. 
\end{align}  
and to the NME $M^{\gamma\gamma}$ of $\gamma\gamma$ transition, respectively. These two quantities are measurable and  in principle can provide a valuable tool to obtain information on the NMEs, even though their measurements are also challenging. Very recently, the correlation relation between $M^{0\nu}$ and $M^{\rm DGT}$ in the shell-model calculation~\cite{Shimizu:2018PRL} was employed by Brase et al. \cite{Brase:2021} to determine  the values of $M^{0\nu}$ in an effective field theory for heavy candidate nuclei. Considering the fact that the  $M^{0\nu}$ was usually calculated in a HO basis with the frequency chosen as $\hbar\omega=41A^{-1/3}$ MeV in EDF calculations~\cite{Rodriguez:2013PLB} or  $\hbar \omega=45 A^{-1/3}-25 A^{-2/3}$ MeV in shell model calculations~\cite{Shimizu:2018PRL}, the  $M^{0\nu\beta\beta}_{\rm GT}$ (proportional to $R_0/b$) is expected to scale as $A^{1/6}$ and the DGT matrix element $M^{\rm DGT}$ is expected to be correlated with $M^{0 \nu}_{\rm GT}\cdot A^{-1/6}$ for all isotopes~\cite{Brase:2021}. Based on this argument, it is reasonable to parametrize the correlation relation between $M^{0\nu}$ and $M^{\rm DGT}$ as follows,
\begin{equation}
\label{eq:correlation}
 |M^{\rm DGT}| = \alpha |M^{0\nu}|\cdot A^{\gamma} + \beta,   
\end{equation}
where the power parameter is fixed to be $-1/6$. The parameters $(\alpha, \beta)$ are free parameters obtained via linear regression.  Using the shell-model results, Brase et al.~\cite{Brase:2021} derived  three lines with the parameters $(\alpha, \beta, \gamma)=(0.447,-0.180, -1/6$), $(0.536, -0.106, -1/6)$ and $(0.699, -0.056,-1/6)$, respectively. 
It was shown there that all the results except for the QRPA results are located inside the two boundary lines. 

The origin of the linear correlation (\ref{eq:correlation}), as  shown in the shell-model study~\cite{Shimizu:2018PRL}, is that both two transitions are dominated by the short-range contribution.  In other words, the net contributions from the intermediate- and long-range regions are negligibly small.  This feature is however not shown in the results of QRPA calculations, where the strong-range contribution to the DGT matrix element is strongly quenched by the intermediate- and long-range  contributions~\cite{Simkovic:2018}. As a result, the total DGT NME $M^{\rm DGT}$ by the QRPA remains small even in the isotopes with large values of $M^{0\nu}$. 

As discussed in Ref.~\cite{Simkovic:2018}, the $2\nu\beta\beta$ decay would be forbidden if the isospin SU(2) and spin-isospin SU(4) symmetries were  exact. Since the isospin symmetry is known to be rather good in nuclei, the double Fermi matrix element is negligibly small. Thus, the main contribution to $2\nu\beta\beta$ decay is given by the DGT matrix
element.  Considering the fact that the NMEs of  $2\nu\beta\beta$ decay deduced from available data are small for nuclei with large $A$, indicating the existence of an (approximate) underlying SU(4) symmetry~\cite{Stefanik:2015}, the QRPA calculations are usually carried out with the isoscalar $g^{T=0}_{pp}$ and isovector $g^{T=1}_{pp}$ particle-particle interaction strength adjusted to impose these symmetries~\cite{Simkovic:2018,Fang:2018}. The $M^{\rm DGT}$ value is proportional to the GT NME of $2\nu\beta\beta$ decay in the closure approximation. This may explain the small $M^{\rm DGT}$ in the QRPA calculation, a consequence of maintaining good spin-isospin SU(4) symmetry.

In the framework of PGCM or MR-EDF, it is challenging to calculate the NMEs of $2\nu\beta\beta$ decay for which the closure approximation is not a good approximation. All intermediate states of neighboring odd-odd nucleus need to be calculated. The cost of this kind of calculation is extremely expensive. In contrast, the DGT NMEs $M^{\rm DGT}$ are much easier to calculate and were computed in the MR-EDF with a Gogny force~\cite{Rodriguez:2010PPNP,Rodriguez:2013PLB}. The results are compared to those of shell model and QRPA calculations in Ref.~\cite{Shimizu:2018PRL}. It was found that the DGT NMEs $M^{\rm DGT}$ from the MR-EDF calculation are generally larger than those by the QRPA, probably indicating a larger violation of the SU(4) symmetries. The values of $M^{\rm DGT}$ and $M^{0\nu\beta\beta}$ by the MR-EDF follow approximately the linear correlation relation (\ref{eq:correlation}) of shell model.  Very recently, three {\em ab initio} methods (IT-NCSM, IM-GCM and VS-IMSRG) are also applied to study the DGT matrix elements in a set of light and medium-mass nuclei~\cite{Yao:2021DGT}. It has been found that the transitions with large values of both  $M^{0\nu}$ and $M^{\rm DGT}$ are always isospin-conserving  processes. In contrast, the DGT NMEs $M^{\rm DGT}$ of isospin-changing processes are generally small and only weakly correlated to $M^{0\nu}$. A further investigation on the  violation of spin-isospin symmetry in each nuclear model is required to understand comprehensively the discrepancy.

 \subsection{Implication on neutrino masses}

 The best-fit values and $3\sigma$ allowed ranges of the neutrino oscillation parameters derived from a global fit of the current neutrino oscillation data by the NuFIT analysis~\cite{Esteban:2020JHEP,NuFIT}  are given below :
\bsub
\begin{align}
 0.269 \leq & s^2_{12}=0.304  \leq 0.343,\\
 0.020 \leq &  s^2_{13}=0.022  \leq 0.024,\\
 6.82\times 10^{-5} {\rm eV^2} \leq &  \Delta m^2_{21}=7.42\times 10^{-5} {\rm eV^2} \leq 8.04\times 10^{-5} {\rm eV^2},\\
 2.431(-2.583)\times 10^{-3} {\rm eV^2}  \leq & \Delta m^2_{31}= 2.514(-2.497)\times 10^{-3} {\rm eV^2} \leq 2.598(-2.412) \times 10^{-3} {\rm eV^2}.
\end{align}
\esub

  \begin{figure}
\centering
\includegraphics[width=7cm]{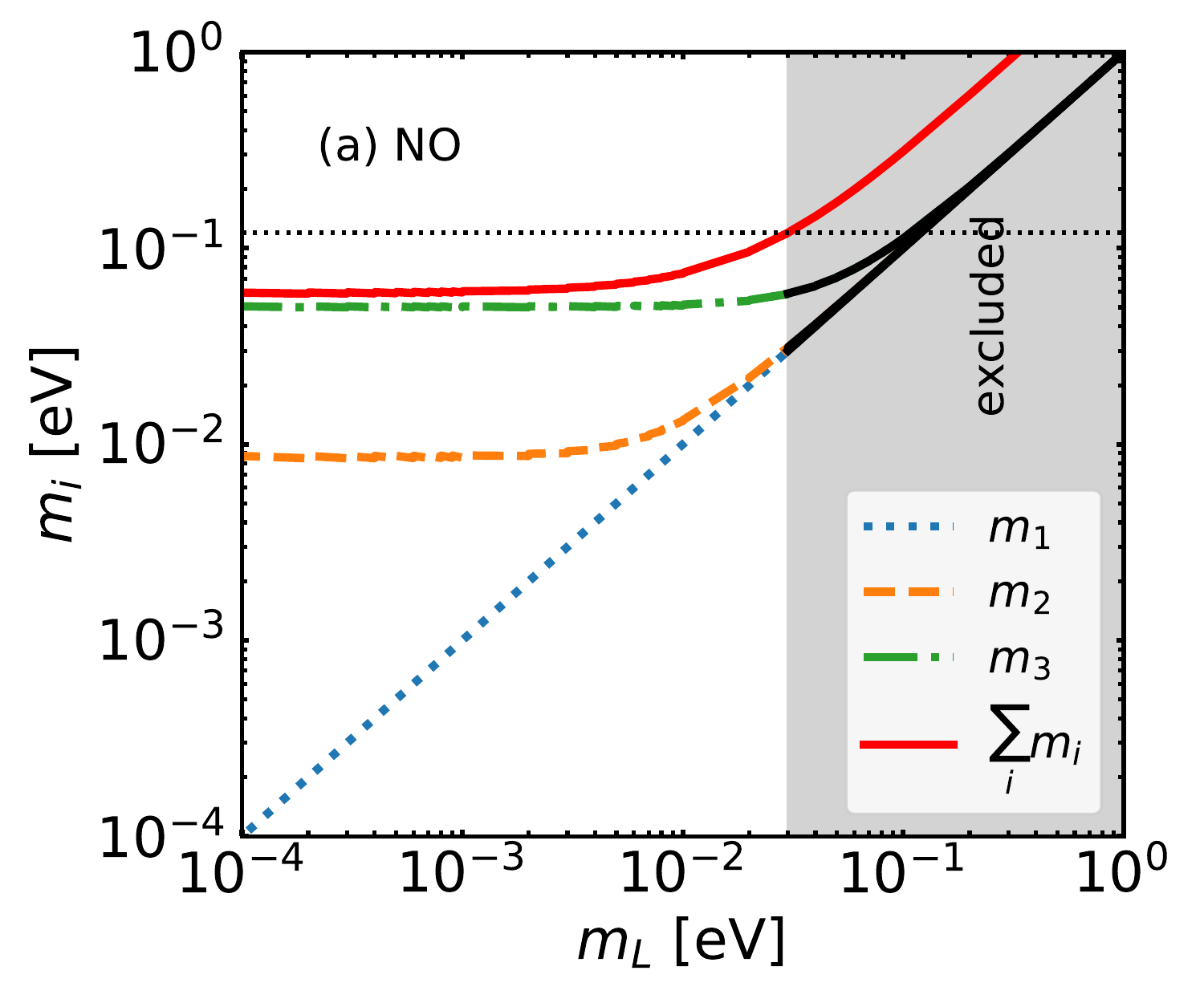}
\includegraphics[width=7cm]{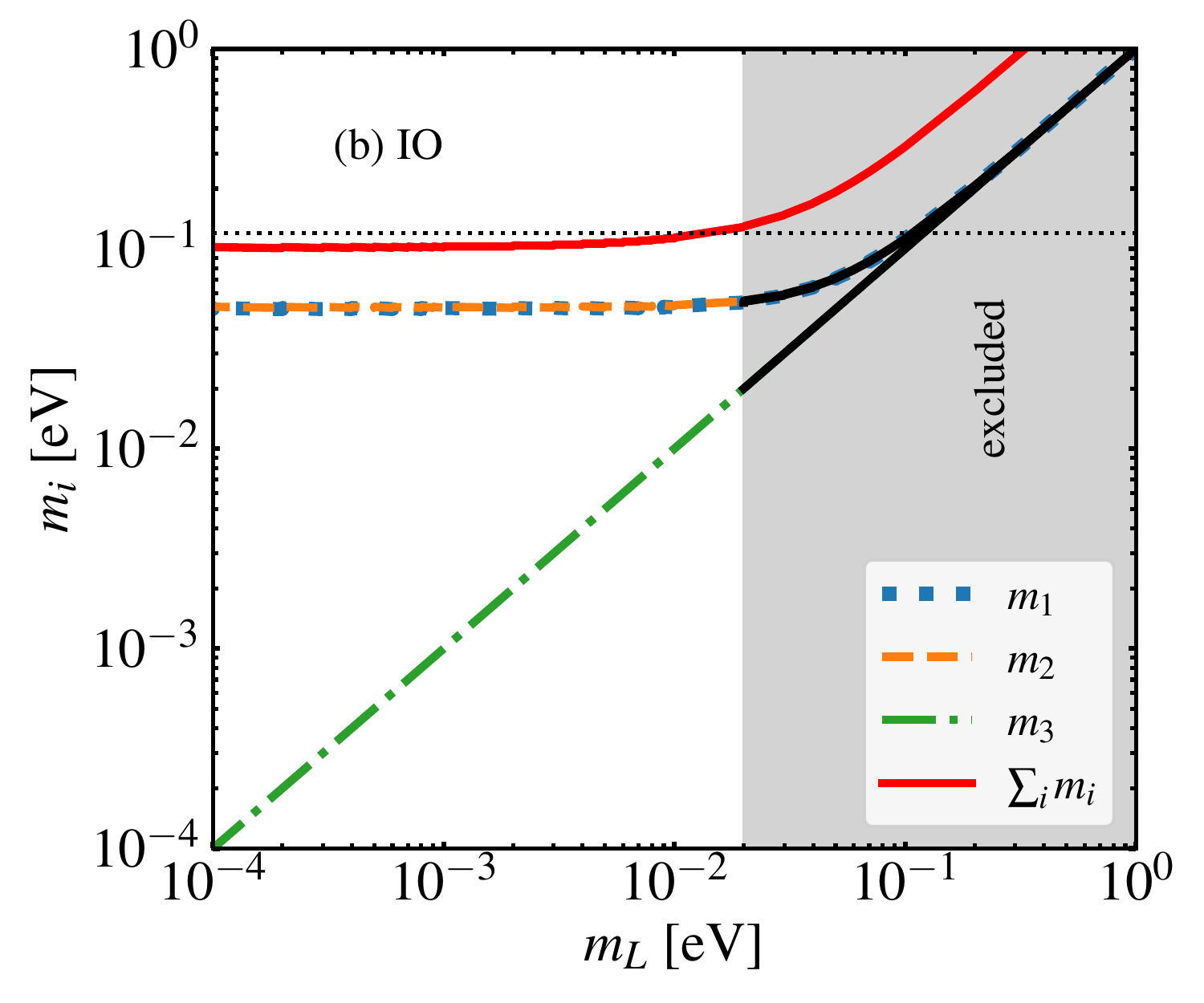}  
\caption{\label{fig:Introd_neutrino-mass} The absolute neutrino masses as a function of the mass of the lightest neutrino in the cases of (a) normal-ordering (NO) and (b) inverted ordering (IO) neutrino mass spectra. The values indicated with black solid lines are ruled out by the upper limit (horizontal dotted line) on the sum of the three neutrino masses $\sum^3_{i=1} m_i <0.12$ eV \cite{Planck:2018} constrained from a Cold Dark Matter model with cosmological observations. In the NO case, $m_1<<m_2 (\simeq \sqrt{\Delta m^2_{21}}=8.6\times 10^{-3}$ eV) $<m_3 (\simeq \sqrt{\Delta m^2_{31}}\simeq 0.05$ eV). In the IO case, $m_3<<m_1\lesssim m_2 (\simeq0.05$ eV). }
\end{figure}

 \begin{figure}
\centering
\includegraphics[width=6cm]{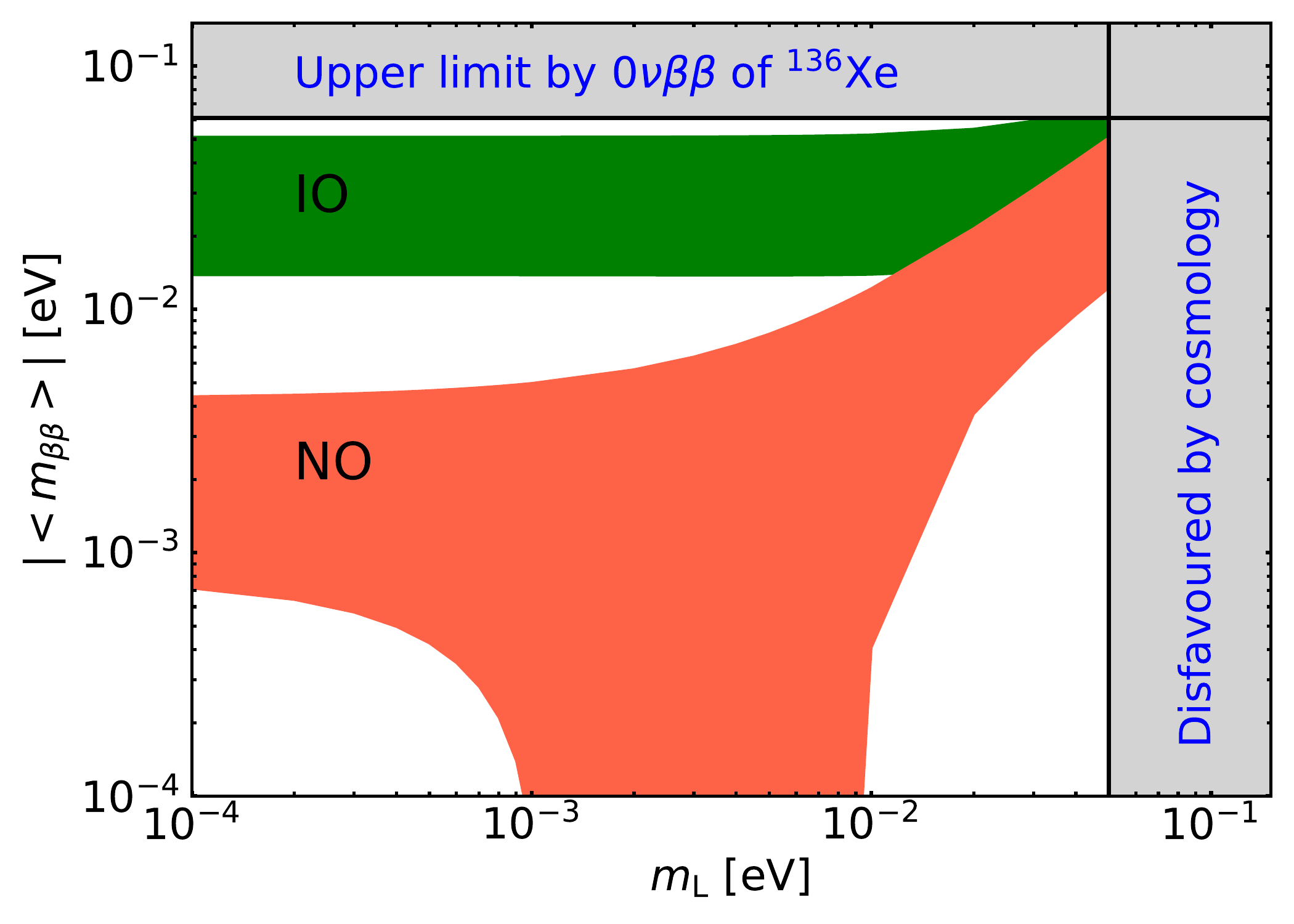} 
\includegraphics[width=6cm]{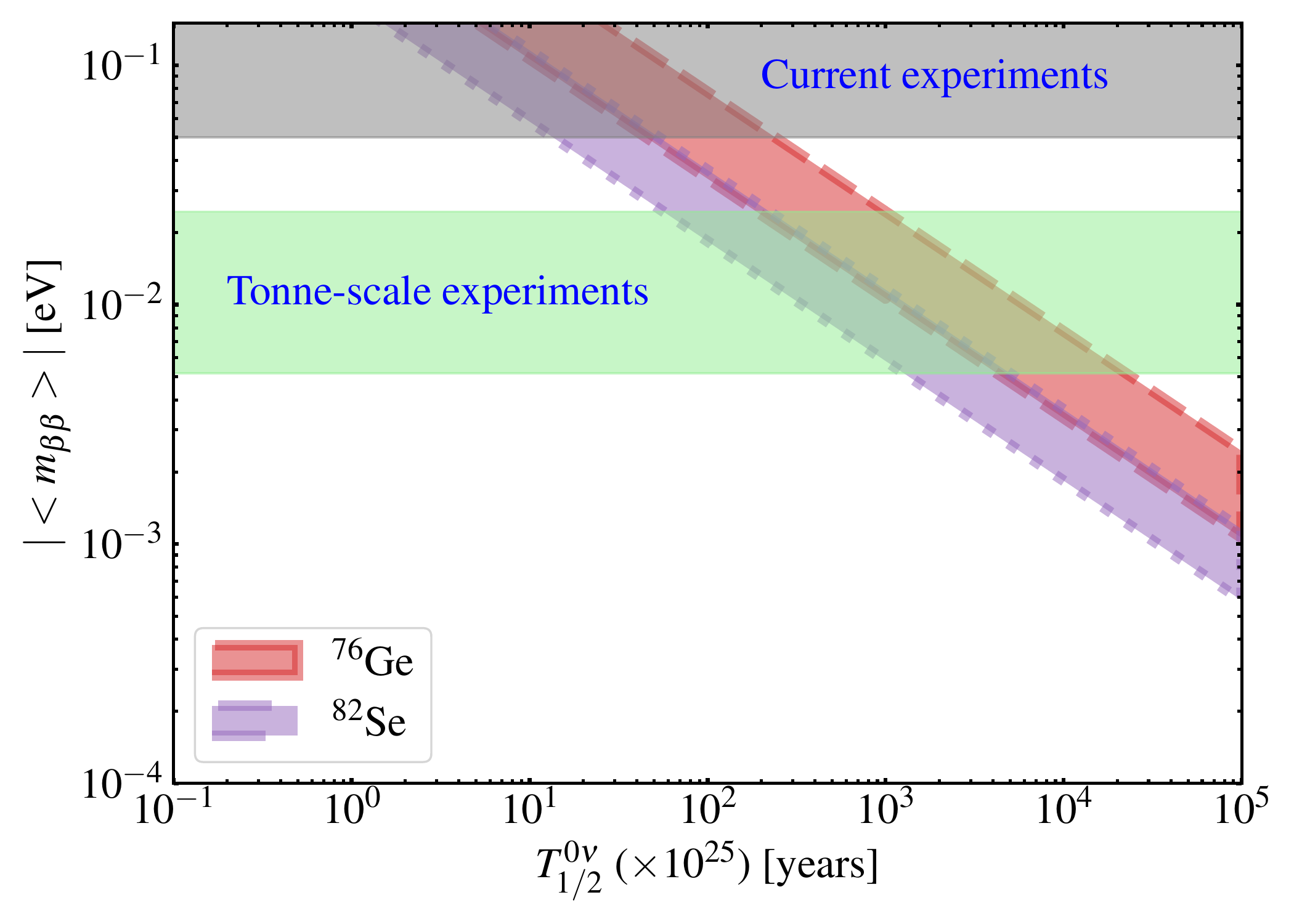}  
\includegraphics[width=6cm]{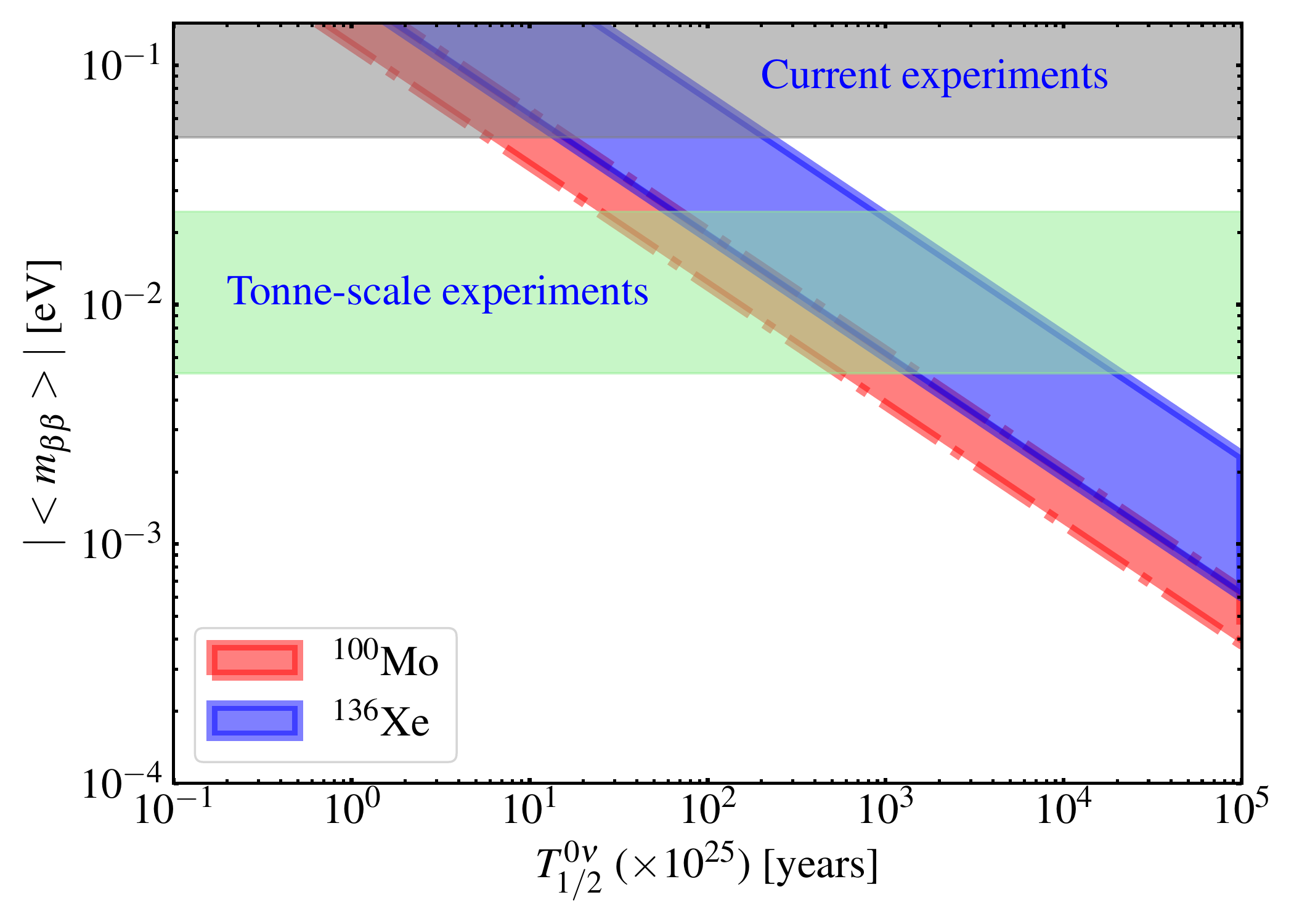}
\caption{\label{fig:effective-neutrino-mass}  (Left) The effective neutrino mass $|\braket{m_{\beta\beta}}|$ in the $0\nu\beta\beta$ decay as a function of the mass $m_L$ of the lightest neutrino. The shaded green and orange areas indicate the allowed values in the case of  inverted-ordering (IO) or normal-ordering (NO) neutrino masses, respectively, determined based on the best-fit values and $3\sigma$ allowed ranges of the neutrino oscillation parameters \cite{PDG:2018}.   The horizontal gray region indicates the upper limit on $|m_{\beta\beta}|<[0.061, 0.165]$ eV from the $0\nu\beta\beta$-decay measurement on $\nuclide[136]{Xe}$ \cite{KamLAND-Zen2016} based on current knowledge on the NME. The vertical gray region marks the area ruled out by the cosmological observations \cite{Planck_Coll:2016}.
(Right) The effective neutrino mass as a function of the half-life of $0\nu\beta\beta$ based on the current knowledge on the NMEs and the phase-space factor with corrections from finite nuclear size and electron screening taken into account in Ref. \cite{Stefanik:2015} for the four candidates. Colored bands are due to the uncertainty in the NMEs.}
\end{figure}

Based on the above data, there are  two possible nonequivalent orderings for the neutrino masses and in each case, one can express all masses in terms of the mass $m_L$ of the lightest neutrino.
\begin{itemize} 
\item  If the neutrino mass spectra follow the normal ordering ($\Delta m^2_{31}>0$), the squared mass of each neutrino in ascending order can be written as
\beq
m^2_1 = m^2_L,\quad m^2_2 = m^2_L+\Delta m^2_{21}, \quad m^2_3 = m^2_L+\Delta m^2_{31}.
\eeq

\item   If the neutrino mass spectra follow the inverted ordering ($\Delta m^2_{31}<0$), the squared mass of each neutrino in ascending order  can be written as
\beq
m^2_3 = m^2_L,\quad m^2_1 = m^2_L-\Delta m^2_{31},\quad m^2_2 = m^2_L - \Delta m^2_{31} + \Delta m^2_{21}.
\eeq

\end{itemize}

Figure~\ref{fig:Introd_neutrino-mass} displays the masses of neutrinos determined from the oscillation experimental data based on the two mass orderings as a function of the unknown mass of the lightest neutrino $m_L$, varying from 0.1 meV to 1.0 eV. It is shown that the minimal value of the total neutrino masses $\sum^3_{i=1} m_i$ is $0.0584$ eV and $0.101$ eV for the normal-ordering and inverted-ordering case, respectively. The latter is only slightly lower than the current constraint $\sum^3_{i=1} m_i<0.12$ eV \cite{Planck:2018}  from  cosmological observations. There are some hints that the normal ordering is favored~\cite{Esteban:2019} which is not a good message for detecting $\znubb$ decay.

 In the $\znubb$ decay, the effective neutrino mass  defined in (\ref{half-life}) can be written explicitly as
 \begin{equation}
     \langle m_{\beta\beta}\rangle
     =m_1c^2_{12}c^2_{13}+m_2c^2_{13}s^2_{12}e^{i\alpha_{21}}+ m_3s^2_{13}e^{i(\alpha_{31}-2\delta)}
     \simeq 0.680m_1+0.297m_2e^{i\alpha_{21}}+0.022m_3e^{i(\alpha_{31}-2\delta)}.
 \end{equation}  
 Following Ref.~\cite{Vissani:1999}, one can plot the  effective neutrino mass $|\braket{m_{\beta\beta}}|$ as a function of the mass of the lightest neutrino $m_L$, where the Majorana phases $\alpha_{21}, \alpha_{31}$ vary within $[0,\pi]$. The result is shown in  Fig.~\ref{fig:effective-neutrino-mass}. The regions allowed in the inverted-ordering and normal-ordering mass spectra are indicated, respectively.  If neutrinos belong to the inverted-ordering, the effective neutrino mass determined by the oscillation parameters is found to be within the interval $2\times 10^{-2} \lesssim |\langle m_{\beta\beta}\rangle| \lesssim 5\times 10^{-2}$ eV \cite{Bilenky:2015}, and in this case the ratio $\left|\langle m_{\beta\beta}\rangle/m_e\right|^2\simeq 10^{-16}$. As discussed before, the typical value of the phase-space factor $G_{0\nu}$ is about $10^{-14} {\rm yr}^{-1}$ which corresponds to the half-life around $10^{28}$ yr if the NME is taken as $\left\lvert M^{0\nu}\right\rvert^2\simeq 10$.  The  current experiments have achieved the half-life sensitivity longer than $10^{25}$ years on $^{136}$Xe~\cite{KamLAND-Zen2016}, $^{76}$Ge~\cite{GERDA2018} and $^{130}$Te~\cite{CUORE2020}.  Next-generation experiments \cite{Dolinski:2019,Xie:2020} aiming to increase detector mass further to tonne-scale size and to reduce backgrounds as much as possible, are expected to reach a discovery potential of half-life $10^{28}$ years after a few years of running, providing a definite answer on the mass hierarchy of neutrinos based on our current knowledge on the NME of $0\nu\beta\beta$ decay~\cite{Engel:2017}. If the  neutrino mass spectra follow the normal-ordering  unfortunately, we need to build a detector with the effective-neutrino-mass sensitivity down to $|\braket{m_{\beta\beta}}|\simeq 1$~meV~\cite{Pascoli:2008PRD,Xing:2015EPJC,Ge:2017PRD,Xing:2017EPJC,Penedo:2018,Cao:2020CPC,Huang:2021JHEP} and the half-life sensitivity up to $10^{30}$ years, which will be a real challenge.

 \jmyr{It is worth mentioning that a quenched $g_A$ is expected to be used  if only one-body current is employed in the calculations of the NME of $\znubb$ decay, as discussed in Sec.~\ref{subsubsec:gA_problem}. The effective $g^{\rm eff}_A$ value should be determined from the corresponding data. However, because of no data on  $\znubb$ decay, different ansatzes were employed to derive the $g^{\rm eff}_A$ value within each nuclear model.  The variation of the derived $g^{\rm eff}_A$ values ranging from 0.30 to the free-space value 1.27 could lead to two orders of magnitude ($\sim 4^4$) less sensitivity for the $0 \nu \beta \beta$ experiments. Of course, this naive analysis of the impact of $g^{\rm eff}_A$ is unfair and turns out to overestimate its impact significantly. It has been pointed out by Suhonen~\cite{Suhonen:2017PRC} that as the  $g^{\rm eff}_A$ increases, the parameters of the adopted nuclear model (such as the pairing strengths in the QRPA) also need to be readjusted to fit the half-life of $2\nu\beta\beta$ decay which alters the predicted NME $M^{0\nu}$ of $0\nu\beta\beta$ decay. With this consideration, it has been shown that the impact of the uncertainty in the $g_A$ on the half-life sensitivity could shrink from $4^4$ to a factor of $2-6$~\cite{Suhonen:2017PRC}.  Besides, the $g_A$ quenching for $0\nu\beta \beta$ decay is expected to be milder than that for $2\nu\beta\beta$ decay. In short, the knowledge of the effective $g^{\rm eff}_A$ value is essential to provide an accurate value on the neutrino effective mass, but is not crucial for the design of $\znubb$-decay experiments.
 }

%% file: 4summary.tex


The search for $0\nu\beta\beta$ decay has become a priority in nuclear physics, particle physics, and astrophysics as its observation would be direct evidence of lepton number violation beyond the standard model, demonstrating that neutrinos are Majorana fermions, shedding light on the mechanism of neutrino mass generation, and probing a key ingredient for generating the matter-antimatter asymmetry in the universe.   The experimental search for the $0\nu\beta\beta$ decay is entering a new era as tonne-scale detectors are planned or under construction for the next-generation experiments, which are expecting to achieve the half-life sensitivity up to $10^{28}$ years. If the $0\nu\beta\beta$ decay is governed by the standard mechanism, the measurements are expected to provide a definite answer on the mass hierarchy of neutrinos based on the current knowledge on the NMEs. 

Accurate NMEs for the $0\nu\beta\beta$ decay of candidate nuclei are important for the design and interpretation of future experiments. In this Review, we have provided an overview of the status of studies on the NMEs with different BMF approaches, including symmetry-projected Hartree-Fock-Bogoliubov theory, quasiparticle random-phase approximation, and symmetry-projected generator coordinate method, using the transition operators derived from the standard mechanism. We have introduced the basic ideas and recent developments in different BMF approaches for atomic nuclei starting from either a valence-space shell model Hamiltonian, a schematic multipole-multipole coupling Hamiltonian, an energy density functional, or a realistic nuclear force. Moreover, we have presented many technical details to tackle the possible difficulties in the BMF approaches and some strategies to improve the description of  nuclear structure properties. In particular, we have provided a detailed derivation of the formulas for computing the NMEs of  $0\nu\beta\beta$ decay.

Here we summarize the main conclusions of the NMEs:

\begin{itemize}
\item Uncertainty from the transition operators of standard mechanism is rather under control. The relativistic effect is within 5\%. The short-range correlation effect is generally shown to be less than 20\%. The specific value depends on the details of the calculations. The two-body weak current may quench the NME by about 10\%, which can be larger and needs further investigation. A recent {\em ab initio} study indicates that a leading-order contact operator may enhance the NME by a factor up to 50\%. One should be very careful while interpreting these numbers as they are obtained within a specific model and are expected to vary with the scheme and energy scale of the adopted nuclear models.

\item Steady progress has been made in understanding the discrepancy in the NMEs attributed to nuclear wave functions. The NMEs of different nuclear models using the same transition operator differ from each other by a factor of up to three. These predictions may be subject to additional uncertainties. The adopted approximations and phenomenological adjustable parameters in the models make it difficult to estimate the associated theoretical errors. The discrepancy has two sources: many-body approximation and nuclear interactions.  Evidence has shown that both need to be taken into account carefully to achieve a full understanding of the difference in the predicted NMEs.

\end{itemize}

 Before concluding this Review, we list some research topics  that are worth being explored soon, including but not limited to

 \begin{itemize}
 
     \item Implementation of neutron-proton isoscalar pairing into MR-EDF frameworks. Adding this new ingredient into the current MR-EDF calculation is expected to lower the NME significantly. However, this is a very challenging task. For one reason, the parameters of nuclear EDF are usually fitted to nuclear global properties, like mass and radius. These properties are usually insensitive to the strength of neutron-proton isoscalar pairing. Therefore, it is difficult to determine the additional parameters for these terms. The observables that are sensitive to neutron-proton pairing, including single-$\beta$ decay rate~\cite{Niu:2013PLB} and $2\nu\beta\beta$ decay, should be included in the optimization of the EDF. In addition, it is known that the MR-EDF based on the popular EDFs usually suffer from the problems of singularity and self-interaction~\cite{Duguet:2009}. Finding an EDF that can provide a good description of nuclear bulk properties and is free from self-interaction, self-pairing, and free from singularities when used at the BMF level is a challenge. There are some works dedicated to this direction~\cite{Sadoudi:2012,Bennaceur:2016}.  
     
     \item Extension of BMF approaches to other weak processes, such as single-$\beta$ decay, $2\nu\beta\beta$ decay, DGT transitions, etc. The validity of PGCM starting from either an EDF, effective nuclear interaction or chiral nuclear force has usually been tested against nuclear structure properties, including the excitation energies and electric multipole transition strengths of nuclear low-lying states, providing a test stone to nuclear models for collective correlations. However, these tests turn out to be insufficient in that two different {\em ab initio} methods produce a similar NME for the $\znubb$ decay of $^{48}$Ca, even though the collective properties are predicted differently in $^{48}$Ti~\cite{Yao:2020PRL,Belley2021PRL}, or two different MR-EDF calculations produce two very different NMEs for the $\znubb$ decay of $^{150}$Nd, but the low-lying states are predicted similarly~\cite{Rodriguez:2010,Song:2014}. It is not a surprise at all because the NME of $\znubb$ decay is sensitive to the overlap of the wave functions of two different nuclei, while the spectroscopic properties are determined by the wave function of one of the two nuclei. From this point of view, the extension of  BMF approaches to $2\nu\beta\beta$ decay would provide a complementary test of the nuclear wave functions. This test has been carried out in the QRPA framework, but not yet in the PGCM/MR-EDF framework, as the computation cost of the latter is much more expensive. As discussed before, the ground-state to ground-state DGT transition can be rather easily studied, compared to the  $2\nu\beta\beta$ decay. A benchmark among different PGCM/MR-EDF calculations for this transition is to be done.

     \item  Assessment of uncertainties originated from many-body approaches, including the error from model-space truncation and the statistic error from nuclear interactions. For the latter, one has to solve the many-body problem and evaluate the NME of $0\nu\beta\beta$ decay starting from a set of Hamiltonians and transition operators with parameters varying around the optimal values. Since each calculation is computational heavy, this kind of calculation is a challenge. To this end, efforts from different aspects to optimize the BMF approach and to speed up the convergence of calculations are worth trying.
     
     On the one hand, previous studies have demonstrated that the choice of single-particle basis in nuclear many-body theory has a significant impact on the convergence behavior of energy. Because of simplicity, the spherical HO basis is employed in most shell models, the convergence of which with the number of basis is very slow. In most BMF approaches, the HF or projected deformed HFB basis which is obtained from the optimal single-particle potential was adopted. Defined in this type of basis, the energy of the BMF approaches converges faster than the shell-model in the HO basis. It has been demonstrated recently in a no-core shell model calculation~\cite{Tichai:2019PRC} and  in-medium similarity renormalization group  \cite{Hoppe:2021PRC} that the use of natural orbitals which are defined as eigenvectors of a correlated one-body density matrix obtained with second-order MBPT provides the fastest model-space convergence and significantly reduced sensitivity to basis parameters. Because of this fact, it is of very interest to build BMF approaches upon the natural orbital basis.

     On the other hand, in recent years, there are a vast of research interests in the application of eigenvector continuation (EC) proposed by Frame et al. \cite{Frame:2018PRL} to nuclear many-body systems. The idea of the EC is analogous to that of GCM in the sense that the trial wave function is expanded in a set of nonorthogonal basis chosen as the eigenvectors for different parameter sets  of the Hamiltonian. The EC  provides a tool of choice to set up an efficient emulator \footnote{The basic idea of setting up an emulator is that one first trains a computer model using the results from full many-body calculations based on Hamiltonians with a representative set of parameters, and then uses the computer model to predict results for the operators with other parameters without carrying out full many-body calculations.} which may provide a much cheaper and quicker way to estimate the  model uncertainty by varying the parameters and truncations of that model.  Effort along this direction is underway, see for instance Ref.~\cite{Melendez:2020PhD}. The combination of EC with PGCM calculation is a direction worth to be explored.

    \item Computing the NMEs of $0\nu\beta\beta$ decay using the transition operators derived fully consistent with the adopted nuclear interactions.  A comprehensive understanding  of the $0\nu\beta\beta$  transition operator necessarily requires the knowledge of physics from the energy  much higher than the electroweak scale  down to the low-energy of nuclear-physics scale. Lattice QCD and chiral EFT provide the tools of choice to match them. Therefore, one of the promising directions is the computation of NMEs in many-body methods with controllable approximations using nuclear interactions and weak transition operators derived consistently from the chiral EFT with the feature of order-by-order convergence. Within this framework, the contribution of two-body weak currents which induce three-nucleon interactions and appear at N$^3$LO need to be examined more carefully. 
    
 \end{itemize}

\jmyr{As the final point we want to stress}, this Review has been focused only on the studies based on the standard mechanism. There are many other possible mechanisms contributing to the $\znubb$ decay, such as the long-range~\cite{Tomoda:1991,Kotila:2021} and short-range~\cite{Graf:2018} nonstandard mechanisms, and the possible mixing between the standard model $W$ boson and its hypothetical heavier right-handed partner $W_R$~\cite{Li:2020PRL}. Once this process is eventually measured, one needs to disentangle the contributions from the standard and nonstandard mechanisms as only the former provides information on the effective neutrino masses. Besides, the phenomenology of $\znubb$ decay may also be significantly impacted with the presence of sterile neutrinos with Majorana mass~\cite{Dekens:2020JHEP}. On the other hand, the NME of the ground-state to ground-state $\znubb$ decay is discussed in this Review. The branching ratio of the decay to the low-lying excited states of the daughter nucleus also needs to be taken into account. Even though this process is strongly quenched by the phase-space factor in the standard mechanism, it could contribute significantly to the NME in the nonstandard mechanism~\cite{Fang:2021PRC}.